\documentclass[twoside,openright,12pt]{CUEDthesisPSnPDF1}
\usepackage{graphicx}
\usepackage{amsmath,amssymb}
\usepackage{amscd,amsfonts,amsthm}
\usepackage{mathtools}
\usepackage[pdftex,bookmarks]{hyperref}
\usepackage{float}
\usepackage{times}
\usepackage{comment}
\usepackage{color}
\usepackage{t1enc}
\usepackage{longtable}
\usepackage{subcaption}
\usepackage{xspace}
\usepackage{psfrag}
\usepackage{latexsym,pstricks,rotating,enumerate}
\usepackage{wrapfig}
\usepackage{color}
\usepackage{multirow}
\usepackage{framed,xcolor}
\usepackage{caption}
\usepackage{tikz}
\usepackage{appendix}
\usepackage{ulem}

\usepackage[numbered]{matlab-prettifier}
\usetikzlibrary{calc,positioning,fit,backgrounds}
\pgfdeclarelayer{background}
\pgfsetlayers{background,main}
\usepackage{watermark}
\newcommand{\ie}{\textit{i.e.~}}

\begin{document}
\renewcommand\baselinestretch{1.2}
\baselineskip=18pt plus1pt
\setcounter{secnumdepth}{3}
\setcounter{tocdepth}{3}
\frontmatter
\thispagestyle{empty}
\baselineskip=18pt
\begin{center}
{\Large \bf Novel techniques for efficient quantum state tomography and quantum process tomography and their experimental implementation} \\
\vspace*{1.5cm}
{\large{\bf Akshay Gaikwad}}\\
\vspace*{1.2cm}
{\textit{A thesis submitted for the partial fulfillment of}}\\
{\textit {the degree of Doctor of Philosophy}}\\
\vspace{0.5cm}

\end{center}
%
\begin{center}
\vspace*{1.5cm}
\hspace*{0cm}
\end{center}
\vspace*{-1cm}
\begin{center}
\includegraphics[scale=0.24]{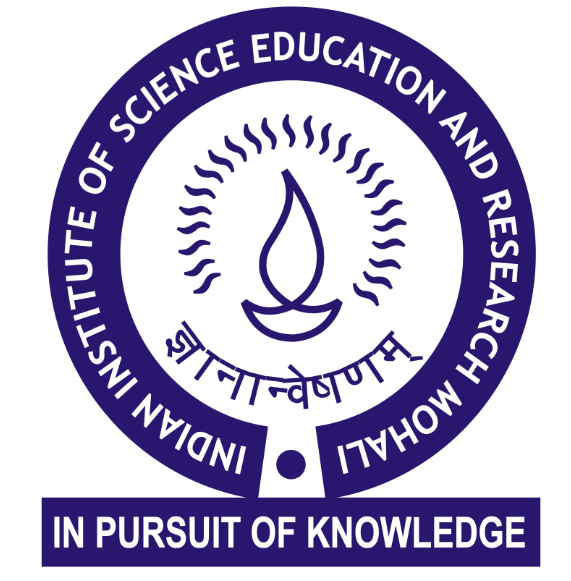}\\
\vspace*{1.5cm}

{Department of Physical Sciences} \\
{Indian Institute of Science Education and Research Mohali} \\

{Knowledge city, Sector 81, SAS Nagar, Manauli PO, Mohali 140306, Punjab, India} \\ 
\vspace*{1.5cm}
December 2023
 
\end{center}

\newpage
\thispagestyle{empty}
\begin{center}
\end{center}
\thispagestyle{empty}
\centerline{\Large \bf Declaration}
\vspace{1cm}
\par 
\noindent The work presented in this thesis has been carried out by me under the guidance of
Prof. Kavita Dorai and Prof. Arvind at the Indian Institute of Science Education and Research Mohali.\\

\noindent This work has not been submitted in part or in full for a degree, diploma or a fellowship to any other University or Institute. Whenever contributions of others are involved, every effort has been made to indicate this clearly, with due acknowledgment of collaborative research and discussions. This thesis is a bonafide record of original work done by me and all sources listed within have been detailed in the bibliography.

\vspace*{1in}

\hspace*{-0.25in}
\parbox{8in}{
\noindent {{\bf Akshay Gaikwad}}
}

\vspace*{0.125in}

\hspace*{-0.25in}
\noindent
\parbox{2.5in}{
\noindent Place~:  \\
\noindent Date~:
}

\vspace*{0.5in}

\noindent
In our capacity as supervisors of the candidate's PhD thesis work, we certify that the above statements by the candidate are true to the best of our knowledge.

\vspace*{2.5cm}

\hspace*{-0.25in}
\parbox{8in}{
\noindent {{\bf Dr. Kavita Dorai} \hspace*{5.5cm}  {\bf Dr. Arvind}} \\
\noindent \textbf{(Supervisor)} \hspace*{6.27cm} \textbf{(Co-supervisor)} \\
\noindent Professor of Physics \hspace*{5.1cm} Professor of Physics\\
\noindent Department of Physical Sciences \hspace*{3cm} Department of Physical Sciences\\
\noindent IISER Mohali \hspace*{6.2cm}  IISER Mohali
}
\vspace*{0.125in}

\hspace*{-0.25in}
\noindent
\parbox{8in}{
\noindent Place~:  \hspace*{7.4cm} Place~: \\
\noindent Date~:   \hspace*{7.55cm} Date~:
}


\newpage
\thispagestyle{empty}

\begin{center}
\end{center}
\thispagestyle{empty}
\centerline{\Large \bf Acknowledgments}
\vspace{10pt}
I wish to begin by expressing my profound gratitude to my thesis supervisors, Prof. Kavita Dorai and Prof. Arvind, for their steadfast support, guidance, and encouragement during my entire PhD journey. Their mentorship not only involved teaching and directing various projects within this thesis but also instilled in me the mindset to comprehend, develop, and accomplish any given task. I am truly thankful for their unwavering assistance and mentorship. I would also like to extend my thanks to Dr. Sandeep Goyal, Dr. K. P. Singh, Dr. Manabendra Nath Bera, and Dr. Vishal Bhardwaj, members of my doctoral committee, for their support and guidance. Special appreciation goes to Dr. Paramdeep Singh Chandi, the scientific officer, for his invaluable help in resolving software issues. Additionally, my gratitude extends to Mr. Balbir Singh, a member of the NMR lab scientific staff, for his generous and consistent support.

I would like to acknowledge the present and past members of the research group, including Rajbinder, Jyotsana, Sumit, Dileep, Akanksha, Vaishali, Krishna, Gayatri, Arshdeep, Matreyee, Shaileyee, Ram Sagar, Rithu, Sneha, Anupama, and past members Amandeep Singh, Harpreet Singh, Shruti Dogra, Satnam Singh, Rakesh Sharma, Navdeep Gogna, for their valuable insights shared during our meetings, which have greatly benefited me. Special thanks are reserved for Amandeep Singh for his guidance in NMR experiments.

My deep gratitude extends to the NMR research facility at IISER Mohali for their tremendous support in facilitating my experimental work. I also appreciate the research fellowship and financial assistance provided by IISER Mohali, enabling me to attend conferences during my PhD.

The accomplishment of obtaining a PhD is not solely attributed to individuals encountered during this journey but also to everyone encountered throughout my life.  I express gratitude to all my teachers, with special thanks to my childhood friends Tushar Pawar and Nitin Shivthare for their steadfast support during my formative years. My gratitude extends to Sameer Shah, my friend from senior secondary education, who played a crucial role in helping me with math during entrance exam preparations. I attribute my success in entering IISER for undergraduate and graduate studies to Sameer's assistance. Subsequently, I made numerous friends during my undergraduate studies, and I express my thanks to Abhinav Kala, Joydeep, Akhil, Love Grover, Rajendra Bhati, Shubham, Gyanendra, Manvendra Rajwanshi, Manvendra Singh, Abhishek Singh for five years of wonderful memories.

During my PhD program at IISER Mohali, I forged new friendships with Sandeep, Mandeep, Mamta, Amreen, Yattu, Sheru, and Chitra. I express my gratitude to Sheru for introducing me to table tennis and Yattu for engaging in humorous chit-chats.  I also spent very good time playing table tennis game with Mamta and Mandeep. Special thanks to Mandeep for tolerating my behavior and imparting valuable habits. I consider myself lucky to have friend like Mandeep in my life. I also appreciate the companionship of my video game and trekking partners Bhati, Jassi, Love, Abhishek, Krishna, and Sandeep, making my PhD journey less stressful.  I cherish the delightful moments I shared with my lovely friend Tarang during my final days at IISER Mohali.

I extend my sincere appreciation to my father Ramdas Gaikwad, my mother Padma, and my brother Ajinkya for their unwavering support and faith in me. I would also like to mention my late grandfather Shankar Gaikwad, whom I called 'Aaba,' as the main source of my inspiration, and I dedicate this thesis to his memory. Their encouragement has been a significant source of motivation throughout my endeavors.

\baselineskip=15pt

\vspace{1cm}

\rightline{\bf \large{Akshay Gaikwad}}

\chapter{Abstract}

The study carried out in this thesis focuses on designing and experimentally implementing various quantum tomography protocols to efficiently characterize and reconstruct unknown quantum states and processes using spin ensemble based nuclear magnetic resonance (NMR) quantum processors and superconducting technology-based IBM quantum processors. The task of reconstructing quantum states is achieved with the help of quantum state tomography (QST) protocols while quantum processes are characterized using quantum process tomography (QPT) protocols. Both QST and QPT are essential to check the reliability and to evaluate the performance of a quantum processor. However, both QST and QPT are cursed with a fundamental difficulty, i.e., the computational complexity increases exponentially with the size of the system which makes them infeasible to perform experimentally, even for smaller dimensional systems. Besides this, having finite size of ensembles and inevitable systematic errors will lead to unphysical density matrices and process matrices. To tackle such issues, numerous QST and QPT protocols have been proposed. However, most of them are yet to be experimentally demonstrated. The prime objective of the study undertaken in this thesis is to design experimental strategies to efficiently implement tomography protocols on NMR and IBM quantum processors. Generalized quantum circuits are proposed to efficiently acquire experimental data to perform QST and QPT and further demonstrated for two- and three-qubit quantum states and quantum processes. 

To tackle the issue of the unphysicality of experimentally reconstructed quantum states and processes using standard tomography techniques, the tasks of QST and QPT are converted into a constrained convex optimization (CCO) problem and the CCO problem is solved to reconstruct valid quantum states and processes which in case of QPT allows us to compute the complete set of Kraus operators corresponding to a given quantum process. Further, the compressed sensing (CS) and artificial neural network (ANN) techniques have also been employed to perform tomography of quantum states and gates from a heavily reduced data set as compared to standard methods. CS and ANN based tomography methods are promising techniques to deal with complexity issue to characterize higher-dimensional quantum gates. Moreover, the problem of selective and direct estimation of desired elements of process matrix characterizing quantum process has also been explored, where partial knowledge about underlying unknown quantum process can be acquired efficiently using selective and efficient quantum process tomography protocol (SEQPT). A generalized quantum algorithm and quantum circuit to perform SEQPT has been proposed and successful experimental demonstration has been shown on NMR and IBM quantum processors. In addition to that, we also proposed an efficient direct QST and QPT scheme based on weak measurement approach and demonstrated experimentally using a three-qubit NMR system. The thesis also investigates the problem of experimentally simulating dynamics of open quantum systems based on dilation techniques. To show the efficacy of above-mentioned quantum tomography and simulation protocols, experimental results are compared with theoretically predicted results in case of several two-and three-qubit quantum systems.
The content of this thesis has been divided into eight chapters as described below:

\subsubsection*{Chapter 1}

The initial section of this chapter presents an overview of quantum computing and information processing. It encompasses essential principles and the physical realization of quantum processors using NMR. Additionally, it provides a brief introduction to various tomography and simulation protocols. Finally, it concludes with closing remarks that outline the objectives and motivations behind the research conducted in this thesis.

\subsubsection*{Chapter 2}
This chapter focuses on the problem of invalid experimental density and process matrices. It introduces the constrained convex optimization (CCO) method for QST and QPT which allow us to reconstruct valid (positive semi-definite) density and process matrices charactering unknown quantum states and processes respectively. It also improves the fidelity of density and process matrix characterization. The chapter discusses the NMR quantum information processor-based implementation of QST and QPT using the CCO method.

\subsubsection*{Chapter 3}
This chapter addresses scalability issues in QST and QPT by employing the application of compressed sensing (CS) algorithms. The CS algorithm allows for full as well as valid QST and QPT from incomplete data sets, resulting in high fidelity estimates of density and process matrices. The chapter also discusses the characterization of small-dimensional quantum gates in higher-dimensional systems. The experimental demonstration of CS based QST and QPT for 2- and 3-qubit system is given using NMR ensemble quantum processor and superconducting based IBM cloud quantum processor.

\subsubsection*{Chapter 4}
This chapter explores the application of artificial neural network (ANN) techniques in QST and QPT to attempt to overcome scalability issues. The Feed- Forward Neural Network (FFNN) architecture is used to reconstruct density and process matrices from noisy experimental data obtained from NMR quantum processor. The results show efficient as well as high fidelity QST and QPT as compared to the standard linear inversion method.

\subsubsection*{Chapter 5}
This chapter introduces a scheme for selective and efficient quantum process tomography (SEQPT) using local measurements without ancilla. The method estimates specific elements of the process matrix by a restrictive set of subsystem measurements, reducing the experimental resources required. The efficacy of the scheme is demonstrated experimentally on NMR and IBM processors for 2- and 3-qubit systems.

\subsubsection*{Chapter 6}
This chapter presents an efficient weak measurement (WM) scheme for direct quantum state tomography (DQST) and direct quantum process tomography (DQPT) without projective measurements. A generalized quantum circuit is proposed and implemented on an NMR ensemble quantum information processor to directly measure multiple selective elements of density and process matrices in a single experiment which enable us to efficiently extract desired information from the system.

\subsubsection*{Chapter 7}
This chapter investigates the experimental simulation of open quantum system dynamics using dilation techniques. The Sz-Nagy's dilation (SND) algorithm is experimentally implemented on an NMR quantum information processor to simulate the action of a 2-qubit pure phase damping channel, correlated amplitude damping channel and magnetic field gradient pulse (MFGP). The algorithm successfully simulates the dynamics using only one ancilla qubit, and the experimental fidelity is assessed using CCO-QPT.

\subsubsection*{Chapter 8}
This chapter describes the summary of the thesis and some future directions.

\thispagestyle{empty}
\pagenumbering{roman}
\setcounter{page}{1}
\thispagestyle{empty}

\begin{center}
\end{center}
\thispagestyle{empty}
{\LARGE \bf List of Publications}
\vspace{1cm}
\vspace*{12pt}
\begin{enumerate}
\addtolength{\itemsep}{12pt} 

\item \textbf{Akshay Gaikwad}, Omkar Bihani, Arvind and Kavita Dorai\\
	\textit{Neural network assisted quantum state and process tomography using limited data sets}\\
	 \href{https://journals.aps.org/pra/abstract/10.1103/PhysRevA.109.012402}{\color{violet}	Phys. Rev. A 109, 012402, (2024)}

\item \textbf{Akshay Gaikwad}, Arvind and Kavita Dorai\\
	\textit{Direct tomography of quantum states and processes via weak measurements of Pauli spin operators on an NMR quantum processor}\\
	  \href{https://epjd.epj.org/articles/epjd/abs/2023/12/10053_2023_Article_791/10053_2023_Article_791.html}{\color{violet} Eur. Phys. J. D (2023) 77: 209, (2023)}

\item \textbf{Akshay Gaikwad}, Krishna Shende, Arvind and Kavita Dorai\\
	\textit{Implementing efficient selective quantum process tomography of superconducting quantum gates on IBM quantum experience}\\
	\href{https://www.nature.com/articles/s41598-022-07721-3}{\color{violet}Scientific reports 12 (1), 1-11, (2022)}
	
	\item \textbf{Akshay Gaikwad}, Arvind and Kavita Dorai\\
	\textit{Experimental simulation of open quantum dynamics using Sz-Nagy's dilation algorithm using NMR}\\
	\href{https://journals.aps.org/pra/abstract/10.1103/PhysRevA.106.022424}{\color{violet}Phys. Rev. A 106, 022424, (2022)}.
	
	\item \textbf{Akshay Gaikwad}, Arvind and Kavita Dorai\\
	\textit{Efficient characterization of quantum processes from reduced data set via compressed sensing using NMR}\\
	\href{https://link.springer.com/article/10.1007/s11128-022-03695-3}
	{\color{violet}Quant. Inf. Proc. 21, (12), (2022)}
	
	\item \textbf{Akshay Gaikwad}, Arvind and Kavita Dorai\\
		\textit{True experimental reconstruction of quantum states and processes via convex optimization using NMR}\\
		\href{https://link.springer.com/article/10.1007/s11128-020-02930-z}{\color{violet}Quant. Inf. Proc. 20, (19), (2021)} 

\item  \textbf{Akshay Gaikwad}, Krishna shende and Kavita Dorai\\
		\textit{Experimental demonstration of optimized quantum process tomography on the IBM quantum experience}\\
		\href{https://www.worldscientific.com/doi/10.1142/S0219749920400043}{\color{violet} Int. J. Quantum Inf. 19 (07), 2040004, (2021)}

\item \textbf{Akshay Gaikwad}, Diksha Rehal, Amandeep Singh, Arvind and Kavita Dorai\\ \textit{Experimental demonstration of selective quantum process tomography on NMR}\\
	    \href{https://journals.aps.org/pra/abstract/10.1103/PhysRevA.97.022311}{\color{violet} Phys. Rev. A 97, 022311, (2018)}

	
\end{enumerate}

\thispagestyle{empty}

\tableofcontents
\listoffigures
\newpage
\thispagestyle{empty}
\pagebreak
\listoftables
\setcounter{page}{1}
\thispagestyle{empty}
\mainmatter
\chapter{Introduction}

\section{Quantum Mechanics as a Computational Paradigm}

Modern computing is based on the laws of classical physics and mathematical logic. Though the functioning of electronic components is based on quantum mechanical principles, the logic they follow is classical. The classical computer is tailored for serial computation, whereby algorithms progress sequentially from one point to another over time. This sequential nature necessitates the completion of specific operation before subsequent ones can commence, thus rendering classical computation time-intensive. It is important to note, however, that parallel computing offers an alternative approach: Parallel computing involves the dissection of computational problems or tasks into autonomous logical units that can be executed concurrently. For instance, the multiplication of two matrices, denoted as $AB=C$. Elements such as $C_{11}$ and $C_{21}$ within matrix $C$ can be computed in parallel fashion. By employing the first row of matrix $A$ and the first column of matrix $B$ to calculate $C_{11}$, and the second row of matrix $A$ and the first column of matrix $B$ for $C_{21}$, a simultaneous and independent computation of all elements $C_{ij}$ within matrix $C$ becomes achievable.

Moreover, classical computers utilize irreversible gates which generate heat during their operational processes. This phenomenon arises when information is erased while operating irreversible gates, leading to the dissipation of energy, as observed by Landauer \cite{land-ibm-1961}. The act of deleting information consistently, necessitates work and the energy consumption \cite{benn-ibm-1973}. In order to render classical computation reversible, supplementary bits must be incorporated, thereby giving rise to the predicament of excess waste commonly referred to as the "garbage problem" in classical reversible computation \cite{fred-ijtp-1982, benn-ibm-1988}. The amount of heat produced by a computer is contingent upon the scale and number of gates (electronic components), and if these components are positioned too closely, the heat generated by one component can potentially harm neighboring components, posing a challenge in the miniaturization of classical computers. This challenge pertaining to the growth of computer hardware was projected and formulated by Intel co-founder Gordon Moore in 1965, famously known as Moore's Law, which states-

\begin{itemize}
\item[-] \textit{The number of transistors on a microchip doubles every two years that we can expect the speed and capability of our computers to increase every couple of years.}
\end{itemize}
 \begin{figure}[t]
\centering
\includegraphics[scale=0.125]{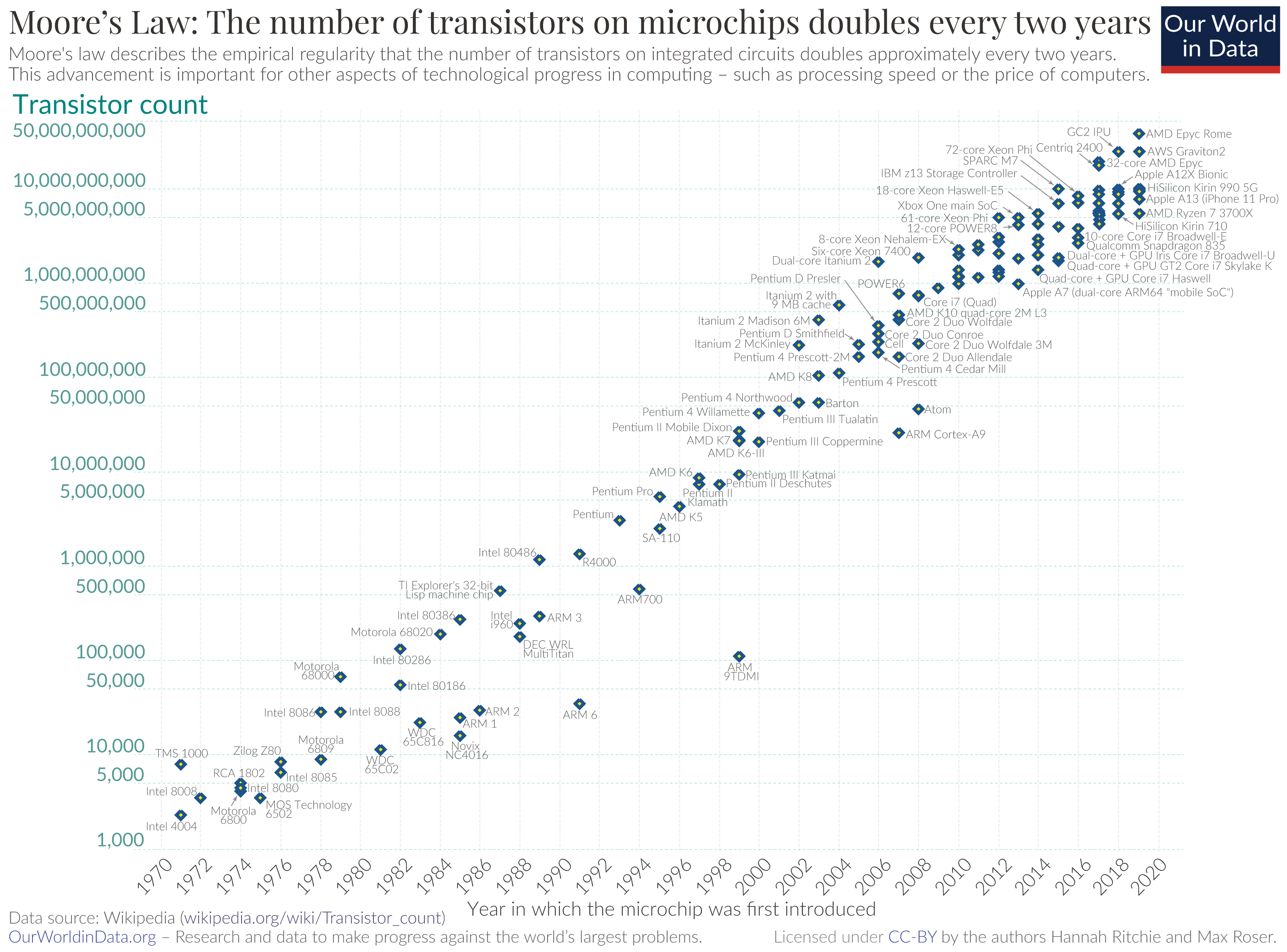}
\caption{A logarithmic graph showing the timeline of how transistor counts in microchips are almost doubling every two years from 1970 to 2020; (Moore's Law). \textit{Source: \href{ https://ourworldindata.org/uploads/2020/11/Transistor-Count-over-time.png}{ourworldindata.org}} }
\label{moors}
\end{figure}
The conventional approaches to developing computer technology will inevitably encounter challenges associated with size limitations. As devices continue to shrink, the impact of quantum effects on their functionality becomes a concern. The question arises: will Moore's Law become obsolete? The answer is no. An alternative solution exists - transitioning to quantum computation! Quantum computation is an immensely captivating and rapidly advancing field of research \cite{bar-cp-1996, ekert-rmp-1996, steame-rpp-1998}. This novel computing paradigm leverages the principles of quantum mechanics instead of classical physics to execute computations, a concept initially conceived by Feynman \cite{feynman-ijtp-1982, feynman-fp-1986} and Deutsch \cite{david-prc-1985}.
 \begin{itemize}
 \item[-] \textit{"How can we simulate the quantum mechanics? Can you do it with a new kind of computer - a quantum computer? It is not a Turing machine, but a machine of a different kind."} - R. P. Feynman, 1981.
 \end{itemize}
 The operation of quantum computers is rooted in the principles of quantum mechanics, harnessing various quantum features such as unitary evolution, quantum superposition, and quantum entanglement. These elements provide a fundamental speed advantage over classical computers \cite{bar-pra-1995, shor-siam-1997}. This speed advantage is so substantial that many researchers believe that no feasible advancements in classical computation could bridge the gap between the power of a classical computer and that of a quantum computer. Quantum computing represents a complete paradigm shift in the functioning and operation of computers. By leveraging these novel quantum properties, we can develop innovative types of software and hardware. It is anticipated that explorations in this field may eventually yield information processing devices with capabilities surpassing those of today's computing and communication systems.

  \section{Quantum Computing and Information Processing}
  
 The domain of quantum computing and quantum information (QCQI) has undergone substantial expansion during the preceding two decades. This field encompasses the study and execution of information processing tasks that can be effectively conducted employing a quantum mechanical framework. Quantum computers possess the capacity to undertake computational operations that lie beyond the capabilities of classical computers. The encoding of $n$-classical information bits mandates no less than $n$ classical resources. However, due to the principle of quantum superposition, quantum mechanical systems can theoretically exhibit a more efficient encoding efficiency compared to classical systems\cite{nielsen-book-10}.

In 1981, R. Feynman introduced the concept of a `quantum computer' and demonstrated that a classical computer would encounter an exponential deceleration while emulating a quantum phenomenon, whereas a quantum computer would not be subject to such constraints\cite{Feynman1982}. Subsequently, in 1985,  D. Deutsch, took Feynman's ideas further and defined two models of quantum computation; he also devised the first quantum algorithm. One of Deutsch's ideas is that quantum computers could take advantage of the computational power present in many \textit{''parallel universes''} and thus outperform conventional classical algorithms\cite{dj-prsl-1985}. In 1994, P. Shor showcased the resolution of two significant problems determining the prime factors of an integer and the discrete logarithm problem both of which could be efficiently solved on a quantum computer, underscoring the prowess of quantum computing\cite{nielsen-book-10, shor-journal-1997}. Furthermore, in 1996, L. Grover demonstrated that a search algorithm for an unsorted database executed on a quantum computer exhibits quadratic speed-up relative to its classical counterpart\cite{grover-prl-1997}.

In 2000, theoretical physicist David P. DiVincenzo proposed a set of prerequisites for the practical realization of an operational quantum computer, later known as \textit{DiVincenzo criteria}\cite{divi-fdp-2000}. These prerequisites encompass a scalable physical framework, the capability to initialize the system to any quantum state, a comprehensive list of quantum gates amenable to implementation, qubit-specific measurement procedures, and sufficiently long coherence times relative to the durations of gate implementations. To this day, no quantum hardware has been found to comprehensively satisfy these specified criteria. Howwver, numerous experiments in the realm of quantum computing have been conducted employing diverse technologies such as optical photons\cite{knill-nature-2001, kok-rmp-2007}, ion traps\cite{bruz-apr-2019,wine-fdp-1998}, superconducting qubits\cite{clarke-nature-2008}, nitrogen-vacancy (NV) centers\cite{casa-prl-2016, child-mrs-2013}, and nuclear magnetic resonance (NMR) techniques\cite{cory-pnas-1997, chuang-sci-1997, chuang-rmp-2005, oliveira-book-07}.

In optical photon-based quantum computers, qubits are encoded in the polarization state of photons. The initial state is prepared by generating single-photon states through light attenuation. Quantum gates are implemented using beam-splitters, phase shifters, and nonlinear Kerr media. Measurement is conducted by detecting individual photons utilizing a photomultiplier tube\cite{knill-nature-2001}. Similarly, trapped ion quantum computers rely on ions that are cooled to a state where their vibrational energy approaches zero, thereby enabling the realization of qubits via the hyperfine state of an atom combined with the lowest-energy vibrational modes of the trapped atoms\cite{wine-fdp-1998}. Quantum gates in this system are constructed using laser pulses, and measurements are derived from population measurements of hyperfine states\cite{bruz-apr-2019}. In the realm of superconducting quantum computers, qubits are denoted by phase, charge, and flux qubits\cite{clarke-nature-2008}. For instance, in the charge qubit, different energy levels correspond to integral numbers of Cooper pairs residing on a superconducting island. Quantum gates for such systems are realized through microwave pulses.

The NV-center, a point defect in diamond, presents a controlled, isolated quantum system that can be manipulated at room temperature. With electronic and nuclear spin components of $S = 1$ and $I = 1$, respectively, the $^{14}N$ NV- center exhibits nine eigenvalue corresponding different spin levels that can be harnessed to realize qubits. Employing resonant microwave pulses, comprehensive quantum control on quantum state is achieved. Measurement methods encompass optical and electrical detection. Although the NV center is susceptible to the effects of absolute temperature and temperature variations, its characteristics at room temperature make it highly appropriate for a diverse range of applications, encompassing quantum sensors and quantum computing, as indicated in \cite{casa-prl-2016, child-mrs-2013}. In May 2016, IBM Corporation introduced a quantum computer comprising five qubits onto the IBM Cloud platform. This initiative aimed to facilitate the execution of algorithms, conduct experiments, and facilitate the exploration of tutorials and simulations to gauge the potential of quantum computing. This five-qubit quantum computer is of a universal nature and is constructed upon superconducting transmon qubits\cite{gaikwad-sr-2022}.  Additionally, IBM is actively engaged in the ongoing enhancement of qubit capabilities. Thus, the concept of a quantum computer has evolved beyond the realm of theory and materialized as a tangible computational apparatus. In the foreseeable future, quantum computers are poised to address genuine problems effectively.

This thesis employs NMR as a tool to undertake tasks associated with quantum information processing. NMR-based quantum computing has established itself as a robust platform for the practical implementation of a diverse array of quantum information processing protocols\cite{dorai-cs-2000}. Within the NMR context, the chemical shifts of different spins are leveraged to individually address these spins in frequency space, and external radio frequency pulses are harnessed for quantum control\cite{oliveira-book-07}.
Quantum information processing necessitates the utilization of pure quantum states. However, the NMR spin system, when operating at room temperature, deviates significantly from this ideal due to the fact that the energy gap between spin levels $\hbar \omega$ is substantially smaller than $k_{\beta}T$. As a result, the initial state of an ensemble of nuclear spins is mixed. Nonetheless, for computational purposes, it becomes possible to initialize the system into a pseudo pure state (PPS) \cite{oliveira-book-07} that emulates a true pure state. By employing radio frequency pulses and leveraging the interactions among spins, any unitary operator can be executed. Furthermore, the mitigation of errors arising from pulse imperfections and offset errors can be achieved through the utilization of numerically optimized pulses employing techniques like Gradient Ascent Pulse Engineering (GRAPE) and genetic algorithms \cite{ khaneja-jmr-2005, manu-pra-2012}. This feature solidifies NMR as an ideal experimental domain for the realization of quantum algorithms\cite{suter-pra-2005, dorai-pra-2000, dorai-pra-2001, xiao-pra-2005}.

The research conducted in this thesis is centred around the designing and empirical implementation of a range of quantum tomography protocols. These protocols are devised to efficiently characterize and reconstruct unknown quantum states and processes, employing both spin ensemble-based NMR quantum information processors and IBM quantum processors built on superconducting technology. The primary goal is to achieve the reconstruction of quantum states through quantum state tomography (QST) protocols and the characterization of quantum processes via quantum process tomography (QPT) protocols. Both QST and QPT play pivotal roles in assessing the reliability and performance of quantum processors. However, the exponential increase in computational complexity with system size poses a fundamental challenge, rendering their experimental execution unfeasible, particularly for systems of greater dimensions. The presence of finite ensemble sizes and inherent systematic errors further leads to the emergence of unphysical density matrices and process matrices. To address these issues, a multitude of QST and QPT protocols have been proposed, albeit many remain untested in experimental settings.

The core objective of this study is to outline pragmatic strategies for the effective implementation of tomography protocols on NMR and IBM quantum processors. To mitigate the unphysicality concern associated with experimentally reconstructed quantum states and processes using standard tomography techniques, the tasks of QST and QPT are reformulated as constrained convex optimization (CCO) problems. This conversion facilitates the reconstruction of valid quantum states and processes. Additionally, for QPT, it permits the computation of a comprehensive set of Kraus operators corresponding to a given quantum process. The application of compressed sensing (CS) and artificial neural network (ANN) techniques is also explored, enabling tomography of quantum states and gates from a significantly reduced dataset compared to conventional methods. CS and ANN-based tomography methods hold promise for handling complexity issues inherent in characterizing higher-dimensional quantum gates.

Furthermore, the investigation delves into the challenge of selectively and directly estimating desired elements within the process matrix that characterizes a quantum process. This exploration results in the formulation of a selective and efficient quantum process tomography protocol (SEQPT), presenting a generalized quantum algorithm and circuit for SEQPT implementation. Successful experimental demonstrations are showcased using NMR and IBM quantum processors. The research also introduces an efficient direct approach to QST and QPT, employing the weak measurement technique, with empirical validation accomplished through a three-qubit NMR system. Moreover, the thesis also studies the simulation and characterization of open quantum system dynamics via dilation techniques. To underscore the efficacy of the proposed quantum tomography and simulation protocols, empirical findings are juxtaposed with theoretically anticipated outcomes across various two- and three-qubit quantum systems.

 \subsection{Quantum bit}
 
A \textit{quantum bit}, generally termed as \textit{Qubit} (this term was coined by Schumacher\cite{sch-pra-1995}) is the quantum analogue of classical \textit{bit} and is the most fundamental and the smallest unit of information used in QCQI .  In the case of qubit, logical states 0 and 1 are represented by $\vert 0 \rangle$ and $\vert 1 \rangle$.  A single qubit is a two-level quantum system represented by a vector in two dimensional Hilbert space spanned by $\vert 0 \rangle = \begin{pmatrix}
1 \\
0 
\end{pmatrix} $ and $\vert 1 \rangle = \begin{pmatrix}
0 \\
1 
\end{pmatrix}$ basis vectors and can be physically realized using the spin-1/2 particle. 
\begin{figure}[t]
\centering
\includegraphics[scale=1]{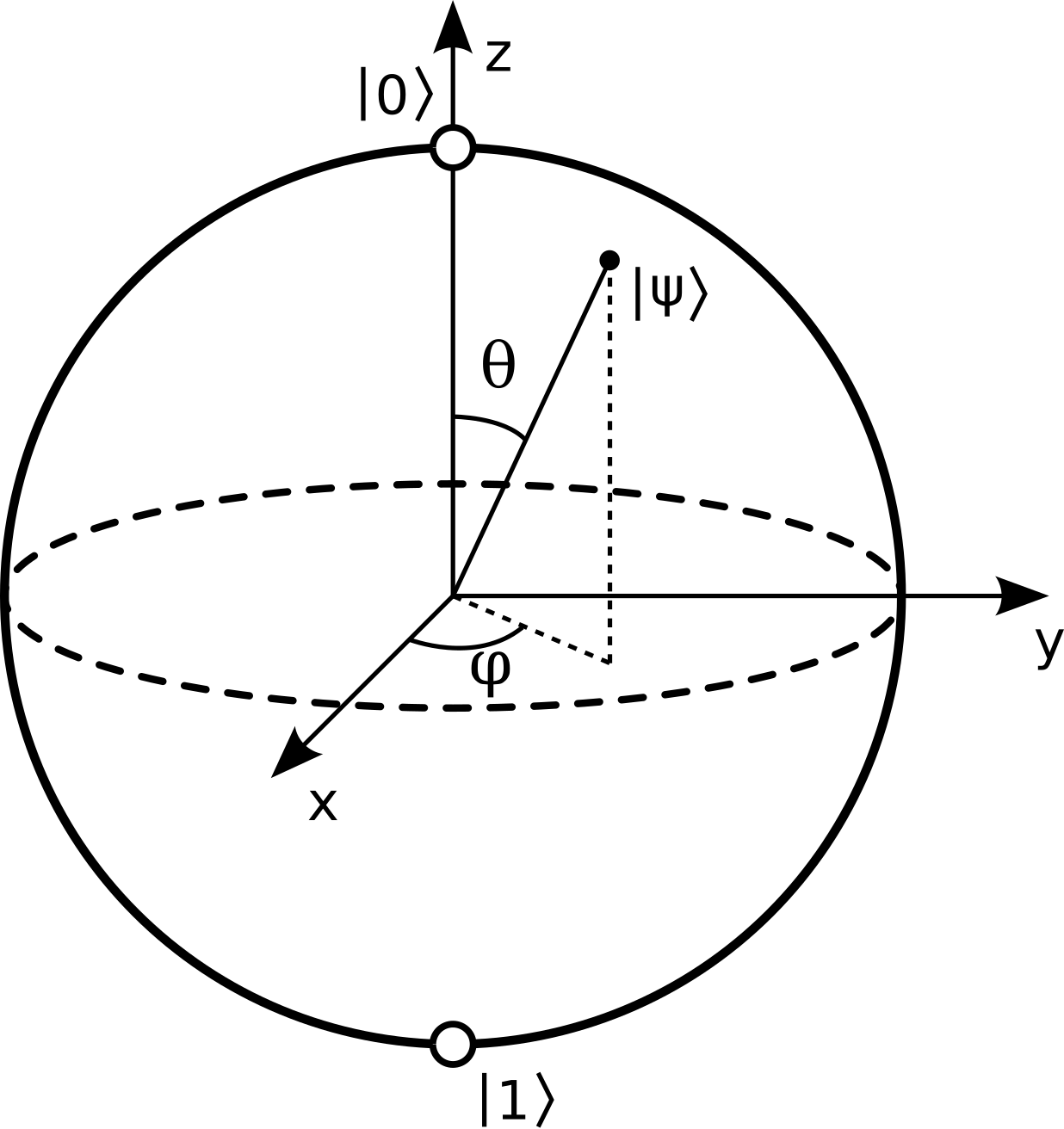}
\caption{Qubit representation on bloch sphere, \textit{source: \href{https://en.wikipedia.org/wiki/Quantum_logic_gate}{Google}} }
\label{bloch}
\end{figure}
 The most general state $\vert \psi \rangle$ of the qubit in polar form, also referred as `Bloch sphere representation of a qubit' is given as,
\begin{equation} 
\label{ch1_eq9}
|\psi\rangle=\cos \left(\frac{\theta}{2}\right)|0\rangle+e^{i \phi} \sin \left(\frac{\theta}{2}\right)|1\rangle
\end{equation}
Note that the global phase is ignored in the above representation of $\vert \psi \rangle$ as it does not have over all observable effect on measurement outcome. The state $\vert \psi \rangle$ given in Eq.\ref{ch1_eq9} can be visualized using Bloch sphere of unit radius as given in Fig.\ref{bloch}.

One can also construct multi-qubit quantum register of length $n$. A $n$-qubit quantum register comprises of $n$ number of qubits and the state of such $n$-qubit composite system can be represented by vector in $2^n$ dimensional Hilbert space. The most general representation of $n$-qubit quantum state is given as,
\begin{equation}\label{ch1_eq10}
\vert \Psi \rangle = \sum_{i=1}^{2^n} c_i \vert e_i \rangle
\end{equation} 
 where $ \vert e_i \rangle  = \lbrace \vert 0 \rangle, \vert 1 \rangle \rbrace^{\otimes n} $ is the $n$-qubit basis vector constructed by taking tensor product of basis vectors of individual qubits and $c_i \in \mathbb{C}$ is the corresponding coefficient such that $\sum_i \vert c_i \vert^2 =1$. As an example, the 2-qubit quantum state is given as,
 \begin{equation}
 \vert \Psi \rangle = c_{00} \vert 00 \rangle + c_{01} \vert 01 \rangle + c_{10} \vert 10 \rangle + c_{11} \vert 11 \rangle 
 \end{equation}
 where $\vert 00 \rangle = \vert 0 \rangle \otimes \vert 0 \rangle$, $\vert 01 \rangle = \vert 0 \rangle \otimes \vert 1 \rangle$, and so on.
 
 If one could able to write the $n$-qubit state given in Eq.\ref{ch1_eq10} as tensor product of individual qubit states $\vert \psi_i \rangle$ as given below,
 \begin{equation}
|\Psi\rangle=\left|\psi_{1}\right\rangle \otimes\left|\psi_{2}\right\rangle \otimes \ldots \otimes\left|\psi_{n}\right\rangle
\end{equation}
then the state $\vert \Psi \rangle$ is said to be \textit{separable} state else it is a \textit{entangled} state. Entangled states don't have any classical analogue and play very crucial role in QCQI\cite{ben-pra-1996, vedral-prl-1997, vedral-jmo-1997, vedral-pra-1998}.  
\begin{figure}[t]
\centering
\includegraphics[scale=0.35]{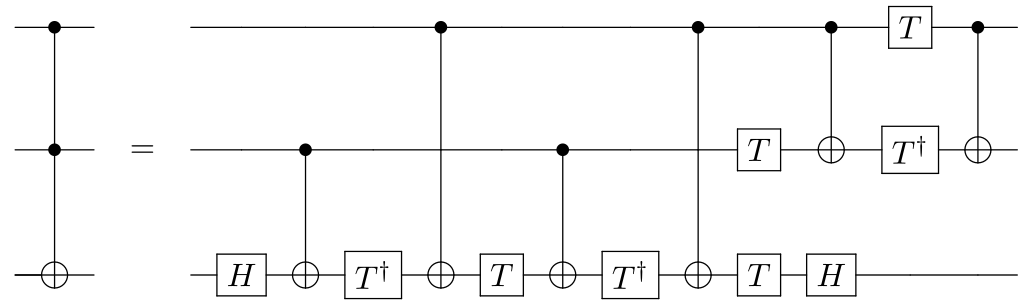} 
\caption{Implementation of 3-qubit \textit{Toffoli} gate using Hadamard \textit{(H)}, phase gate \textit{(S)}, $\pi/8$ gate \textit{(T)} and \textit{CNOT} gate.  }
\label{fig14}
\end{figure}

 \subsection{Quantum gates}
 
\begin{figure}[h!]
\centering
\includegraphics[scale=1.1]{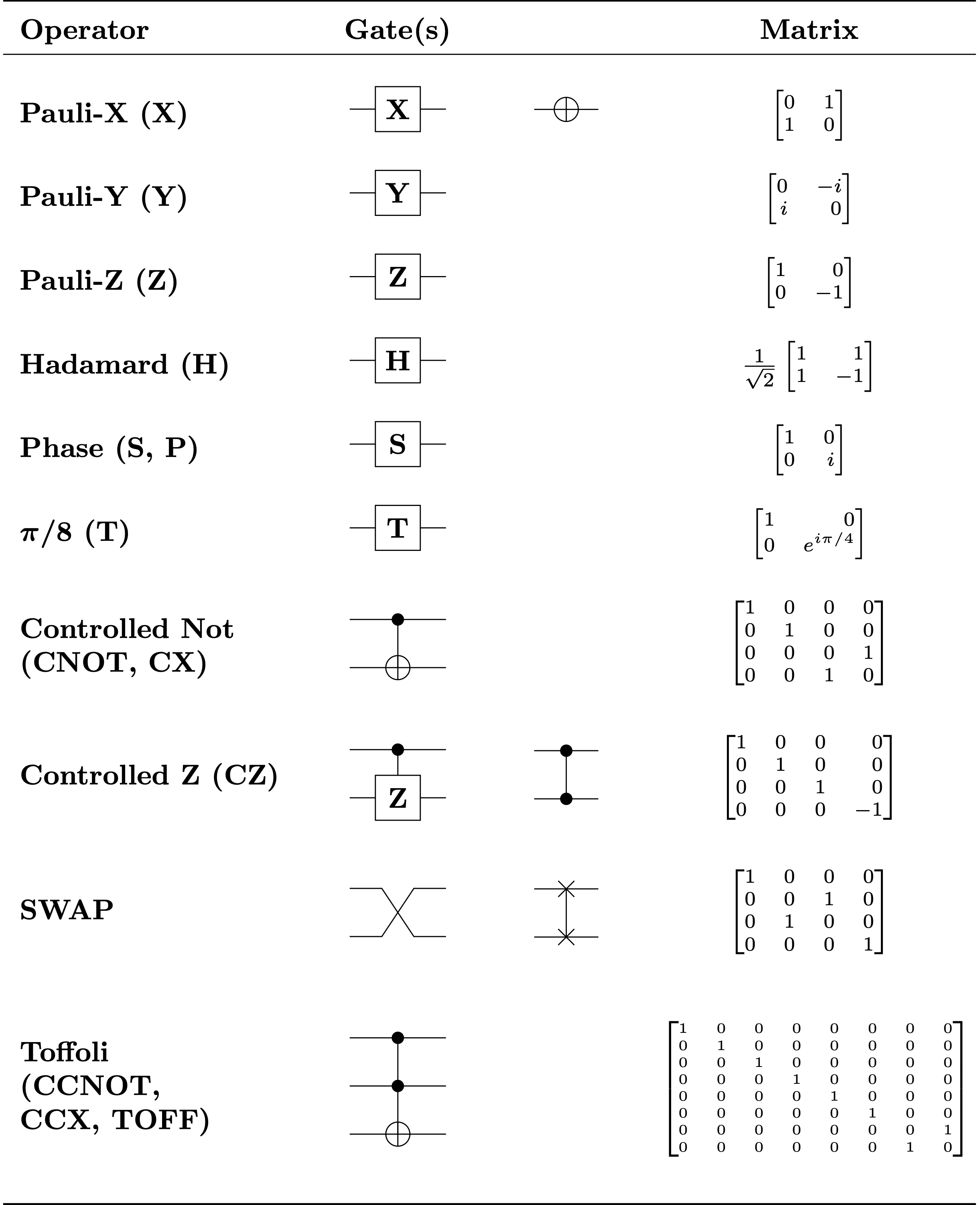} 
\caption{Common quantum logic gates by name, circuit form(s) and matrices. \textit{source: \href{https://en.wikipedia.org/wiki/Quantum_logic_gate}{Google}} }
\label{fig13}
\end{figure}

 An $n$-qubit quantum gate is represented by $2^n \times 2^n$ dimensional unitary matrix which acts on $n$-qubit input state vector and gives $n$-qubit output state vector. The following are some of the important quantum gates (matrix representation is given in Fig.\ref{fig13}):
 \begin{align}
\text{1-qubit QGs}&: \textit{X, Y, Z, H, S, T} \nonumber \\
\text{2-qubit QGs}&: \textit{CNOT, CZ, SWAP} \nonumber \\
\text{3-qubit QGs}&: \textit{Toffoli} \nonumber
\end{align}
 Arbitrary $n$-qubit quantum gate can be decomposed into set of 1 and 2-qubit quantum gates called \textit{universal quantum gates}. One of the widely used universal set of quantum gates is: $\lbrace \textit{H, S, T, CNOT} \rbrace$.
As an example, the decomposition of 3-qubit \textit{Toffoli} gate into universal set of quantum gates is given in the Fig.\ref{fig14}. However, in general, finding decomposition using universal quantum gates in order to create a given unitary matrix that can be implemented efficiently is itself a challenging task. 

\subsection{Quantum Measurement}

 Consider the Hermitian observable represented by operator $A$ with eigenvectors $\vert \lambda_n \rangle$ and corresponding eigenvalues $\lambda_n$. Using spectral theorem, operator $A$ can be rewritten as,
 \begin{equation}
 A = \sum_n \lambda_n P_n
 \end{equation}
 where $P_n = \vert \lambda_n \rangle \langle \lambda_n \vert$ is a projection operator onto eigenstate $\vert \lambda_n \rangle$ satisfying following properties: i) $P_m P_n = \delta_{mn} P_n$, ii) $P_n^{\dagger} = P_n$ and iii) $\sum_n P_n = I$.
 
 If a measurement of $A$ is made on quantum state $\vert \Psi \rangle$ given in Eq.\ref{ch1_eq10} then the state of the system will collapse onto one of the eigenstates of $A$ and the result will be the corresponding eigenvalue $\lambda_n$ with probability $\vert c_n \vert^2=\langle \Psi \vert P_n \vert \Psi \rangle$. The post-measurement state $\vert \Psi \rangle_{final}$ of the system is given as,
 \begin{equation}
  \vert \Psi \rangle_{final} = \frac{P_n \vert \Psi \rangle }{\sqrt{\langle \Psi \vert P_n \vert \Psi \rangle}}
 \end{equation}
 
 The expectation value $\langle A \rangle$ of observable $A$ is defined as,
\begin{equation} \label{ch1_eq24}
\langle A \rangle = \langle \Psi \vert A \vert \Psi \rangle
\end{equation}

\subsection{Quantum ensemble and density matrix formalism}
 A collection of independent, isolated (not interacting) and identical systems is called an \textit{ensemble}. For example, the collection of proton spins in tube of water can be treated (up to a very good approximation) as \textit{ensemble of spin-$\frac{1}{2}$ particles}. Consider the ensemble of size $N$ out of which $n_1$ number of individual systems are prepared in the state $\vert \psi_1 \rangle$, $n_2$ number of individual systems are prepared in the state $\vert \psi_2 \rangle$, and so on such that $\sum_i n_i = N$. The density matrix $\rho$ describing the state of such ensemble system is given as,
 \begin{equation}
 \rho  = \sum_i p_i \vert \psi_i \rangle \langle \psi_i \vert
 \end{equation}
 where $p_i = \frac{n_i}{N}$ is classical probability of having individual system in the state $\vert \psi_i \rangle$ such that $\sum_i p_i = 1$.  The $\rho$ has following three properties: i) $\rho$ is Hermitian matrix, \ie $\rho = \rho^{\dagger}$, ii) $\rho$ is positive semi-definite, \ie $\rho \geq 0$. In other words, all the eigenvalues of $\rho$ are non-negative and iii) $\rho$ has a unit trace, \ie $\rm{Tr}(\rho) = 1$. It turns out that for pure ensemble we have $\rm{Tr}(\rho^2) = 1$ while in the case of mixed ensemble we have $\rm{Tr}(\rho^2) < 1$. The density matrix elements in given basis set $\lbrace \vert e_i \rangle \rbrace$ can be written as,
\begin{equation}
\rho_{mn} = \langle e_m \vert \rho \vert e_n \rangle
\end{equation}
The diagonal elements $\rho_{mm}$ represents the \textit{Born probability} of getting state $\vert e_m\rangle $ or sometimes also interpreted as the \textit{population} of state $\vert e_m \rangle$  while the off-diagonal elements $\rho_{mn}$ ($m \neq n$) are called \textit{coherences} between state $\vert e_m \rangle$ and $\vert e_n \rangle$.

\section{Theory of NMR}\label{nmr_basic}
The term 'NMR' stands for 'Nuclear Magnetic Resonance'. It is a natural phenomenon observed in the atomic nuclei having non-zero nuclear spin (or magnetic dipole moment). If such nucleus is concomitantly placed in the presence of a static magnetic field and an oscillating electromagnetic field with appropriate frequency then an absorption or emission can occur, the phenomenon is termed as 'nuclear magnetic resonance'. In NMR, the Zeeman Hamiltonian of single nuclear spin having magnetic dipole moment $\boldsymbol{\mu}$ placed in external magnetic field ($\mathbf{B} = B_0 \hat{z}$) is given as,
\begin{equation}\label{ch1_eq26}
H =  -\boldsymbol{\mu}.\mathbf{B}  = -\mu_z B_0 = -\gamma \hbar B_0 I_z = -\omega_0 \hbar I_z
\end{equation}
 where $\gamma$ is called the gyromagnetic ratio of the nucleus, $I_z$ is a dimensionless operator representing \textit{z}-component of angular momentum or nuclear spin and $\omega_0 = \gamma B_0$ is known as 'Larmor frequency'. The eigenvectors of $I_z$ are denoted by $\vert m \rangle$ where $m = -I,-I+1,-I+2,...,I-1,I$.  The eigenvalues of the Hamiltonian are directly proportional to eigenvalues of $I_z$, given as,
\begin{equation}\label{ch1_eq29}
E_m = -m \hbar \omega_0
\end{equation}

In the case where given molecule has more than one NMR active nuclei placed in magnetic field then the total NMR Hamiltonian $H_{\rm tot}$ is given as,
\begin{equation} \label{ch1_eq30}
H_{\rm tot} = \sum_i H_z^i + H_{\rm int}
\end{equation}
 where $H_z^i$ is Zeeman Hamiltonian of $i$th nucleus and $H_{\rm int}$ is inter-spin interaction Hamiltonian. The general form of interaction Hamiltonian $H_{\rm int}$ for a system of coupled spins is given as,
 \begin{equation}
 H_{\rm int} = H_{\rm cs} + H_{\rm J} + H_{\rm Q} + H_{\rm D}
 \end{equation}
 where  $H_{\rm cs}$ is chemical shift interaction, $H_{\rm J}$ is scalar or J-coupling,  $H_{\rm Q}$ is quadrupolar interaction and $H_{\rm D}$ is dipolar interaction. The $H_{\rm cs}$ is due to orbital motion of the surrounding electrons, $H_{\rm J}$ is the electron-mediated interaction between nuclei, $H_{\rm Q}$ is the quadrupolar interaction between a nucleus with spin $>$ 1/2 and the electric field gradient at the nuclear position and $H_{\rm D}$ is the direct (through space) dipolar interaction between nuclei. However, in liquid state NMR, $H_{\rm Q}$ and $H_{\rm D}$ get averaged to zero and only surviving terms are $H_{\rm cs}$ and $H_{\rm J}$.

\subsubsection*{Interaction with radio frequency field: The NMR phenomenon}

The transitions between energy eigenstates of Zeeman Hamiltonian can be induced using a radio-frequency pulse. The Hamiltonian associated with RF pulse is known as RF Hamiltonian $H_{RF}$. For RF pulse applied along x direction with oscillating magnetic field $\pmb{B}_1(t)=2B_1 \cos(\Omega t + \phi)\hat{x}$, RF Hamiltonian has the form\cite{oliveira-book-07},
\begin{equation}
H_{RF}=-\pmb{\mu}\cdot\pmb{B}_1(t)=-\hbar \gamma I_x [2B_1\cos(\Omega t + \phi)]
\end{equation}
 On resonance ($\omega_0 \approx \Omega$), the transition rate between spin state $\vert m \rangle$ and  $\vert n \rangle$ is given by \textit{Fermi golden rule}\cite{oliveira-book-07},
\begin{equation}
P_{m \longrightarrow n} \propto \gamma^2 \hbar^2 B_1^2 \vert \langle m \vert I_x \vert n \rangle \vert^2.
\end{equation}
 The allowed transitions are given by selection rule: $\triangle m =\pm 1$. The amplitude of oscillating magnetic field $B_1$ is very small compared to $B_0$. So, $H_{RF}$ can be treated as a small perturbation to the Zeeman Hamiltonian and the evolution of nuclear spin in presence of RF pulse can be obtained using time-dependent perturbation theory. However, for simplicity the evolution of spin states under $H_{RF}$ can also be visualized using rotating frame approximation via semi-classical treatment known as 'vector model'. The effective Hamiltonian $H_{eff}$ of single spin in rotating frame of frequency $\Omega$ can be approximated as\cite{oliveira-book-07},
 \begin{equation} \label{ch1_eq35}
  H_{e f f}=\omega^{0} I_{z}+w_{1}\left[\cos (\phi) I_{x}+\sin (\phi) I_{y}\right] 
 \end{equation}
where $\omega^{0}  = (\omega_0 - \Omega)$ is called resonance offset and $\omega_1 = \gamma B_1$ is known as \textit{Nutation frequency}. The resonance occurs at $\omega_0 = \Omega$.
%
\subsubsection*{Thermal density matrix}      

In reality, no system is completely isolated, there is always some interaction present with environment which leads system to go to thermal equilibrium with it's environment. The thermal density matrix $\rho_{0}$    is simply related to system's Hamiltonian $H$ as,
\begin{equation}
\rho_{0}=\frac{e^{-H / k_{B} T}}{\sum_{m} e^{-E_{m} / k_{B} T}}
\end{equation}
where sum in denominator is called the \textit{partition function} $Z$ of the system extended over all Hamiltonian eigenstates and $E_m$ represents the eigenvalues of $H$.  In the basis formed by Hamiltonian eigenstates, the thermal density
matrix is given as,
\begin{equation} \label{ch1_eq37}
\rho_{0}=\frac{1}{Z}\left(\begin{array}{cccccc}
e^{\frac{-E_{1}}{k_{B} T}} & 0 & 0 & \cdot & \cdot & \\
0 & e^{\frac{-E_{2}}{k_B^{T}}} & 0 & 0 & \cdot & \cdot \\
0 & 0 & e^{\frac{-E_{3}}{k_{B} T}} & 0 & 0 & \cdot \\
\cdot & \cdot & \cdot & \cdot & & \\
\cdot & \cdot & \cdot & & \cdot & \\
\cdot & \cdot & \cdot & & & .
\end{array}\right)
\end{equation}
where $E_m = m \hbar \omega_o$. In \textit{high-temperature} limit ($k_{B} T \gg \hbar \omega_0$) we get,
\begin{equation} \label{ch1_eq38}
\begin{aligned}
&e^{m \hbar \omega_{0} / k_{B} T} \cong 1+\frac{m \hbar \omega_{0}}{k_{B} T}  \\
&\sum_{s=-I}^{I} e^{s \hbar \omega_{0} / k_{B} T} \cong 2 I+1 
\end{aligned}
\end{equation}
Using Eq.\ref{ch1_eq37} and \ref{ch1_eq38}, thermal density matrix $\rho_0$ for spin-1/2 ensemble can be written as,
\begin{equation}\label{ch1_eq39}
\rho_{0}=\left(\frac{1}{2 I+1}\right) \mathbf{1}+\left(\frac{1}{2 I+1} \Delta\right) I_{z}
\end{equation}
where $\mathbf{1}$ is the identity matrix and $ \Delta = \frac{\hbar \omega_{0}}{ k_{B} T}$. The first term in the Eq.\ref{ch1_eq39} is treated as uniform background hence it does not contribute to NMR signal. On the other hand the second term is called as \textit{deviation density matrix} ($\Delta \rho_0$) which constitutes the starting point for all NMR experiments.

\subsubsection*{Relaxation phenomenon}

NMR spin ensemble regain the thermal equilibrium state through mainly two types of relaxation processes:  i) Transverse relaxation ($T_2$) affecting $M_{x,y}$, and ii) Longitudinal relaxation ($T_1$) impacting $M_z$. The transverse and longitudinal damping rates, $\gamma_2$ and $\gamma_1$, are defined by $\gamma_2 = 1/T_2$ and $\gamma_1 = 1/T_1$ respectively. These processes are governed by the Bloch equations, encompassing the decay of transverse components ($M_{x,y}$) and recovery of the thermal magnetization along z-axis ($M_z=M_0$).

\begin{align}
\text{Transverse relaxation}&: \frac{d M_{x,y}}{dt} = \frac{-M_{x,y}}{T_2} \Longrightarrow M_{x, y}=M_{0} e^{-t / T_{2}}  \nonumber \\
\text{Longitudinal relaxation}&: \frac{d M_{z}}{dt} = \frac{M_0-M_z}{T_1} \Longrightarrow M_{z}=M_{0}\left(1-e^{-t / T_{1}}\right) \nonumber 
\end{align}

\subsubsection*{NMR spectrum}

In signal acquisition, an RF pulse flips (also referred as \textit{readout} pulse),  the magnetization vector from equilibrium to the transverse plane which precesses at Larmor frequency in the xy-plane. This precession induces a magnetic field change, leading to electromagnetic induction and voltage signal emergence, known as free induction delay (FID). The total FID signal is generally represented by complex number $S(t)$ as,

\begin{equation}
\begin{aligned}
S(t) &=(S_{x}(t)+i S_{y}(t))e^{-t/T_2} \\
&=(S_{0} \cos \omega_0 t+i S_{0} \sin \omega_0 t)e^{-t/T_2} \\
&=S_{0} \exp (i \omega_0 t)e^{-t/T_2} .
\label{ch1_eq40}
\end{aligned}
\end{equation}
where $S_0$ is total magnetization while $S_x = S_{0} \cos \omega_0 t$ and $S_y = S_{0} \sin \omega_0 t $ are the $x$ and $y$ components
of the total magnetization and $e^{-t/T_2}$ is transverse magnetization damping factor. Fourier transformation (FT) of $S(t)$ gives the frequency domain signal $S(\omega)$, a Lorentzian function, called `NMR spectrum' given as,
 \begin{equation}
\begin{aligned}
S(t) &\xrightarrow[\text{}]{\text{FT}}S(\omega) \\
S_{0} \exp (i \omega_0 t)e^{-\gamma_2 t}  &\xrightarrow[\text{}]{\text{FT}} \underbrace{\frac{S_{0} \gamma_2}{\gamma_2^{2}+(\omega-\omega_0)^{2}}}_{\text {real }}+\mathrm{i} \underbrace{\frac{-S_{0} (\omega-\omega_0)}{\gamma_2^{2}+(\omega-\omega_0)^{2}}}_{\text {imaginary }} \label{ch1_eq41}
\end{aligned}
\end{equation}
 In the NMR language, the real part of spectrum is called `\textit{absorption mode}' while the imaginary part of spectrum is called `\textit{dispersion mode}'. Generally, an NMR spectrum is a combination of the absorption and dispersion modes. 
 
 \section{NMR quantum computing}
 
 In the year 1997, D. G. Cory and I. L. Chuang independently introduced a quantum computer model based on Nuclear Magnetic Resonance (NMR), capable of being programmed akin to conventional quantum computers\cite{cory-1997, neil-sci-1997}. Their model involves an ensemble quantum computer framework where measurement outcomes manifest as the anticipated values of observables. This computational paradigm can be actualized through NMR spectroscopy on extensive ensembles of nuclear spins. The NMR has been used to experimentally demonstrate several quantum algorithms like Grover's search algorithm\cite{naka-pra-2004}, Deutsch-Jozsa algorithm\cite{dorai-pra-2000}, Order-Finding algorithm\cite{vand-prl-2000}, Shor's algorithm\cite{vand-nature-2001} and many more\cite{oliveira-book-07}. 

The NMR spectrometer encompasses a superconducting magnet inducing a robust magnetic field along the z-direction, along with RF coils for both exciting spins and capturing NMR signals from the relaxing spin ensemble. Upon sample placement within the magnetic field, spin interactions trigger energy level splitting contingent on the spin system size. At room temperature, energy level populations adhere to the Boltzmann distribution, yielding a mixed state at thermal equilibrium. This scenario presents a challenge for quantum computing, which mandates pristine initial quantum states. This challenge is surmounted in NMR quantum computing by initiating a "pseudopure" state that mimics a pure state. Quantum gates are executed using RF pulses and inter-spin interactions, with computation outcomes recorded as NMR signals representing average magnetization in the \textit{x} and \textit{y} directions. These signals correspond directly to expectation values of certain qubit basis set elements. By applying RF pulses on individual spins, the anticipations for all basis set elements can be computed, enabling density matrix reconstruction.

Recent advancements in NMR, specifically in the domain of spin dynamics control through RF pulses, have facilitated the high-fidelity implementation of quantum gates for NMR quantum computing. Nuclear spins exhibit extended coherence times due to their isolation from the environment. Despite these merits, a notable constraint of liquid state NMR quantum computers remains scalability. Subsequent sections delve into topics including state initialization, quantum gate implementation, and measurement within NMR quantum computing.

%

 \subsection{Well characterized qubits in NMR}
 
 A \textit{Qubit} is a two level quantum system which can be realized in NMR by spin-$\frac{1}{2}$ nucleus placed in static magnetic field with corresponding Zeeman Hamiltonian $H = \nu I_z$, yielding two energy eigenstate: $\vert \frac{1}{2} \rangle$ and $\vert -\frac{1}{2} \rangle$ with corresponding energy eigenvalues $+\frac{1}{2}\nu$ and $-\frac{1}{2}\nu$ respectively where $\nu = \frac{\gamma B_0}{2 \pi}$ is Larmor frequency (in Hertz (Hz)) of given nucleus. 
 
The Hamiltonian for a system of $n$-interacting spin-1/2 nuclei in a magnetic field, is given as: 
 
\begin{equation}
H_0=\sum_{i=1}^n \omega_i I_z^i+2 \pi \sum_{i<j}^n J_{i j} I^i \cdot I^j
\end{equation}
where $J_{ij}$ is scalar coupling between the $i$th and $j$the coupling. Multi-qubit system can be realized using such $n$-interacting spin-1/2 nuclei.

\begin{figure}[t]
\centering
\includegraphics[scale=0.34]{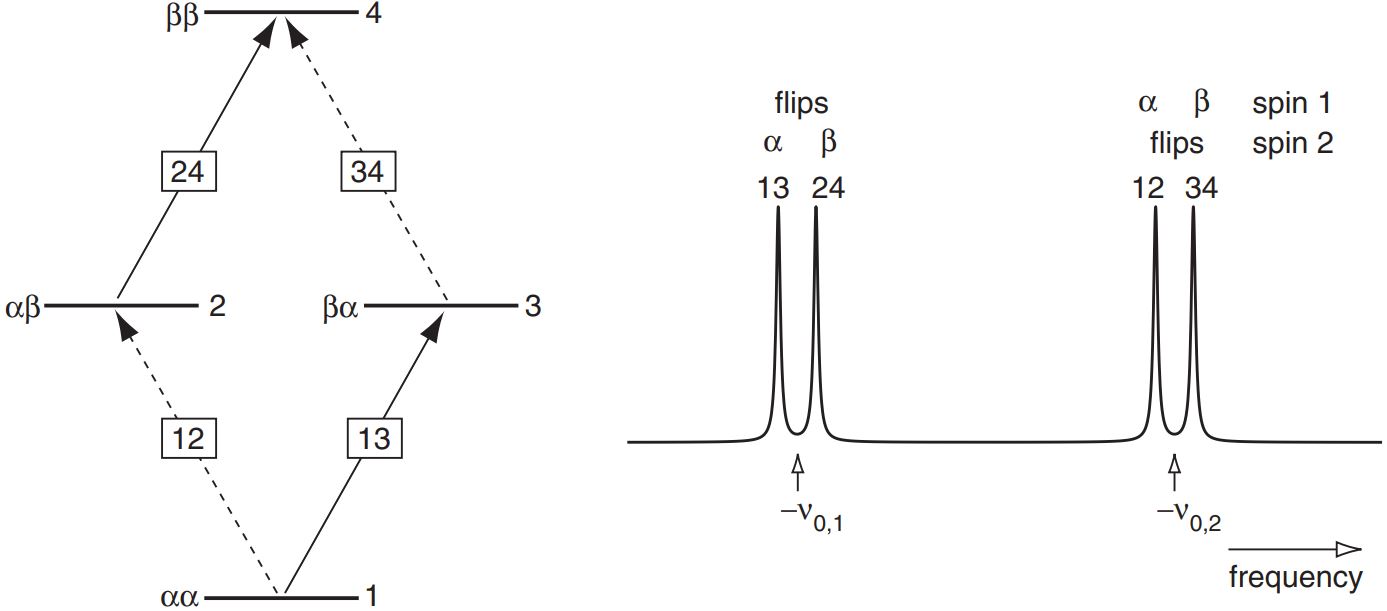} 
\caption{Allowed transitions and NMR spectrums in coupled two spin-1/2 nuclei. }
\label{fig15}
\end{figure}
Apart from qubit one can also have other units of information like:
\begin{align}
\text{\textit{Qutrit}} &: \text{3-level quantum system} \quad (\text{spin-1 nuclei like: $^2H$, $^{14}N$}) \nonumber \\
\text{\textit{Ququart}} &: \text{4-level quantum system}  \quad (\text{spin-3/2 nuclei like: $^{11}B$, $^{75}As$, $^{79}Br$}) \nonumber \\
\text{\textit{Qudit}} &: \text{d-level quantum system} \quad (\text{A quadrupolar nuclei}) \nonumber \nonumber 
\end{align}  
Nuclei with spin $>1/2$ are called \textit{quadrupolar} nuclei.


\subsection{Initializing quantum circuits in NMR}

The NMR apparatus operates with a spin ensemble at ambient temperature. In liquid-state NMR experiments, the sample volume is typically 400-600 $\mu l$, containing approximately $10^{17}-10^{18}$ nuclear spins. At thermal equilibrium, Boltzmann distribution governs energy eigenstate population, resulting in a mixed thermal density matrix of the form:
\begin{equation}
\rho_{\mathrm{eq}} \approx \frac{1}{2^{\mathrm{N}}}\left({I}+\epsilon \Delta \rho_{\mathrm{eq}}\right)
\end{equation}
where $I$ is ${2^n \times 2^n}$ identity matrix , $\epsilon$ is associated with NMR signal strength, The first term in the $\Delta \rho_{\mathrm{eq}}$ is deviation density matrix. Using the fact that NMR is only responsive to deviation part of the density matrix which provides the base of the concept of Pseudo-pure state (PPS)\cite{oliveira-book-07}.
The PPS ($\rho_{pps}$) of an n-qubit system corresponding to $\vert \psi \rangle$ is given as,
\begin{equation}
\rho_{pps}=\frac{1-\epsilon}{N} I+\epsilon|\psi\rangle\langle\psi|
\end{equation}   
  The first term in the $\rho_{pss}$ corresponds to uniform background while the second term is the deviation density matrix. All the computational tasks and calculations are analysed by considering only deviation term.

In NMR, the preparation of PPS from thermal density matrix is done using several techniques, out of which two methods used in study are briefly described here: i) Temporal averaging method and ii) Spatial averaging method. 
\subsubsection*{Temporal averaging method}
In temporal averaging method, results from multiple experiments are added together to get PPS. E.g, in the case of two-qubit PPS state preparation, three unitary gates $\lbrace U_0, U_1, U_2 \rbrace $ are applied on initial thermal density matrix $\rho_{th}$ given as,
   \begin{equation}
\rho_{th}=\left(\begin{array}{llll}
a & 0 & 0 & 0 \\
0 & b & 0 & 0 \\
0 & 0 & c & 0 \\
0 & 0 & 0 & d
\end{array}\right)
\end{equation}
The three unitary matrices are given as,
\begin{equation}
U_{0}=\left(\begin{array}{llll}
1 & 0 & 0 & 0 \\
0 & 1 & 0 & 0 \\
0 & 0 & 1 & 0 \\
0 & 0 & 0 & 1
\end{array}\right), U_{1}=\left(\begin{array}{llll}
1 & 0 & 0 & 0 \\
0 & 0 & 1 & 0 \\
0 & 0 & 0 & 1 \\
0 & 1 & 0 & 0
\end{array}\right) \text { and } U_{2}=\left(\begin{array}{llll}
1 & 0 & 0 & 0 \\
0 & 0 & 0 & 1 \\
0 & 1 & 0 & 0 \\
0 & 0 & 1 & 0
\end{array}\right)
\end{equation}
After adding all output density matrices $\rho_i = U_i \rho_{th} U_i^{\dagger}$, the PPS can be obtained as:
\begin{equation}
\rho_{pps}=\rho_{0}+\rho_{1}+\rho_{2}=\left(\begin{array}{cccc}
3 a & 0 & 0 & 0 \\
0 & b+c+d & 0 & 0 \\
0 & 0 & b+c+d & 0 \\
0 & 0 & 0 & b+c+d
\end{array}\right)
\end{equation}

%
The above equation can be rewritten as,
\begin{equation}
\rho_{pps}=(1-a) \mathbf{1}+(4 a-1)|00\rangle\langle 00|
\end{equation}
The $\rho_{pps}$ represents the PPS corresponding to state $\vert 00 \rangle$ and will evolve exactly as pure state $\vert 00 \rangle \langle 00 \vert$. 

\subsubsection*{Spatial averaging method}
The spatial averaging method is based on spatially separated sub-ensembles. In NMR, sub-ensemble can be accessed using combination of RF pulses and pulsed magnetic field gradients (helps to dephase spins so that off-diagonal terms vanishes). The PPS is then average over all sub-ensembles. So, using both RF and gradient pulses together the PPS can be prepared in a single experiment. In case of two-qubit system, the pulse sequence used in spatial averaging method to prepare PPS corresponding to $\vert 00 \rangle$ from thermal density matrix is given below,
\begin{equation}
\left[G_{z}(\tau)\left(\frac{\pi}{4}\right)_{-y}^{1} U_{J}\left(\frac{1}{2 J}\right)\left(\frac{-\pi}{4}\right)_{x}^{1} G_{z}(\tau)\left(\frac{\pi}{3}\right)_{x}^{2}\right]
\end{equation}
where $G_{z}(\tau)$ represents a gradient pulse of duration $\tau$, and $U_{J}\left(\frac{1}{2 J}\right)$ is evolution under J-coupling and $\theta^i_{\pm x(y)}$ is local rotation about $\pm x(y)$ on $i$th qubit. Generally spatial averaging method of PPS preparation suffer magnetization loss due to non-unitary evolution achieved by gradient pulses.

\subsection{Quantum gate implementation in NMR}

The quantum gates are unitary operators which are used to process the quantum information encoded in the state of the quantum system. They are broadly classified as: single qubit gates and multi-qubit gates.  As mentioned earlier in this chapter, any $n$-qubit arbitrary quantum gate can be decomposed using universal set of quantum gates comprising the set of single qubit gates and two-qubit CNOT gate. In this section we briefly describe the NMR implementation of universal quantum gates which further can be used to implement arbitrary high dimensional gates.

\subsubsection*{Single qubit gates}

In NMR rotation operators are implemented using appropriate radio frequency (RF) pulses. From Eq.\ref{ch1_eq35}, the effective RF Hamiltonian (on resonance) is given as: 
\[
H_{e f f}=w_{1}\left[\cos (\phi) I_{x}+\sin (\phi) I_{y}\right] 
\]
If RF pulse of length $t$ is applied with appropriate phase $\phi$, then rotation operators can be implemented as follows:
\begin{equation}
\begin{aligned}
&R_{x}(\theta) = e^{-i \theta I_x}=  (\theta)_x, \quad \text{such that $\phi = 0$ and $\theta = \omega_1 t$}  \nonumber \\
&R_{y}(\theta) = e^{-i \theta I_y}= (\theta)_y, \quad \text{such that $\phi = \frac{\pi}{2}$ and $\theta = \omega_1 t$}  \nonumber \\
&R_{z}(\theta) = e^{-i \theta I_z} = (\frac{\pi}{2})_x(\theta)_y(\frac{-\pi}{2})_x  \nonumber 
\end{aligned}
\end{equation}

It has been shown that, an arbitrary single qubit unitary matrix can be decomposed and implemented on NMR using following pulse sequence:

\begin{equation}
U=e^{i \alpha}\left[\begin{array}{cc}
e^{-i \beta / 2} & 0 \\
0 & e^{i \beta / 2}
\end{array}\right]\left[\begin{array}{cc}
\cos \frac{\gamma}{2} & -\sin \frac{\gamma}{2}  \\
\sin \frac{\gamma}{2} & \cos \frac{\gamma}{2}
\end{array}\right]\left[\begin{array}{cc}
e^{-i \delta / 2} & 0 \\
0 & e^{i \delta / 2}
\end{array}\right] = e^{i\alpha} (\beta)_z (\gamma)_y (\delta)_z 
\end{equation}
where $\alpha$ is overall phase and $\beta$, $\gamma$ and $\delta$ are real numbers whose values depend on given unitary matrix $U$.

\subsubsection*{2-qubit CNOT gate}

The two-qubit CNOT gate is in general entangling gate which can be implemented with set of local rotations and J-coupling ($U_{J}$). The exact pulse sequence is given below:
\begin{equation}
\text{CNOT} = (\frac{\pi}{2})^1_x (\frac{\pi}{2})^1_y (\frac{\pi}{2})^1_{\bar{x}} (\frac{\pi}{2})^2_x (\frac{\pi}{2})^2_{\bar{y}}  U_J(\frac{1}{2J}) (\frac{\pi}{2})^2_y   
\end{equation} 
where $U_J(\frac{1}{2J}) = e^{-i 2\pi J I_{1z}I_{1z} t}$ is free evolution under J-coupling for time $t = \frac{1}{2J}$.
Note that all pulse sequences are to be read from right to left. 

\subsubsection*{Gate time vs decoherence time}
The typical gate implementation time lies in the range of microseconds (local rotation gates) to milliseconds (high dimensional entangling gates like CNOT or Toffoli) in contrast to decoherence time which lies in the range of few seconds, in some cases it is in hours. 

\subsection{Measurement in NMR}

\begin{figure}[t] 
\centering
\includegraphics[scale=0.29]{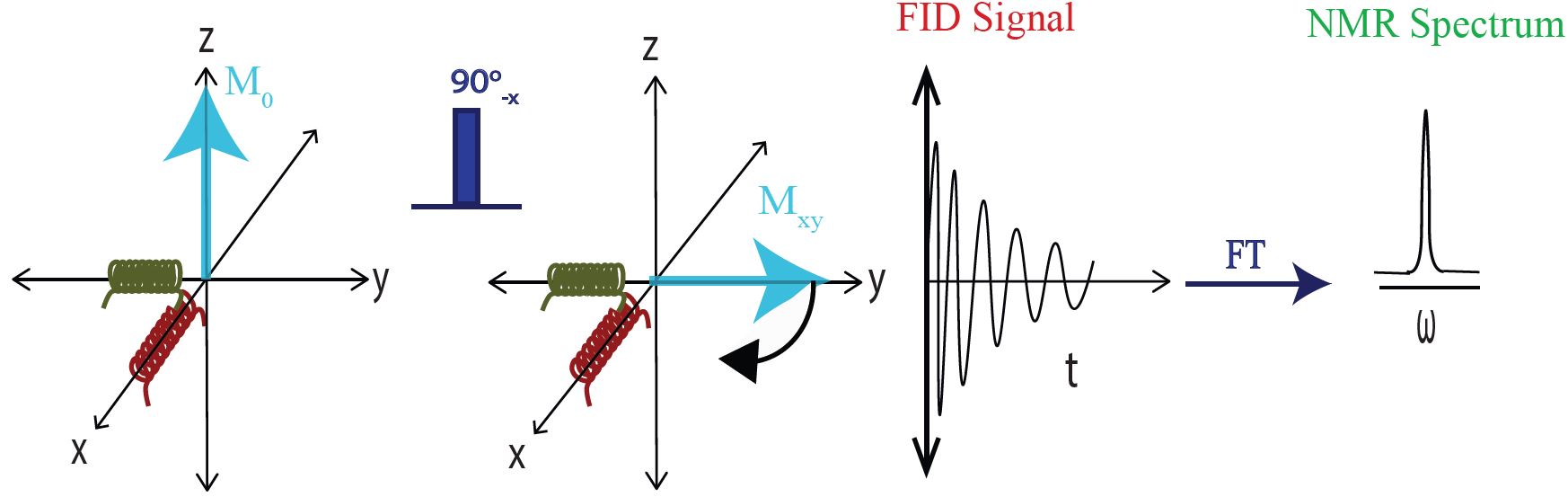} 
\caption{Signal acquisition process in NMR. }
\label{fig16}
\end{figure}

The conventional NMR signal detection involves an ensemble weak measurement, wherein the spin's interaction with the radio-frequency coil does not considerably alter the quantum state during the measurement of total spin magnetization.  As detailed in Sec.\ref{nmr_basic}, when nuclear spins are placed in magnetic field $B_0$ along \textit{z}-axis, the bulk magnetization is produced alined along \textit{z}-axis. Application of \textit{rf} pulse rotates this magnetization to the \textit{xy} plane, initiating precession around the \textit{z}-axis at a Larmor frequency $\omega_0$. The resulting precessing magnetization introduces flux changes in the rf coils, generating a signal voltage (as depicted in Fig.\ref{fig16}). This precession induces a magnetic field change, leading to electromagnetic induction and voltage signal emergence, known as free induction delay (FID), illustrated in Fig.\ref{fig16}. The temporal transverse magnetization signal in the time domain is expressed as follows:
\begin{equation}
S(t) =(S_{x}(t)+i S_{y}(t))e^{-t/T_2} \propto S(t) \propto \operatorname{Tr}\left\{\rho(t) \sum_k\left(\sigma_{k x}+i \sigma_{k y}\right)\right\}
\end{equation}
where $\sum_k (\sigma_{k x}+i \sigma_{k y})$ is detection operator, $\sigma_{kx}$ and $\sigma_{ky}$ are Pauli spin operators proportional to \textit{x} and \textit{y} component of \textit{k}th spin. Fourier transformation (FT) of $S(t)$ gives the frequency domain signal $S(\omega)$ and the spectral intensity gives the expectation values of detection operators which further can be used to perform QST.


\subsection{Open quantum dynamics}

In the case of closed quantum system, the time evolution of a density matrix $\rho$ is given by \textit{Liouville-Von Neumann equation}:
\begin{equation}
\frac{\partial \rho}{\partial t} =-\frac{i}{\hbar}[H, \rho] \label{ch1_eq19}
\end{equation}
where $[H, \rho] = H \rho - \rho H$, is the commutator between operators $H$ and $\rho$.

However, in more general scenario where the system under consideration may not be isolated and may have interaction with it's surrounding, generally referred as \textit{system's environment}, in such case the total Hamiltonian is given as,
\begin{equation}
H_{tot} = H_{sys} + H_{env} +H_{int}
\end{equation}
where $H_{sys}$ and $H_{env}$ represent the system's and environment's Hamiltonian while $H_{int}$ represents system-environment interaction Hamiltonian. One can still treat such composite system (main system + environment) as whole and the time evolution of corresponding composite density matrix $\rho_{comp}$ is still given by Eq.\ref{ch1_eq19}. However, if one is only interested in the time evolution of reduced density matrix $\rho_{sys}$ describing the dynamics of system alone then one has to trace out the environment, \ie $\rho_{sys}(t) = {\rm Tr}_{env}(\rho_{comp}(t))$. In this scenario system is generally termed as \textit{open quantum system} and the differential equation describing the time evolution of such system is given by the \textit{'Lindblad master equation'}\cite{daniel-aip-2020}, 
\begin{equation} \label{ch1_eq22}
\frac{\mathrm{d} \rho_{sys}}{\mathrm{d} t}=-i[H_{sys}, \rho_{sys}]+\frac{1}{2} \sum_{k} \left(\left[L_{k} \rho_{sys}, L_{k}^{\dagger}\right]+\left[L_{k}, \rho_{sys} L_{k}^{\dagger}\right]\right)
\end{equation}
 where $L_k$ are called Lindblad operators generally describing the effect of environment on the system. The Lindblad master equation can be separated into components which cause unitary and non-unitary
evolution of the system:
\begin{align}
\text{Unitary evolution}&: -i[H_{sys}, \rho_{sys}] \nonumber \\
\text{Non-unitary evolution}&: \frac{1}{2} \sum_{k} \left(\left[L_{k} \rho_{sys}, L_{k}^{\dagger}\right]+\left[L_{k}, \rho_{sys} L_{k}^{\dagger}\right]\right) \nonumber
\end{align}

Another equivalent representation describing the dynamics of open quantum system is given via super operator method which generally referred as \textit{Kraus operator representation}. Consider a quantum system undergoing a general quantum evolution represented by a completely positive (CPTP) map. The
action of such a map on a quantum state $\rho_{sys}$ via the superoperator
$\Lambda$ in the Kraus operator representation is given as\cite{sudarshan-pr-61,kraus-book-83},
\begin{equation} \label{ch1_eq23}
\rho_{sys}(t) = \Lambda(\rho_{sys}(0))=\sum_{i} A_{i}(t) \rho_{sys}(0) A_{i}^{\dagger}(t), \quad s.t \quad\sum_{i} A_{i}(t) A_{i}^{\dagger}(t)=I \quad \forall \hspace{2mm} t
\end{equation}
where $ \rho_{sys}(0) $ is initial density matrix at $t = 0$ and $A_i(t)$'s are Kraus operators at particular time instant $t$. The Kraus operator representation is usually helpful when dealing with system evolution for fixed time interval. 
The utilization of Kraus operator representation becomes evident during the execution of quantum process tomography (QPT) task which will be elaborated later in the subsequent chapters.  Note that the time evolution of density matrix given in Eq.\ref{ch1_eq22} and \ref{ch1_eq23} are equivalent and equations can be inter-converted using techniques given in \cite{tong-pra-2004, lidar-cp-2001}. 


\section{Quantum state and process tomography}

The reliability of quantum devices primarily relies on two factors: the initial states in which information is encoded and the quantum gates used to process that information. Additionally, the measurement process plays a crucial role in extracting information from an unknown quantum state. Therefore, it is essential to characterize quantum states and processes (gates) using efficient measurement techniques. This evaluation is typically achieved through QST and QPT methods.

QST allows us to estimate the unknown density matrix, while QPT enables us to estimate a quantity known as the process matrix, which characterizes the unknown quantum process. Both QST and QPT are statistical processes that consist of two key components: a set of measurements and an estimator that maps measurement outcomes to an estimate of the unknown state or process. Due to the finite size of the ensemble and the presence of systematic errors, there is always some degree of ambiguity associated with estimating an experimentally created or implemented state or process. This ambiguity often results in unphysical density or process matrices.

Moreover, both tasks face a common fundamental challenge: as the size of the quantum system grows, the required resources and computational complexity increase exponentially. This exponential growth makes the implementation of QST and QPT protocols infeasible, even for systems with only a few qubits. Therefore, it is crucial to design efficient QST and QPT protocols that yield physically valid density or process matrices.

On an ensemble quantum computer such as NMR, standard QST is carried out by measuring the expectation values of a fixed set of basis operators, with the n-qubit density operator $\rho$ being represented in the tensor product of the Pauli basis:
\begin{equation} 
\rho=\sum_{i=0}^{3} \sum_{j=0}^{3} \ldots \sum_{n=0}^{3} c_{i j \ldots n} \sigma_{i} \otimes \sigma_{j} \otimes \ldots \sigma_{n} 
\label{ch1_eq57}
\end{equation}
where $c_{00...0} = 1/2^n$ and $\sigma_0$ denotes the $2 \times 2$ identity matrix and $\sigma_i$, $i = 1,2,3$  are single qubit Pauli operators. The standard procedure to estimate all the coefficients $c_{ij...n}$  
typically involves solving linear system of equations of the form\cite{gaikwad-qip-2021}:
\begin{equation}
\mathcal{A}x = \mathcal{B}
\end{equation}
where matrix $\mathcal{A}$ is referred to as a fixed coefficient matrix, the vector $x$ contains elements of the density matrix which needs to be reconstructed and vector $\mathcal{B}$ contains
actual experimental data. 

To formulate the QPT task mathematically, consider the quantum state $\rho$ is undergoing through general quantum evolution given by completely positive trace preserving (CPTP) map $\Lambda$. The action of such map on
input state $\rho$ is described by operator sum representation given in Eq.\ref{ch1_eq23} as:
\begin{equation} \label{ch1_eq59}
\Lambda(\rho)=\sum_{i=1}^{4^n} A_{i} \rho A_{i}^{\dagger}
\end{equation}
In the fixed set of basis operator $\lbrace E_i \rbrace$ the Kraus operators can be expanded as $A_{i}=\sum_{m} a_{i m} E_{m}$ and Eq.\ref{ch1_eq59} can be rewritten as,
\begin{equation}
\Lambda(\rho)=\sum_{m, n} \chi_{m n} E_{m} \rho E_{n}^{\dagger}
\end{equation}
where $\chi_{m n}=\sum_{i} a_{i m} a_{i n}^{*}$. The quantities $\chi_{mn}$ are the elements of $\chi$ matrix which characterize the given quantum process $\Lambda$. The $\chi$ matrix is also referred as process
matrix. The $\chi$ matrix satisfies the following three properties:
\begin{enumerate}
\item $\chi$ should be hermitian. \ie, $\chi = \chi^{\dagger}$
\item Trace($\chi$) = 1
\item $\chi$ should be positive. \ie, $\chi \ge 0$
\end{enumerate}
The standard procedure of estimating full $\chi$ matrix involves solving linear system of equations of the form\cite{chuang-jmo-09}:
\begin{equation}
\mathcal{\beta} \chi = \mathcal{\lambda}
\end{equation} 
where matrix $\beta$ is the coefficient matrix $\mathcal{A}$, column matrix $\chi$ contains the elements $\chi_{mn}$ which are to be determined and column matrix $\lambda$ is actual experimental data.

%

\subsection{Dealing with invalid density and process matrices via convex optimization methods}

Utilizing the standard linear inversion technique for reconstructing quantum states and processes often leads to unphysical density and process matrices respectively. This issue is evident in scenarios like QPT, where non-completely positive maps emerge as result of experimental reconstruction of processes using linear inversion method. This signifies that inversion is not the optimal fit for raw tomography outcomes. Incorrect or unphysical reconstructions could mislead assessments of quantum state and channel. Approaches such as convex optimization, a fundamental concept in machine learning, strive to identify parameters within a model that align best with prior information. Convex optimization yields a global optimum that provides the most accurate fit to raw data. By employing convex optimization, one can extract comprehensive and accurate information from measurement results ensuring the positivity of quantum states and processes. Variety of convex optimization techniques are used in the past to perform valid QST and QPT. Most of the convex optimization methods are based on least square (LS) optimization problem (${\Vert \mathcal{A}x-\mathcal{B}\Vert}_2$) subject to positivity condition ($\rho \geq 0$)\cite{gaikwad-qip-2021, gaikwad-ijqi-2020, huang-sb-2020, qi-quantum-inf-2017}. However, computation complexity grow exponentially as the size of the system which make these methods computationally expensive in terms of time required to reach global optimal solution. To address this issue, these methods are further refined using more advance algorithm like hyperplane intersection projection algorithm which results in projected least-squares QPT protocol\cite{surawy-quantum-2022} and projected gradient descent algorithms for QST\cite{Bolduc2017}.  These convex optimization methods offer several distinct advantages, including its independence from prior knowledge about the system and its ability to operate without the need for additional ancillary qubits. Specifically, in the context of QPT, the convex optimization method allows for the computation of a complete set of Kraus operators. 

Moreover, compressed sensing (CS) optimization method has also been extensively used to perform QST as well as QPT in which the $\Vert . \Vert_{l_1}$ ($l_1$ norm of variable vector) is the main objective function with multiple constraints\cite{rodionov-prb-2014, yang-pra-2017, Steffens_2017}. It tunes out that CS optimization method not only deals with unphysicality issue but also with complexity issue. That is, CS method allows to reconstruct valid density and processes matrix using heavily reduced data set and in some cases outperforms LS method \cite{rodionov-prb-2014, gaikwad-qip-2022}. However, the applicability of CS method is limited and required prior knowledge about the sparsity and noise present in the measurement data. The CS optimization method turns out to be promising while characterizing high dimensional quantum gates.

\subsection{Data driven based QST and QPT via machine learning methods}

Machine learning methods primarily address the issue of complexity and allow to perform full QST as well as QPT using incomplete measurement set. Majority of such protocols are primarily based on data driven approach which includes protocols like self guided tomography in which quantum state is iteratively learned by optimizing a projection measurement without any data storage or post-processing\cite{ferrie-prl-2014, ferrir-prl-2016}, adaptive quantum tomography method using Bayesian approach\cite{straupe-jetp-2016} and via linear regression estimation\cite{qi-quantum-inf-2017, houlsby-pra-2012}. Though self guided and adaptive protocols are efficient as compared to standard method, it requires large number of projective measurements which are hard to implement on ensemble quantum system. 

\subsubsection*{Artificial Neural Networks (ANNs)} 

ANN is a sub-branch of machine learning method and the framework  is inspired from neurons in human brain. ANN consists of many artificial neurons which are
connected to each other. The neuron is activated when its value is greater than
threshold value called bias. The Feed-Forward-Neural network (FFNN) is one the type of ANN model where the architecture consists of three layers: input layer,
hidden layer and output layer. Data is fed into input layer which is passed on
layer by layer till it arrives at the output. Data is divided into two parts i.e. train data-set and
test data-set. As name implies, train data-set is used to train the model
(update network parameters, weights and biases) and test data-set is used to
evaluate the network. Consider '$m$' training elements $\lbrace
(\vec{x}^{(1)},\hat{\vec{y}}^{(1)}),(\vec{x}^{(2)},\hat{\vec{y}}^{(2)}),\cdots ,(\vec{x}^{(m)},\hat{\vec{y}}^{(m)}) \rbrace$ where
$\vec{x}^{(i)}$ is $i^{th}$ input and $\hat{\vec{y}}^{(i)}$ is corresponding labelled output.
Feeding these inputs to the network produces outputs
$[\tilde{\vec{y}}^{(1)},\tilde{\vec{y}}^{(2)},...,\tilde{\vec{y}}^{(m)}]$. Since network
parameters are initialized randomly, predicted output is not equal to expected
output. Training of this network can be achieved by minimizing the
mean-squared-error \textit{cost function}, $J(w,b)=\frac{1}{m}\sum_{i=1}^{m}
||\hat{\vec{y}}^{(i)}-\tilde{\vec{y}}^{(i)}||^2 $ with respect to network parameters by stochastic gradient descent,

\begin{align}
w_{ij} \rightarrow w'_{ij} =& w_{ij}-\frac{\eta}{m'} \sum_{i=1}^{m'}
\frac{\partial}{\partial  w_{ij}} \mathcal{L}(\vec{x}^{(i)}) \\ b_{i} \rightarrow
b'_{i} =& b_{i}-\frac{\eta}{m'} \sum_{i=1}^{m'} \frac{\partial}{\partial  b_{i}}
\mathcal{L}(\vec{x}^{(i)})
\end{align}

where $\mathcal{L}(x^{(i)})= ||\hat{\vec{y}}^{(i)}-\tilde{\vec{y}}^{(i)}||^2 $ is the cost function
of randomly chosen $m'$ training inputs $x^{(i)}$ and $\eta$ is the learning
rate. $w'_{ij}$ and $b'_{i}$ are updated weights and biases. Once the model is trained then it can be applied to test data set which can be a experimentally obtained data set.  In the recent years, such ANN models are applied to perform efficient QST as well as QPT\cite{carleo-rmp-2019, carleo-science-2017, torlai-prl-2018, palmieri-npj-2020, torlai-np-2018, quek-npj-2021}.

\subsection{Selective and direct QST and QPT protocols}

The selective and direct estimation of desired element of density and process matrices is referred as direct QST (DQST) and QPT (DQPT) respectively. The objective of DQST and DQPT is very well explored in the past where plenty of algorithms and protocols are proposed to selectively compute element of density and process matrix\cite{paz-prl-08, ekert-prl-2022, feng-pra-2021}.  Although the traditional methods like adaptive, self guided, maximum likelyhood estimation (MLE) and ancilla assisted methods offer
some advantages over standard QPT, they still are not very
useful when only certain elements of the density or process matrix need to
be estimated. The methods like selective and efficient quantum process
tomography (SEQPT) based on quantum 2-design states\cite{perito-pra-2018,pears-pra-2021, gaikwad-pra-2018} turns out to efficient while performing DQPT whereas weak measurement (WM) based techniques have been used to perform efficient DQST as well as DQPT where in certain special
cases WM method outperform projective measurements\cite{lundeen-prl-2012, arvind-pra-2014}. 
\subsection{Simulation and characterization of open quantum dynamics}

 One of the main building
blocks of a quantum computer is the underlying physical
system and its time evolution under a given
Hamiltonian~\cite{divi-fdp-2000}, while the main obstacle in
building such a quantum computer is its unwanted and
inevitable interaction with its environment, generally
referred to as decoherence.  Efforts
to mitigate decoherence led to studies of open quantum
dynamics, whereby the time evolution of a quantum system is
studied using a master equation
approach~\cite{Breuer2007,Rotter-rpp-2015}.  In real
situations, the physical system under consideration is
continuously interacting with its environment, causing its
time evolution to be non-unitary, contributing
significantly to errors in the computational output and
reducing the quality of the quantum
device~\cite{zuniga-pra-2012}.  Hence the task of designing
quantum algorithms to simulate open quantum dynamics is
important from a fundamental as well as a practical point of
view.

 The duality quantum algorithm for simulating evolution of an
open quantum system was proposed \cite{Wei-sr-2016, Zheng-sr-2021} where the time evolution of the open quantum system is
realized by using Kraus operators. In duality algorithm the evolution operator
is a linear combination of unitary operators resulting in desired non-unitary
evolution. However, the method is experimentally expensive and requires ancilla system of dimension equal to number of Kraus operators characterizing given open quantum system. One the other hand, a class of promising quantum algorithms to
simulate arbitrary non-unitary evolutions on quantum devices
have been reported, which are primarily based on the dilation
technique namely, the Stinespring dilation
algorithm~\cite{shirokov-jmp-2020} and Sz.-Nagy's dilation
algorithm~\cite{prineha-prr-2021}.  The basic
tenet of these algorithms is, to construct a unitary
operation in a higher-dimensional Hilbert space, which
simulates the desired non-unitary evolution in a
lower-dimensional Hilbert space.  The Stinespring dilation
algorithm requires a larger Hilbert space dimension, which
makes it computationally and experimentally expensive, as
compared to the Sz.-Nagy algorithm.  The Sz.-Nagy algorithm
has been used to experimentally simulate a single-qubit
amplitude damping channel on the IBM quantum
processor~\cite{Hu-sr-2020}.

\section{Organization of the thesis}

Chapter 2 presents overview of standard QST and QPT methods based on the linear inversion technique. A detailed description of the associated non-physicality issue concerning states and processes is given. Additionally, the CCO-based QST and QPT protocols are introduced to effectively address and resolve the non-physicality problem.

Chapters 3 and 4 are dedicated to the consideration of the scalability concern of QST and QPT protocols using the CS algorithm and techniques based on artificial neural networks, respectively. The performance of these methods is demonstrated on both NMR and IBM quantum processors. In Chapters 5 and 6, the challenge of selectively and directly estimating desired elements of unknown density and process matrices is addressed. The modified selective quantum process tomography (MSQPT) protocol is proposed, and the weak measurement method is applied on NMR and IBM processors, respectively.

Chapter 7 focuses on the simulation and characterization of arbitrary open quantum dynamics through the utilization of Sz-Nagy's dilation algorithm on an NMR processor. Chapter 8 provides a summary of the main results obtained throughout the study and outlines potential directions for future research.
\chapter{Achieving valid quantum states and processes on an NMR quantum information processor through convex optimization}\label{chap2} 

\section{Introduction}\label{ch2_intro}

In recent years, much research has been conducted to design high-fidelity quantum devices based on quantum technology. This requires the characterization of quantum states and processes, which are essential in studying the behavior of quantum processors and validating quantum devices. The characterization is usually done through quantum state tomography (QST) and quantum process tomography (QPT). Both QST and QPT are statistical processes that consist of two elements: a set of measurements and an estimator that maps the measurement outcomes to an estimate of the unknown state or process. Since the sample size is finite and systematic errors are inevitable, there is always some uncertainty or error associated with the estimated state, which can sometimes result in unphysical density matrices\cite{james-pra-2001}. Therefore, it is crucial to have an estimation protocol that produces valid quantum states (processes).

To date, several tomography protocols have been proposed and successfully applied to physical systems, such as the state of nuclear spin ensembles\cite{long-job-2001, singh-pla-2016, vind-pra-2014}, photon polarization states\cite{qi-quantum-inf-2017, miranowicz-pra-2014}, and infinite-dimensional coherent states of light. These protocols are typically based on least-square linear inversion\cite{long-job-2001, miranowicz-pra-2014, bartkiewicz-sr-2016, miranowicz-prb-2015, james-pra-2001, adamson-prl-2010}. However, there are also other strategies for QST, such as maximum likelihood estimation\cite{singh-pla-2016}, hedged likelihood function estimation\cite{robin-prl-2010}, model averaging\cite{ferrie-njp-2014}, adaptive bayesian state estimation\cite{houlsby-pra-2012}, linear regression\cite{qi-quantum-inf-2017}, gradient approach(self guided)\cite{ferrie-prl-2014}, numerical strategies\cite{kaznady-pra-2009}, weak measurement\cite{lundeen-prl-2016}, and controlled-SWAP quantum network\cite{ekert-prl-2002}. Similar protocols have also been proposed for QPT, including ancilla-assisted QPT\cite{altepeter-prl-2003}, simplified QPT\cite{kosut-njp-2009}, selective QPT using quantum 2-design states\cite{perito-pra-2018}, self-consistent QPT\cite{merkel-pra-2013}, compressed sensing QPT\cite{rodionov-prb-2014}, adaptive measurement-based QPT\cite{pogorelov-pra-2017}, and selective QPT via sequential weak value measurement of incompatible observables\cite{kim-natcom-2018}. These protocols have been implemented on various platforms, such as NMR\cite{maciel-njp-2015, gaikwad-pra-2018, singh-pla-2016, long-job-2001}, superconducting qubits\cite{neeley-nature-2008, chow-prl-2009, steffen-science-2006}, nitrogen vacancy centers\cite{zhang-prl-2014}, linear optics\cite{pogorelov-pra-2017, Chapman-prl-2016}, and ion-trap quantum processors\cite{riebe-prl-2006}. Despite the availability of many tomography protocols, most of them do not produce valid density (process) matrices. Protocols such as adaptive measurements and self-guided tomography that do produce valid states/processes often involve a large number of projective measurements, which can be challenging to implement on ensemble quantum computers. Additionally, in the MLE protocol, knowledge of the noise distribution in the system is crucial for constructing the likelihood function. Care should be taken when estimating special states, such as entangled states, using MLE\cite{ schwemmer-prl-2015, silva-pra-2017}.

In this chapter, a method is presented to address the issue of unphysical experimentally reconstructed density matrices and process matrices. The method involves optimizing a least square objective function while taking into consideration the positivity condition as a nonlinear constraint and the unit trace condition as a linear constraint\cite{gaikwad-qip-2021}. This transforms the standard linear inversion-based tomography problem into a constrained convex optimization (CCO) problem, which ensures the positivity of the reconstructed state/process and produces valid quantum states and quantum processes. The CCO tomography method does not require any prior knowledge about the system or any additional ancillary qubits. The advantages of the CCO method over the standard method are demonstrated by characterizing unknown quantum states and quantum processes for two- and three-qubit quantum systems. Additionally, for QPT, the complete set of valid Kraus operators for a given quantum process can be efficiently computed via the unitary diagonalization of the experimentally reconstructed positive process matrix, which was not previously possible using other QPT techniques.



\section{QST via standard and CCO method}\label{ccotomo_sec2}
\subsection{Linear inversion based standard QST on NMR }\label{lsqst}

The general density matrix corresponding to n-qubit system has $4^n-1$ independent real parameters and can be reconstructed experimentally by determining all these $4^n-1$ numbers\cite{singh-pla-2016}. QST is usually done by performing set of repeated projective measurements on multiple copies of identically prepared states in different measurement bases. On an ensemble quantum computer like NMR, it is very hard to perform projective measurements, in that case QST is carried out by measuring expectation values of certain fixed set of basis operators which are further related to single quantum coherence (off-diagonal elements $\langle m \vert \rho \vert n \rangle$, such that total \textit{z}-angular momentum quantum numbers between $ \vert m \rangle $ and $ \vert n \rangle $ differ by one) terms in the underlying density matrix. This is generally done by rotating the state via different unitary transformations before performing the measurement
to collect information about different elements of the density matrix\cite{long-job-2001}.

%

\subsubsection*{2-qubit standard QST on NMR}
 For a system of two qubits, the general form of  density matrix $\rho$ is given as,

 \begin{equation}
   \rho=\begin{bmatrix} 
x_1 & x_2+i x_3 & x_4+i x_5 & x_6+i x_7 \\
x_2-i x_3 & x_8 & x_9+i x_{10} & x_{11}+i x_{12} \\
x_4-ix_5 & x_9-ix_{10} & x_{13} & x_{14}+ix_{15} \\
x_6-ix_7 & x_{11}-ix_{12} & x_{14}-ix_{15} & x_{16} 
\end{bmatrix}   
\label{ch1_eq68}
 \end{equation}
 where all $x_i$'s are real numbers. 
 \begin{figure}
  \centering
  \includegraphics[width=.65\linewidth]{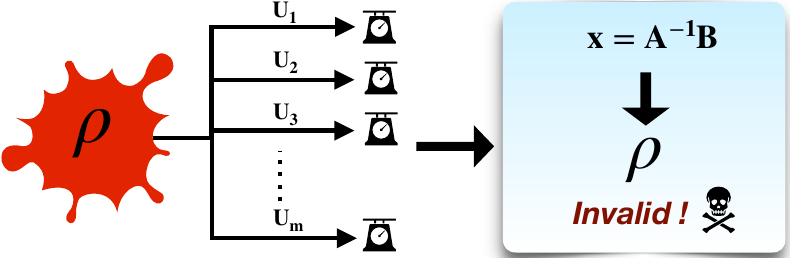}
  \caption{Schematic of standard QST based on linear inversion.}
  \label{toc_qst}
  \end{figure}
   The NMR signal corresponding to 1st spin is associated with $\rho_{24} = x_{11} + ix_{12}$ and $\rho_{13} = x_{4} + ix_{5}$ while signal corresponding to 2nd spin is associated with $\rho_{34} = x_{14} + ix_{15}$ and $\rho_{12} = x_{2} + ix_{3}$. These elements are generally referred to as readout positions of density matrix and are also interpreted as transition amplitudes between two energy eigenstates:
 \begin{align}
\rho_{24} &: \vert 01 \rangle   \longleftrightarrow \vert 11 \rangle \nonumber \\
\rho_{13} &: \vert 00 \rangle   \longleftrightarrow \vert 10 \rangle \nonumber \\
\rho_{34} &: \vert 10 \rangle   \longleftrightarrow \vert 11 \rangle \nonumber \\
\rho_{12} &: \vert 00 \rangle   \longleftrightarrow \vert 01 \rangle \nonumber 
\end{align}
 Note that these are the only four allowed transitions possible in two qubit system as, $\Delta m = \pm 1$ (see Fig.\ref{fig15}). All other transitions are forbidden by selection rule. In the case of 2-qubit system, the complete NMR signal from two spins consists of 4 spectrums (2 spectrum per nuclear spin). Each spectrum corresponds to given transition. The absorption part of given spectrum is associated with real part of corresponding density matrix element whereas dispersion part corresponds to imaginary part of density matrix element. For example, the intensity of absorption part of spectrum (integrated area) corresponding to element $\rho_{24}$ is proportional to $x_{11}$ likewise intensity of dispersion part of the same spectrum is proportional to $x_{12}$. So the readout from two spins will give the four density matrix elements or in more general sense it gives rise to 8 linear equations. To extract other elements of density matrix we apply certain unitary gates on density matrix and then perform measurement.  One of the possible complete set of unitary gates for 2-qubit QST is given as: $\lbrace I I, R_x^2, R_y^2, R_x^1 R_x^2 \rbrace$ where $I$ is single qubit identity matrix and $R_{x(y)}^i$ denotes $90^{\circ}$ rotation about $x(y)$ axis on $i$th qubit. Unitary gate, $II = I \otimes I$ is identity operation on both qubits and so on. The complete set of unitary gates followed by measurements is referred as \textit{tomographically complete set of measurements}. For a given tomographically complete set of measurements a system of linear equations can be formed. Each tomographic unitary gate yields 8 linear equations, so in total we get 32 equations. An extra equation is added as unit trace condition: $x_1 + x_8 + x_{13} + x_{16} = 1$. For this system of 33 linear equations, the coefficient matrix $\mathcal{A}$ is of dimension $33 \times 16$ and column matrix $\mathcal{B}$ is of dimension $33 \times 1$. The column matrix $x$ containing the variables $x_i$s is of dimension $16 \times 1$. Solving linear inversion problem given below will yield the column matrix $x$ and so do given unknown density matrix $\rho$ as,
 
 \begin{align}
\mathcal{A}x &=\mathcal{B} \nonumber \\
\mathcal{A}^{T} \mathcal{A}x &=\mathcal{A}^{T} \mathcal{B} \nonumber \\
x &=(\mathcal{A}^{T} \mathcal{A})^{-1} \mathcal{A}^{T} \mathcal{B}   \label{ch1_eq69}
\end{align}
 
The elements of column matrix $\mathcal{B}$ are the intensities of NMR spectrums obtained experimentally and matrix $(\mathcal{A}^{T} \mathcal{A})^{-1} \mathcal{A}^{T}$ can be computed analytically. Computing $x$  using Eq.\ref{ch1_eq69} and constructing $\rho$ is referred as \textit{standard QST} based on linear inversion method. 
 
\subsubsection*{3-qubit standard QST on NMR}

The $8 \times 8$ dimensional density matrix for 3-qubit system will have 64 real independent parameters (trace condition excluded), \ie column matrix $x$ will be of dimension $64 \times 1$. The NMR signal corresponding to given spin is associated with density matrix elements as follows:
\begin{align}
\text{1st spin or 1st qubit} &: \rho_{48}, \rho_{26}, \rho_{37}, \rho_{15} \nonumber \\
\text{2nd spin or 2nd qubit} &: \rho_{57}, \rho_{13}, \rho_{68}, \rho_{24} \nonumber \\
\text{3rd spin or 3rd qubit} &: \rho_{56}, \rho_{12}, \rho_{78}, \rho_{34} \nonumber 
\end{align}
These readout elements are associated with transition amplitudes between two energy eigenstates: 
{\footnotesize	
\begin{align}
\rho_{48} &: \vert 011 \rangle   \longleftrightarrow \vert 111 \rangle,\quad \rho_{26}: \vert 001 \rangle   \longleftrightarrow \vert 101 \rangle,\quad \rho_{37}: \vert 010 \rangle   \longleftrightarrow \vert 110 \rangle,\quad \rho_{15}: \vert 000 \rangle   \longleftrightarrow \vert 100 \rangle \nonumber  \\
\rho_{57} &: \vert 100 \rangle   \longleftrightarrow \vert 110 \rangle,\quad \rho_{13}: \vert 000 \rangle   \longleftrightarrow \vert 010 \rangle,\quad \rho_{68}: \vert 101 \rangle   \longleftrightarrow \vert 111 \rangle,\quad \rho_{24}: \vert 001 \rangle   \longleftrightarrow \vert 011 \rangle \nonumber  \\
\rho_{56} &: \vert 100 \rangle   \longleftrightarrow \vert 101 \rangle,\quad \rho_{12}: \vert 000 \rangle   \longleftrightarrow \vert 001 \rangle,\quad \rho_{78}: \vert 110 \rangle   \longleftrightarrow \vert 111 \rangle,\quad \rho_{34}: \vert 010 \rangle   \longleftrightarrow \vert 011 \rangle \nonumber 
\end{align}}
The complete NMR signal from three spins will have 12 spectrums (4 spectrums per spin). A given tomographic unitary operation will yield 24 linear equations. The complete set of tomographic unitary gates for 3-qubit QST is given as:
\[
\lbrace I I I, R_y^3, R_y^1, R_y^2 R_y^3, R_x^1 R_y^2 R_x^3, R_x^1 R_x^2 R_y^3, R_x^1 R_x^2 R_x^3 \rbrace \] where $III = I \otimes I \otimes I$ and so on. For given 7 tomographic operations and including unit trace condition we get total of 169 linear equations. So for 3-qubit case, $\mathcal{A}$ will be $169 \times 64$ dimensional matrix and $\mathcal{B}$ will be $64 \times 1$ dimensional column matrix. Using Eq.\ref{ch1_eq69}, column matrix $x$ can be determined and $\rho$ can be constructed. 

\subsubsection*{n-qubit standard QST on NMR}

For $n$-qubit system, one can efficiently construct tomographically complete optimal set of unitary rotations using integer programming method\cite{li-pra-2017}. For example, in case of 4-qubit system the cardinality of such optimal set turns out to be 15. That is, it is necessary to execute 15 tomographic rotation operations to reconstruct the complete density matrix of a 4-qubit system, which entails solving a total of 961 linear equations. In the case of 4-qubit QST, one possible set is given below:

\[
\begin{gathered}
\lbrace  IIII,  R_x^4, R_x^1 R_x^4, R_x^1 R_y^4, R_y^1 R_y^2, R_x^2 R_y^3 R_y^4, R_y^2 R_x^3 R_y^4, \\
R_y^2 R_y^3 R_y^4, R_x^1 R_x^2 R_x^3, R_x^1 R_y^2 R_y^3, R_y^1 R_x^3 R_x^4, R_y^1 R_y^2 R_y^3 \\
R_x^1 R_x^2 R_x^3 R_x^4, R_y^1 R_x^2 R_x^3 R_x^4, R_y^1 R_x^2 R_y^3 R_y^4 \rbrace
\end{gathered}
\]

Similarly, for 5-qubit system tomographically complete optimal set of unitary rotations of cardinality 33 is given below:

\[
\begin{gathered}
\lbrace IIIII, R_x^5, R_x^4 R_y^5, R_y^4 R_y^5, R_x^3 R_y^5, R_y^3 R_y^5, R_x^2 R_x^3 R_x^4, \\
R_y^2 R_y^3 R_y^4, R_x^1 R_y^3 R_x^4, R_x^1 R_x^2 R_y^4, R_x^1 R_y^2 R_y^5, R_x^1 R_y^2 R_x^3 \text {, } \\
R_y^1 R_x^3 R_y^4, R_y^1 R_y^3 R_y^5, R_y^1 R_x^2 R_y^5, R_y^1 R_x^2 R_y^3, R_y^1 R_y^2 R_y^5, \\
R_y^1 R_y^2 R_x^4, R_x^2 R_x^3 R_x^4 R_x^5, R_y^2 R_y^3 R_x^4 R_x^5, R_y^2 R_y^3 R_y^4 R_x^5, \\
R_x^1 R_y^3 R_x^4 R_x^5, R_x^1 R_x^2 R_y^4 R_x^5, R_x^1 R_y^2 R_x^3 R_x^5, R_y^1 R_x^3 R_y^4 R_x^5 \text {, } \\
R_y^1 R_x^2 R_y^3 R_x^5, R_y^1 R_y^2 R_x^4 R_x^5, R_x^1 R_x^2 R_x^3 R_x^4 R_y^5, \\
R_x^1 R_x^2 R_x^3 R_y^4 R_y^5, R_x^1 R_x^2 R_y^3 R_x^4 R_y^5, R_x^1 R_x^2 R_y^3 R_y^4 R_y^5, \\
R_y^1 R_y^2 R_x^3 R_x^4 R_y^5, R_y^1 R_y^2 R_x^3 R_y^4 R_y^5  \rbrace
\end{gathered}
\]   
The QST of 5-qubit density matrix involves solving 5281 number of linear equations.

One can see that, as the system size increases the complexity of QST task increases exponentially and to address this scalability issue one need to design more efficient algorithms. The schematic representation of standard QST  is given in Fig.{\ref{toc_qst}}.


\subsection{Fidelity measure for quantum states }\label{fidqst}

Fidelity serves as a fundamental concept in the realm of quantum information, offering a mathematical framework for quantifying the \textit{degree of similarity} between two quantum states. In practical applications, numerous scenarios arise where such comparisons prove beneficial. For instance, due to inherent imperfections and noise in experimental preparations of quantum states, there arises a genuine interest in determining the closeness between the actual state produced and the intended target state. This concern commonly arises in quantum communications and quantum computing, where the goal is to either generate or transmit precisely defined quantum states amidst the challenges posed by noise and other sources of error. Consequently, fidelity for mixed states can be regarded as a more practical measure, given its generic nature, as opposed to pure state fidelity, which represents an idealized scenario. Nevertheless, defining fidelity for mixed states lacks a unique, clear-cut definition, leading to the existence of several different approaches\cite{jozsa-jmo-1994}.


The Uhlmann-Jozsa (UJ) fidelity and normalized trace distance are two of the most widely used state fidelity measures in the literature. 
 The UJ-fidelity $\mathcal{F}_{UJ}(\rho,\sigma)$ between two density matrices $\rho$ and $\sigma$ is defined as, the maximal transition probability between the purification of a pair of density matrices $\rho$ and $\sigma$. The mathematical formula for UJ-fidelity is given as\cite{jozsa-jmo-1994, uhlman-rmp-1976},
 
\begin{equation}
\mathcal{F}_{UJ}(\rho, \sigma):=\max _{|\psi\rangle,|\varphi\rangle}|\langle\psi \mid \varphi\rangle|^2=(\operatorname{Tr} \sqrt{\sqrt{\rho} \sigma \sqrt{\rho}})^2
\label{ujfid}
\end{equation}
where $\vert \psi \rangle$ and $ \vert \phi \rangle $ are the purifications of $\rho $ and $\sigma$ respectively. Similarly, the normalized trace norm $\mathcal{F}_{tr}(\rho,\sigma)$ between two density matrices $\rho$ and $\sigma$ is defined as\cite{liang-rpp-2019},

\begin{equation}
{\mathcal F}_{tr}(\rho,\sigma)=
\frac{|{\rm Tr}(\rho \sigma^\dagger)|}
{\sqrt{{\rm Tr}(\rho^\dagger \rho)
{\rm Tr}(\sigma^\dagger \sigma )}}
\label{ch2_eq71}
\end{equation} 

Both the fidelity measures given in Eq.\ref{ujfid} and \ref{ch2_eq71} satisfy a list of fidelity axioms (the most basic requirements to be satisfied by any generalization of fidelity measure) proposed by Jozsa in order to be a suitable fidelity measure between pair of mixed states\cite{liang-rpp-2019}:
\begin{itemize}

\item[-] Normalization: $ 0 \leq \mathcal{F}(\rho, \sigma) \leq 1$

\item[-] Symmetric: $\mathcal{F}(\rho, \sigma) = \mathcal{F}(\sigma, \rho )$

\item[-] Unitary invariance: $\mathcal{F}(\rho, \sigma) = \mathcal{F}( U \rho U^{\dagger}, U \sigma U^{\dagger})$

\item[-] Orthogonality support: $\mathcal{F}(\rho, \sigma) = 0 $ iff $\rho \sigma = 0$

\item[-] $\mathcal{F}(\rho, \sigma) = \rm{Tr}( \rho, \sigma )$ if either $\rho$ or $\sigma$ is a pure state.
\end{itemize}

When employing the Uhlmann-Jozsa (U-J) fidelity measure, it is imperative that the density matrices $\rho$ and $\sigma$ be positive semi definite - a condition that may not hold true for experimentally reconstructed density matrices. To circumvent this concern, in this thesis, we have opted to utilize the normalized trace norm as a state and process fidelity measure, unless explicitly stated otherwise. One can also find the application of normalized trace norm as a valid fidelity measure in References \cite{liang-rpp-2019, weinstein-prl-2001, yang-pra-2017}.

	\begin{figure}
\begin{subfigure}{.5\textwidth}
  \centering
  \includegraphics[width=.85\linewidth]{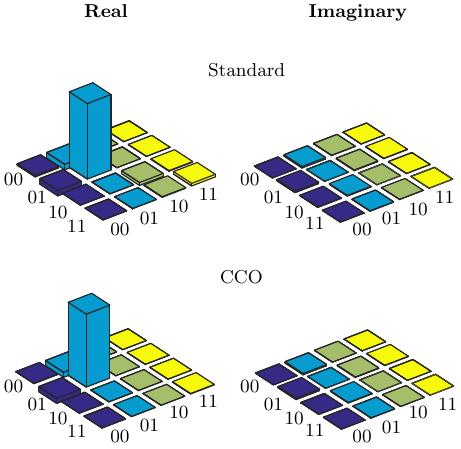}
  \caption{$\vert 01 \rangle$}
  \label{fig:01}
\end{subfigure}%
\begin{subfigure}{.5\textwidth}
  \centering
  \includegraphics[width=0.85\linewidth]{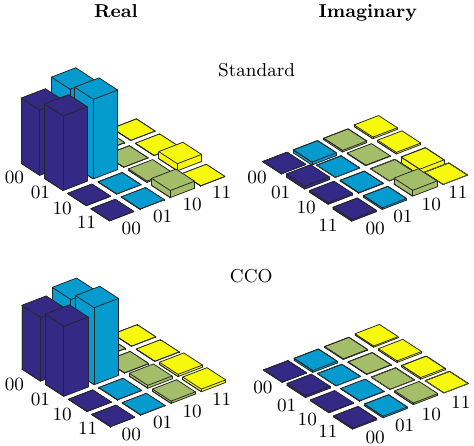}
\caption{$\vert 0+ \rangle$}
  \label{fig:0+}
\end{subfigure}
\begin{subfigure}{.5\textwidth}
  \centering
  \includegraphics[width=0.85\linewidth]{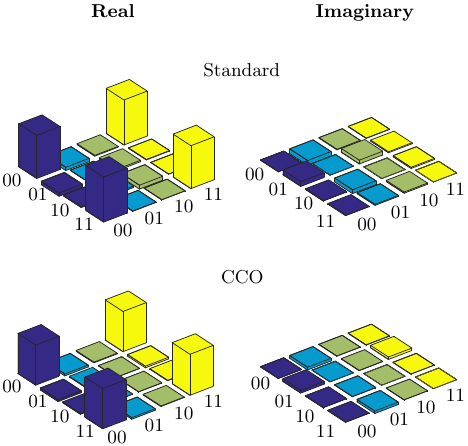}
\caption{$(\vert 00 \rangle + \vert 11 \rangle)/\sqrt{2}$}
  \label{fig:bell1}
\end{subfigure}
\begin{subfigure}{.5\textwidth}
  \centering
  \includegraphics[width=0.85\linewidth]{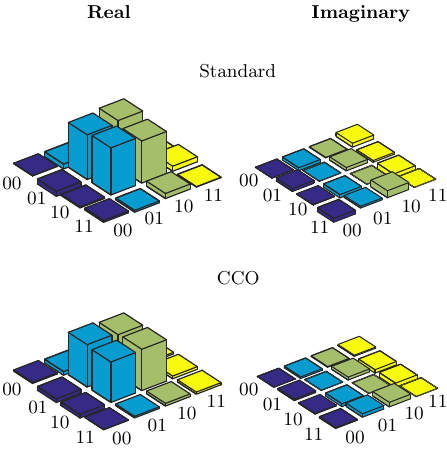}
\caption{$(\vert 01 \rangle + \vert 10 \rangle)/\sqrt{2}$}
  \label{fig:bell2}
\end{subfigure}
\caption{Real (left) and imaginary (right)
parts of experimental tomographs of two qubit states,
constructed via standard (1st row) and CCO (2nd row)
method are shown.}
\label{ccoqst}
\end{figure}
\subsection{Convex optimization based QST }

       The reconstructed $\rho$ by standard protocol is hermitian in nature and has trace 1 but there is no guarantee that it will be positive since the positivity constrain is not explicitly included in standard protocol. And in general it is very hard to incorporate positivity condition in any general state estimation protocol. To address this situation, the linear inversion-based standard QST problem has been reformulated into a constrained convex optimization (CCO) problem. The standard mathematical form of CCO problem in context of QST is given as follows,
       
\begin{subequations}
\begin{alignat}{2}
&\!\min_{x}        &\qquad& {\Vert \mathcal{A} x- \mathcal{B} \Vert}_2\label{ch1_eq70}\\
& s.t. &      & \rho \geq 0,\label{ch1_eq70:c1}\\
&                  &      & \mathrm{Tr}(\rho) = 1 \label{ch1_eq70:c2}.
\end{alignat}
\end{subequations}       
       

where $\Vert v \Vert_2$ denotes the $l_2$ norm of vector $v$. The least square objective function given in Eq.\ref{ch1_eq70} is defined in the paper\cite{gaikwad-qip-2021}. The CCO problem stated in Eq.\ref{ch1_eq70} is formulated using YALMIP MATLAB package\cite{lofberg-2004} which employed SeDuMi\cite{sturm-oms-1999} as solver. Upon solving the aforementioned CCO problem, the valid density matrix can be obtained. This matrix provides the least square fit to the experimental data and reveals the actual quantum state.

In order to assess the effectiveness of QST using the CCO method, various two- and three-qubit quantum states were experimentally prepared and subjected to tomography. Subsequently, their fidelities and corresponding eigenvalues were calculated using both the standard method and the CCO method. Some of the experimentally constructed tomographs for 2-qubit states via standard and CCO method are depicted in the Figs.\ref{ccoqst}. The fidelity between the theoretically expected ($\rho^{}_{\rm theo}$) 
and the experimentally reconstructed ($\rho^{}_{\rm expt}$) 
quantum state has been computed using Eq.\ref{ch2_eq71}. The computed fidelities for several different quantum states  were slightly improved for the CCO method as compared to the standard method. However,  the main advantage of the CCO method is that the experimentally reconstructed density matrix is always positive semi-definite and hence represents a valid quantum state.


\begin{table}[h!]
\centering
\label{qsteig}
\caption{Eigenvalues for the
two-qubit density matrix, obtained from experimentally 
reconstructed  density matrices via standard and CCO QST.}
\begin{tabular}{p{3cm}|p{5.5cm}|p{3.4cm}}
\hline
 Quantum state & Standard &  CCO \\
 \hline
 $\vert 00 \rangle$   & -0.0488, -0.0171, 0.0499, 1.0160    & 0, 0, 0.0225, 0.9775\\
  \hline
 $\vert 01 \rangle$   & -0.0429, -0.0222, 0.0364, 1.0287    & 0, 0, 0.0067, 0.9933 \\
  \hline
   $\vert 10 \rangle$   & -0.1486, -0.0911, 0.1915, 1.0482    & 0, 0, 0.0807, 0.9193 \\
 \hline
 $\vert 11 \rangle$   & -0.1457, -0.0955, 0.1933, 1.0480 & 0, 0, 0.0808, 0.9192\\
  \hline
  $(\vert 00 \rangle + \vert 11 \rangle)/\sqrt{2} $  & -0.0822, -0.0456, 0.0508, 1.0778   & 0, 0, 0.0105, 0.9895\\
   \hline
  $(\vert 01 \rangle - \vert 10 \rangle)/\sqrt{2} $  & -0.0950, -0.0370, 0.0624, 1.0696   & 0, 0, 0.0142, 0.9858\\
   \hline
  $(\vert 00 \rangle - \vert 11 \rangle)/\sqrt{2} $  & -0.1315, -0.0455, 0.1180, 1.0591   & 0, 0, 0.0592, 0.9408\\
   \hline
  $(\vert 01 \rangle + \vert 10 \rangle)/\sqrt{2} $  & -0.1175, -0.0278, 0.0910, 0.0543  & 0, 0, 0.0397, 0.9603\\
 \hline
  $(\vert 01 \rangle + \vert 11 \rangle)/\sqrt{2} $  & -0.0892, -0.0493, 0.1060, 1.0326   & 0, 0, 0.0255, 0.9745\\
  \hline
 $(\vert 00 \rangle + \vert 01 \rangle)/\sqrt{2} $   & -0.0587, -0.0166, 0.0683, 1.0070  & 0, 0, 0.0375, 0.9625 \\
 \hline
  $(\vert 10 \rangle + \vert 11 \rangle)/\sqrt{2} $  & -0.1017, -0.0730, 0.1209, 1.0538 & 0, 0, 0.0381, 0.9619 \\
  \hline
   $(\vert 00 \rangle + \vert 10 \rangle)/\sqrt{2} $   & -0.0884, -0.0469, 0.1093, 1.0260  & 0, 0, 0.0303, 0.9697 \\
   \hline
   $(\vert 01 \rangle + i\vert 11 \rangle)/\sqrt{2} $  & -0.0936, -0.0436, 0.0987, 1.0385   & 0, 0, 0.0267, 0.9733\\
   \hline
   $(\vert 10 \rangle + i\vert 11 \rangle)/\sqrt{2} $   & -0.1122, -0.0962, 0.1549, 1.0536    & 0, 0, 0.0544, 0.9456 \\
   \hline
   $(\vert 00 \rangle + i\vert 10 \rangle)/\sqrt{2} $   & -0.0898, -0.0420, 0.1028, 1.0290  & 0, 0, 0.0304, 0.9696\\
   \hline
  $(\vert 00 \rangle + i\vert 01 \rangle)/\sqrt{2} $   & -0.0862, -0.0379, 0.0837, 1.0405  & 0, 0, 0.0329, 0.9671 \\
 \hline
 $ (\vert 00 \rangle + \vert 01 \rangle + \vert 10 \rangle + \vert 11 \rangle)/2  $   & -0.0823, -0.0293, 0.0974, 1.0142  & 0, 0, 0.0293, 0.9707\\
 \hline
  $ (\vert 00 \rangle + i\vert 01 \rangle + \vert 10 \rangle + i\vert 11 \rangle)/2  $   & -0.0917, -0.0619, 0.1120, 1.0416  & 0, 0, 0.0298, 0.9702\\
 \hline
 $ (\vert 00 \rangle + \vert 01 \rangle + i\vert 10 \rangle +i \vert 11 \rangle)/2  $   & -0.0728, -0.0110, 0.0770, 1.0068   & 0, 0, 0.0298, 0.9702\\
 \hline
 $ (\vert 00 \rangle + i\vert 01 \rangle + i\vert 10 \rangle - \vert 11 \rangle)/2  $  & -0.0828, -0.0347, 0.0904, 1.0271  & 0, 0, 0.0234, 0.9766\\
\hline
\end{tabular}
\end{table}

\begin{table}[h!]
\centering
\caption{\label{qstfid}
Two-qubit quantum state fidelity obtained from standard and CCO method.}
\begin{tabular}{c c c}
\hline\hline
Quantum state &
Standard&
CCO ~~~\\
\hline
$\vert 00 \rangle $ & 0.9969 & 0.9993~~~\\
$\vert 01 \rangle $ & 0.9926 & 0.9948~~~\\ 
$\vert 10 \rangle $ & 0.9690 & 0.9947~~~\\
$\vert 11 \rangle $ & 0.9665 & 0.9928~~~\\
$ (\vert 00 \rangle + \vert 11 \rangle)/\sqrt{2}  $  & 0.9921 & 0.9968~~~\\ 
$ (\vert 01 \rangle - \vert 10 \rangle)/\sqrt{2}  $  & 0.9887 & 0.9942~~~\\ 
$ (\vert 00 \rangle - \vert 11 \rangle)/\sqrt{2}  $  & 0.9796 & 0.9911~~~\\ 
$ (\vert 01 \rangle + \vert 10 \rangle)/\sqrt{2}  $  & 0.9786 & 0.9956~~~\\ 
$ (\vert 01 \rangle + \vert 11 \rangle)/\sqrt{2}  $  & 0.9832 & 0.9946~~~\\ 
$ (\vert 00 \rangle + \vert 01 \rangle)/\sqrt{2}  $  & 0.9929 & 0.9966~~~\\     
$ (\vert 10 \rangle + \vert 11 \rangle)/\sqrt{2}  $  & 0.9834 & 0.9966~~~\\      
$ (\vert 00 \rangle + \vert 10 \rangle)/\sqrt{2}  $  & 0.9883 & 0.9982~~~\\   
$ (\vert 01 \rangle + i\vert 11 \rangle)/\sqrt{2}  $  & 0.9853 & 0.9952~~~\\  
$ (\vert 10 \rangle + i\vert 11 \rangle)/\sqrt{2}  $  & 0.9771 & 0.9953~~~\\
$ (\vert 00 \rangle + i\vert 10 \rangle)/\sqrt{2}  $  & 0.9895 & 0.9980~~~\\
$ (\vert 00 \rangle + i\vert 01 \rangle)/\sqrt{2}  $  & 0.9920 & 0.9984~~~\\  
$ (\vert 00 \rangle + \vert 01 \rangle + \vert 10 \rangle + \vert 11 \rangle)/2  $  & 0.9884 & 0.9964~~~\\ 
$ (\vert 00 \rangle + i\vert 01 \rangle + \vert 10 \rangle + i\vert 11 \rangle)/2  $  & 0.9856 & 0.9958~~~\\        
$ (\vert 00 \rangle + \vert 01 \rangle + i\vert 10 \rangle + i\vert 11 \rangle)/2  $  & 0.9919 & 0.9969~~~\\ 
$ (\vert 00 \rangle + i\vert 01 \rangle + i\vert 10 \rangle - \vert 11 \rangle)/2  $  & 0.9898 & 0.9966~~~\\
\hline
\end{tabular}
\end{table}

\section{QPT via standard and CCO method}\label{ccotomo_sec3}
\subsection{Standard QPT}

Quantum process tomography (QPT) can quantitatively characterize an unknown quantum process. Any quantum state $\rho$ undergoing a physically valid process can described by a completely positive (CP) map, and an unknown process $\Lambda$  can be described in the operator-sum representation of Kraus operators\cite{chuang-jmo-09}:
\begin{equation} 
\Lambda (\rho) = \sum_{i=1}^{d^2} A_i \rho A_i^\dagger 
\label{ch1_eq72}
\end{equation}
where $A_i$'s are the Kraus operators satisfying $\sum_{i} A_i A_i^\dagger =
I$. The Kraus operators can be expanded using a fixed complete set of basis operators $\lbrace E_i \rbrace $ as $A_i = \sum_m a_{im} E_m$, then Eq.\ref{ch1_eq72} can be rewritten as,
\begin{equation} 
\Lambda (\rho) = \sum_{m,n=1}^{d^2} \chi_{mn} E_m \rho E_n^\dagger
\label{ch1_eq73} 
\end{equation} 
where $\chi_{mn} = \sum_{i}a_{im}a_{in}^*$ is called the
process matrix in $\lbrace E_i \rbrace$ basis operators and is a positive Hermitian matrix satisfying the
trace preserving constraint $\sum_{m,n} \chi_{mn} E_{n}^{\dagger}
E_m = I$. The $n$-qubit quantum process will be uniquely characterized by $4^n \times 4^n$ dimensional $\chi$ matrix. 

The idea behind the standard QPT is to estimate the complete $\chi$ matrix by preparing the system in different quantum states then letting it to evolve under given quantum process and then measuring the set of observables. The full data set for QPT can be acquired using tomographically complete sets of input states $\lbrace \rho_1, \rho_2,....,\rho_k \rbrace $ and measurement operators $\lbrace M_1, M_2,..., M_l \rbrace$. For given $\rho_i$, the measurement of all $\lbrace M_i \rbrace$ is equivalent to performing QST of given $\rho_i$. Suppose the system is prepared in the state $\rho_i$, evolved  under a given quantum process $\Lambda$ and then measurement is performed correspond to measurement operator $M_j$ then using Eq.\ref{ch1_eq73} we will have 
 \begin{equation}
 \lambda^i_j=\text{Tr}(M_j\Lambda(\rho_i)) = \sum_{m,n} \chi_{mn} \text{Tr}(M_j E_m \rho_i E_n^{\dagger})
 \label{ch1_eq74}
 \end{equation}
 For all input states $\lbrace \rho_i \rbrace$ and measurement operators $\lbrace M_j \rbrace$, the  Eq.\ref{ch1_eq74} can be rewritten in the compact form as\cite{chuang-jmo-09},
\begin{equation}
\overrightarrow{\lambda}= \beta \overrightarrow{\chi}
\label{ch1_eq75}
\end{equation} 
 where $\overrightarrow{\lambda}$ and $\overrightarrow{\chi}$ are vectorized form of $\lambda^i_j$ and $\chi_{mn}$ respectively. The matrix $\beta$ is coefficient matrix with entries given as $\beta_{ji,mn} = \text{Tr}(M_j E_m \rho_i E_n^{\dagger})$. For full data set Eq.\ref{ch1_eq75} allows us to estimates complete $\chi$ by simple linear inversion technique. This is known as standard QPT. In practice, using standard QPT, the experimentally constructed $\chi$ matrix may not be positive semi definite due to experimental uncertainties.
 
\begin{figure}[t]
  \centering
  \includegraphics[width=.75\linewidth]{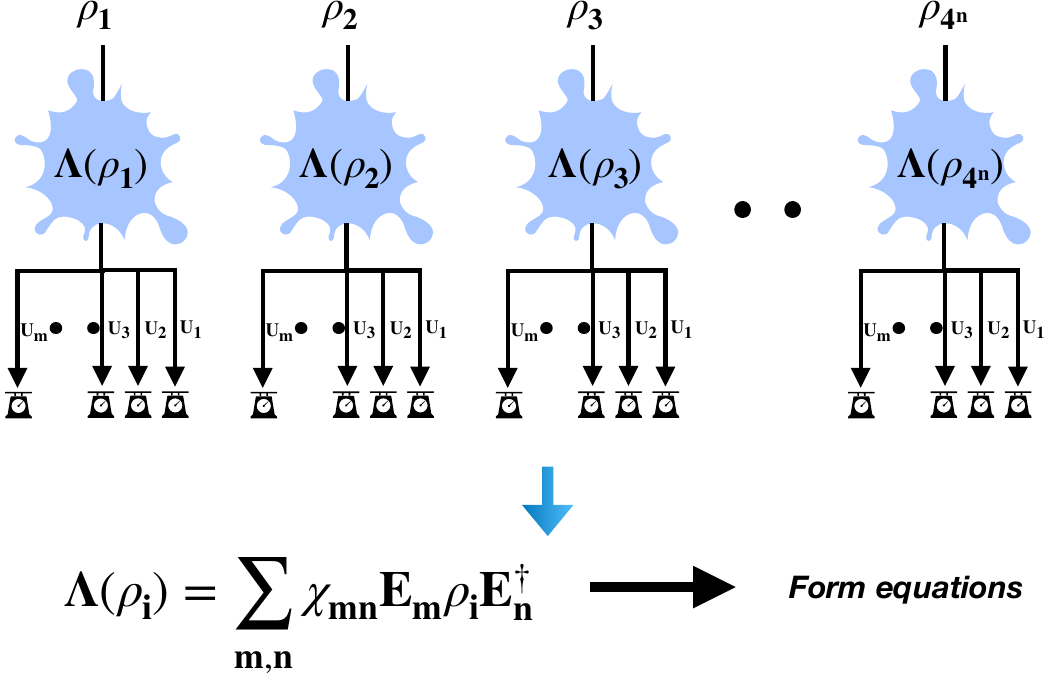}
  \caption{Schematic of standard QPT based on linear inversion.}
  \label{toc_qpt}
  \end{figure}

 For $n$-qubit case tomographically complete set of input states is given as:
 \[ \lbrace \vert 0 \rangle, \vert 1 \rangle, \vert + \rangle, \vert - \rangle \rbrace ^{\otimes n} \]
  where $\vert + \rangle = (\vert 0 \rangle + \vert 1 \rangle)/\sqrt{2} $ and $\vert - \rangle = (\vert 0 \rangle + i\vert 1 \rangle)/\sqrt{2} $. In NMR, the tomographic measurements are carried out by applying tomographically complete set of unitary rotations followed by signal acquaisition. Here the signal is recorded as the free induction decay (FID) (time domain) then by applying Fourier Transform we get spectrum (frequency domain) which effectly measures the net magnetization in transverse plane (x-y plane). For the case of two- and three-qubit system tomographically complete set of unitary rotations is given as:  \[ \lbrace I I, R_x^2, R_y^2, R_x^1 R_x^2 \rbrace \quad \text{and} \quad \lbrace I I I, R_y^3, R_y^1, R_y^2 R_y^3, R_x^1 R_y^2 R_x^3, R_x^1 R_x^2 R_y^3, R_x^1 R_x^2 R_x^3 \rbrace \] respectively. The general schematic of standard QPT based on linear inversion is depicted in the Fig.\ref{toc_qpt}.
\subsection{Convex optimization based QPT }

The $\chi$ matrix obtained from standard QPT protocols
is Hermitian and has unit trace, but there is no assurance that it will
be positive. 
Standard QPT methods could
lead to an unphysical density matrix which implies that the inversion
was not able to optimally fit the experimental data, and hence 
more constraints should be used
while performing the reconstruction of the
$\chi$ matrix.  
One viable alternative is the CCO method of 
reconstruction, which always leads to a valid process matrix. 
The mathematical formulation of the CCO method for QPT is
given by\cite{gaikwad-qip-2021}:
\begin{subequations}
\begin{alignat}{2}
&\!\min_{\chi}        &\qquad& {\Vert \Phi \overrightarrow{\chi}- \overrightarrow{\lambda} \Vert}_2\label{ch1_eq76}\\
& s.t. &      & \chi \geq 0,\label{ch1_eq76:c1}\\
&                  &      & \sum_{m,n} \chi_{mn} E_{n}^{\dagger}
E_m = I \label{ch1_eq76:c2}
\end{alignat}
\end{subequations}       
The CCO problem given in Eq.~\ref{ch1_eq76} is formulated using YALMIP MATLAB package which employed SeDuMi as solver. Once the $\chi$ matrix is determined, it can be unitarily diagonalized and the Kraus operators can be
determined from this diagonalized $\chi$ matrix. However, the reconstruction of the full set of Kraus operators 
only works if the experimentally determined $\chi$
matrix is positive semidefinite i.e. if the $\chi \ge 0$. As a demonstration, the CCO based QPT has been applied to characterize two-qubit unitary quantum gates: Identity, CNOT and C-$R_{x}^{\pi}$ and natural decoherence process (relaxation phenomenon) occurred in spin ensemble system. Furthermore, a complete set of Kraus operators was also computed, which was not achievable using the standard method.

\begin{figure}
\begin{subfigure}{.63\textwidth}
  \includegraphics[width=1.3\linewidth]{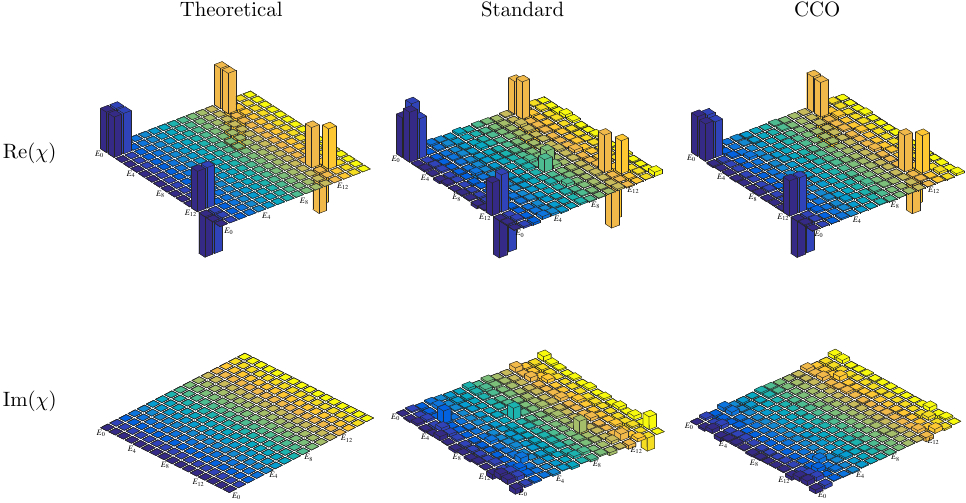}
  \caption{CNOT gate}
  \label{fig:cnot}
\end{subfigure}%
\newline
\par\bigskip
\par\bigskip
\begin{subfigure}{.65\textwidth}
  \includegraphics[width=1.3\linewidth]{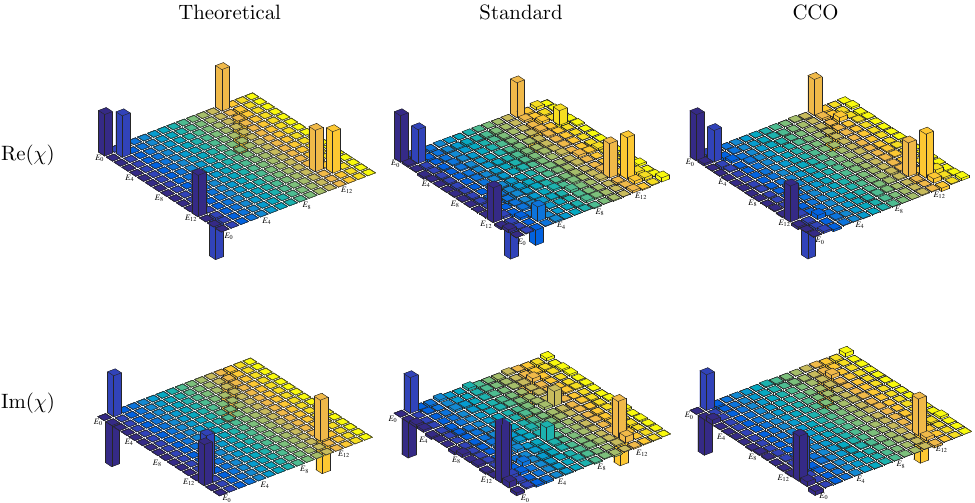}
\caption{C-$R_x^{\pi}$ gate}
  \label{fig:cx}
\end{subfigure}
\caption{The tomographs in the row denote the real and imaginary parts of the process matrix for 2-qubit quantum gates. The tomographs in the first column represents the theoretically constructed process matrix while
those in second and third represent the experimentally measured linear inversion based and CCO method based process matrices
respectively.}
\label{ccoqpt}
\end{figure}

For illustration, theoretically constructed and experimentally tomographed (via standard as well as CCO method) $\chi$ matrices for
 CNOT and control-$R^{\pi}_{x}$ operators are
depicted in Fig.\ref{fig:cnot} and \ref{fig:cx},
respectively. Upper panel in Fig.\ref{fig:cnot} (\ref{fig:cnot}) denotes real part of $\chi$ matrix and lower panel denotes imaginary
part of $\chi$ matrix for CNOT (control-$R^{\pi}_{x}$) gate. The
tomographs in the first column represents the theoretically
constructed $\chi$ matrix while those in second and third
columns represent the experimentally measured linear
inversion based and CCO method based $\chi$ matrices
respectively. The
eigenvalues of experimentally constructed $\chi$ matrices
via standard method and CCO method for all three operations
are calculated in the Table.\ref{qpteig}. 

\begin{table}[h!]
\centering
\caption{\label{qpteig} Eigenvalues
obtained from experimental $\chi$ matrices constructed via
standard and CCO QPT.}
\begin{tabular}{ p{1.6cm}|p{6.1cm}|p{4.3cm} } \hline Quantum
operation& Standard QPT &  CCO QPT\\ \hline CNOT   & 1.0117,
0.1331, -0.1421, 0.1247, 0.0934, 0.0860, 0.0716, 0.0541,
0.0668, -0.1135, -0.0935, -0.0838, -0.0315, -0.0672,
-0.0598, -0.0503    & 0.0077, 0.0201, 0.0245, 0.0438,
0.9038, 0, 0, 0, 0, 0, 0, 0, 0, 0, 0, 0\\ \hline
C-$R^{\pi}_{x}$   & 0.9972, 0.1435, -0.1305, 0.1198, 0.1061,
0.0971, 0.0837, 0.0746, 0.0553, -0.01187, -0.1044, -0.0838,
-0.0415, -0.0767, -0.0578, -0.0639  & 0.0077, 0.0166,
0.0315, 0.0397, 0.9045, 0, 0, 0, 0, 0, 0, 0, 0, 0, 0, 0\\
\hline Identity   & 1.0087, 0.1205, -0.0547, 0.0581, 0.0355,
-0.0441, -0.0122, -0.0385, -0.0338, -0.0271, -0.0213,
-0.0151, 0.0019, 0.0006, -0.0067  & 0.0166, 0.0357, 0.9477,
0, 0, 0, 0, 0, 0, 0, 0, 0, 0, 0, 0, 0 \\ \hline
\end{tabular}
\end{table}

One can see that experimentally estimated $\chi$ matrix via
linear inversion based standard method has some negative
eigenvalues which make it unphysical and does not
correspond to any valid quantum operation. On the other hand
all the eigenvalues of experimentally estimated $\chi$
matrix via CCO method are the positive which make them physical. The fidelity of the experimentally
constructed $\chi_{\rm{expt}}$ matrix with reference to
the 6theoretically expected $\chi_{\rm{theo}}$ matrix was
calculated using the measure\cite{zhang-prl-2014}:
\begin{equation} {\mathcal F}(\chi^{}_{\rm
expt},\chi^{}_{\rm theo})= \frac{|{\rm Tr}[\chi^{}_{\rm
expt}\chi_{\rm theo}^\dagger]|} {\sqrt{{\rm Tr}[\chi_{\rm
expt}^\dagger\chi^{}_{\rm expt}] {\rm Tr}[\chi_{\rm
theo}^\dagger\chi^{}_{\rm theo}]}} \label{ch1_eq77}
\end{equation} Fidelity measure $\mathcal F$ is normalized
in the sense that as $\chi_{\rm{expt}} \rightarrow
\chi_{\rm{theo}} $ \ie ~experimentally constructed $\chi$
matrix approaches theoretically expected $\chi$ matrix leads
to $\mathcal F$ $\rightarrow$ 1.  The calculated fidelities
via standard and CCO method are given in Table.\ref{qptfid}:

\begin{table}[h!]
\centering 
\caption{\label{qptfid} Two-qubit gate
fidelities obtained via standard QPT and CCO QPT.}
\begin{tabular}{ p{3.6cm} p{2.1cm} p{2.5cm}}
\hline \hline
Quantum gate & Standard QPT & CCO\\
\hline Identity & 0.9809 & 0.9959~~~\\ 
CNOT & 0.9313 & 0.9817~~~\\
Control-$R^{\pi}_{x}$  & 0.9269 & 0.9831~~~\\
\hline \end{tabular}
\end{table}

In all three cases, the fidelity $\mathcal{F}$ obtained via CCO method is greater than 0.98, which shows the efficacy of CCO based QPT protocol. 

Moreover, by utilizing the constructed positive $\chi$ matrix based on CCO, one can effectively calculate the complete set of legitimate Kraus operators using the following approach:\cite{gaikwad-qip-2021}: 
\begin{itemize}
\item Step 1: Find the unitary diagonalization of $\chi$ matrix. i.e $\chi_{exp} = VDV^{\dagger}$, where columns of $V$ are eigenvectors of $\chi_{exp}$ and $D$ is diagonal matrix with eigenvalues of $\chi_{exp}$ matrix as diagonal elements ($d_i$).
\item Step 2: Compute Kraus operators ($A_i$) using above decomposition,
\begin{equation}
A_i = \sqrt{d_i} \sum_p V_{pi}E_p
\label{e13}
\end{equation}
where $E_p$ are basis operators.
\end{itemize}
From the Table.\ref{qpteig} one can see that for CNOT and control-$R^{\pi}_{x}$ operations there are a total of 5 non-zero eigenvalues of $\chi$ matrix (constructed via CCO method) which will produce 5 Kraus operators and there will be 3 Kraus operators corresponding to identity operation. The complete set of Kraus operators for all three gates are given below.
\begin{itemize}
\item Kraus operators corresponding to Identity gate
{\footnotesize
\[
A_1= {\begin{bmatrix}
    -0.0308 + 0.0457i  & -0.0028 - 0.0077i &  0.0626 + 0.1056i &  0.0022 + 0.0078i \\
   0.0070 + 0.0095i & -0.0393 + 0.0633i &  -0.0055 - 0.0060i &  0.0550 + 0.1068i \\
   0.0755 - 0.0678i &  0.0203 + 0.0042i &  0.0279 - 0.0575i &  0.0052 + 0.0001i \\
   0.0395 + 0.0187i &  0.0850 + 0.0079i &  0.0153 + 0.0005i &  0.0451 - 0.0399i
  \end{bmatrix}}
\]
\[
A_2={ \begin{bmatrix}
    0.0571 + 0.0943i &  -0.0133 + 0.0201i & -0.1932 - 0.0604i & -0.0069 - 0.0085i \\
  -0.0071 + 0.0271i & -0.0082 + 0.0968i &  0.0173 - 0.0019i & -0.1724 - 0.0574i \\
   0.0189 - 0.1154i &  0.0269 - 0.0103i & -0.0281 - 0.1005i &  0.0091 - 0.0038i \\
   0.0481 - 0.0087i &  0.0723 - 0.0485i & -0.0040 + 0.0067i & -0.0220 - 0.0980i
  \end{bmatrix}}
\]
\[
A_3={ \begin{bmatrix}
    -0.0442 - 0.9758i & -0.0014 + 0.0418i &  0.0103 + 0.0259i & -0.0100 - 0.0007i \\
   0.0005 - 0.0271i & -0.0550 - 0.9813i &  0.0095 + 0.0029i &  0.0129 + 0.0272i \\
   0.0101 - 0.0239i &  0.0088 - 0.0002i & -0.0190 - 0.9617i &  0.0233 + 0.0404i \\
  -0.0096 + 0.0017i &  0.0101 - 0.0205i &  0.0242 - 0.0412i &  0.0021 - 0.9671i \\
  \end{bmatrix}}
\]
\item Kraus operators corresponding to CNOT gate
\[
A_1={\begin{bmatrix}
    0.0344 - 0.0042i &  0.0389 + 0.0130i & -0.0068 + 0.0035i & -0.0691 - 0.0003i \\
  -0.0039 - 0.0038i & -0.0208 - 0.0054i & -0.0494 + 0.0543i &  0.0125 + 0.0320i \\
   0.0548 + 0.0194i & -0.0023 - 0.0251i & -0.0714 + 0.0137i & -0.0094 - 0.0117i \\
   0.0208 + 0.0094i &  0.0654 + 0.0264i & -0.0162 + 0.0148i &  0.0221 + 0.0124i 
  \end{bmatrix}}
\]
\[
A_2={\begin{bmatrix}
   0.0124 + 0.0245i &  0.0065 - 0.0008i & -0.0508 - 0.1283i & -0.0079 + 0.0205i \\
   0.0727 + 0.0709i & -0.0132 - 0.0155i & -0.0941 + 0.0168i &  0.0494 - 0.0326i \\
   0.0323 - 0.0020i & -0.0552 + 0.0537i &  0.1017 + 0.0139i & -0.0577 - 0.0248i \\
   0.0400 + 0.0811i & -0.0112 + 0.0281i &  0.0603 + 0.0204i & -0.0381 - 0.0261i
  \end{bmatrix}}
\]
\[
A_3={\begin{bmatrix}
   0.0907 - 0.0140i & -0.0599 + 0.0491i &  0.0581 + 0.0467i &  0.0292 + 0.0058i \\
  -0.0567 + 0.0142i & -0.0978 - 0.0171i &  0.0109 + 0.0093i & -0.0310 - 0.1036i \\
   0.0267 + 0.0135i & -0.0221 + 0.0546i & -0.0700 + 0.0595i & -0.0752 - 0.0404i \\
  -0.0269 - 0.0463i & -0.0340 + 0.0427i &  0.0765 + 0.0205i & -0.1294 + 0.0564i 
  \end{bmatrix}}
\]
\[
A_4={ \begin{bmatrix}
   0.1786 + 0.0344i &  0.1327 - 0.0629i &  0.0228 + 0.0866i & -0.0018 - 0.0201i \\
   0.0052 - 0.0290i & -0.1264 - 0.0353i &  0.0174 - 0.0797i & -0.0397 + 0.0932i \\
  -0.0346 + 0.0199i &  0.0383 - 0.0361i &  0.1008 - 0.0337i & -0.1466 - 0.0222i \\
  -0.0024 - 0.0169i & -0.0415 - 0.0293i &  0.1058 - 0.0050i &  0.1214 - 0.1034i
  \end{bmatrix}}
\]
\[
A_5={\begin{bmatrix}
   0.0706 + 0.9517i & -0.0369 + 0.0847i &  0.0250 + 0.0166i & -0.0245 - 0.0130i \\
  -0.0139 - 0.1052i & -0.1412 + 0.9412i & -0.0442 - 0.0280i & -0.0077 - 0.0040i \\
  -0.0187 + 0.0169i & -0.0218 - 0.0073i & -0.0414 + 0.0215i & -0.0410 + 0.9380i \\
  -0.0065 - 0.0224i & -0.0537 + 0.0269i &  0.0297 + 0.9390i & -0.0516 + 0.0110i
  \end{bmatrix}}
\]

\item Kraus operators corresponding to control-$R^{\pi}_{x}$ gate
\[
A_1={ \begin{bmatrix}
   0.0012 + 0.0210i &  0.0165 + 0.0134i & -0.0744 + 0.0001i & -0.0121 - 0.0364i \\
   0.0359 + 0.0089i &  0.0251 - 0.0080i & -0.0654 - 0.0831i & -0.0097 - 0.0251i \\
  -0.0234 + 0.0153i & -0.0354 - 0.0211i & -0.0461 + 0.0131i &  0.0004 + 0.0068i \\
   0.0365 + 0.0369i & -0.0233 + 0.0167i & -0.0068 + 0.0143i &  0.0125 - 0.0159i
  \end{bmatrix}}
\]
\[
A_2={ \begin{bmatrix}
   0.0153 + 0.0286i &  0.0001 - 0.1142i & -0.0595 - 0.0245i &  0.0153 - 0.0085i \\
  -0.0141 + 0.0290i & -0.0039 - 0.0323i & -0.0097 + 0.0340i &  0.1150 + 0.0166i \\
   0.0058 + 0.0004i & -0.0092 - 0.0963i &  0.0556 + 0.0204i & -0.0167 - 0.0196i \\
  -0.0192 + 0.0136i & -0.0308 + 0.0298i &  0.0407 + 0.0222i & -0.1054 - 0.0405i
  \end{bmatrix}}
\]
\[
 A_3={ \begin{bmatrix}
   0.1537 + 0.0345i &  0.0717 + 0.0257i &  0.0169 - 0.0805i & -0.0006 - 0.0094i \\
  -0.0074 + 0.0235i & -0.1425 - 0.0207i & -0.0232 + 0.0145i &  0.0001 - 0.0276i \\
   0.0178 + 0.0428i & -0.0375 - 0.0168i &  0.0243 - 0.0328i & -0.0154 + 0.1688i \\
  -0.0239 - 0.0132i & -0.0232 + 0.0146i &  0.0017 - 0.1398i &  0.0512 - 0.0880i
  \end{bmatrix}}
\]
\[
 A_4={ \begin{bmatrix}
   0.0686 - 0.0160i & -0.1101 - 0.0036i & -0.0419 - 0.0764i & -0.0257 - 0.0232i \\
  -0.1221 - 0.0570i & -0.0450 + 0.0133i & -0.0657 + 0.0230i &  0.0491 + 0.0496i \\
   0.0651 + 0.0241i &  0.0643 - 0.0430i & -0.1377 + 0.1614i & -0.0061 + 0.0345i \\
  -0.0239 + 0.0021i & -0.0537 - 0.0392i & -0.0594 - 0.0210i &  0.0213 + 0.1907i
  \end{bmatrix}}
\]
\[
 A_5={\begin{bmatrix}
   0.1841 + 0.9399i & -0.0704 + 0.1026i &  0.0143 + 0.0058i &  0.0012 + 0.0004i \\
  -0.0949 - 0.0906i &  0.0979 + 0.9445i &  0.0084 + 0.0134i & -0.0049 + 0.0210i \\
   0.0077 - 0.0108i & -0.0075 + 0.0042i & -0.0249 + 0.0765i &  0.9336 - 0.0790i \\
  -0.0076 - 0.0216i & -0.0092 - 0.0086i &  0.9304 - 0.0835i &  0.0338 + 0.0811i
  \end{bmatrix}}
\]
}
\end{itemize}
 Given that the computation of Kraus operators necessitates the positive eigenvalues of the $\chi$ matrix according to the procedure, an advantage can be gained through the utilization of the CCO method over standard and other QPT methods. Additionally, the fidelity between the ideal and predicted gate outputs (derived from the experimentally constructed $\chi$ matrix) has been calculated for 16 arbitrary input states across all three quantum operations using Eq.\ref{ch2_eq71}, as presented in Table \ref{qptfid2}. 

\begin{table}[h!]
\centering
\caption{\label{qptfid2}
Difference in fidelity between ideal and 
predicted gate output (from experimentally constructed 
$\chi$ matrix) for all 16 input states, obtained via 
standard QPT and CCO QPT.}
\begin{tabular}{ c|c|c|c|c|c }
\hline
\multicolumn{2}{c|}{CNOT} & \multicolumn{2}{c|}{Control-$R^{\pi}_{x}$} & 
\multicolumn{2}{c}{Identity} \\
Standard & CCO & Standard & CCO & Standard & CCO \\ 
\hline
0.9803 & 0.9862 & 0.9703 & 0.9799 & 0.9969 & 0.9983 \\
0.9831 & 0.9870 & 0.9792 & 0.9825 & 0.9926 & 0.9981 \\
0.9687 & 0.9722 & 0.9808 & 0.9839 & 0.9929 & 0.9976 \\
0.9709 & 0.9768 & 0.9835 & 0.9884 & 0.9920 & 0.9997 \\
0.9898 & 0.9907 & 0.9887 & 0.9905 & 0.9690 & 0.9953 \\
0.9868 & 0.9920 & 0.9834 & 0.9904 & 0.9665 & 0.9959 \\
0.9921 & 0.9960 & 0.9949 & 0.9961 & 0.9834 & 0.9958 \\
0.9817 & 0.9904 & 0.9806 & 0.9876 & 0.9771 & 0.9974 \\
0.9560 & 0.9813 & 0.8834 & 0.9872 & 0.9883 & 0.9967 \\
0.9863 & 0.9958 & 0.9143 & 0.9892 & 0.9832 & 0.9960 \\
0.9592 & 0.9812 & 0.8801 & 0.9912 & 0.9884 & 0.9955 \\
0.9323 & 0.9711 & 0.8743 & 0.9867 & 0.9856 & 0.9983 \\
0.9003 & 0.9872 & 0.9726 & 0.9835 & 0.9895 & 0.9981 \\
0.9269 & 0.9939 & 0.9854 & 0.9862 & 0.9853 & 0.9972 \\
0.8984 & 0.9839 & 0.9759 & 0.9896 & 0.9919 & 0.9965 \\
0.8779 & 0.9807 & 0.9760 & 0.9867 & 0.9898 & 0.9986 \\
\hline
\end{tabular}
\end{table} 
 Also the output state (density matrix) obtained from experimentally constructed $\chi$ matrix via CCO method satisfies all three properties of a valid density matrix. So CCO based QPT method allows us to accurately predict the action of the quantum process on arbitrary input quantum states. Using Table\ref{qptfid2} one can easily calculate the \textit{average gate fidelity} which is defined as state fidelity between actual and ideal gate outputs, averaged over all input states. In all three cases the average gate fidelity calculated using CCO method is higher with respect to standard method. Furthermore, an additional measure, referred to as the \textit{average state deviation} $\Delta_{avg}$, has been employed to evaluate the effectiveness of the CCO method in comparison to the standard method. This metric involves a element-wise comparison between two density matrices, offering a valuable means to quantify the closeness between the two matrices. The state deviation $\Delta$ is defined as\cite{huang-sb-2020},
\begin{equation}
\Delta = \sum_{ij}\frac{(\rm{abs}(\overline{\rho}_{ij}-{\rho_{ij}}))^2}{d^2}
\label{delta}
\end{equation} 
where $\rm{abs}(z)$ denotes the absolute value of complex number $z$ and $\lbrace \overline{\rho}_{ij} \rbrace$ are elements of the predicted density matrix using experimentally constructed $\chi$ matrix while $\lbrace \rho_{ij} \rbrace $ are elements of ideal gate output.
$\Delta_{avg}$ can be calculated by averaging over all the input states. The smaller the value of $\Delta_{avg}$ better is the performance of QPT protocol. The state deviation $\Delta$ for each input states for all three quantum gates are listed in Table.\ref{std}

\begin{table}[h!]
\centering
\caption{\label{std}
Calculated state deviation for each input state 
(using experimentally reconstructed $\chi$ matrix) via
standard QPT and CCO QPT.}
\begin{tabular}{ c|c|c|c|c|c }
\hline
\multicolumn{2}{c|}{CNOT} & 
\multicolumn{2}{c|}{Control-$R^{\pi}_{x}$} & \multicolumn{2}{c}{Identity} \\
 Standard & CCO & Standard & CCO & Standard & CCO \\ 
 \hline
0.0025 &  0.0017 & 0.0037 & 0.0025 & 4.1733e-04 & 2.5963e-04 \\
0.0021 & 0.0017 & 0.0026 & 0.0023 & 0.0010 & 2.5509e-04 \\
0.0039 & 0.0036 & 0.0024 & 0.0021 & 9.0279e-04 & 3.6542e-04  \\
0.0038 & 0.0031 & 0.0021 & 0.0016 & 0.0012 & 5.9867e-05 \\
0.0017 & 0.0015 & 0.0017 & 0.0016 & 0.0046 & 7.4988e-04 \\
0.0019 & 0.0013 & 0.0014 & 0.0015 & 0.0049 & 6.2527e-04 \\
0.0010 & 0.0006 & 0.0008 & 0.0008 & 0.0025 & 6.5192e-04 \\
0.0029 & 0.0018 & 0.0027 & 0.0020 & 0.0034 & 4.6806e-04 \\
0.0058 & 0.0024 & 0.0368 & 0.0017 & 0.0016 & 5.0433e-04 \\
0.0017 & 0.0019 & 0.0223 & 0.0014 & 0.0023 & 5.6461e-04 \\
0.0051 & 0.0025 & 0.0354 & 0.0012 & 0.0015 & 6.7930e-04 \\
0.0083 & 0.0039 & 0.0373 & 0.0019 & 0.0021 & 2.8602e-04 \\
0.0317 & 0.0016 & 0.0034 & 0.0022 & 0.0015 & 3.1576e-04 \\
0.0198 & 0.0008 & 0.0019 & 0.0019 & 0.0021 & 4.0010e-04 \\
0.0310 & 0.0020 & 0.0030 & 0.0016 & 0.0010 & 5.3135e-04 \\
0.0320 & 0.0026 & 0.0033 & 0.0021 & 0.0014 & 2.2969e-04 \\
\hline
\end{tabular}
\end{table} 
For all three quantum gates, one can see that from Table.\ref{std} the state deviation of each input state is smaller in case of CCO method w.r.t. the standard method. Furthermore the average state deviation $\Delta_{avg}$ is given in the Table.\ref{avg}. From Table\ref{avg} one can see that the performance of CCO method is better than the standard method as $\Delta_{avg}^{std} > \Delta_{avg}^{cco} $ for all three quantum gates.
\begin{table}[h!]
\centering
\caption{\label{avg}
Average state deviation computed from standard ($\Delta_{avg}^{std}$) 
and from CCO ($\Delta_{avg}^{cco}$) methods.}
\begin{tabular}{ p{3.6cm} p{2.1cm} p{2.1cm}}
\hline
Quantum process &
$\Delta_{avg}^{std}$&
$\Delta_{avg}^{cco}$ ~~~\\
\hline
Identity & 0.0020 & 4.3414e-04~~~\\
CNOT & 0.0097 & 0.0021~~~\\ 
control-$R^{\pi}_{x}$  & 0.0101 & 0.0018~~~\\   
\hline
\end{tabular}
\end{table}

\subsection{Markovian Quantum Process Tomography}
	\begin{figure}
\begin{subfigure}{.5\textwidth}
  \centering
  \includegraphics[width=.85\linewidth]{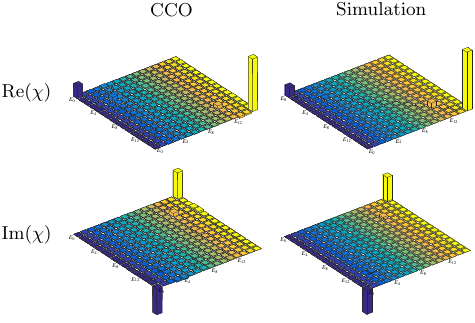}
  \caption{$t=0.05s$}
  \label{fig:d1}
\end{subfigure}%
\begin{subfigure}{.5\textwidth}
  \centering
  \includegraphics[width=0.85\linewidth]{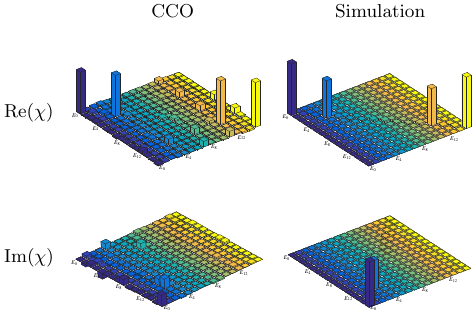}
\caption{$t=0.5s$}
  \label{fig:d2}
\end{subfigure}
\par\bigskip
\par\bigskip
\begin{subfigure}{.5\textwidth}
  \centering
  \includegraphics[width=0.85\linewidth]{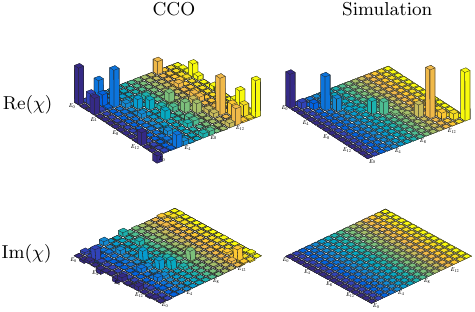}
\caption{$t=5s$}
  \label{fig:d3}
\end{subfigure}
\begin{subfigure}{.5\textwidth}
  \centering
  \includegraphics[width=0.85\linewidth]{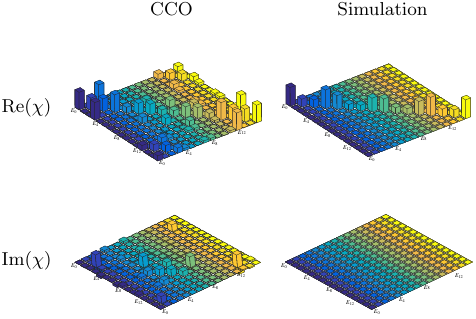}
\caption{$t=15s$}
  \label{fig:d4}
\end{subfigure}
\caption{The tomographs in the row denote
the real and imaginary parts of the process matrix for the decoherence process at various times. The tomographs in the
first and second column represents the experimentally constructed process matrix via CCO method and process matrix obtained by
numerically simulating the decoherence model
respectively.}
\label{ccodeco}
\end{figure}
The CCO based QPT method has been used to characterize non-unitary decoherence processes present in NMR. For NMR
systems, the relevant time scales are
$1/(2J) \approx 0.002325$ sec, $\overline{T}_1 \approx 15$
sec and $\overline{T}_2 \approx 0.5$ sec where $J$ is scaler
coupling constant while $\overline{T}_1$ and
$\overline{T}_2$ are in the range of longitudinal ($T_1$)
and transverse ($T_2$) relaxation parameters. Four distinct time intervals were selected: $t_1 = 0.05$ sec, $t_2 = 0.5$ sec,
$t_3 = 5$ sec, $t_4 = 15$ sec to capture the nature of
decoherence and calculated the $\chi$ matrix corresponding
to these 4 time intervals. The experimental results were
compared to $\chi_{num}$ i.e. $\chi$ matrix obtained by
numerically simulating the decoherence model (model includes
the internal Hamiltonian of the system plus phase
damping and generalized amplitude damping acting
independently on each subsystem). The phase damping channel $\mathcal{E}_t^{\mathrm{PD}}$ acts on a qubit as:

\begin{equation}
\mathcal{E}_t^{\mathrm{PD}}(\rho)=\left(\begin{array}{cc}
\rho_{00} & e^{-\gamma t} \rho_{01} \\
e^{-\gamma t} \rho_{10} & \rho_{11}
\end{array}\right)
\end{equation}
for some damping rate $\gamma$. Similarly the action of generalized amplitude damping channel $\mathcal{E}_t^{\mathrm{GAD}}$ on a qubit is given as:

\begin{equation}
\mathcal{E}_t^{\mathrm{GAD}}(\rho)=\left(\begin{array}{cc}
k_1 \rho_{00}+k_2 \rho_{11} & e^{-\Gamma t / 2} \rho_{01} \\
e^{-\Gamma t / 2} \rho_{10} & k_3 \rho_{00}+k_4 \rho_{11}
\end{array}\right)
\end{equation}
where $k_2 =(1-\bar{n})\left(1-e^{-\Gamma t}\right), k_3 = \bar{n}\left(1-e^{-\Gamma t}\right), k_1 = 1-k_3$, and $k_4 = 1-k_2$, with $\Gamma$ a damping rate and $\bar{n}$ a temperature parameter. The detailed construction
of decoherence model is given in the paper\cite{childs-pra-2001}.
\begin{table}[t]
\centering
\caption{\label{bell_decoherence} The fidelity 
difference between an
actual evolved Bell state (denoted by $\vert \rm{B_i}
\rangle$) computed via CCO QPT,
and predicted output state QPT.
The
first column represents the 
different time intervals for which evolution under the
decoherence process was considered.}
\begin{tabular}{ c|c|c|c|c|c }
\hline
Process & & $\vert \rm{B_1} \rangle$ 
& $\vert \rm{B_2} \rangle$ & 
$\vert \rm{B_3} \rangle$ & $\vert \rm{B_4} \rangle$
\\
\hline
\multirow{ 2}{*}{t=0.05 sec} & CCO & 0.9672 & 0.9808 & 0.9767 & 0.9902 \\
& Numerical & 0.9822 & 0.9921 & 0.9844 & 0.9952  \\ \hline
\multirow{ 2}{*}{t=0.5 sec} & CCO & 0.9884 & 0.9899 & 0.9891 & 0.9757 \\
& Numerical & 0.9785 & 0.9795 & 0.9770 & 0.9831 \\ \hline
\multirow{ 2}{*}{t=5 sec} & CCO & 0.9925 & 0.8410 & 0.9946 & 0.8866 \\
& Numerical & 0.6658 & 0.7193 & 0.6642 & 0.7177 \\ \hline
\multirow{ 2}{*}{t=15 sec} & CCO & 0.9964 & 0.9228 & 0.9959 & 0.9031\\
& Numerical & 0.6060 & 0.7121 & 0.6069 & 0.7126 \\ 
\hline
\end{tabular}
\end{table} 

The evolution of Bell entangled states under the natural decoherence processes is further examined through QST. Subsequently, a comparison is made between the QST results and the states predicted using CCO-based QPT, along with the utilization of a numerically simulated decoherence model. The $\chi$ matrices for decoherence process at various
times are given in Fig.\ref{ccodeco}. Note that all the $\chi$ matrices constructed via
CCO method represent valid quantum map implying true
characterization of decoherence process and from the
decoherence results one can see that the decoherence model
fits with the experimental data upto a good extent (for
small time intervals) although there are extra peaks in
experimental $\chi$ matrices which can not be explained
using the decoherence model under consideration. To
quantitatively characterize the correctness of decoherence
model we have calculated process fidelity between
experimentally constructed $\chi$ matrix and $\chi_{num}$
for each time interval. For the time intervals $t=0.05$
sec, $0.5$ sec, $5$ sec and $15$ sec the calculated
fidelities are 0.9901, 0.8441, 0.7245 and 0.6724
respectively which implies that for small time intervals the
process can be modeled with the proposed decoherence model
while for large time intervals the decoherence model needs
to be modified by including more decoherence
terms.
 
Furthermore, the dynamics of maximally entangled Bell states are studied: $\vert \rm{B_1} \rangle = (\vert 00 \rangle +
\vert 11 \rangle)/\sqrt{2}$, $\vert \rm{B_2} \rangle = (\vert
01 \rangle + \vert 10 \rangle)/\sqrt{2}$, $\vert \rm{B_3}
\rangle = (\vert 00 \rangle - \vert 11 \rangle)/\sqrt{2}$
and $\vert \rm{B_4} \rangle = (\vert 01 \rangle - \vert 10
\rangle)/\sqrt{2}$ under decoherence process. These states were experimentally prepared with fidelities of 0.9968, 0.9956, 0.9911, and 0.9942, respectively. The fidelity between
actual evolved state (constructed using CCO based QST
method) and output state predicted via QPT (both using
experimental and numerical $\chi$ matrix) is given in the
Table\ref{bell_decoherence}. One can see from
Table\ref{bell_decoherence} that for short time intervals
(upto $t \approx O(10^{-1})$ sec) the decoherence model is
able to predict the dynamics of maximally entangled Bell
states with fidelities more than 0.90 while CCO based QPT is
able to predict the true dynamics at all time stages with
good fidelity.

\subsection{Comparison of CCO QPT with standard Protocols}
\label{compare}
\begin{itemize}
 
\item Standard tomography protocols in general do not produce valid density and process
matrices, while CCO based tomography method always produces valid density and process
matrices.
 
\item For QST and QPT, the fidelity obtained via the CCO method
is better than standard methods.
 
\item The \textit{Average state deviation} $\delta_{avg}^{cco}$ given in Eq.\ref{delta} obtained via CCO method is much smaller than standard method which shows that CCO method has better performance than standard method.
 
\item  In the context of QPT, the experimentally constructed $\chi$ matrix enables the efficient computation of all Kraus operators. On the other hand, standard QPT does not yield valid Kraus operators.
 
\item The operation of a quantum gate on any arbitrary input state can be accurately predicted using CCO-based QPT, whereas output state predicted using standard QPT does not correspond to a valid quantum state.
 
\item Both the CCO method and the standard method employ identical experimental data for conducting QST and QPT. Therefore, the experimental complexity remains same for the two approaches.

 \item The fidelities of CNOT and Control-$R^{\pi}_{x}$ gates using the CCO method are improved by 5.41\% and 6.06\% respectively.
 
\item As a result of statistical and systematic errors, the experimentally reconstructed states and processes can occasionally deviate from being positive and physically valid. However, the enhanced fidelity of quantum states and processes, while maintaining the positivity constraint, distinctly indicates that the CCO method significantly mitigates these errors.
 
\end{itemize}

\section{Conclusions}\label{ccotomo_sec4}

This study utilizes a constrained convex optimization (CCO) method to comprehensively characterize different quantum states and quantum processes of two qubits on an NMR quantum information processor. Convex optimization serves as a search procedure, ensuring that the obtained solutions satisfy both experimental and mathematical constraints, resulting in globally optimal outcomes. To assess the effectiveness of the CCO method, the results of QST and QPT using CCO are compared with those obtained from standard linear inversion-based methods. The experiments demonstrate that the CCO method generates physically valid density and process matrices, closely resembling the reconstructed quantum state or the mapped evolution of the quantum process, respectively. Furthermore, the fidelity attained using the CCO method exceeds that of the standard method. Additionally, the experimentally constructed process matrix is utilized to calculate a complete set of Kraus operators corresponding to given quantum process.

In situations where high-fidelity quantum states are prepared, any deviations between the experimental data and the reconstructed process matrix can be attributed to noise. Consequently, CCO-based QPT proves to be a robust method for investigating the nature of noise processes within the quantum system. Assuming system Markovian dynamics, the CCO method was employed to characterize inherent decoherence processes in the NMR system. These results represent a significant advancement in estimating noise and enhancing the fidelity of quantum devices.  The results presented in this chapter have been published in \href{https://link.springer.com/article/10.1007/s11128-020-02930-z}{Quant. Inf. Proc 20(1), 19 (2021)}.

\chapter{Scalable characterisation of noisy quantum gates using compressed sensing and reduced data set}\label{csqpt_chap}

\section{Introduction}

This chapter focuses on addressing the scalability issue in QPT. As previously stated, QPT and QST are standard methods for characterizing the evolution and state of a system. Although both methods are easily demonstrated in experiments, they share a common challenge of exponential growth in required resources and computational complexity with the size of the system, making them impractical even for small numbers of qubits (as seen in Sec.\ref{lsqst} for 1 to 5 qubit systems). To overcome this challenge, various strategies and protocols have been proposed and tested on various physical platforms, as listed in Chap.\ref{chap2}. The performance and experimental efficiency of QPT protocols depend on the specific system being considered. The choice of method depends on the problem being solved. For instance, if one is looking to find a specific element of the process matrix in QPT, the selective QPT protocol would be a better choice than other techniques. In this chapter, the objective is to address a similar issue, focusing on achieving a thorough and precise characterization of quantum processes in situations where the process matrix is nearly sparse. This implies that the process matrix can be effectively approximated by an s-sparse process matrix within a known basis. By taking the sparsity of the process matrix into account, some measurements may be redundant, making QPT experimentally efficient even for high-dimensional systems\cite{shabani-prl-2011}.
\begin{figure}
\centering
\includegraphics[angle=0,scale=1]{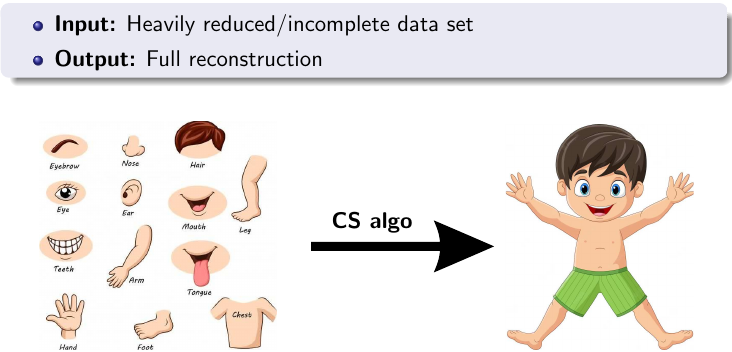} 
\caption{ Graphical depiction illustrating the operational functionality of the compressed sensing algorithm. }   
\label{toc_csqpt}
\end{figure}
To address the scalability problems in QPT, methods such as Monte-Carlo process certification\cite{silva-prl-2011} and randomized benchmarking\cite{knill-pra-2008} have been developed. However, these methods only calculate the fidelity of the quantum gate, which limits their ability to identify gate errors and improve gate quality. Adaptive and self-guided tomography protocols are efficient and can compute the full process matrix, but require a large number of projective measurements that are difficult to perform on ensemble quantum computers. The ancilla-assisted QPT uses an extra clean qubit as an ancilla and exploits the correlation between the system and ancilla qubit, which significantly reduces the experimental complexity of full reconstruction of the quantum process\cite{altepeter-prl-2003}. The simplified QPT approach uses system-environment interaction to simplify the QPT task\cite{kosut-njp-2009}. However, having an extra clean qubit is already a scalability issue, and having prior knowledge about the system-environment is not always easy for high-dimensional systems due to the added interaction terms. This is where the Compressed Sensing (CS) algorithm helps to perform a complete and accurate characterization of the quantum process from a reduced data set without performing actual projective measurements, and without requiring any extra resources like an ancilla\cite{gaikwad-qip-2022}. The CS algorithm was originally proposed for signal processing and later adapted for QCQI tasks such as QST and QPT. Successful experimental demonstrations of CS-based QST and QPT have been conducted using NMR\cite{yang-pra-2017, gaikwad-qip-2022}, linear optics\cite{shabani-prl-2011}, and superconducting qubits\cite{rodionov-prb-2014}. The CS-QPT method is expected to have an exponential speed-up over the standard QPT method. For a d-dimensional Hilbert space system, the method requires only $O$(\textit{s} log$(d)$) experimental probabilities to construct a good estimate of the process matrix $\chi$ if $\chi$ is known to be $s$-compressible in a known basis.

In this chapter, the CS algorithm is used to perform QPT on 2- and 3-qubit quantum gates using an NMR ensemble quantum information processor and IBM cloud experience. We experimentally estimate the full process matrix, $\chi_{CS}$, from a heavily reduced data set compared to standard full QPT. The sparsity, $s$, of the process matrix depends on the operator basis used to estimate $\chi$. The algorithm's performance has been tested for two different basis sets: the usual Pauli basis (PB), where $\chi$ is approximately sparse for most quantum gates, and the Pauli error basis (PEB), where $\chi$ is maximally sparse. In both cases, it is determined that the CSQPT method performs well, particularly in the latter scenario, resulting in a fidelity exceeding 0.9 with a data set size approximately 5-6 times smaller than that of a complete data set. A comparison between the performance of the LS method and the CS method on a reduced data set is conducted as well. Both approaches yield viable solutions and are capable of reconstructing the corresponding $\chi$ matrix. However, it is reported that the LS method does not provide a feasible solution if the size of the reduced data set is below a certain threshold. The quantum circuit and NMR implementation of the CSQPT method for a standard two-qubit gates and a three-qubit gates are provided as examples. The study also includes the characterization of reduced dynamics of a 2-qubit subsystem embedded in a 3-qubit system. The NMR implementation is designed to efficiently acquire experimental data that is compatible with the CS algorithm. The quantum circuit is also valid for other experimental platforms and can be extended to higher-dimensional systems.


\section{Compressed sensing algorithm for QPT}\label{csqpt_sec2}

The CS-QPT method provides the way to reconstruct complete and true $\chi$ matrix of given quantum process from heavily reduced data set provided that $\chi$ matrix is sufficiently sparse in some known basis i.e. the number of non-zero entries in $\chi$ matrix are very less. Specifically in the case of quantum gates which are trace preserving unitary quantum processes, one can always find the proper basis in which the corresponding $\chi$ matrix is maximally sparse. In such scenarios the CS-QPT method helps to verify the quality of quantum gates for large scale quantum computers where standard and other QPT protocols are infeasible. The CS algorithm is typically based on constrained convex optimization problem which minimizes the $l_1$-norm of variable vector, in this case $\overrightarrow{\chi}$ and take the $l_2$-norm of measurement error as a constraint. 

For trace-preserving maps, the complete
convex optimization problem for CS-QPT is formulated as
follows\cite{gaikwad-qip-2022}: 
\begin{subequations} \begin{alignat}{2}
&\!\min_{\chi}        &\qquad&
\Vert{\overrightarrow{\chi}}\Vert_{l_1}\label{ch3_eq80}\\
&\text{subject to} &      & \Vert
\overrightarrow{B}^{exp}-\Phi \overrightarrow{\chi}
\Vert_{l_2} \leq \epsilon,\label{ch3_eq80:c1}\\ &
&      & \chi \geq 0,\label{ch3_eq80:c2}\\ &
&      & \sum_{m,n}\chi_{mn}E_m^{\dagger}E_n =
I_d.\label{ch3_eq80:c3} \end{alignat} \end{subequations}
where Eq.~\ref{ch3_eq80} is the main objective function which is
to be minimized and Eq.~\ref{ch3_eq80:c1} is the
standard constraint involved in the CS algorithm;
Eqs.~\ref{ch3_eq80:c2} and \ref{ch3_eq80:c3} denote
the positivity and trace preserving constraints of the
process matrix, respectively. The parameter $\epsilon$
quantifies the level of uncertainty in the measurement, \ie
the quantity $\overrightarrow{B}^{{\rm exp}}=\Phi
\overrightarrow{\chi}_0+\overrightarrow{z}$ is observed,
with $\Vert \overrightarrow{z} \Vert_{l_2}  \leq \epsilon$,
where $\overrightarrow{\chi}_0$ is the vectorized form of
true process matrix and $\overrightarrow{z}$ is an unknown
noise vector. The general $l_p$- norm of
given vector $ {\overrightarrow{x}}$ is defined as:
$\|x\|_{p}=\left(\sum_{i}\left|x_{i}\right|^{p}\right)^{1 /
p}$. If the process matrix is sufficiently sparse and the
coefficient matrix $\Phi$ satisfies the restricted isometry
property (RIP) condition then by solving the optimization
problem delineated in Eq.~\ref{ch3_eq80}, one can accurately
estimate the process matrix.
The RIP condition is satisfied if the coefficient
matrix $\Phi$ satisfies the following
conditions\cite{shabani-prl-2011, rodionov-prb-2014}:
\begin{itemize} \item[(i)] \begin{equation} 1-\delta_s \leq
\frac{\Vert \Phi \overrightarrow{\chi}_1 - \Phi
\overrightarrow{\chi}_2 \Vert_{l_2}^2}
{\Vert\overrightarrow{\chi}_1 - \overrightarrow{\chi}_2
\Vert_{l_2}^2}\leq 1+\delta_s \label{rip1} \end{equation}
for all $s$-sparse vectors $\overrightarrow{\chi}_1$ and
$\overrightarrow{\chi}_2$.  
An $N \times 1$ dimensional
vector $\overrightarrow{x}$ is $s$-sparse, if only $s < N$
elements are non-zero.  
\item [(ii)] The value of the
isometry constant $\delta_s < \sqrt{2}-1$. 
The restricted isometry constants (RIC) of a
matrix $A$ measures how close to an isometry is the action of
$A$ on vectors with few a nonzero entries, measured in the
$l_2$-norm\cite{emmanuel-crm-2008}. 
Specifically, the upper and lower RIC of a
matrix $A$ of size $n \times N$ is the maximum and the minimum
deviation from unity (one) of the largest and smallest,
respectively, square of singular values of $\text { all
}\left(\begin{array}{c} N \\ k \end{array}\right) \text {
matrices }$ formed by taking $k$ columns from $A$. 
\item[(iii)] The size of the data set is sufficiently large
\ie $m_{\text{conf}}\geq C_0 s  \text{log}(d^4/s)$ where
$C_0$ is a constant,  
$m_{\text{conf}}$ is the size of the data set, $s$ is 
the sparsity of
the process matrix and $d$ is the dimension of 
the Hilbert space. 
\end{itemize} 

Once the basis operators $\lbrace E_{\alpha}
\rbrace$ and the configuration space  $\lbrace \rho_i,
M_j\rbrace $ are chosen, the coefficient matrix
$\Phi_{\text{full}}$ corresponding to the entire data set is
fully defined and does not depend on the measurement
outcomes. It has been shown that if $\Phi_{m}$ is built by
randomly selecting $m$ rows (\ie $m$ number of random
configurations) from $\Phi_{\text{full}}$ then it is most
likely to satisfy the RIP
conditions.  
Hence the sub-matrix
$\Phi_m \in \Phi_{\text{full}} $ together with the
corresponding observation vector $\overrightarrow{B}^{exp}_m
\in \overrightarrow{B}^{exp}_\text{full}$ can be used to
estimate the process matrix by solving the optimization
problem (Eq.\ref{ch3_eq80}). 

In this study, two different operator basis sets are used:
the standard Pauli basis (PB) and the Pauli error
basis (PEB).  In order to compare process $\Lambda$ with desired unitary operation $U$ (i.e with the map $\mathcal{U}(\rho)= U \rho U^{\dagger}$), let's apply inverse operation $\mathcal{U}^{-1}$ after process $\Lambda$. The resulting composed process will be $\tilde{\Lambda} = \mathcal{U}^{-1}\circ \Lambda$ which characterizes the error. That is, if $\Lambda$ is closed to desired $\mathcal{U}$ then $\tilde{\Lambda}$ is closed to identity operation. So the process matrix $\tilde{\chi}$ corresponding to $\tilde{\Lambda}$ in the Pauli basis is referred as process matrix $\chi$ in Puali error basis. To mathematically prove that, consider the action of $\tilde{\Lambda}$ represented in pauli basis as follows,
\begin{equation}
\sum_{m,n=1}^{d^2}\tilde{\chi}_{mn} P_m \rho P_n^{\dagger} = U^{-1} (\sum_{m,n=1}^{d^2}\chi_{mn} P_m \rho P_n^{\dagger}) U
\end{equation} 
which can be rewritten as,
\begin{equation} \label{eq_peb}
\sum_{m,n=1}^{d^2}\tilde{\chi}_{mn} (UP_m) \rho ({UP_n})^{\dagger} = \sum_{m,n=1}^{d^2}\chi_{mn} P_m \rho P_n^{\dagger}
\end{equation}
So one can see that from eq.\ref{eq_peb}, the error matrix $\tilde{\chi}$ is basically the process matrix of original map $\Lambda$ expressed in the operator basis $E_l = UP_l$.

 For both bases, the orthogonality condition is
given by $ \langle E_{\alpha} \vert E_{\beta} \rangle = d
\delta_{\alpha \beta} $.  
For
an $n$-qubit system, 
the basis operators $P_i$ in the PB set are 
$P_i = \lbrace I, \sigma_x, \sigma_y, \sigma_z \rbrace
^{\otimes n}$, while the basis operators $E_i$ in 
the PEB set are:
$E_i = UP_i$ where $U$ is the desired unitary matrix for
which the process matrix needs to be estimated.  Furthermore,
the process matrix in PEB corresponding to the desired $U$,
is always maximally sparse, \ie it contains only one
non-zero element.  The convex
optimization problems involved in LS-QPT and CS-QPT
(Eq.\ref{ch1_eq76} and \ref{ch3_eq80}, respectively) can be solved
efficiently using the YALMIP MATLAB
package, which employs SeDuMi as a solver.

\section{Experimental implementation of CS-QPT on NMR}\label{csqpt_sec3}
\subsection{CS-QPT of two-qubit gates}

 The CS-QPT protocol was applied to
two-qubit quantum gates namely, the CNOT gate and a
controlled rotation gate.  The controlled rotation gate is a
nonlocal gate which rotates the state of the second qubit
via $R_x(\theta)$ if the first qubit is in the state $\vert
1 \rangle $. 

For two qubits the tomographically complete set of input
states is given by: \[\lbrace \vert 0 \rangle, \vert 1
\rangle, \vert + \rangle, \vert - \rangle \rbrace ^{\otimes
2} \] where $\vert + \rangle = (\vert 0 \rangle + \vert 1
\rangle)/\sqrt{2} $ and $\vert - \rangle = (\vert 0 \rangle
+ i\vert 1 \rangle)/\sqrt{2} $. 
In NMR, tomographic measurements are carried out by applying
a set of unitary rotations followed by signal
acquisition. The time-domain NMR signal
is recorded as a free induction decay and then Fourier
transformed to obtain the frequency spectrum, which
effectively measures the net magnetization in the transverse
($x-y$) plane.  For two NMR qubits, the tomographically
complete set of unitary rotations is given
by: $\lbrace I I, R_x^2, R_y^2, R_x^1 R_x^2 \rbrace$ 
\begin{figure}
\centering
\includegraphics[angle=0,scale=1.2]{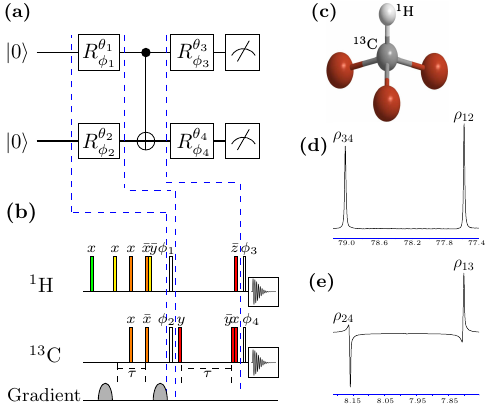} 
\caption{(Color online) (a) Quantum circuit comprising all settings to acquire data to perform CS-QPT of a CNOT gate. (b) NMR implementation to perform CS-QPT of CNOT gate. All rectangles filled with different colours denotes hard NMR pulses to perform various single qubit rotation gates.
(c) ${}^{13}$C-enriched chloroform molecule
with ${}^{1}$H and ${}^{13}$C labeling the
first and second qubits, respectively. (d) and (e) depict
the
NMR spectra of ${}^{13}$C and ${}^{1}$H, respectively,
corresponding to the configuration $\lbrace \vert ++ \rangle
\langle ++ \vert, IX \rbrace$.  }   
\label{ckt1}
\end{figure}

As an illustration, the quantum circuit and corresponding
NMR implementation of CS-QPT for the two-qubit CNOT gate is
given in Fig.~\ref{ckt1}. To implement CS-QPT of any
other quantum gate, the CNOT gate (third block of
Fig.~\ref{ckt1}) needs to be replaced with the desired gate,
while the remaining circuit is unaltered.  Fig.\ref{ckt1}(a)
depicts the general quantum circuit to acquire data for
CS-QPT and contains all possible settings corresponding to a
tomographically complete set of input quantum states and
measurements. The first two blocks in Fig.\ref{ckt1}(a)
prepare the desired initial input state from $\vert 00
\rangle$. In the second block the quantum process (CNOT gate
in this case) which is to be tomographed, is applied to the
system qubits and in the fourth block, tomographic unitary
rotations are applied followed by measurements on each
qubit. 
The first block in Fig.\ref{ckt1}(b) represents the NMR
pulse sequence which prepares the spin ensemble in the pseudo
pure state (PPS) $\vert 00 \rangle$.
In the second block, spin-selective rf pulses 
are applied to prepare the desired input
state from the $\vert 00 \rangle $ state. In the
third block the pulse sequence corresponding to the CNOT gate
(quantum process which is to be tomographed) is given and
finally in the last block the desired set of tomographic pulses are
applied and the NMR signal is acquired.

The $^{13}C$-enriched chloroform molecule
(Fig.\ref{ckt1}(c)) dissolved in acetone-D6 is used to
physically realize a two-qubit system, with the
${}^{1}$H and ${}^{13}$C spins denoting the 
first and second qubits, respectively. 
The NMR
Hamiltonian in the rotating
frame is given by: 
\begin{equation}
\mathcal{H} = - \sum_{i=1}^2 \nu_i I_{iz} +  J_{\rm CH} I_{1z} I_{2z} \label{eq7} 
\end{equation}
where $\nu_1$,
$\nu_2$ are the chemical shifts, $I_{1z}$,
$I_{2z}$ are the $z$-components of the spin angular
momentum operators of the ${}^{1}$H and ${}^{13}$C spins
respectively, and $J_{{\rm CH}}$ is the scalar coupling
constant;
The spatial averaging technique is utilized to
initialize the system in the PPS corresponding to
$\vert 00 \rangle$, with the density matrix
$\rho_{00}$ given by\cite{oliveira-book-07}: 
\begin{equation}
\rho_{00}=\frac{1}{4}(1-\eta)I_4+\eta \vert 00\rangle \langle
00 \vert \label{eq8} \end{equation} where $\eta$ 
corresponds to the net spin magnetization 
at thermal equilibrium and $I_4$ is a $4 \times 4$ identity
operator.

Figs.~\ref{ckt1}(d) and \ref{ckt1}(e) depict the
NMR spectra corresponding to carbon and hydrogen
respectively, obtained for the configuration $\lbrace \vert
++ \rangle, R_x^2  \rbrace $, where $\lbrace \vert ++ \rangle$
refers to the initial state and $R_x^2 $ denotes the 
tomographic pulse set used. 
The system is prepared in
the initial input state $\vert ++ \rangle$, 
a CNOT gate is applied, and
finally the tomographic pulse $R_x^2 $ is
applied to obtain the
NMR spectrum.  For the first qubit, the area under 
the spectrum
is related to 
the density matrix elements 
$\rho_{24}$ and $\rho_{13}$, while for the second
qubit, the area under the spectrum is related to
the 
density matrix elements 
$\rho_{34}$ and $\rho_{12}$.
In general, the four readout elements 
of the density
matrix are complex numbers; in NMR the imaginary part of the
density matrix can be calculated by applying a $90^{\circ}$
phase shift to the spectrum (post-processing) and then
measuring the area. 
Hence a given configuration 
comprises four data
points (two for each qubit).
Since the size of the full configuration space is 64 (16 states
$\times$ 4 tomographic rotations), the size of
full data set is $64 \times 4 = 256$. The vector
$\overrightarrow{B}^{exp}_\text{full}$ (256$\times$1
dimensional)  can be constructed by computing the area under
the spectrum for the full configuration space. One 
can hence construct
$\overrightarrow{B}^{exp}_m $ and 
the corresponding sub-matrix
$\Phi_{m}$ by randomly selecting $m$ rows from
$\overrightarrow{B}^{exp}_\text{full}$ and
$\Phi_\text{full}$ respectively, solving the optimization
problem (Eq.~\ref{ch3_eq80}) for a reduced data set of size
$m$, and estimating the process matrix. 
\subsection{CS-QPT of three-qubit gates}

The CS-QPT protocol is implemented to characterize the
three-qubit Controlled-NOT-NOT ($U_{{\rm CNN}}$) gate with
multiple targets, with the first qubit being the control,
while the other two qubits are the target qubits.  The gate
can be decomposed using two CNOT gates :$U_{{\rm CNN}}$ =
CNOT$_{13}.$CNOT$_{12}$ and is widely used in encoding
initial input states in error correction codes\cite{shor-pra-1995}, fault
tolerant operations\cite{egan-arxiv-2021} and in the preparation of three-qubit maximally entangled states\cite{dogra-pra-2015}.

\begin{figure}
\centering
\includegraphics[angle=0,scale=1.3]{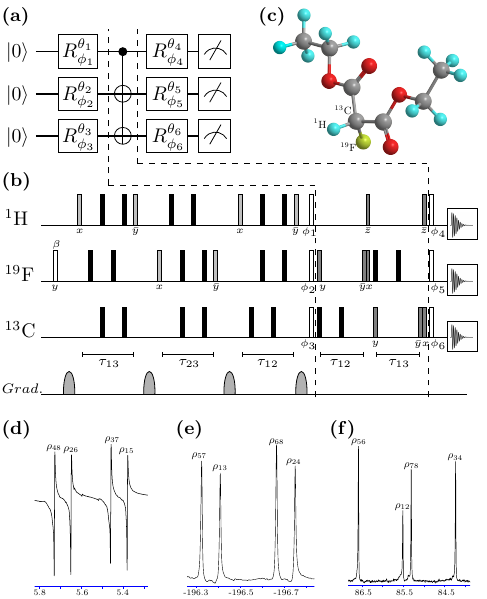}
\caption{(Color online) General quantum circuit to obtained data for
implementation of CS-QPT of 3-qubit Control-NOT-NOT ($U_{\rm {CNN}}$) gate.
(b) NMR implementation of CS-QPT of 3-qubit Control-NOT-NOT ($U_{\rm {CNN}}$) gate. (c)
$^{13}C$-labeled diethyl fluoromalonate with $^1H$, $^{19}F$
and $^{13}C$ nuclei labeled as the first, second and third
qubits, respectively. NMR spectra depicted in (d), (e) and (f)
correspond to ${}^{1}$H, ${}^{19}$F and ${}^{13}$C nuclei
respectively, for the configuration $\lbrace \vert 11+
\rangle \langle 11+ \vert,R_x^3 R_y^2 R_x^1 \rbrace$. 
}
\label{ckt3}
\end{figure}

The NMR Hamiltonian for three qubits
in the rotating frame is given by:
\begin{equation}
\mathcal{H} = - \sum_{i=1}^3 \nu_i I_{iz} + \sum_{i,j=1
(i \ne j)}^3 J_{ij} I_{iz} I_{jz} \label{ham3q} 
\end{equation}
where the indices $i$, $j$ label the qubit and
$\nu_i$ denotes the respective chemical offset. The 
quantities $J_{ij}$ denote the scalar coupling strengths
between the $i$th and $j$th qubits, while $I_{iz}$ represents
the $z$-component of the spin angular momentum of the $i$th qubit. The ${}^{13}$C-labeled diethyl fluoromalonate molecule
(Fig.\ref{ckt3}(c)) dissolved in acetone-D6 is used to physically
realize a three-qubit system, with the
$^1H$, $^{19}F$ and $^{13}C$ spin-1/2 nuclei labeled as the first,
second and third qubits, respectively. State initialization is
performed by preparing the system in the PPS
corresponding to $\vert 000 \rangle$ via the spatial averaging
technique\cite{singh-pra-2019} where the corresponding density matrix is given
by: \begin{equation} \rho_{000} = (\frac{1-\epsilon}{8})I_8
+ \epsilon \vert 000 \rangle \langle 000 \vert \label{pps3}
\end{equation} where $\epsilon \approx 10^{-5}$ represents
the net thermal magnetization and $I_8$ is the 8$\times$8 identity
operator.

For a three-qubit system, the tomographically complete set
of input states is given by: $\lbrace \vert 0 \rangle, \vert
1 \rangle, \vert + \rangle, \vert - \rangle \rbrace
^{\otimes 3}$ where $\vert + \rangle = (\vert 0 \rangle +
\vert 1 \rangle)/\sqrt{2} $ and $\vert - \rangle = (\vert 0
\rangle + i\vert 1 \rangle)/\sqrt{2} $ and the tomographically
complete set of unitary rotations is given
by:~\[\lbrace I I I, R_y^3, R_y^1, R_y^2 R_y^3, R_x^1 R_y^2 R_x^3, R_x^1 R_x^2 R_y^3, R_x^1 R_x^2 R_x^3 \rbrace\]
The quantum circuit and the corresponding NMR
pulse sequence to perform CS-QPT for the three-qubit gate $U_{\rm
CNN}$ is given in Fig.~\ref{ckt3}. The first block in
Fig.\ref{ckt3}(a) represents the input state
preparation while the second block represents the
application of quantum gate $U_{\rm{CNN}}$ (i.e. quantum
process which is to be tomographed),
and tomographic unitary rotations are
applied in the last block, followed by measurement on each qubit. 
Fig.\ref{ckt3}(b) represents the corresponding NMR
implementation of quantum circuit given in Fig.\ref{ckt3}(a).  The spatial averaging
techniques are used in the first block 
to initialize system in the desired PPS,
followed by the application of spin-selective rf
pulses to prepare the desired input state. 
In the second
block the pulse sequence corresponding to $U_{\rm{CNN}}$ is
applied on the input state and in the last block after
application of tomographic pulses, the signal 
of the desired nucleus is recorded.

The NMR spectra corresponding to ${}^1$H, ${}^{19}$F and
${}^{13}$C are given in
Figs.~\ref{ckt3}(d), (e) and (f), respectively,
for the configuration $\lbrace \vert
11+ \rangle \langle 11+ \vert , R_x^3 R_y^2 R_x^1 \rbrace $, \ie the
input state  $ \vert 11+ \rangle \langle 11+ \vert $ is
prepared, evolved under the quantum process corresponding
to $U_{\rm{CNN}}$, the tomographic set of pulses $R_x^3 R_y^2 R_x^1$ is
applied, and finally the NMR signal is recorded. 
For the first qubit (${}^{1}$H)
the area under the four spectral lines 
correspond to the
density matrix elements
$\rho_{48}$, $\rho_{26}$, $\rho_{37}$ and $\rho_{15}$,
for the second qubit (${}^{19}$F)
the area under the four spectral lines 
correspond to the
density matrix elements
$\rho_{57}$, $\rho_{13}$, $\rho_{68}$ and $\rho_{24}$, while
for the third qubit (${}^{13}$C)
the area under the four spectral lines 
correspond to the
density matrix elements
$\rho_{56}$, $\rho_{12}$, $\rho_{78}$ and $\rho_{34}$. 
For a three-qubit
system there are 12 experimental data points (4 per
qubit) for a given configuration and the total number
configurations are  = 448 (64 input states $\times$ 7
tomographic unitary operations) which yields the
$\overrightarrow{B}^{exp}_\text{full}$ of size = 5376 (448
configurations $\times $ 12 data points per configuration).
One can construct $\overrightarrow{B}^{exp}_\text{m}$ by
randomly selecting $m$ number of rows from
$\overrightarrow{B}^{exp}_\text{full}$, and using the
corresponding coefficient matrix $\Phi_{m}$, one can solve
the optimization problem (Eq.\ref{ch3_eq80}) and construct 
the process matrix for a reduced data
set of size $m$.

\subsection{CS-QPT of two-qubit processes in 
a three-qubit system}

In order to experimentally implement a two-qubit CNOT gate
in a multi-qubit system, one needs to allow the two system
qubits to interact with each other \ie, let them evolve under
the internal coupling Hamiltonian for a finite time. In
reality, this is non-trivial to achieve experimentally, as
during the evolution time the other qubits are also
continuously interacting with system qubits, and one has to
``decouple'' the system qubits from the other qubits. In the
language of NMR, this is  referred to as refocusing of the
scalar $J$-coupling. 

The two-qubit CNOT gate is realized using four single-qubit
rotation gates and one free evolution under the internal
coupling Hamiltonian (Fig.~\ref{ckt1}).  The single-qubit
rotation gates are achieved by applying very short duration
rf pulses of length $\approx 10^{-6}$ s, while the time
required for free evolution under the coupling Hamiltonian
is $\approx 10^{-3}$ s. The quality of the experimentally
implemented quantum gate depends on the time required for
gate implementation, which for the two-qubit CNOT gate, is
primarily determined by the free evolution under the
coupling Hamiltonian.
The CS-QPT protocol is used to
efficiently characterize three coupling evolutions
corresponding to $U^J_{ij}$ of the form: 
\begin{equation}
U^J_{ij}(t) = e^{-i 2 \pi J_{ij} I_{iz} I_{jz} t} \label{hf}
\end{equation} 
where the indices $i$ and $j$ label the qubit
and $J_{ij}$ is the 
strength of the scalar
coupling between the $i$th and the $j$th qubit. 
For the CNOT gate,
$t = \vert \frac{1}{2 J_{ij}} \vert $. 
A three-qubit system is continuously evolving
under the three $J_{ij}$ couplings, so in order to let a
subsystem of two qubits effectively evolve under one 
of these
couplings, one has to refocus all the other $J$-couplings.
For example, consider the subsystem of the $i$th and  $j$th
qubit with the effective evolution $U_{ij}^{J}(t)$ given by:
\begin{equation}
U_{ij}^{J}(t) = U_{\rm{int}}(\frac{t}{2}) R_{x}^k(\pi) U_{\rm{int}}(\frac{t}{2}) R_{x}^k(-\pi)
\end{equation}
where $R_{x}^k(\pm \pi)$ is an $x$-rotation on the $k$th
qubit by an angle $\pm \pi$ and $U_{{\rm int}}(\frac{t}{2})$
is the unitary operator corresponding to free
evolution for a duration $\frac{t}{2}$ under the internal Hamiltonian
$\mathcal{H}_{{\rm int}} = \sum_{i,j=1, i>j}^3 J_{ij} I_{iz}
I_{jz}$.
The procedure for tomographic reconstruction
of the reduced two-qubit density matrix from the
full three-qubit density matrix is given in
Table~\ref{redqst}.
All three $U_{ij}^{J}(t)$ were successfully characterized using the CS-QPT method. The corresponding process matrices were constructed using a significantly reduced data set of approximately size 20, achieving experimental fidelities greater than 0.94.
 Note that using information given in Table \ref{redqst} one can efficiently characterize general quantum state as well as dynamics of subsystem of 2-qubits in 3-qubit system. Also note that, in this case the experimental data is acquired by measuring only 2-qubits under consideration, so the complete set of input states and tomographic rotations are same as case of 2-qubit system. 
 
\begin{table}
\centering
 \caption{\label{redqst} Relation
between readout positions $\rho_{ij}'$ of subsystem's
reduced density matrix with readout positions $\rho_{mn}$ of
system's full density matrix is given.} 
\begin{tabular}{c | c c c c}
\hline \hline
~Subsystem ~& \multicolumn{4}{c}{Readout positions of reduce density
matrix}~ \\
~~ & ~$\rho_{24}'$ & ~$\rho_{13}'$ & ~$\rho_{34}'$ & ~$\rho_{12}'$ ~~~~\\ \hline 

~$^1H + ^{19}F$~ & ~$\rho_{48}+\rho_{37}$ & ~$\rho_{26}+\rho_{15}$ & ~$\rho_{57}+\rho_{68}$ &~ $\rho_{13}+\rho_{24}$ ~~~ \\

~$^1H+^{13}C$~ & ~$\rho_{48}+\rho_{26}$ & ~$\rho_{37}+\rho_{15}$ & ~$\rho_{56}+\rho_{78}$ &~ $\rho_{12}+\rho_{34}$ ~~~ \\

~$^{19}F+^{13}C$~ & ~$\rho_{68}+\rho_{24}$ &~ $\rho_{57}+\rho_{13}$ & ~$\rho_{78}+\rho_{34}$ &~ $\rho_{56}+\rho_{12}$ ~~\\

\hline \hline
\end{tabular}
\end{table}
\begin{figure*}
\centering
\includegraphics[angle=0,scale=1.1]{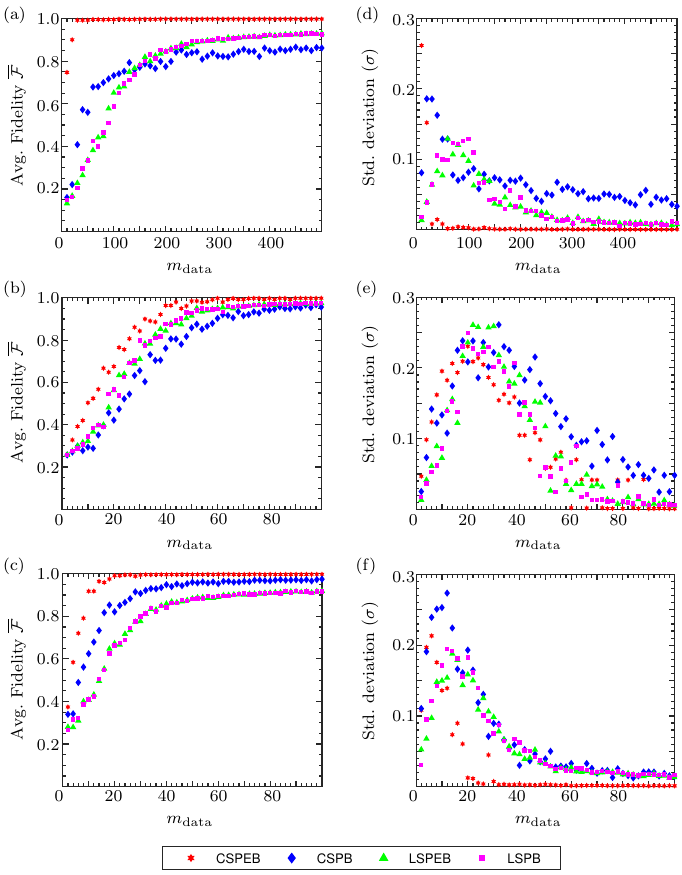} 
\caption{(Color online)The first column represents the average gate fidelity $\overline{\mathcal{F}}$ corresponding to (a) a three-qubit $U_{\rm {CNN}}$ gate, (b) a CNOT gate and (c) $U^J_{23}$
 vs number of data points
$m_{\text{data}}$ while second column represents standard deviation ($\sigma$) in average fidelity corresponding to (d) a three-qubit $U_{\rm {CNN}}$ gate, (e) a CNOT gate and (f) $U^J_{23}$ vs number of data points
$m_{\text{data}}$.
The plots in red, blue, green and magenta 
correspond to CS-PEB, CSPB, LSPEB and LSPB methods, 
respectively. 
} 
\label{plots}
\end{figure*}
\begin{table}[h!]
\centering
\caption{\label{complexity3}
For given quantum gate, the minimum value of $m_{\rm{data}}$
at which the experimental average gate fidelity
$\overline{\mathcal{F}}$ turns to be $\geq$ 0.9 computed via
CS and LS algorithm with corresponding standard deviation
$\sigma$ for PEB and PB is given.}
\setlength{\tabcolsep}{5pt} 
\renewcommand{\arraystretch}{1.1} 
\begin{tabular}{cc|cccccc}
 \hline \hline 
& & $U_{\mathrm{CNN}}$ & CNOT & $\mathrm{C}-R_{x}^{\pi}$ & $U_{12}^{J}$ & $U_{23}^{J}$ & $U_{13}^{J}$ \\
 \hline
 \multirow{3}{4em}{CSPEB} & $m_{\text {data }}$ & 30 & 44 & 48 & 14 & 14 & 18 \\
 & $\overline{\mathcal{F}}$ & 0.9920 & 0.9798 & 0.9728 & 0.9549 & 0.9641 & 0.9417 \\
 & $\sigma$ & 0.0081 & 0.0701 & 0.0797 & 0.0963 & 0.0734 & 0.0980 \\
  \hline
  \multirow{3}{4em}{CSPB} & $m_{\text {data }}$ & - & 62 & 58 & 24 & 28 & 38 \\
 & $\overline{\mathcal{F}}$ & - & 0.9203 &0.9068 & 0.9464 & 0.9145 & 0.9067 \\
 & $\sigma$ & - & 0.0905 & 0.0746 & 0.0468 & 0.0710 & 0.0695 \\
 \hline
 \multirow{3}{4em}{LSPEB} & $m_{\text {data }}$ & 320 & 52 & 48 & 32 & 66 & - \\
  & $\overline{\mathcal{F}}$ & 0.9109 & 0.9514 & 0.9332 & 0.9075 & 0.9019 & - \\
  & $\sigma$ & 0.0123 & 0.0263 & 0.0805 & 0.0459 & 0.0217 \\
  \hline
 \multirow{3}{4em}{LSPB} & $m_{\text {data }}$ & 290 & 48 & 52 & 34 & 68 & - \\
 & $\overline{\mathcal{F}}$ & 0.9006 & 0.9308 & 0.9503 & 0.9071 & 0.9048 & - \\
 & $\sigma$ & 0.0147 & 0.0475 & 0.0504 & 0.0561 & 0.0198 & - \\

\hline \hline
\end{tabular}
 \end{table}

\subsection{Comparison of CS-QPT and LS-QPT protocols}

The fidelity of the experimentally estimated $\chi_{{\rm exp}}$
is computed using Eq.\ref{ch1_eq77}. QPT of various two- and three-qubit quantum gates was conducted utilizing both the CS-QPT and LS-QPT protocols on a reduced data set.  The CS-QPT method is implemented for two
different basis sets, namely PB and PEB.  For a two-qubit
system $m_{\text{data}}^{\rm full} = 256$,  while for a
three-qubit system, $m_{\text{data}}^{\rm full} = 5376$ where the $m_{\text{data}}^{\rm full}$ denotes the size of full data set obtained using complete set of input states and tomographic rotation operators for respective spin system.  In
the PEB set, $\chi_{\text{ideal}}$ is maximally sparse for
all unitary quantum gates, while in the PB set,
$\chi_{\text{ideal}}$ corresponding to the two- qubit CNOT,
control-$R_{x}^{\pi}$ and $U^J_{ij}$ gates have 16, 16 and 4
non-zero elements, respectively (out of a total of 256
elements).  For the three-qubit gate $U_{\rm CNN}$,
$\chi_{\text{ideal}}$ has 16 non-zero elements (out of a
total of 4096 elements).
The performance of the CS-QPT method is compared  with the LS-QPT method for six different quantum processes corresponding to: (i) a three-qubit $U_{\rm CNN}$ gate, (ii) a two-qubit CNOT gate, (iii) a control-$R_x^{\pi}$ rotation (iv) $U^J_{23}$, (v) $U^J_{13}$ and (vi) $U^J_{12}$ out of which results of three quantum process corresponding to (a)
a three-qubit $U_{\rm CNN}$ gate, (b) a two-qubit CNOT gate and (c) $U^J_{23}$ are displayed in Fig.\ref{plots}. The first column in Fig.\ref{plots} represents the average gate fidelity $\overline{\mathcal{F}}$ vs $m_{\text{data}}$ while second column represents standard deviation $\sigma$ in average gate fidelity $\overline{\mathcal{F}}$ vs $m_{\text{data}}$. The average gate fidelity is obtained using average process matrix estimated via CS and LS algorithm in PEB and PB. The plots in red and blue color represent the case of CS-QPT performed in PEB and PB respectively while the plots in green and magenta color represent the case of LS-QPT performed in PEB and PB respectively. The average
fidelity and the value of $\sigma$ is computed by 
implementing the
CS-QPT and LS-QPT protocols 50 times for randomly selected
$m_{\text{data}}$ number of data points,
and $\sigma$ is calculated from: \begin{equation}
\sigma = \sqrt{\frac{\sum_{i=1}^{N}
(\mathcal{F}_i-\overline{\mathcal{F}})^2}{N-1}} \label{eq10}
\end{equation} where $N=50$ and $\overline{\mathcal{F}}$ is
the average fidelity.
\begin{table} 
\centering
\caption{\label{fid} Experimental quantum gate
fidelities obtained via CS and LS method using full data set $m^{\rm{full}}_{\rm{data}}$.}
\setlength{\tabcolsep}{6.3pt} 
\renewcommand{\arraystretch}{1.2} 
\begin{tabular}{c c c c c}
\hline 
\hline
Process & ~~CS-PEB~~& ~~CSPB~~ & ~~LSPEB~~& ~~LSPB\\
\hline $U_{\rm{CNN}}$ & 0.9980 & 0.8877 & 0.9542 &~ 0.9542\\ 
CNOT & 0.9984 & 0.9843 & 0.9817 &~ 0.9817\\ 
C-$R^{\pi}_x$ & 0.9980 & 0.9744 & 0.9831 &~ 0.9831\\
$U^J_{12}$ & 0.9967 & 0.9894 & 0.9819 &~ 0.9819 \\
$U^J_{23}$ & 0.9976 &  0.9793 & 0.9273 &~ 0.9273 \\
$U^J_{13}$ & 0.9895 & 0.9710 & 0.8942 &~ 0.8942 \\
\hline \end{tabular}
\end{table}

 The plots in the first row given in Fig.\ref{plots} corresponds to three-qubit gate $U_{\rm{CNN}}$ where plot (a) gives the information about the accuracy in characterizing $U_{\rm{CNN}}$ gate for given value of $m_{\text{data}}$ while the plot (d) gives the information about the precision in characterizing $U_{\rm{CNN}}$ for given value of $m_{\text{data}}$. Similarly second and third rows represent the experimental results corresponding to CNOT gate and $U^J_{23}$ process, respectively. The plots corresponding to two-qubit control rotation gate ($C-R_{x}^{\pi}$) is similar to CNOT gate given in second row while plots corresponding to $U^J_{13}$ and $U^J_{12}$ are similar to  $U^J_{23}$ given given in third row.
As seen from Fig.~\ref{plots}, the CS-QPT method implemented
using the PEB set performs better than the LS-QPT and the
CS-QPT method implemented using the PB set in all cases.
The performance of LS-QPT method is independent of the
choice of basis operators.  On the other hand, the
performance of the CS-QPT may yield a lower fidelity as
compared to LS-QPT method if the basis operators are not
properly chosen.  For a reduced data set, the overall
performance for the three-qubit gate $U_{\rm{CNN}}$ is
CS-PEB $>$ CSPB $>$ LSPEB $\approx$ LSPB while for the
two-qubit CNOT and C-$R_x^{\pi}$ gates, CS-PEB $>$
LSPEB$\approx$ LSPB $>$ CSPB. 
For all the two-qubit quantum processes
corresponding to $U_{ij}^J$, CS-PEB $>$
CSPB $>$ LSPEB $\approx$ LSPB.  Note that in the case of CNOT and C-$R_x^{\pi}$ gates the LS algorithm works better than CS algorithm for CSPB case for all values of $m_{\rm{data}}$ while in case of three-qubit $U_{\rm{CNN}}$ gate,    the performance of LS algorithm beats the CS algorithm in case of CSPB for $m_{\rm{data}} \geq 160$  which clearly shows the importance of selecting appropriate operator basis set for CS algorithm. In all the cases examined, it is observed that the standard deviation ($\sigma$) in average fidelity does not exhibit a monotonic behaviour. That is due to, for very small value of $m_{\rm {data}}$ the process of randomly selecting $m_{\rm {data}}$ data points to estimate process matrix will more likely to yield less fidelity and for sufficiently large value of $m_{\rm {data}}$ the process of randomly selecting $m_{\rm {data}}$ data points to estimate process matrix is more likely to yield high fidelity using both CS and LS algorithm. In such extreme cases one will have small value of standard deviation $\sigma$, \ie maximum precision. However, for intermediate values of $m_{\rm {data}}$ the process of randomly selecting data points of given size to estimate process matrix may or may not yield high fidelity in such cases the $\sigma$ will have higher value. For the
two-qubit CNOT and control-$R_x^{\pi}$ gates, $\sigma$ has a maximum  around $m_{\rm {data}}\approx 20 $, while for the
$U_{CNN}$, $U^J_{23}$, $U^J_{13}$ and $U^J_{12}$ 
quantum processes, $\sigma$ is maximum around $m_{\rm {data}}\approx 10 $. However, note that, in all cases the CS-PEB yields better precision compared to CSPB, LSPEB and LSPB. In brief, what is experimentally observed is that for exceedingly small values of $m_{\rm {data}}$, lower accuracy and higher precision are obtained. With an increase in $m_{\rm {data}}$ up to a certain threshold, accuracy rises but precision decreases. Subsequently, upon further increase in $m_{\rm {data}}$, both accuracy and precision are enhanced.

The experimentally obtained minimum
value of  $m_{\rm {data}}$  at which the
experimentally computed average gate fidelity is
$\geq$ 0.9, 
is given in 
Table~\ref{complexity3}, 
for all the quantum processes.  

For the three-qubit $U_{\rm{CNN}}$ gate, a result of $\overline{\mathcal{F}}_{\rm{CSPEB}} = 0.9920 \pm 0.0081$ was achieved using a reduced data set of size $m{\rm{data}} = 30$. On the other hand, the two-qubit CNOT and control-$R_x^{\pi}$ gates exhibited results of $\overline{\mathcal{F}}_{\rm{CSPEB}} \geq 0.9790 \pm 0.0701$ and $\overline{\mathcal{F}}_{\rm{CSPEB}} \geq 0.9729 \pm 0.0797$ for data sets of size $m_{\rm{data}} \geq 44$ and $m_{\rm{data}} \geq 48$, respectively. The
reduced data set is $\approx 5$
times smaller than the full data set, which implies
that the experimental
complexity is reduced by $\approx 80 \%$ as compared to
the standard QPT method. Furthermore, for all the two-qubit quantum
processes corresponding to $U_{ij}^J$, 
$\overline{\mathcal{F}}_{\rm{CSPEB}} \geq  0.9417 \pm 0.0980
$ for $m_{\rm{data}} \geq 18$. This reduced
data set is $\approx 12$ times
smaller than the full data set which implies the experimental
complexity in these cases is reduced by $\approx 92 \%$ 
as compared to
the standard QPT method. In short, the plots given in Fig.\ref{plots} also provides the information about experimental complexity of CS and LS algorithm, \ie, number of experiments required to characterize given quantum process depending upon desired accuracy and precision. 

\section{CS-QPT on IBM cloud quantum computer}\label{csqpt_sec4}

The IBM quantum processor represents a quantum operating system based on superconducting transmon qubits. It offers free accessibility and incorporates a Python-based software framework called \textit{QISKit}, which facilitates the implementation of theoretical protocols \cite{santos-2017}.
In this study, the five-qubit IBM QX2 processor was specifically employed, designating the first two qubits as $q[0]$ and $q[1]$ as system qubits. The objective was to showcase the QPT of different two-qubit IBM quantum gates using two distinct methods, namely LS-QPT and CS-QPT. Experimental parameters, including those related to the Hamiltonian and relaxation times specific to the IBM QX2 processo is given in Table.\ref{qx2}. The more details about IBM QX2 processor can be found in References\cite{devitt-pra-2016, shukla-pla-2018}

\begin{table} 
\centering
\caption{\label{qx2} The first row represents the qubit index q[i] within the 5-qubit IBM QX2 quantum processor. The second row displays the resonance frequencies $\omega_R$ of the corresponding read-out resonators. Qubit frequencies $\omega$ are provided in the third row. The anharmonicity $\delta$ is given in the fourth row, gauges the extent of information leakage from the computational space. The fifth and sixth rows correspond to the qubit-cavity coupling strengths $\mathcal{C}$ and the coupling of the cavity to the environment $\kappa$ for the respective qubits. The seventh and eighth rows present the longitudinal relaxation time $T_1$ and transverse relaxation time $T_2$, respectively.}
\setlength{\tabcolsep}{6.3pt} 
\renewcommand{\arraystretch}{1.2} 
\begin{tabular}{c c c c c c}
\hline 
\hline
  & $\mathrm{q}[0]$ & $\mathrm{q}[1]$ & $\mathrm{q}[2]$ & $\mathrm{q}[3]$ & $\mathrm{q}[4]$ \\
\hline 
$\omega_R / 2 \pi(\mathrm{GHz})$ & 6.530350 & 6.481848 & 6.436229 & 6.579431 & 6.530225 \\
$\omega / 2 \pi(\mathrm{GHz})$ & 5.2723 & 5.2145 & 5.0289 & 5.2971 & 5.0561 \\
$\delta / 2 \pi(\mathrm{MHz})$ & -330.3 & -331.9 & -331.2 & -329.4 & -335.5 \\
$\mathcal{C} / 2 \pi(\mathrm{kHz})$ & 476 & 395 & 428 & 412 & 339 \\
$\kappa / 2 \pi(\mathrm{kHz})$ & 523 & 489 & 415 & 515 & 480 \\
$\mathrm{~T}_1(\mu \mathrm{s})$ & 53.04 & 63.94 & 52.08 & 51.78 & 55.80 \\
$\mathrm{~T}_2(\mu \mathrm{s})$ & 48.50 & 35.07 & 89.73 & 60.93 & 84.18 \\
\hline \end{tabular}
\end{table}

\begin{figure}
\centering
\includegraphics[angle=0,scale=1.4]{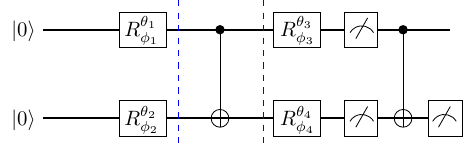} 
\caption{Two-qubit quantum circuit to acquire experimental data for
QPT of CNOT gate.} 
\label{ckt}
\end{figure}

\begin{figure}
\centering
\includegraphics[angle=0,scale=1]{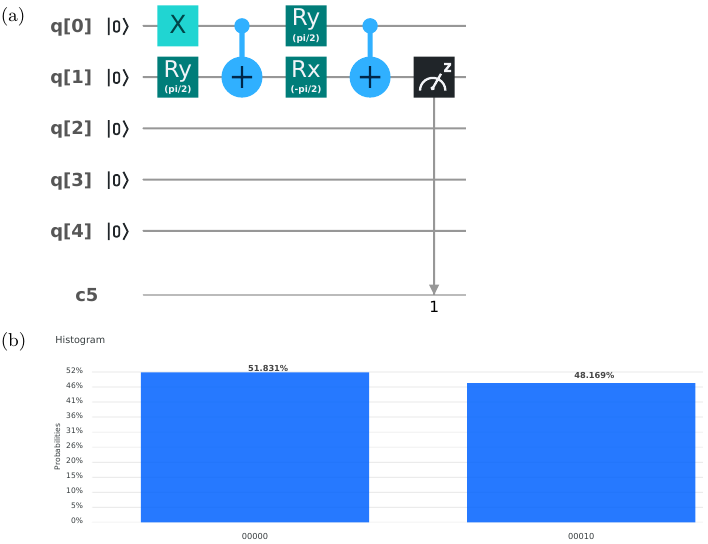} 
\caption{ (a) IBM quantum circuit for implementing QPT of a
CNOT gate, corresponding to configuration $\lbrace \vert 1+
\rangle \langle 1+ \vert, \sigma_{1x}\sigma_{2y} \rbrace$.
Bit flip gate $X \equiv R_{y}(\pi)$. (b) Histogram of the probabilities
$p_0$ and $p_1$, of obtaining the two-qubit states $ \vert 0
\rangle $ and $\vert 1 \rangle$ respectively, after running
the quantum circuit (given in (a)) 4096 times. We obtain
$p_0 = 0.51831 $ and $p_1 = 0.48169$.}
\label{ibm} 
\end{figure}

In the case of IBM like quantum processor, experimental data point corresponding to a given configuration $\lbrace \rho_i, M_j \rbrace $ can be expressed using Eq.\ref{ch1_eq74}. For an $n$-qubit system, complete set of input states is same as NMR case, however, the tomographic measurements are carried out by directly computing expectation values of measurement operators: $\lbrace I, \sigma_x, \sigma_y, \sigma_z \rbrace^{\otimes n} $ where $\sigma_x$, $\sigma_y$ and $\sigma_z$ are single-qubit Pauli spin operators. The entries in $\overrightarrow{\lambda}$ are experimentally computed expectation values $\text{Tr}(M_j \Lambda(\rho_i))$ where the elements of the coefficient matrix $\Phi$ being the coefficients $\text{Tr}(M_j E_m \rho_i E_n^{\dagger})$. Quantum circuit to acquire experimental data for
QPT is given in Fig.\ref{ckt}. As an example, the IBM quantum circuit for a configuration corresponding to $\lbrace \vert 1+ \rangle \langle 1+ \vert,
\sigma_{1x}\sigma_{2y} \rbrace$ to perform QPT of a CNOT
gate is given in Fig.~\ref{ibm}. The proposed quantum
circuit requires detection only on a single qubit {\ie}
the measurement outcomes $\lbrace \lambda_i^j \rbrace$ of all
configurations are mapped to either $\langle \sigma_{1z}
\rangle$ or $\langle\sigma_{2z}\rangle$.  For two qubits the
complete list of unitary operations that map the Pauli
measurement outcomes to either $\langle \sigma_{1z} \rangle$
or $\langle\sigma_{2z}\rangle$ is given in
Reference\cite{gaikwad-pra-2018}. Since on the IBM QX2
processor, measurements are done in the $\sigma_z$ basis,
the probabilities $p_0$ and $p_1$ of observing $\vert 0
\rangle$ and $\vert 1 \rangle$ respectively are calculated
by implementing the same circuit 4096 times. The
experimentally computed probabilities are used to compute
the expectation values: 
\begin{equation} 
\lambda^j_i= \sum_{m,n=1}^{d^2} \chi_{mn}
{\rm Tr}(M_j E_m \rho_i E_n^\dagger) = e_0 p_0 +
e_1 p_1 
\label{e8} 
\end{equation} 
where $e_0=1$
and $e_1=-1$ are eigenvalues of the $\sigma_z$ operator.


The QPT of three quantum gates, specifically the Identity, CNOT, and SWAP gates, was performed on the IBM QX2 processor using CS and LS optimization techniques. The first two qubits ($q[0]$ and $q[1]$) were designated as system qubits. In all cases, $\chi_{{\rm exp}}^{{\rm CS}}$ and $\chi_{{\rm exp}}^{{\rm LS}}$ were experimentally constructed using both CS and LS methods. Additionally, quantum circuits were simulated on the IBM simulator, and $\chi_{{\rm sim}}^{{\rm CS}}$ and $\chi_{{\rm sim}}^{{\rm LS}}$ were computed. The gate fidelity was evaluated using Equation \ref{ch1_eq77}.

\begin{figure}
\centering
\includegraphics[angle=0,scale=1.05]{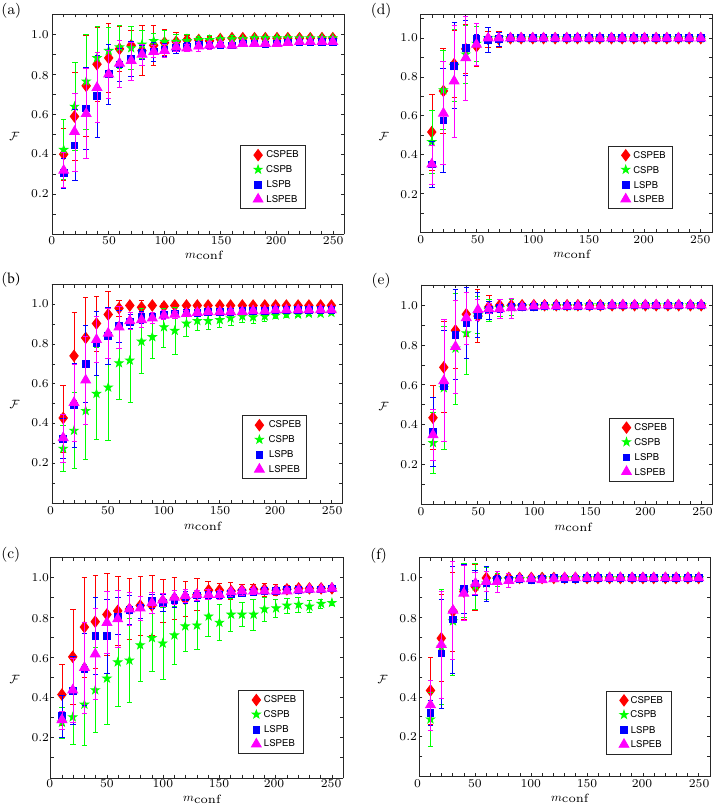} 
\caption{Experimental plots (a), (b), and (c) were obtained using the IBM QX2 processor, while corresponding simulation results (d), (e), and (f) are provided for the QPT of the Identity gate, CNOT gate, and SWAP gate, respectively. The average gate fidelity $\mathcal{F}$ is represented on the $y$-axis, while the size of the data set $m_{{\rm conf}}$ is depicted on the $x$-axis. The CS method is represented by red diamonds and green stars, while the LS method is indicated by pink triangles and blue squares for the PEB and PB basis, respectively. Error bars in all cases are determined by calculating the standard deviation $\sigma$ through the random selection of $m_{{\rm conf}}$ configurations, repeated 50 times.} 
\label{fig2}
\end{figure}
\subsection{IBM experimental results and analysis }
Figure~\ref{fig2} contains the 
the experimental and simulation results of performing QPT
via LS and CS methods 
corresponding to Identity, CNOT and SWAP gates,
respectively.
The $y$-axis denotes the average
fidelity $\mathcal F$ of a given quantum gate and the $x$-axis
denotes the number of configurations (size of reduced data
set) $m_{conf}$, used to construct the average process matrix.
The average process matrix is constructed by randomly
choosing $m_{\rm {conf}}$ number of configurations from
the full configuration space for 50 times. The
error bars are calculated using Eq.\ref{eq10}

For the identity gate the CS method works better than the LS
method.  The plots in Fig.~\ref{fig2} demonstrate that for a
data set of size $m_{{\rm conf}} > 60$, the average fidelity
of the Identity gate is $> 0.9$ for randomly chosen settings
using the CS method.  Since for the Identity gate PEB = PB,
the performance of the optimization algorithms can be ranked
as CSPEB $\approx$ CSPB $>$ LSPEB $\approx$ LSPB. When employing the complete data set consisting of 256 entries, the experimental gate fidelities obtained through CSPEB, CSPB, LSPEB, and LSPB methods are 0.98247, 0.98215, 0.96342, and 0.96308, respectively.

For the CNOT gate, the CS method in general performs better than
the LS method. However, unlike for the identity gate,
choosing the PEB basis for both methods gives better results
as compared to the PB basis.  The performance of the
optimization algorithms as evidenced by the plots in
Fig.~\ref{fig2} can be ranked as CSPEB $>$ LSPEB $\approx$
LSPB $>$ CSPB. For the CSPEB case, we found that for a data
set of size $m_{{\rm conf}}>50$ the average gate fidelity is
$> 0.9$ for randomly chosen configurations. Using the full
data set of size 256, it is found the experimental gate
fidelities obtained via CSPEB, CSPB, LSPEB and LSPB methods
are 0.99558, 0.95573, 0.97282 and  0.97319, respectively. Similar in the CSPEB scenario, it is observed that with a data set size of $m_{{\rm conf}}>100$, the average gate fidelity surpasses 0.9 for randomly selected configurations. Utilizing the complete data set of size 256, the experimental gate fidelities obtained through CSPEB, CSPB, LSPEB, and LSPB methods are 0.94671, 0.8715, 0.94323, and 0.94367, respectively.

 For all three quantum gates, it was observed that the CS method performs more effectively and generates a complete and valid process matrix with reasonably high accuracy. This is achieved using a heavily reduced data set when conducted in the PEB. The simulation results in the Fig.~\ref{fig2} are obtained by theoretically simulating all quantum circuits on IBM simulator. It is a ideal and noise free theoretical simulation where high fidelity simulation results ensure that all the quantum circuits are correct and can be used further in the experiments. The simulation
results shown in Fig.~\ref{fig2} verify that the quantum
circuits that we implemented are correct and that the errors
in the experimentally constructed process matrices are
mainly due to gate imperfections and decoherence effects.
Furthermore, no significant improvement in fidelity was achieved by increasing the number of experiments beyond a certain threshold value of $m^{{\rm thrs}}_{{\rm conf}}$. In case of the
Identity, CNOT and SWAP gates, this threshold value
$m^{{\rm thrs}}_{{\rm conf}}$ turns out to be $\approx$ 60, 50 and
100 respectively. Both intuitively and through experimental observation, it becomes evident that there exists a point, denoted as $m_{conf}^{thrs}$, beyond which the fidelity remains relatively constant, rendering further experimentation less valuable. This phenomenon might be attributed to the repetition of information obtained after a certain number of experiments. 
Nevertheless, one might anticipate that within the PEB framework, all three quantum processes (Identity, CNOT, and SWAP) would be equivalent to the identity operation and should exhibit the same threshold $m_{conf}^{thrs}$. However, in practice, the implementation of quantum gates is not flawless and is influenced by environmental noise at varying degrees. Because of this, the corresponding process matrices have different sparsity and since the performance of CS method depends on sparsity, the $m_{conf}^{thrs}$ will vary for different gates even in the PEB basis.  Also note that all process matrices obtained via the CS and LS method
are physically valid and can hence be used to compute a complete set
of valid Kraus operators via a unitary diagonalization
procedure (see Eq.\ref{e13}).

\section{Conclusions}\label{csqpt_sec5}
A general quantum circuit was implemented to gather experimental data that is compatible with the CS-QPT algorithm. This was carried out on both an NMR ensemble processor and an IBM cloud processor. The
proposed quantum circuit can also be used for other
experimental platforms and can be extended to
higher-dimensional systems. The efficacy of the CS-QPT protocol for various quantum processes, including the $U_{\rm{CNN}}$ gate, CNOT gate, controlled rotation gate, and $U_{ij}^J$ unitary operations, was successfully demonstrated. Additionally, solutions corresponding to the process matrices were obtained with significant accuracy using both CS-QPT and LS-QPT, while utilizing a substantially reduced data set compared to the full data set required for standard QPT. The results indicate that the CS-QPT protocol is notably more efficient than the LS-QPT protocol, particularly when the process matrix is maximally sparse and an appropriate operator basis is selected. Since the CS-QPT method is
experimentally feasible it can be used to characterize
higher-dimensional quantum gates and to validate the
performance of large-scale quantum devices. The results presented in this chapter have been published in \href{https://link.springer.com/article/10.1007/s11128-022-03695-3}{Quantum Inf. Process. 21, 388 (2022)} and \href{https://www.worldscientific.com/doi/abs/10.1142/S0219749920400043}{Int. J. Quantum. Inf. 19,
2040004 (2021)}.
\chapter{Predicting quantum states and processes with incomplete measurements using feed-forward neural network}

\section{Introduction}
\label{sec1}

  In recent years, there has been a proposal for a new category of tomography methods that are more flexible and efficient. These methods primarily rely on a data-driven approach, inspired by machine learning (ML) techniques\cite{carleo-rmp-2019, carleo-science-2017, torlai-prl-2018}. ML advancements have contributed to various areas such as language translations\cite{popel-nc-2020}, self driven cars\cite{bachute-mlwa-2021},
vaccine development\cite{yang-sr-2021}, etc. Moreover, ML has found applications in quantum-related tasks, including classifying phases of
matter\cite{carrasquilla-np-2017}, representation of many body quantum
states\cite{gao-nc-2017}, classification of quantum
states\cite{harney-njp-2020}, quantum error
correction\cite{nautrup-quantum-2019}, quantum state tomography
(QST)\cite{torlai-arcm-2020} and quantum process tomography (QPT). 
 In this chapter, the emphasis is on utilizing machine learning for tasks associated with QST and QPT. Several methods has been proposed to carry
out QST using ANNs. A simulation of QST using Restricted Boltzmann Machine (RBM)
based ANN model was done on highly entangled states\cite{torlai-np-2018}.
Experimental implementation of this RBM based QST was studied for two qubit
state on an optical system\cite{neugebauer-pra-2020}.  ML based adaptive QST was
performed which adapts to current experiment and suggests suitable next
measurement to gain maximum knowledge\cite{quek-npj-2021}. QST using attention
based generative network which learns noisy density matrix was realized
experimentally on IBMQ quantum computer\cite{cha-mlst-2021}. Study of NN
enhanced QST was carried out where state preparation and measurement (SPAM)
errors was minimized when reconstructing the state on photonic quantum
set\cite{palmieri-npj-2020}. A convolutional neural network model was employed
to reconstruct quantum states with tomography measurements in presence of
simulated noise\cite{lohani-mlst-2020}. Furthermore, local-measurement-based
quantum state tomography via neural networks has been demonstrated on
NMR\cite{xin-npj-2019}. Most of the ML based protocols mentioned have only been applied to state tomography and yet to be demonstrated for process tomography.

In this study, the Feed-Forward Neural Network (FFNN) architecture is employed, where information flows in the forward direction. This architecture is used to conduct quantum state and process tomography using a set of noisy experimental data acquired from an NMR quantum information processor\cite{gaikwad-pra-2024}. The model is trained and tested on computationally generated states/processes and then validated using experimental data obtained from NMR spectra. Furthermore, the FFNN model is assessed on an incomplete measurement dataset, where a fraction of the total measured dataset is randomly selected. It is observed that the model provides predictions for states and processes with a notably high fidelity, even when considering a substantially reduced subset of the total measured dataset randomly.  


\section{ANN based QST and QPT }\label{sec2}
\subsection{Artificial Neural Networks (ANNs)}

\begin{figure*}[t]
\centering
\includegraphics[angle=0,scale=0.8]{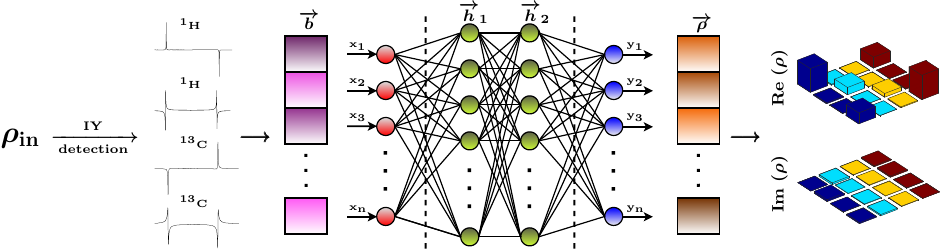} 
\caption{ Schematics representation of 2-qubit QST using FFNN model.} 
\label{flowchart}
\end{figure*}

Artificial Neural Network (ANN) consists of many artificial neurons which are
connected to each other. The neuron is activated when its value is greater than
threshold value called bias. 
In this study, the task of characterizing quantum states and processes is undertaken by employing a multilayer perceptron model, also known as a Feed-Forward Neural Network (FFNN). The FFNN architecture mainly consists of three layers: input layer,
hidden layer and output layer. Data is fed into input layer which is passed on
layer by layer till it arrives at the output. (See third step in
Fig.\ref{flowchart}). Data is divided into two parts i.e. train data-set and
test data-set. As name implies, train data-set is used to train the model
(update network parameters, weights and biases) and test data-set is used to
evaluate the network. Consider '$m$' training elements $\lbrace
(\vec{x}^{(1)},\hat{\vec{y}}^{(1)}),(\vec{x}^{(2)},\hat{\vec{y}}^{(2)}),\cdots ,(\vec{x}^{(m)},\hat{\vec{y}}^{(m)}) \rbrace$ where
$\vec{x}^{(i)}$ is $i^{th}$ input and $\hat{\vec{y}}^{(i)}$ is corresponding labeled output.
Feeding these inputs to the network produces outputs
$[\tilde{\vec{y}}^{(1)},\tilde{\vec{y}}^{(2)},...,\tilde{\vec{y}}^{(m)}]$. Since network
parameters are initialized randomly, predicted output is not equal to expected
output. Training of this network can be achieved by minimizing the
mean-squared-error \textit{cost function}, $J(w,b)=\frac{1}{m}\sum_{i=1}^{m}
||\hat{\vec{y}}^{(i)}-\tilde{\vec{y}}^{(i)}||^2 $ with respect to network parameters by stochastic
gradient descent,

\begin{align}
w_{ij} \rightarrow w'_{ij} =& w_{ij}-\frac{\eta}{m'} \sum_{i=1}^{m'}
\frac{\partial}{\partial  w_{ij}} \mathcal{L}(\vec{x}^{(i)}) \\ b_{i} \rightarrow
b'_{i} =& b_{i}-\frac{\eta}{m'} \sum_{i=1}^{m'} \frac{\partial}{\partial  b_{i}}
\mathcal{L}(\vec{x}^{(i)})
\end{align}

where $\mathcal{L}(x^{(i)})= ||\hat{\vec{y}}^{(i)}-\tilde{\vec{y}}^{(i)}||^2 $ is the cost function
of randomly chosen $m'$ training inputs $x^{(i)}$ and $\eta$ is the learning
rate. $w'_{ij}$ and $b'_{i}$ are updated weights and biases.

\subsection{Training FFNN for QST and QPT}

 \subsubsection{Data generation and training FFNN for QST and QPT}
 
In the case of QST, the training of FFNN is done using Eq.\ref{ch1_eq69}. The training data set is constructed by generating pure and mixed quantum states as follows,
\begin{equation} \label{e7}
c_i  = N(0,1)+i\,N(0,1)
\end{equation} 
\begin{equation} \label{ann_e8}
R = N(0,1,\left[2^n,2^n \right]) + i\, N(0,1,\left[2^n,2^n \right])
\end{equation}  
 where $N(0,1)$ in Eq.\ref{e7} is random number generated normally with zero mean and unit variance. For '$n$'-qubit pure states we generate $2^n$ random complex numbers $c_i$ given in Eq.\ref{e7} and normalize it. In Eq.\ref{ann_e8} $ N(0,1,\left[2^n,2^n \right])$ is  $2^n \times 2^n$  random matrix distributed normally with zero mean and unit variance. The corresponding mixed state density matrix $\rho_{\rm{mix}}$ is constructed as $\rho_{\rm{mix}} = \frac{R R^{\dagger}}{\rm{Tr}(R R^{\dagger})}$. After generating density matrices $\rho_i$, corresponding $\vec{b}_i$ are computed.  Training elements $\lbrace \vec{b}_{i}, \vec{\rho}_{i} \rbrace$ will then used to train the FFNN model given in Fig.\ref{flowchart} where $\vec{b}_{i}$ acts as input to FFNN and $\vec{\rho}_{i}$ is corresponding labeled output.
 
\begin{table}[t]
\centering
\caption{The average state fidelity over 3000 2-qubit test states for  $M_{data} = 20$ for different training size and different numbers of epochs. }
\setlength{\tabcolsep}{15pt} 
\renewcommand{\arraystretch}{1}
\begin{tabular}{c | c | c | c}
\hline 
\hline
Training data & \multicolumn{3}{c}{Epochs} \\
size & 50 & 100 & 150\\
\hline
500 & 0.8290 & 0.9176 & 0.9224 \\
1000 & 0.9244 & 0.9287 & 0.9298 \\
5000 & 0.9344 & 0.9379 & 0.9389 \\
10000 & 0.9378 & 0.9390 & 0.9400 \\
20000 & 0.9394 & 0.9414 & 0.9409 \\
80000 & 0.9426 & 0.9429 & 0.9422 \\
\hline \end{tabular}
 \label{2qst_epoch}
\end{table}
\begin{table}[t]
\centering
\caption{The average state fidelity over 3000 3-qubit test states for  $M_{data} = 120$ for different training size and different numbers of epochs.}
\setlength{\tabcolsep}{15pt} 
\renewcommand{\arraystretch}{1}
\begin{tabular}{c | c | c | c}
\hline 
\hline
Training data & \multicolumn{3}{c}{Epochs} \\
size & 50 & 100 & 150\\
\hline
 500 & 0.6944 & 0.8507 & 0.8716 \\
1000 & 0.8793 & 0.8994 & 0.9025 \\
5000 & 0.9231 & 0.9262 & 0.9285 \\
10000 & 0.9278 & 0.9321 & 0.9332 \\
20000 & 0.9333 & 0.9362 & 0.9393 \\
80000 & 0.9413 & 0.9433 & 0.9432 \\
\hline \end{tabular}
\label{3qst_epoch}
\end{table}
\begin{table}[t]
\centering
\caption{The average process fidelity over 3000 2-qubit test processes for  $M_{data} = 200$ for different training size and different numbers of epochs.}
\setlength{\tabcolsep}{15pt} 
\renewcommand{\arraystretch}{1}
\begin{tabular}{c | c | c | c}
\hline 
\hline
Training data & \multicolumn{3}{c}{Epochs} \\
size & 50 & 100 & 150\\
\hline
 500 & 0.4904 & 0.5421 & 0.5512 \\
2000 & 0.6872 & 0.7090 & 0.7202 \\
15000 & 0.7947 & 0.8128 & 0.8218 \\
20000 & 0.8047 & 0.8203 & 0.8295 \\
50000 & 0.8305 & 0.8482 & 0.8598 \\
80000 & 0.8441 & 0.8617 & 0.8691 \\
\hline \end{tabular}
\label{2qpt_epoch}
\end{table}

 Similarly, in case of n-qubit QPT, training of FFNN  is done using Eq.\ref{ch1_eq73} and \ref{ch1_eq75}. The training data set is constructed by randomly generating set of unitary operators. After generating the operator, it is then acted on the input states $ \rho_{in} = \lbrace |0\rangle, |1\rangle, \frac{\vert 0 \rangle + \vert 1 \rangle} {\sqrt{2}},\frac{\vert 0 \rangle + i\vert 1 \rangle} {\sqrt{2}} \rbrace^{\otimes n}$ and get $\rho_{out} = U \rho_{in} U^{\dagger}$. Stacking all $\rho_{out}$s we form $\vec{\lambda}$. Then using Eq.\ref{ch1_eq75}, $\vec{\chi}$ is calculated characterizing given unitary operator $U$. Training elements $\lbrace \vec{\lambda}_{i}, \vec{\chi}_{i} \rbrace$ will then used to train the FFNN model where $\vec{\lambda}_{i}$ acts as input to FFNN and $\vec{\chi}_{i}$ is corresponding labeled output.

For carrying out Quantum State Tomography (QST) and Quantum Process Tomography (QPT) using an FFNN with a reduced data set of size $m$, fractional input vectors $\vec{b}{m}$ and $\vec{\lambda}{m}$ are formulated. This is achieved by randomly selecting $m$ elements from the input vectors, while the remaining elements are assigned a value of 0 (zero padding). Subsequently, the reduced input vectors, along with their corresponding labeled output vectors, are employed for training the FFNN.

 For both the tasks of QST and QPT, the activation function used was LeakyReLU ($\alpha=0.5$) for hidden layers,
  \begin{equation} 
\begin{split}
\text{LeakyReLU(x)} = & x\, ; \, x>0 \\
 = & \alpha x\, ; \, x<0
\end{split}
\end{equation}
and linear activation function was used for output layer. Cosine similarity loss function,
$\mathcal{L} = \arccos \left( \frac{\hat{\vec{y}} . \tilde{\vec{y}}}{ ||\hat{\vec{y}}||.||\tilde{\vec{y}}|| } \right)$
was used and adagrad($\eta=0.5$) optimizer with learning rate, $\eta$ was used to train the network. Adagrad optimizer adapts the learning rate  relative to how frequently a parameter gets updated during training.
The average fidelity of 3000 2 and 3-qubit test states and 2-qubit processes for different training size and different number of epochs are shown in the Tables.\ref{2qst_epoch}, \ref{3qst_epoch} and \ref{2qpt_epoch} respectively. For both the tasks of QST and QPT, the training data size has been set to 80000, and 150 epochs have been selected.

\section{Experimental results and analysis} \label{sec3}
This study involves applying FFNN to experimentally measured noisy data obtained from NMR. The objective is to perform comprehensive QST for both 2-qubit and 3-qubit quantum states, along with full QPT for various 2-qubit quantum processes. These tasks are executed utilizing a diminished data set. In all the cases, performance of FFNN is evaluated by computing average state (process) fidelity. 
%

\subsection{2- and 3-qubit QST on NMR via FFNN} \label{secqst}

\begin{figure}
\centering
\includegraphics[angle=0,scale=1]{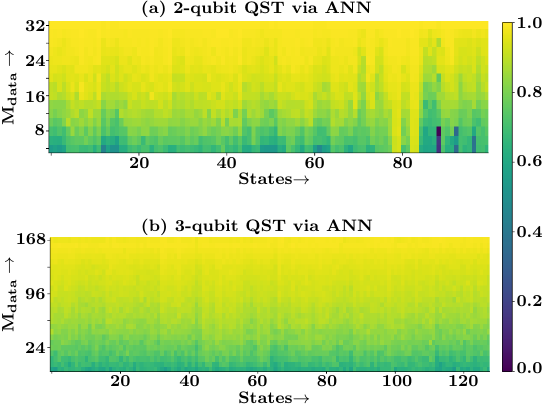} 
\caption{ (Color online) The graph illustrates the experimental state fidelity ($\bar{\mathcal{F}}$) of the FFNN model and the linear inversion method with respect to the dataset size ($M_{data}$).  The panel (a) represents 100 2-qubit states, while (b) represents 128 3-qubit states. The x-axis represents the numbering of the states, while the y-axis corresponds to the dataset size ($M_{data}$). The legend on the graph provides the fidelity values for reference.  } 
\label{qst}
\end{figure}

\begin{figure}
\centering
\includegraphics[angle=0,scale=1.1]{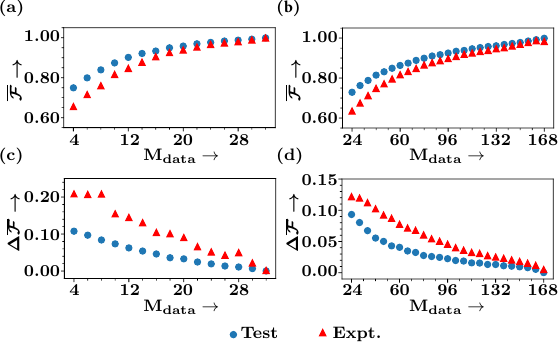} 
\caption{ (Color online) In the first row graphs represents average state fidelity ($\bar{\mathcal{F}}$) obtained via FFNN method vs size of dataset ($M_{data}$) while in second row corresponding  standard deviation $\Delta \bar{\mathcal{F}}$ are given. Plots (a) and (c) given in first column correspond to 2-qubit system while plots (b) and (d) correspond to 3-qubit system. For test data-set, average is calculated on 3000 states while for experimental data-set $\bar{\mathcal{F}}$ and $\Delta \bar{\mathcal{F}}$ are calculated on randomly considering $M_{data}$ elements 50 times for 100 2-qubit states and 128 3-qubit states each.  }
\label{avgqst}
\end{figure}

\begin{figure}
\centering
\includegraphics[angle=0,scale=1.1]{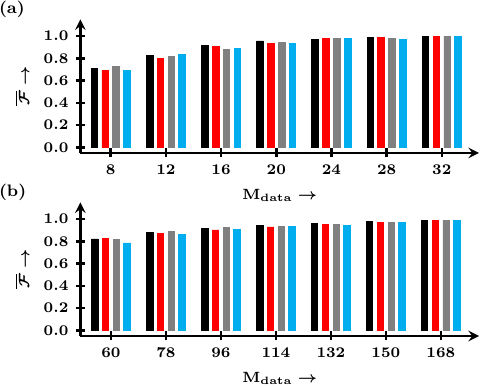} 
\caption{(Color online)  Experimental state fidelity ($\bar{\mathcal{F}}$) vs size of dataset ($M_{data}$) of entangled states is given. Part (a) corresponds to 2-qubit bell states while part (b) corresponds to 3-qubit GHZ and bi-separable states. $\bar{\mathcal{F}}$ is calculated between respective density matrices predicted via FFNN and density matrices obtained via standard linear inversion for reduced data set of a size $M_{\rm data}$ }
\label{bell_ghz}
\end{figure}

As described in previous chapters, the QST data is obtained by application various tomographic pulses followed by detection. In the case of 2-qubit system, each measurement produces eight elements of vector $\vec{b}$. For full QST, the dimension of vector $\vec{b}$ is $33 \times 1$ (32 from tomographic pulses and 1 is from unit trace condition). To demonstrate the performance of the FFNN model, 100 two-qubit states were experimentally prepared using different preparation settings. Full QST was then conducted using the FFNN model, followed by the computation of the average fidelity between the density matrix predicted through FFNN and the one obtained using the standard linear inversion method.


Similarly, three-qubit QST is done using seven tomographically complete unitary rotations. Each measurement produces 12 spectrums (4 per nucleus). Since two elements can be extracted from single spectrum, the size of full input vector $\vec{b}$ will be $169 \times 1$. 
To demonstrate the performance of the FFNN model, 128 three-qubit states were experimentally prepared using different preparation settings. Full QST was subsequently conducted using the FFNN model. The average fidelity between the density matrix predicted by the FFNN and that computed through the standard linear inversion method was calculated. Additionally, full QST of maximally entangled 2-qubit Bell states and 3-qubit GHZ and bi-separable states was carried out using a reduced data set, employing the FFNN model.

As a illustration, flowchart representing quantum state tomography of 2-qubit bell state $\vert B_1 \rangle = (\vert 00 \rangle + \vert 11 \rangle)/\sqrt{2} $ via FFNN using  experimental data obtained on NMR quantum information processor is given in Fig.\ref{flowchart}. In the first step, initial state $\rho_{in}$ corresponding to the bell state $ \vert B_1 \rangle$ is experimentally prepared. In second block step the NMR signals (frequency domain) obtained by applying tomographic pulse $IY$ on initial state $\rho_{in}$ followed by detection are given where first and second NMR signals correspond to $^1$H nucleus and third and forth NMR signals correspond to $^{13}$C nucleus. Each signal consists of two spectra giving total of 8 spectra for given tomographic measurement. In the third step, reduced vector $\vec{b}$ of a size $8 \times 1$ is constructed using spectral intensities of 8 NMR spectra given in the second block. Then the elements of vector $\vec{b}$ are fed into the input layer of FFNN as $\vec{b}_1 = x_1$, $\vec{b}_2 = x_2$,... and so on, where $x_i$ is numerical value assign to $i$th neuron in input layer. At the output layer of the FFNN, the elements $y_i$ are obtained, which then form the vector $\vec{\rho}$. In the final stage, the density matrix $\rho$ is constructed in the last block. Note that each output element $y_i$ is a real number and the imaginary values in the $\rho$ are stored into the two output neurons and further can be extracted as $\rho_{mn} = y_p + i y_q$.  The tomographs $\rm{Re}(\rho)$ and $\rm{Im}(\rho)$ in the last block represent the real and imaginary part of density matrix predicted by FFNN using data set of a size $M_{data} = 8$ obtained on NMR for given tomographic pulse $IY$.  The flowchart given in Fig.\ref{flowchart} also works for QPT task where in the first step we will have initial state $\Lambda(\rho)$ and rest of the procedure will be same.

In the case of QST the FFNN model is trained on 80,000 states. 
For both the tasks of 2-qubit and 3-qubit QST, three hidden layers have been employed. These layers consist of 100, 100, and 50 neurons for the 2-qubit case, and 300, 200, and 100 neurons for the 3-qubit scenario.  The performance of trained FFNN is shown in Fig.\ref{qst} and \ref{avgqst}. In the part (a) of Fig.\ref{qst}, the fidelity between density matrices obtained via FFNN and standard linear inversion method of individual 100 two-qubit experimental quantum states is given. Y-axis denotes the reduced input vector of size $M_{data}$ and on X-axis the quantum states are numbered in the step of 20. Similarly, in the part (b) of Fig.\ref{qst}, the fidelity between density matrices obtained via FFNN and standard linear inversion method of individual 128 three-qubit experimental quantum states is given.

Furthermore, in Fig.\ref{avgqst},  the performance of FFNN is evaluated by means of average state fidelity $\mathcal{\bar{F}}$ calculated over set of states (test/experimental). Plots (a) and (c) given in the first column correspond to 2-qubit QST while plots (b) and (d) given in the second column correspond to 3-qubit QST. In the plots (a) and (b) given in Fig.\ref{avgqst}, y-axis denotes the average fidelity $\mathcal{\bar{F}}$ calculated over set of states (test/experimental) and the x-axis denotes reduced size $M_{data}$ of input vector which was fed into FFNN. For a given value of $M_{data}$, the average fidelity $\mathcal{\bar{F}}_i=\frac{1}{50}\sum_{n=1}^{50} \mathcal{F}_n$ of given quantum state $\rho_i$ predicted via FFNN is calculated by randomly selecting $M_{data}$ elements from corresponding full input vector $\vec{b}$ for 50 times. In the case of test data set denoted by solid blue circle, the performance of FFNN is evaluated by means of average state fidelity $\mathcal{\bar{F}} = \frac{1}{3000}\sum_{n=1}^{3000} \mathcal{\bar{F}}_n$ calculated over 3000 2-qubit and 3-qubit test states while in the case of experimental data set denoted by solid red triangle, $\mathcal{\bar{F}}$ is calculated over 100 two-qubit and 128 three-qubit experimental states.  In the plots (c) and (d), the y-axis denotes standard deviation $\sigma$ in average state fidelity $\mathcal{\bar{F}}$ computed as, 
\begin{equation}
 \sigma = \sqrt{\frac{\sum_{i=1}^{N} (\mathcal{\bar{F}}_i-\bar{\mathcal{F}})^2}{N-1}}
 \label{e10}
 \end{equation}
where $N=3000$ for test states for both two- and three-qubit case while for experimental data set, $N = 100$ for two-qubit case and $N = 128$ for three-qubit case.
\begin{figure}[t]
\centering
\includegraphics[angle=0,scale=1.1]{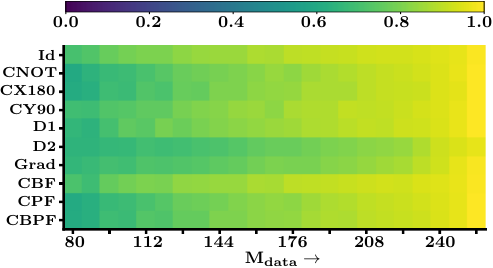} 
\caption{ (Color online) Experimental process fidelity ($\bar{\mathcal{F}}$) vs data elements ($M_{data}$). $M_{data}$ is given on x-axis while experimental quantum processes/gates are given on y-axis. The legend represents the fidelity value. } 
\label{qpt}
\end{figure}

From Fig.\ref{avgqst} one can see that, for test data set, the FFNN model is able to predict the 2-qubit unknown test quantum state with average fidelity $\mathcal{\bar{F}} \geq 0.8392 \pm 0.084$ for reduced data set of a size $M_{data} \geq 8$ while in the case of 3-qubit unknown test quantum state average fidelity turned out to $\mathcal{\bar{F}} \geq 0.8630 \pm 0.0407$ for reduced data set of a size $M_{data} \geq 60$. Similarly, for experimental quantum states, the FFNN model is able to predict the 2-qubit quantum state with the average fidelity $\mathcal{\bar{F}} \geq 0.8466 \pm 0.1450$ for reduced data set of a size $M_{data} \geq 12$ while in 3-qubit case the FFNN is able to predict the unknown quantum state with average fidelity $\mathcal{\bar{F}} \geq 0.8327 \pm 0.0716$ for reduced data set of a size $M_{data} \geq 60$. For full input vector $\vec{b}$, the average fidelity calculated over 3000 two and three-qubit test quantum states turns out to be $\mathcal{\bar{F}} = 0.9993 $ and $\mathcal{\bar{F}} = 0.9989 $ respectively 
 whereas average fidelity calculated over 100 two-qubit and 128 three-qubit experimental quantum states turns out to be $\mathcal{\bar{F}} = 0.9983 $ and $\mathcal{\bar{F}} = 0.9833 $ respectively.

Furthermore, the FFNN model was also utilized for the analysis of two-qubit maximally entangled Bell states, as well as three-qubit GHZ and bi-separable states. In Fig.\ref{bell_ghz} the experimental fidelities $\mathcal{F(\rho_{\rm \small FFNN},\rho_{\rm \small STD})}$ of two-qubit Bell states and three-qubit GHZ and bi-separable states are calculated between respective density matrices predicted via FFNN and density matrices obtained via standard linear inversion for reduced data set of a size $M_{\rm data}$. In the part (a) of Fig.\ref{bell_ghz} black, red, gray and blue bars correspond to Bell states $\vert B_1 \rangle = (\vert 00 \rangle + \vert 11 \rangle)/\sqrt{2} $, $\vert B_2 \rangle = (\vert 01 \rangle - \vert 10 \rangle)/\sqrt{2} $, $\vert B_3 \rangle = (\vert 00 \rangle - \vert 11 \rangle)/\sqrt{2} $ and $\vert B_4 \rangle = (\vert 01 \rangle + \vert 10 \rangle)/\sqrt{2} $ respectively. Similarly in part (b) black and red bars correspond to three-qubit GHZ states $\vert \psi_1 \rangle = (\vert 000 \rangle + \vert 111 \rangle)/\sqrt{2}$ and $\vert \psi_2 \rangle = (\vert 010 \rangle + \vert 101 \rangle)/\sqrt{2} $ while gray and blue bars correspond to three-qubit bi-separable states $\vert \psi_3 \rangle = (\vert 000 \rangle + \vert 001 \rangle + \vert 110 \rangle + \vert 111 \rangle)/2 $ and $\vert \psi_4 \rangle = (\vert 000 \rangle + \vert 010 \rangle + \vert 101 \rangle + \vert 111 \rangle)/2 $ respectively. The bar plots given in Fig.\ref{bell_ghz}, clearly shows that our FFNN model is able to predict the two and three-qubit entangled states with very high fidelity for reduced data set.

\subsection{2-qubit QPT on NMR via FFNN} \label{qptsec}

\begin{figure}[t]
\centering
\includegraphics[angle=0,scale=1]{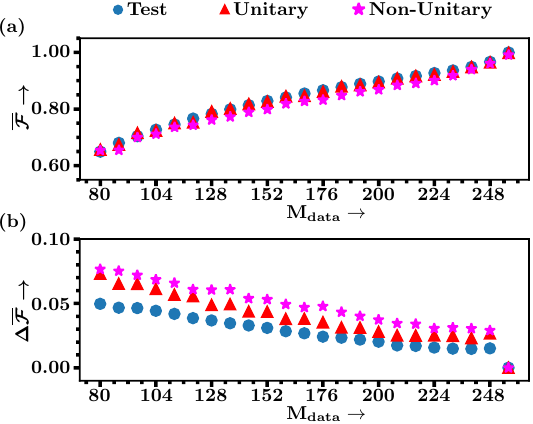} 
\caption{ (Color online) Panel (a) represents average process fidelity ($\bar{\mathcal{F}}$) obtained via FFNN method vs size of dataset ($M_{data}$) while panel (b) corresponds to standard deviation $\Delta \bar{\mathcal{F}}$. For test data-set, denoted by solid blue circle, average is calculated on 3000 processes while for experimental unitary and non-unitary processes, denoted by solid red triangle and solid magenta star respectively, $\bar{\mathcal{F}}$ and $\Delta \bar{\mathcal{F}}$ are calculated on randomly considering $M_{data}$ elements 300 times for 4 unitary quantum gates and 6 non-unitary processes. } 
\label{avgqpt}
\end{figure}
The standard QPT of given quantum process uses Kraus operator representation given in Eq.\ref{ch1_eq23}.  For 2-qubit QPT on NMR, the complete set of linearly independent input quantum states which need to be prepared is given by: $ \rho_{in} = \lbrace |0\rangle, |1\rangle, \frac{\vert 0 \rangle + \vert 1 \rangle} {\sqrt{2}},\frac{\vert 0 \rangle + i\vert 1 \rangle} {\sqrt{2}} \rbrace^{\otimes 2}$. After preparing given input state, the quantum process which is to be tomographed is applied on input state. Then QST of output state is performed using tomographic complete set of unitary rotation operators given in subsection.\ref{secqst}. The input vector $\vec{\lambda}$ is then constructed using tomographic data of output states. For all input quantum states $\rho_{in}$, the element wise comparison of output density matrix $\Lambda(\rho_{in})$ and matrices $E_m \rho_{in} E_n^{\dagger}$ yields the column vector $\vec{\lambda}$ of size $256 \times 1$.   
\begin{table}[H]
\centering
\caption{\label{ann_qptfid} Experimental fidelities $\mathcal{F(\chi_{\rm \small FFNN},\chi_{\rm \small STD})}$ are given. $\chi_{\rm \small FFNN}$ is process matrix corresponding given 2-qubit quantum process predicted via FFNN using full data set and $\chi_{\rm \small STD}$ is process matrix corresponding given quantum process obtained via standard linear inversion method. }
\setlength{\tabcolsep}{3.5pt} 
\renewcommand{\arraystretch}{1}
\begin{tabular}{c c c c}
\hline 
\hline
Non-unitary process & ~~$\mathcal{\bar{F}}$~~& ~~Unitary process~~ & ~~$\mathcal{\bar{F}}$~~~~\\
\hline 
D1  & 0.9987 & Test & 0.9997  ~~~\\ 
D2 & 0.9635  & Identity & 0.9943  ~~~\\ 
Grad & 0.9917 & CNOT & 0.9996  ~~~\\
CBF & 0.9943 & CX180 & 0.9996  ~~~\\
CPF & 0.9996 & CY90 & 0.9996 ~~~\\
CBPF & 0.9996 &  &   ~~~\\

\hline \end{tabular}
\end{table}
 
In the case of 2-qubit QPT, the FFNN model is trained on 80,000 synthetic quantum processes using reduced data set. For 2-qubit QPT, three hidden layers have been employed, consisting of 600, 400, and 300 neurons, respectively. The performance of trained FFNN is evaluated using 3000 test and 10 experimentally implemented quantum processes on NMR. The FFNN results for QPT of various 2-qubit experimental quantum processes are shown in Fig.\ref{qpt}. The quality of FFNN is evaluated by means of average process fidelity $\mathcal{\bar{F}(\chi_{\rm \small FFNN},\chi_{\rm \small STD})}$ of given quantum process where $\chi_{\rm \small FFNN}$ is process matrix predicted by FFNN using reduced data set of given size $M_{data}$ and $\chi_{\rm \small STD}$ is process matrix obtained via standard linear inversion method using full data set. In the part (a), the performance of FFNN is evaluated on 3000 2-qubit test quantum processes where y-axis denotes the average fidelity $\mathcal{\bar{F}} = \frac{1}{3000} \sum_{n=1}^{3000} \mathcal{\bar{F}}_n$, where $\mathcal{\bar{F}}_n = \frac{1}{300} \sum_{i=1}^{300} \mathcal{\bar{F}}_i $ is average fidelity of $n$th test quantum process calculated by randomly constructing input vector of given size for 300 times. The standard deviation in average fidelity $\mathcal{\bar{F}}$ is calculated using Eq.\ref{e10} over 3000 quantum processes. For experimental data set, the FFNN model is applied to perform two-qubit QPT for three different scenarios: i) unitary quantum gates implemented on NMR, ii) intrinsic non-unitary processes in NMR involving natural decoherence processes (free evolution for $D1 = 0.05$ sec and $D2 = 0.5$ sec) and magnetic field gradient pulse operation and iii) experimentally simulated correlated bit flip, correlated phase flip and correlated bit+phase flip error channels using duality algorithm on NMR.  For first two scenarios, the QPT data is collected using standard QPT procedure mentioned above while data collection for third scenario is briefly described in the following subsection.

\subsubsection{Experimental simulation of fully correlated error channels and their  reconstruction via ANN on NMR}

Here, a brief description of the duality simulation algorithm (DSA) is provided for the case of fully correlated two-qubit error channels: CBF, CPF and CBPF. Later the FFNN model is employed to fully characterize these channels. DSA allows to simulate arbitrary dynamics of open quantum system in single experiments where the ancilla system is required having dimension equal to the total number of Kraus operators characterizing the given quantum channel which is to be simulated. The arbitrary quantum channel having $d$ number of Kraus operators can be simulated via DSA using unitary operations, $V$, $W$, and control operation $U_c = \sum_{i=0}^{d-1} \vert i \rangle \langle i \vert \otimes U_i $ such that the following condition is satisfied,
\begin{equation}
E_{k}=\sum_{i=0}^{d-1} W_{k i} V_{i 0} U_{i}  \quad (k = 0,1,2,...,d-1)
\label{dsa}
\end{equation} 
where $E_k$ is Kraus operator.  $V_{i0}$ and $W_{ki}$ are the elements of $V$ and $W$ respectively. The generalized quantum circuit for DSA is given in the paper\cite{xin-pra-2017} where initial state of the system is encoded as $ \vert 0 \rangle_{a} \otimes \vert \psi \rangle_{s} $ which is then acted upon by $V \otimes I$ followed by $U_c$ and then $W \otimes I$ and at last measurement is performed on system qubits. 

\begin{figure}[t]
\centering
\includegraphics[angle=0,scale=1.1]{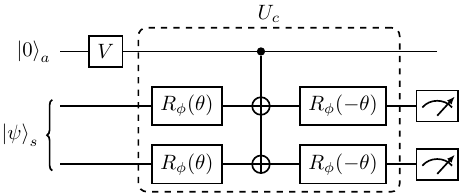} 
\caption{ (Color online) General quantum circuit to simulate the action of error channels: correlated bit flip, correlated phase flip  and correlated bit+phase flip.  } 
\label{dsackt}
\end{figure}

The two-qubit CBF, CPF and CBPF channels are characterized using two Kraus operators given as follows:
\begin{align}
\text{CBF} &: E_0 = \sqrt{1-p} I^{ \otimes 2}, \quad E_1 = \sqrt{p} \sigma_x^{ \otimes 2}  \nonumber \\
\text{CPF} &: E_0 = \sqrt{1-p} I^{ \otimes 2}, \quad E_1 = \sqrt{p} \sigma_z^{ \otimes 2}  \nonumber \\
\text{CBPF} &: E_0 = \sqrt{1-p} I^{ \otimes 2}, \quad E_1 = \sqrt{p} \sigma_y^{ \otimes 2}  \nonumber 
\end{align}
where $p$ is the error parameter or can be interpreted as probability with which the state of the system is affected by given error channel. For $p = 0$ the state of the system is unaffected and for $p=1$ the state of the system is maximally affected by the given error channel. Since all three channels have only two Kraus operators, they can be simulated using single ancilla qubit. For all three channels, one can set $V = \left(\begin{array}{cc}
\sqrt{1-p} & -\sqrt{p} \\
\sqrt{p} & \sqrt{1-p}
\end{array}\right)$, $W = I$ and $U_0  = I \otimes I$. And $U_1$ for CBF, CPF and CBPF are set to be $\sigma_x \otimes \sigma_x$, $\sigma_z \otimes \sigma_z$ and $\sigma_y \otimes \sigma_y$ respectively such that the condition given in Eq.\ref{dsa} is satisfied.  Note that $V$ can be interpreted as rotation about y-axis by angle $\theta$ such that, $p = Sin^2(\frac{\theta}{2})$. The single generalized quantum circuit for DSA to simulate all three error channels is given in the Fig.\ref{dsackt}. For CBF channel $U_c$ turns out to be Control-Not-Not gate (C-$(\sigma_x \otimes \sigma_x)$) where value of $\theta$ in Fig.\ref{dsackt} is 0. For CPF the value of $\theta$ is $\frac{\pi}{2}$ and $\phi$ is $-y$ (axis of rotation) while for CBPF channel $\theta$ is $\frac{\pi}{2}$ and $\phi$ is $z$ (axis of rotation). The output from the tomographic measurements on system qubits forms the column vector 
$\overrightarrow{\lambda}$. For given value of $p$, the full vector $\overrightarrow{\lambda}$ can be constructed by preparing the system qubits into complete set of linearly independent input states given in Sec.\ref{qptsec}. 
The experimental fidelity obtained via FFNN of individual quantum process is given in the Fig.\ref{qpt} where the data $M_{data}$ is given on x-axis and the experimental quantum process is given on y-axis. The legend color denotes the average process fidelity $\mathcal{\bar{F}} = \frac{1}{300} \sum_{i=1}^{300}\mathcal{\bar{F}}_i $ which is calculated by randomly constructing reduced input vector of given size $M_{data}$ for 300 times. In the case of experimental unitary quantum gates: Identity, CX180, CNOT and CY90 the FFNN is able to predict the corresponding process matrix with average fidelity 
 $\mathcal{\bar{F}} = 0.8767 \pm 0.0356, 0.8216 \pm 0.0463, 0.8314 \pm 0.0387 $ and $0.8489 \pm 0.0315$ using the reduced data set of size 160 respectively. Similarly in the case of intrinsic non-unitary quantum processes: D1, D2 and gradient pulse, the FFNN is able to predict corresponding process matrix with average fidelity of
$\mathcal{\bar{F}} = 0.8373 \pm 0.0381, 0.7607 \pm 0.0690$ and $0.7858 \pm 0.0703$ using the reduced data set of size 160 respectively. Furthermore, for experimentally simulated correlated error channels CBF, CPF and CBPF the average fidelity turned out to be
 $\mathcal{\bar{F}} = 0.8738 \pm 0.0366, 0.8272 \pm 0.0403$ and $ 0.8273 \pm 0.0416 $ using the reduced data set of size 160 respectively. For full data set exact numerical values of process fidelity obtained via FFNN are shown in the Table.\ref{ann_qptfid}. 

In addition to that, Fig.\ref{avgqpt} also represents the performance of FFNN model where the average fidelity is calculated over set of processes for given value of $M_{data}$. For test data set the $ \mathcal{\bar{F}} $ is calculated over 3000 test processes where as in the case of experimental data set, $ \mathcal{\bar{F}} $ is computed over 4 unitary  and 6 non-unitary processes. Plot (a) given in Fig.\ref{avgqpt} clearly shows that the FFNN model is able to predict the unitary as well as non-unitary quantum processes from noisy experimental reduced data set. In the part (b) of Fig.\ref{avgqpt}, the standard deviation is calculated using Eq.\ref{e10}. E.g.,
for $M_{data} = 160$, FFNN is able to predict the arbitrary test process with $ \mathcal{\bar{F}} = 0.8411 \pm 0.0284 $ whereas experimental unitary and non-unitary process can be obtained with $ \mathcal{\bar{F}} = 0.8447 \pm 0.03803$ and $0.8187 \pm 0.04932$ respectively. That is, depending upon the desired accuracy and precision the value of $M_{data}$ can be set to perform QPT of given quantum process via FFNN.

\section{Conclusions}\label{annqpt_sec4}
The application of the Feed Forward Neural Network architecture to experimentally measured data from NMR was performed. This was done to reconstruct the density and process matrix characterizing the given quantum state and process, respectively. These reconstructions were carried out using a reduced data set in various scenarios. For both QST and QPT, the FFNN is able to predict the given quantum state and process with very high fidelity using a sufficiently small input data set. The experimental results indicate that the FFNN can be experimentally more efficient than the standard linear inversion method as it requires sufficiently less number of experiments to perform full QST and QPT. The results presented in this chapter are available here: \href{https://journals.aps.org/pra/abstract/10.1103/PhysRevA.109.012402}{Phys. Rev. A 109, 012402 (2024)}
\chapter{Generalized algorithm and experimental demonstration of selective quantum process tomography on NMR and IBM quantum processor}\label{chapter_3QEntDet}

\section{Introduction} 

This chapter deals with the problem of the selective estimation of desired element of an unknown process matrix with required precision.  Till date, numerous QPT protocols have been proposed which certainly offer some advantages over standard
QPT but they still are not very useful when only certain
elements of the $\chi$ matrix need to be estimated. Typically most QPT protocols for selective estimation of an element of the process matrix require extra ancilla qubits which makes them experimentally expensive. Hence
much effort has recently focused on achieving a selective
estimation of elements of the $\chi$ matrix via a technique
called selective and efficient quantum process tomography
(SEQPT) without using any ancilla qubits\cite{paz-prl-08,paz-pra-09,paz-prl-10}.
The SEQPT without ancilla method interprets the elements of
the $\chi$ matrix as an average of the survival
probabilities of a certain quantum map; while the method
certainly has advantages over other existing schemes, it
still requires a large number of state preparations and
experimental settings to carry out element-wise complete process tomography.

In this study, a generalization of the SEQPT without ancilla method is proposed. This approach requires significantly fewer experimental settings compared to the SEQPT or standard QPT protocols. The utilization of local measurements (LM)\cite{singh-pra-16} enables the selective estimation of the quantum process matrix to a desired precision. The scheme is referred to as local measurement-based selective and efficient quantum process tomography or modified SEQPT (MSQPT)\cite{gaikwad-pra-2018}. The scheme simplifies the QPT protocol as, in this local measurement-based approach, the detection settings remain unchanged while estimating different elements of the $\chi$ matrix. Efficiency is a key feature of this scheme, as it involves calculating the expectation values of specific Hermitian observables through the local in a predetermined set of quantum states. Experimental demonstration of the scheme is carried out by implementing it on a two-qubit NMR system. For demonstration, the no operation, controlled-NOT, and controlled-Hadamard gates are tomographed.


\section{Protocol for selective and efficient quantum process tomography 
using local measurements}\label{seqpt_sec1}

In this section, the standard SEQPT protocol is initially described, along with the obstacles and complexity associated with it, as well as its conventional method of implementation. Subsequently, the `\textit{modified} SEQPT (MSQPT)' protocol is introduced. 

 \subsection{Standard SEQPT protocol}
Consider a quantum system undergoing a general quantum
evolution represented by a completely positive (CP) map.
The action of such a  map on a quantum state $\rho$ via the
superoperator $\Lambda$ in the fixed set of basis operators $\lbrace E_i \rbrace$ is given in the Eq.\ref{ch1_eq73} as,
\[
    \Lambda(\rho)=\sum_{a,b=0}^{D^2-1} \chi_{ab} {E_a} \rho
{E_b}^\dagger 
\]
where the quantities $\chi_{ab}$ are the elements of a
matrix $\chi$ characterizing the given CP map $\Lambda$. For the SEQPT purpose, consider a
complete set of $D^2-1$ basis operators $\lbrace E_i\rbrace$
for a $D$-dimensional Hilbert space, 
satisfying the 
orthogonality conditions:
\begin{equation}
\textbf{Tr}(E_m E_n^\dagger)=D\delta_{mn} \quad {\rm and} 
\quad \textbf{Tr}(E_m)=D\delta_{m0}
\label{ortho}
\end{equation}

A major step towards the determination of $\chi_{ab}$
elements is to relate them to the quantities $F_{ab}$ called
the \textit{average survival probabilities} of a special quantum map\citep{paz-pra-09,paz-prl-11}:

\begin{equation}
F_{a b} \equiv \int d|\phi\rangle\left\langle\phi\left|\Lambda\left(E_{a}^{\dagger}|\phi\rangle\langle\phi| E_{b}\right)\right| \phi\right\rangle=\frac{D \chi_{a b}+\delta_{a b}}{(D+1)}
\label{asp}
\end{equation}
The average is taken over entire Hilbert space of the average survival probability of the special map $\Lambda_{ab}$ defined as $\Lambda_{ab}(\rho) = \Lambda(E_a^{\dagger} \rho E_b)$. (\ie the transformation $\rho\rightarrow \Lambda_{ab}(\rho)$ is obtained by first mapping $\rho$ to $E_a^{\dagger} \rho E_b$ and then applying the given channel $\Lambda$). From Eq.\ref{asp}, one can see that the the efficient estimation of $\chi_{ab}$ is equivalent to that of $F_{ab}$.

However, there are two major obstacles one has to deal with while estimating $F_{ab}$: i) The average over entire Hilbert space requires infinite number of state preparations and measurements and ii) the special quantum map $\Lambda_{ab}$ is not valid quantum process and can not be implemented in the lab. It has been shown that the first obstacle can be surmounted using special quantum states called \textit{`quantum 2-design states'}. Using set of quantum 2-design states one can transform continuous integral over entire Hilbert space into a discrete sum over finite set of 2-design states as follows\cite{dankert-pra-09}:

\begin{equation}
\label{e_3} F_{ab} = \frac{1}{K}
\sum_{j} \langle \phi_j \vert \Lambda (E_a^\dagger \vert
\phi_j \rangle \langle \phi_j \vert E_b)\vert \phi_j \rangle
\end{equation} 
Here a quantum 2-design set $S=\{ \vert \phi_j\rangle:
j=1,....,K \}$ of cardinality $K$ has been used to provide
a way to discretely sample the system Hilbert space so as to
avoid continuous integration over the entire space. In spite of that, the size of $S$ increases exponentially with the dimension of the Hilbert space, \ie $K = O(D^2)$. However, by randomly sampling over subset of size $N$, one can still estimate the $F_{ab}$ with the error that scales as, $\Delta F_{a b} \propto \sqrt{\frac{1}{N}\left(1-\frac{N-1}{K-1}\right)}$. For the case of $N \ll K$, the error approximately scale as, $1/\sqrt{N}$. Thus,
the precision fixes the required number of experiments.

The second obstacle can be surpassed by decomposing special map $\Lambda_{ab}$ into two CP maps $\Lambda_{ab}^{\pm}$.  To estimate real part of $F_{ab}$, consider the average survival probabilities $F_{ab}^{\pm}$ of two special quantum maps as\cite{paz-prl-11},
\begin{equation}
F_{a b}^{\pm}=\frac{1}{K}\sum_{j} \langle \phi_j \vert \Lambda ((E_a \pm E_b)^{\dagger} \vert \phi_j \rangle \langle \phi_j \vert (E_a \pm E_b)) \vert \phi_j \rangle
\end{equation}
 where $\Lambda_{ab}^{\pm} = \Lambda ((E_a \pm E_b)^{\dagger} \vert \phi_j \rangle \langle \phi_j \vert (E_a \pm E_b))$. One can see that, the states corresponding to $(E_a \pm E_b)^{\dagger} \vert \phi_j \rangle \langle \phi_j \vert (E_a \pm E_b)$ (represents valid density matrix upto some normalization) can be prepared in the lab. The real part of $F_{ab}$ can be calculated as, 
 \begin{equation}
 {\text {Re}}(F_{ab}) = \frac{F_{ab}^{+}-F_{ab}^{-}}{2}
 \label{reseqpt}
 \end{equation}
Similarly, for imaginary part of $F_{ab}$ the average survival probabilities $F_{ab}^{\pm}$ of corresponding CP maps $\Lambda_{ab}^{\pm}$ are given as,
\begin{equation}
F_{a b}^{\pm}=\frac{1}{K}\sum_{j} \langle \phi_j \vert \Lambda ((E_a \pm i E_b)^{\dagger} \vert \phi_j \rangle \langle \phi_j \vert (E_a \pm iE_b)) \vert \phi_j \rangle
\end{equation}
 where $\Lambda_{ab}^{\pm} = \Lambda ((E_a \pm i E_b)^{\dagger} \vert \phi_j \rangle \langle \phi_j \vert (E_a \pm i E_b))$. One can see that, the states corresponding to $(E_a \pm i E_b)^{\dagger} \vert \phi_j \rangle \langle \phi_j \vert (E_a \pm i E_b)$ (represents valid density matrix upto some normalization) can be prepared in the lab. The imaginary part of $F_{ab}$ can be calculated as, 
 \begin{equation}
 {\text{Im}}(F_{ab}) = \frac{F_{ab}^{+}-F_{ab}^{-}}{2}
 \label{iMSQPT}
 \end{equation}
Using Eq.\ref{reseqpt} and \ref{iMSQPT} $F_{ab}$ can be calculated and with it $\chi_{ab}$ can be estimated. However, one can clearly see that these
procedures involve using different experimental settings for
different values of $a$ and $b$ to prepare the required
state and a large number of experiments have to be performed in
order to achieve a high precision. Further, constructing and
implementing the corresponding unitary operators is a
challenging task. This procedure was implemented on an optics set-up demonstrating the first experimental demonstration of selective QPT\cite{paz-prl-11}. So we refer to it as standard SEQPT protocol. 

\subsection{Modified SEQPT}

A different approach is adopted to implement SEQPT, wherein a method involving the weighted average results of various experiments is utilized, analogous to the temporal averaging scheme used to obtain a pseudopure state\cite{vandersypen-prl-00,leung-rsl-98}. The computation of the expectation values of basis operators is accomplished through local measurements.
Eq.~(\ref{e_3}) can be rewritten in terms of density
operators corresponding to the quantum 2-design states $\rho_j=\vert
\phi_j \rangle \langle \phi_j\vert$ as:
\begin{equation}
\label{e_2.5}
F_{ab} \equiv \frac{1}{K} \sum_{j} \textbf{Tr} [ \rho_j
\Lambda (E_a^\dagger \rho_j E_b)]=\frac{
D\chi_{ab}+\delta_{ab}}{D+1}
\end{equation}
The basis operators  $\lbrace E_i \rbrace$ can be used to 
decompose the operator $E^{\dagger}_a \rho_j E_b$:
\begin{equation}
\label{e_2.6} 
E_a^\dagger \rho_j E_b=
\sum_{i} {^jc}_i^{ab}E_i 
\end{equation}  
where the coefficients $^jc_i^{ab}\in\mathbb{C}$ are
independent of the quantum process characterized by $\Lambda$,
and can be computed analytically using the orthogonality
condition: 
\begin{equation}
{^jc_i}^{ab}=
\frac{1}{D}\textbf{Tr}[(E_a^\dagger \rho_j E_b) E_i]
\label{coefficients}
\end{equation}
The superoperator $\Lambda$ is linear and hence can be
expanded as:
\begin{equation}
\label{e_2.7}
\Lambda (E_a^\dagger \rho_j E_b)=  \sum_{i} {^jc}_i^{ab} \Lambda(E_i)
\end{equation}
Using the above decomposition, Eq.(\ref{e_2.5}) can be rewritten as
\begin{equation}\label{e_2.8}
F_{ab}= \frac{1}{K} \sum_{j} \textbf{Tr}\left[\rho_j \sum_{i}
{^jc}_i^{ab} \Lambda(E_i)\right] 
\end{equation} 
Every  
basis operator $E_i$  (other than the first one which we
take proportional to identity) is a  
Hermitian operator with zero trace and can be interpreted as a
deviation density operator in NMR system\cite{oliveira-book-07}, and thus it can be
experimentally prepared in the lab.

The quantum process $\Lambda$ is permitted to act on the basis operator state $E_i$, resulting in $\Lambda (E_i)$ for each basis operator. Consequently, if the state $\Lambda (E_i)$ is experimentally tomographed, then one can use theoretically calculated coefficients
${^jc_i}^{ab}$ as per the Eq.~(\ref{coefficients}) and compute
$F_{ab}$ in Eq.~(\ref{e_2.8}). 
The
results from individual $E_i$'s weighted by $^jc_i^{ab}$ are added to
obtain the final result. However, the aim is to avoid the full state tomography of the state
$\Lambda(E_i)$. So, further decomposing $\rho_j$ as $\rho_j=\sum_{k}
{^je_k E_k}$ (with $^je_k\in\mathbb{R}$), and using the
linearity of trace, Eq.~(\ref{e_2.8}) reduces to 
\begin{equation}\label{e_2.10}
F_{ab}= \frac{1}{K} \sum_{j} \sum_{i,k} {}^je_k\, {}^jc^{ab}_i
\textbf{Tr} [E_k \Lambda(E_i)]
\end{equation}
where the coefficients ${}^je_k \,{}^jc^{ab}_i$ are process 
independent and can
be computed analytically. Rewriting them as $^j\beta^{ab}_{ki}$,
Eq. (\ref{e_2.10}) takes a simple form\cite{gaikwad-pra-2018}:
\begin{equation}\label{e_2.11}
F_{ab}= \frac{1}{K} \sum_{i,j,k} {^j\beta^{ab}_{ki}} \textbf{Tr}
\left[E_k \Lambda(E_i)\right]
\end{equation}
where $\textbf{Tr}[E_k \Lambda(E_i)] \equiv \langle E_k^i
\rangle$ is the expectation value of basis operator $E_k$ in
the state $\Lambda(E_i)$.  The information about the quantum
process is now
stored in the output state $\Lambda(E_i)$. To calculate a selective
element of $\chi$ matrix, all one needs to do is to calculate
expectation values of $E_k$ and take the weighted average of
these expectation values using the theoretically calculated 
coefficients $^j\beta^{ab}_{ki}$.
\begin{table}
\centering
\caption{\label{complex}
Comparison of experimental resources for different
protocols for the determination of a specific element of the $\chi$ matrix
for a two-qubit system.}
\begin{tabular}{c c c c }
\hline \hline
 &
QPT & 
SEQPT &
MSQPT~~~\\
\hline \hline
Preparations & $15$ & $80$ & $15$~~~\\
Readouts & $120$ & $240$ & $60$~~~\\
\hline
\end{tabular}
\end{table}
To determine $F_{ab}$,there is no necessity to conduct a comprehensive quantum state tomography of the output state $\Lambda(E_i)$ which constitutes a highly resource-intensive operation.  The expectation values $\langle
E^i_k \rangle$ can be determined by mapping them to local
expectation values of appropriate operators.  To demonstrate
this, standard Pauli basis set is chosen as $\lbrace E_i\rbrace$ which
for $N-$qubits is given as: $\{I^j, \sigma_x^j,
\sigma_y^j, \sigma_z^j \}$ for the $j$th qubit  and
taking all possible tensor products to form the set
$\{E_i\}$. The measurements of elements of the Pauli basis can
be measured via local measurements and in fact can be 
mapped to measurements of various $\sigma_z^j$ by applying 
certain fixed operations before measurement. This is
particularly suitable for NMR where such measurements can be
readily accomplished.
For a two-qubit system 
this map is given in Table~\ref{Map}
where the measurement of each member of the Pauli basis set is 
mapped to a measurement of certain local $z$ magnetizations.
This significantly simplifies the experimental complexity
of the SEQPT scheme and further named as MSQPT.

\begin{table}[t]
\centering
\caption{\label{Map}
Fifteen observables for two qubits mapped
to local \textit{z}-magnetization of
one of the qubit.
$\rho_k=U_k \Lambda(E_i)U_k^\dagger$ in order to calculate
$\langle E_k\rangle=\textbf{Tr}\left[E_k\Lambda(E_i)\right]$.}
\begin{tabular}{rp{12pt}l}
\hline \hline
\textrm{Observable Expectation}&&
\textrm{Unitary operator $U_k$}\\
\hline \hline
$\langle \sigma_{2x} \rangle$ = Tr[$\rho_{1}.\sigma_{2z}$]
&&
$U_{1}=\overline{Y}_2$  \\
$\langle \sigma_{2y} \rangle$ = Tr[$\rho_{2}.\sigma_{2z}$]
&&
$U_{2}=X_2$ \\ 
$\langle \sigma_{2z} \rangle$ = Tr[$\rho_{3}.\sigma_{2z}$] &&
$U_{3}$ = Identity  \\ 
$\langle \sigma_{1x} \rangle$ =
Tr$\left[\rho_{4}.\sigma_{1z}\right]$ &&
$U_{4}=\overline{Y}_1$ \\ 
$\langle \sigma_{1x}\sigma_{2x} \rangle$ = Tr[$\rho_{5}.
\sigma_{2z}$] && $U_5={\rm{CNOT~  }}Y_2Y_1$  \\ 
$\langle \sigma_{1x}\sigma_{2y} \rangle$ =
Tr[$\rho_{6}.\sigma_{2z}$] && $U_6={\rm{CNOT~  }}\overline{X}_2Y_1$
\\ 
$\langle \sigma_{1x}\sigma_{2z} \rangle$ =
Tr[$\rho_{7}.\sigma_{2z}$] && $U_7={\rm{CNOT~  }}\overline{Y}_1$ \\ 
$\langle \sigma_{1y} \rangle$ = Tr[$\rho_{8}.\sigma_{1z}$]
&&
$U_{8}=X_1$ \\ 
$\langle \sigma_{1y}\sigma_{2x} \rangle$ =
Tr[$\rho_{9}.\sigma_{2z}$] && $U_9={\rm{CNOT~  }}\overline{Y}_2X_1$
\\ 
$\langle \sigma_{1y}\sigma_{2y} \rangle$ =
Tr[$\rho_{10}.\sigma_{2z}$] &&
$U_{10}={\rm{CNOT~  }}\overline{X}_2\overline{X}_1$  \\ 
$\langle \sigma_{1y}\sigma_{2z} \rangle$ =
Tr[$\rho_{11}.\sigma_{2z}$] && $U_{11}={\rm{CNOT~  }}X_1$  \\ 
$\langle \sigma_{1z} \rangle$ = Tr[$\rho_{12}.\sigma_{1z}$]
&& $U_{12}$ = Identity \\ 
$\langle \sigma_{1z}\sigma_{2x} \rangle$ = 
Tr[$\rho_{13}.\sigma_{2z}$] && $U_{13}={\rm{CNOT~  }}\overline{Y}_2$  \\ 
$\langle \sigma_{1z}\sigma_{2y} \rangle$ = 
Tr[$\rho_{14}.\sigma_{2z}$] && $U_{14}={\rm{CNOT~  }}X_2$ \\ 
$\langle \sigma_{1z}\sigma_{2z} \rangle$ = 
Tr[$\rho_{15}.\sigma_{2z}$] && $U_{15}=\rm{CNOT~  }$ \\ 
\hline
\end{tabular}
\end{table}
A stepwise description of the 
experimental implementation of the MSQPT protocol 
to selectively determine the 
element $\chi_{ab}$ of the process matrix is as follows:
\begin{enumerate}
\item[(i)] Choose any state $\rho_j=\vert \phi_j\rangle
\langle \phi_j\vert$ from the
set of quantum 2-design and find the decomposition  of
$E_a^\dagger \rho_j E_b $ in terms of basis operators $E_i$.
\item[(ii)]
Experimentally prepare the quantum system in one of the  basis
states having non-vanishing coefficients ${}^jc^{ab}_i$ as per
Equation~(\ref{e_2.6}).
\item[(iii)] Apply the quantum channel $\Lambda$ to $E_i$ to get
the output state $\Lambda(E_i)$. 
\item[(iv)] Find the decomposition of the chosen state $\rho_j$
in terms of basis operators analytically and then
experimentally determine the expectation values of all those
$E_k$'s which have
non-vanishing coefficients, ${}^je_k$, using the local
measurement technique. 
\item[(v)] Repeat the procedure for all the states in the
chosen  quantum
2-design set.
\end{enumerate}
\begin{figure}[h]
\centering
\includegraphics[angle=0,scale=1.5]{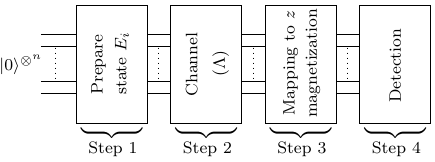}
\caption{The figure presents a block diagram illustrating the Multi-Step Quantum Process Tomography (MSQPT) protocol. The protocol involves several steps. Step 1 involves the preparation of the basis operator $E_i$ state.
In Step 2, a quantum channel $\Lambda$ is applied. Step 3 consists of mapping the basis operator to local measurements, followed by the measurement of the expectation values of Pauli $z$-operators on the subsystems. Finally, in Step 4, local detection takes place.}
\label{SEQPTlm}
\end{figure} 

The MSQPT protocol is schematically depicted in
Fig.~\ref{SEQPTlm}: the first step is to
prepare the basis state, followed by the action of the quantum
process. After the quantum process has acted on the basis
state the next step is to map the required measurements to
single-spin magnetization measurements, and finally
the single spin magnetization detection is carried out. The MSQPT scheme has two advantages: first, it is simpler
than the original scheme as one does not have to choose different
experimental settings for the estimation of each element of $\chi$ matrix, and, second, it involves fewer experiments. The
comparison of experimental resources required by different
protocols to determine a specific element of the $\chi$ matrix for
two-qubit systems is given in Table.\ref{complex}. The standard QPT
method implemented on two NMR qubits relies on the channel
action on number of input states. This requires state preparation settings ($\rho_{in}$) = 15, with the
number of tomographs required being 15. Since each tomograph requires eight readouts, the total number of readouts
required is $15 \times 8 = 120$. 

In the standard SEQPT protocol, the states to be prepared for estimating the real part of $\chi_{ab}$ are $(E_a \pm E_b)^{\dagger} \rho_j (E_a \pm E_b)$, where $j = 1$ to $20$ (2-design states).  The states to be prepared for
estimating the imaginary part of $\chi_{ab}$ are $(E_a \pm i E_b)^{\dagger} \rho_j (E_a \pm i E_b)$. The number of state preparation settings required to
obtain $\chi_{ab} = 80$ [20 for $(E_a + E_b)$ + 20 for $(E_a - E_b)$ + 20 for $(E_a + i E_b)$ + 20 for $(E_a - i E_b)$]. This method requires three readouts (the number of nonvanishing coefficients in the expansion of $\rho_j$) for each state, in order to obtain survival probabilities $\textbf{Tr}\lbrace \rho_j \Lambda [(E_a \pm E_b)^{\dagger} \vert \phi_j \rangle \langle \phi_j \vert (E_a \pm E_b)]  \rbrace$ and $\textbf{Tr}\lbrace \rho_j \Lambda [(E_a \pm i E_b)^{\dagger} \vert \phi_j \rangle \langle \phi_j \vert (E_a \pm i E_b)] \rbrace$. Hence the total number of readouts required required for the SEQPT method is $3 \times 80 = 240$. For the MSQPT protocol, the number of state preparation settings are $15$, while the number of readouts
required for each state preparation is four. Hence the total
number of readouts required for the MSQPT method is $4 \times 15 = 60$.

\subsubsection*{Quantum 2-design set using mutually unbiased basis}
\label{2-design}

To enable the experimental implementation of MSQPT, one of the essential prerequisites is a quantum 2-design set denoted as $S$. Fortunately, there are available algorithms that can be utilized to construct such a set\cite{dankert-pra-09,paz-pra-09}. One approach involves identifying a complete set of mutually unbiased basis (MUBs) states. In a system with a state space of dimension $D$, if $D$ is a prime number or a power of a prime number, it will possess ($D+1$) MUBs\cite{bandyopadhyay-alg-02,lawrence-pra-02,
Klappenecker-book-04}. In the case of our two-qubit system with $D=2^2$, we can construct the quantum 2-design set by employing a complete set of MUBs, which amounts to five MUBs in this instance. The MUBs states satisfy the relation ${\vert \langle \phi_p^{B_k} \vert \phi_q^{B_l} \rangle\vert}^2=\frac{1}{D}$ for all $k \neq l$, where $B_k$ represents the labels of the basis sets and $\vert \phi_p \rangle$ denotes the elements within the basis set.

For a 2-qubit system with $D=4$, the complete set of MUBs constituting the states in the quantum 2-design set, in the computational basis, is provided below\cite{Klappenecker-book-04}:

\begin{equation*}
B_1=\left \lbrace \begin{pmatrix}
1\\0\\0\\0
\end{pmatrix},\begin{pmatrix}
0\\1\\0\\0
\end{pmatrix},\begin{pmatrix}
0\\0\\1\\0
\end{pmatrix},\begin{pmatrix}
0\\0\\0\\1
\end{pmatrix}\right \rbrace 
\end{equation*}
\begin{equation*}
B_2=\frac{1}{2}\left \lbrace \begin{pmatrix}
1\\1\\1\\1
\end{pmatrix},\begin{pmatrix}
1\\1\\-1\\-1
\end{pmatrix},\begin{pmatrix}
1\\-1\\-1\\1
\end{pmatrix},\begin{pmatrix}
1\\-1\\1\\-1
\end{pmatrix}\right \rbrace 
\end{equation*}
\begin{equation*}
B_3=\frac{1}{2} \left \lbrace \begin{pmatrix}
1\\i\\i\\-1
\end{pmatrix},\begin{pmatrix}
1\\-i\\i\\1
\end{pmatrix},\begin{pmatrix}
1\\i\\-i\\1
\end{pmatrix},\begin{pmatrix}
1\\-i\\-i\\-1
\end{pmatrix}\right \rbrace  
\end{equation*}
\begin{equation*}
B_4=\frac{1}{2} \left \lbrace\begin{pmatrix}
1\\-1\\-i\\-i
\end{pmatrix},\begin{pmatrix}
1\\-1\\i\\i
\end{pmatrix},\begin{pmatrix}
1\\1\\-i\\-i
\end{pmatrix},\begin{pmatrix}
1\\1\\-i\\i
\end{pmatrix}\right \rbrace 
\end{equation*}
\begin{equation}
B_5=\frac{1}{2}\left \lbrace \begin{pmatrix}
1\\-i\\-1\\-i
\end{pmatrix},\begin{pmatrix}
1\\-i\\1\\i
\end{pmatrix},\begin{pmatrix}
1\\i\\-1\\i
\end{pmatrix},\begin{pmatrix}
1\\i\\1\\-i
\end{pmatrix}\right \rbrace
\end{equation}
For example $\vert \phi_3^{B_1} \rangle $ is the third
element of $B_1$ basis set and the state is $\vert 10 \rangle$.
Also $B_1$ is the commonly used computational basis. All the
twenty states in the above defined MUBs comprise the quantum
2-design set $S$ for the present study. 
\begin{figure}
\centering
\includegraphics[angle=0,scale=1.5]{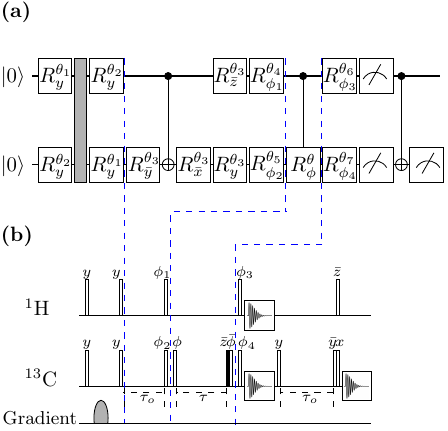}
\caption{(a)The provided quantum circuit depicts the implementation of the MSQPT. The circuit incorporates local unitary operations denoted as $R_{\phi}^\theta$, which are accomplished through rotations by an angle $\theta$ and a phase $\phi$. This particular quantum circuit utilises non-unitary operation  illustrated by rectangle filled with gray color.
(b)NMR implementation of MSQPT. The non-unitary operation in implemented using magnetic field gradient pulse.  } 
\label{ckt+seq}
\end{figure}
\section{NMR implementation of MSQPT}
\label{NMRseqpt}

A demonstration of the MSQPT protocol was conducted using an NMR quantum information processor. This demonstration involved three distinct 2-qubit unitary processes: an identity operation (referred to as 'no operation'), a controlled-NOT (CNOT) gate, and a controlled-Hadamard (CH) gate. The entangling CNOT gate is a well-studied nonlocal unitary quantum process. It performs a controlled bit flip on the target qubit if the control qubit is in the state $\vert 1 \rangle$. On the other hand, the controlled Hadamard gate corresponds to applying a Hadamard (or pseudo-Hadamard) gate to the target qubit when the control qubit is in the state $\vert 1 \rangle$. Figure \ref{ckt+seq} illustrates a general rotation operation on a qubit, characterized by an angle $\theta$ and a phase $\phi$, and represented by the unitary operator $R_{\phi}^{\theta}$. The corresponding values for $\theta$ and $\phi$ employed in the quantum circuit (depicted in Figure \ref{ckt+seq}(a)) to achieve the desired unitaries are provided in Table \ref{Gates}. In the context of the circuit, the identity operation signifies 'no operation', while the CNOT gate performs a state flip on the target qubit (accompanied by a phase factor of $e^{-\iota\frac{\pi}{2}}$) when the control qubit is in the state $\vert 1\rangle$. The controlled-Hadamard (CH) gate generates a superposition state of the target qubit, transforming $\vert 0 \rangle$ to $\vert - \rangle$ and $\vert 1 \rangle$ to $\vert + \rangle$, when the control qubit is in the state $\vert 1 \rangle$. Here, $\vert \pm \rangle = \frac{1}{\sqrt{2}}(\vert 0 \rangle \pm \vert 1 \rangle)$, and a negative phase is indicated by a bar over the phase term.
\begin{figure}[t]
\centering
\includegraphics[angle=0,scale=1.3]{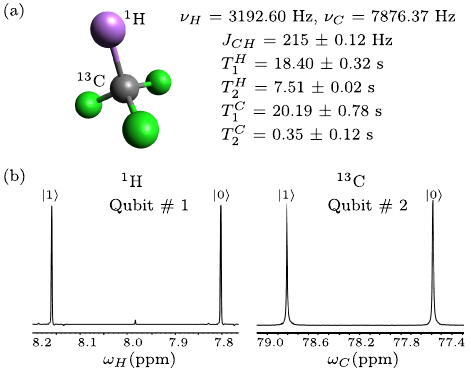}
\caption{(Color online) (a) Molecular structure of ${}^{13}$C labeled chloroform, which serves as a two-qubit quantum system. In this system, the first qubit is encoded as the nuclear spin of ${}^{1}$H, while the second qubit is encoded as the nuclear spin of ${}^{13}$C. The values of the scalar coupling $J_{\rm CH}$ (in Hz), relaxation times $T_{1}$ and $T_{2}$ (in seconds), and chemical shifts $\nu_i$ are indicated alongside.  (b) Thermal
equilibrium NMR spectrum after a $\frac{\pi}{2}$ detection
pulse.}
\label{molecule}
\end{figure}
The Pauli operators are selected as the basis operators for the analysis and further denoted as $\lbrace E_i\rbrace$. It is important to note that the choice of basis operators does not affect the quantum process being tomographed, and alternative choices of basis operators are equally valid as long as the orthogonality conditions stated in Eq.\ref{ortho} are satisfied. For two-qubit system, the set of sixteen product operators is defined as follows (Ernst et al., book reference): $I$, $\sigma_{2x}$, $\sigma_{2y}$, $\sigma_{2z}$, $\sigma_{1x}$, $\sigma_{1x}\sigma_{2x}$, $\sigma_{1x}\sigma_{2y}$, $\sigma_{1x}\sigma_{2z}$, $\sigma_{1y}$, $\sigma_{1y}\sigma_{2x}$, $\sigma_{1y}\sigma_{2y}$, $\sigma_{1y}\sigma_{2z}$, $\sigma_{1z}$, $\sigma_{1z}\sigma_{2x}$, $\sigma_{1z}\sigma_{2y}$, $\sigma_{1z}\sigma_{2z}$. Here, $I$ represents a $4 \times 4$ identity matrix, and $\sigma$ denotes the Pauli matrices. Terms such as $\sigma_{1x}\otimes\sigma_{2z}$ are simplified as $\sigma_{1x}\sigma_{2z}$ for convenience. Table \ref{Map} represents the quantum mapping for experimentally measuring the expectation values of product operators using LM (Linear Measurement). To illustrate, if we wish to determine the expectation value of $\langle \sigma_{1x}\sigma_{2y} \rangle$ in the state $\Lambda(E_i)$, we map $\Lambda(E_i)$ to $\rho_6$ with $\rho_6=U_6 \Lambda(E_i) U_6^\dagger$. According to Table \ref{Map}, $U_6={\rm CNOT~}\overline{X}_2 Y_1$, indicating that the system needs to undergo a local $\frac{\pi}{2}$ rotation of the first qubit with a phase $y$ and of the second qubit with a phase $\overline{x}$, followed by a CNOT gate. Consequently, $\langle \sigma_{2z} \rangle$ in the state $\rho_6$ is equivalent to $\langle \sigma_{1x}\sigma_{2y} \rangle$ in the state $\Lambda(E_i)$. In the context of NMR, it is convenient to compute the expectation values for Pauli $z$-operators, as they correspond to the $z$ magnetizations of the nuclear spins.
\begin{table}
\centering
\caption{\label{Gates}
Parameters chosen to implement
different unitary quantum processes.}
\begin{tabular}{c c c}
\hline \hline
Quantum process &
Phase $\phi $&
Angle $\theta$~~~\\
\hline \hline
Identity & $x$, $y$ & 0~~~\\
CNOT & $x$ & $\pi$~~~\\ 
CH & $\overline{y}$ & $\frac{\pi}{2}$~~~\\   
\hline
\end{tabular}
\end{table}
The two NMR qubits are realized using a molecule of
${}^{13}$C-enriched chloroform dissolved in acetone-D6, with
the nuclear spins ${}^{1}$H and ${}^{13}$C labeled as
`Qubit~1' and `Qubit~2', respectively. The molecular
structure, experimental parameters and the NMR spectrum
obtained at thermal equilibrium after a $\frac{\pi}{2}$
detection pulse are shown in Fig.~\ref{molecule}. 
All the experiments were performed at ambient temperature on
a Bruker Avance III 400 MHz FT-NMR spectrometer equipped
with a BBO probe. 
\begin{figure}[t]
\centering
{\includegraphics[scale=1.3]{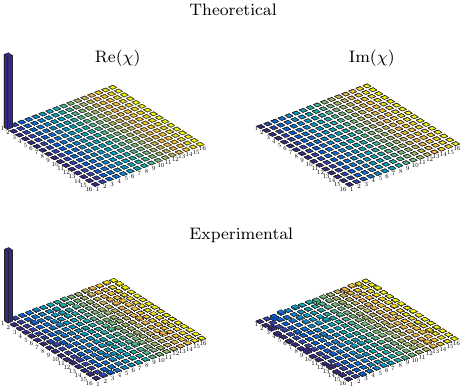}} 
\caption{(Color online). The tomographs in the columns
denote the real and imaginary parts of the $\chi$ matrix
respectively, for the identity operator. The tomographs on
the top represent the theoretically constructed while those
on the bottom represent the experimentally measured $\chi$
matrix of the identity operator. The fidelity of the
identity operator turned out to be 0.98.  }
\label{tomo-id}
\end{figure}
\begin{figure}[t]
\centering
{\includegraphics[scale=1.3]{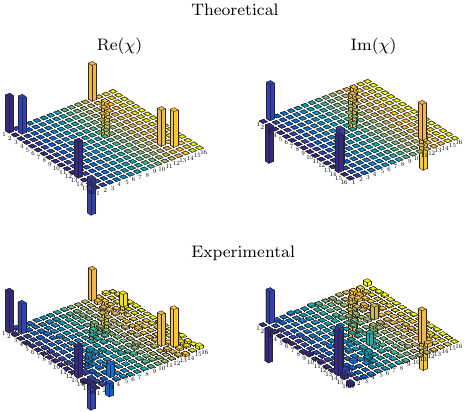}}
\caption{(Color online). The tomographs in the columns
denote the real and imaginary parts of the $\chi$ matrix
respectively, for the CNOT gate (control-$R_x^{\pi}$).  The
tomographs on the top represent the theoretically
constructed while those on the bottom represent the
experimentally measured $\chi$ matrix of the CNOT operator.
The fidelity of the CNOT operator turned out to be 0.93.}
\label{tomo-cnot} \end{figure}
\begin{figure}[t]
\centering
{\includegraphics[scale=1.3]{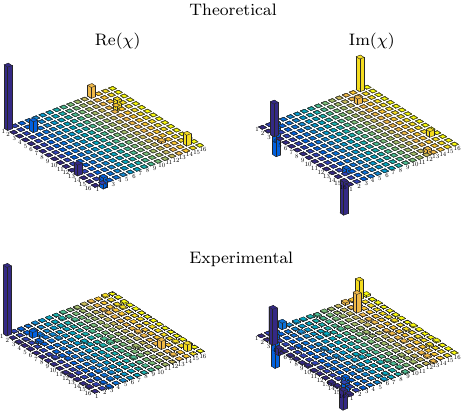}}
\caption{(Color online). The tomographs in the columns
denote the real and imaginary parts of the $\chi$ matrix
respectively, for the controlled-Hadamard gate
(control-$R_y^{-\frac{\pi}{2}}$). The tomographs on the top
represent the theoretically constructed while those on the
bottom represent the experimentally measured $\chi$ matrix
of the CH operator. The fidelity of the CH operator turned
out to be 0.92.}
\label{tomo-ch}
\end{figure}

The Hamiltonian for a two-qubit system in the rotating frame is given by 
\begin{equation}\label{e_2.13} 
{\mathcal{H}}=2 \pi \left[
(\nu_{{\rm H}}-\nu_{{\rm H}}^{{\rm rf}}) I_z^{{\rm H}} + 
(\nu_{{\rm C}}-\nu_{{\rm C}}^{{\rm rf}}) I_z^{{\rm C}} + 
J_{{\rm CH}} I_z^{{\rm H}} I_z^{{\rm C}}\right]
\end{equation}
where $\nu_{{\rm H}}$, $\nu_{{\rm C}}$ are the 
chemical shifts 
and $I_z^{{\rm H}}$, $I_z^{{\rm C}}$ are the
z-components of the spin angular momentum operators of
the ${}^{1}$H and ${}^{13}$C spins respectively, and J$_{{\rm
CH}}$ is the scalar coupling constant; 
$\nu_{{\rm H}}^{{\rm rf}}$ and
$\nu_{{\rm C}}^{{\rm rf}}$ are the rotating
frame frequencies.
The spatial
averaging technique is used to prepare the spins in an initial
pseudopure state\citep{cory-physicad,oliveira-book-07}:
\begin{equation} \rho_{00}=\frac{1}{4}(1-\epsilon)I+\epsilon
\vert 00\rangle \langle 00 \vert \end{equation}
where $\epsilon$ is proportional to spin polarization and
can be evaluated from the ratio of magnetic and thermal
energies of an ensemble of magnetic moments $\mu$ in a
magnetic field $B$ at temperature $T$; 
$\epsilon \thicksim
\frac{\mu B}{k_B T}$ and at room temperature  
and for a $B \approx$ 10 Tesla, $\epsilon \approx
\rm{10^{-5}}$. 

Fig.\ref{ckt+seq}(a) and \ref{ckt+seq}(b) display the quantum circuit and the corresponding NMR pulse sequence for implementing the MSQPT scheme, respectively. The circuit is divided into four modules, indicated by dashed blue lines, each serving a distinct purpose. In the circuit, the unitary operation $R_{\phi} ^{\theta}$ represents a local rotation with an angle $\theta$ and phase $\phi$. The rotation angles in the circuit are either zero or $\frac{\pi}{2}$, and the circuit is designed accordingly. The first two modules of the circuit are responsible for preparing the required basis state $E_i$. In the first part of the quantum circuit, the shaded rectangle represents a non-unitary quantum process that eliminates unwanted quantum coherences. The second module is executed only when a nonzero $\theta_3$ is required based on the experimental settings. The third module corresponds to the unitary quantum process $\Lambda$, which transforms $E_i$ to $\Lambda(E_i)$. Finally, the last module performs the quantum mapping, as specified in Table \ref{Map}, of the desired operator $E_i$ to local Pauli $z$-operators. The meter symbol represents an NMR measurement, and only one of the three measurements occurs in each experimental setting.

The quantum gates (including local rotations) are implemented using highly accurate radio frequency (rf) pulses and free evolution periods under the system Hamiltonian. Spin-selective hard pulses with the desired phase are utilized for local rotations. For $\rm{^1H}$, a $\frac{\pi}{2}$ hard pulse corresponds to an rf pulse duration of 12.95 $\mu$s at a power level of 20.19 W, while for $\rm{^{13}C}$, the pulse duration is 8.55 $\mu$s at a power level of 74.67 W. Unfilled rectangles represent $\frac{\pi}{2}$ hard pulses, while the filled rectangle represents a $\frac{\theta}{2}$ hard pulse according to the unitary quantum process $\Lambda$. The phases of all hard pulses are indicated above the respective pulses. To eliminate undesired coherences during basis state preparation, a $z$-gradient is employed. The measurement boxes represent the time-domain NMR signal, which is proportional to the expectation value of $\sigma_z$ after a Fourier transformation.

The fidelity between the experimentally constructed $\chi_{\rm{expt}}$ and the theoretically expected $\chi_{\rm{theo}}$ was calculated using Eq. \ref{ch1_eq77}. Fig.\ref{tomo-id} to \ref{tomo-ch} display the theoretically constructed and experimentally tomographed $\chi$ matrices for the 'no operation' (identity), the CNOT gate, and the CH gate, respectively. The calculated fidelities $(\mathcal{F})$ for the Identity, CNOT, and CH operators were found to be 0.98, 0.93, and 0.92, respectively. In Fig.\ref{tomo-id}, the upper panel illustrates the theoretically expected $\chi$ matrix, while the lower panel shows the experimentally constructed $\chi$ matrix (real and imaginary parts) for the Identity operator. The axes of the $\chi$ matrix are labeled with the indices of the product basis operators $E_i$. Similarly, Fig.\ref{tomo-cnot} and \ref{tomo-ch} represent the $\chi$ matrices for the CNOT and CH operators, respectively. In all three cases, the fidelity $\mathcal{F}$ was greater than 0.92, indicating the successful experimental implementation of the MSQPT protocol.

It is important to note that the NMR implementation of the MSQPT protocol involves the utilization of non-unitary evolution through the application of a magnetic field gradient pulse. This type of evolution is generally not feasible on conventional quantum processors, as they only allow for the use of unitary operators. In the subsequent section, a more generalized quantum circuit is proposed for MSQPT that exclusively relies on unitary gates, making it compatible with other physical platforms.
\section{IBM implementation of MSQPT protocol}\label{IBMseqpt}

In recent years, significant advancements have been made by researchers worldwide in the development of controllable quantum systems with higher dimensions. Particularly noteworthy is the successful construction of a large-scale 20-qubit quantum processor by IBM, utilizing superconducting technology. Previous studies have reported achievements in NMR and linear optical photonic systems, reaching capacities of up to 12 and 10 qubits, respectively \cite{negrevergne-prl-2006,gao-nature-phyics-2010}. Additionally, ion-trap-based quantum processors have demonstrated experimental control over as many as 14 qubits \cite{monz-prl-2011}. These progressions indicate a positive trajectory in addressing scalability challenges, especially for physical systems like superconducting and ion-trap technologies. Inspired by these developments, the MSQPT technique was employed to characterize superconducting quantum gates by incorporating additional ancilla qubits.

This study presents a demonstration of the MSQPT protocol using the IBM QX2 quantum information processor, specifically focusing on selectively characterizing various superconducting quantum gates. A general quantum algorithm and the corresponding quantum circuit are provided for preparing the initial input states, enabling efficient implementation of the MSQPT protocol on an n-qubit system. The demonstration primarily concentrates on 2-qubit and 3-qubit quantum gates. The experimental results exemplify the usefulness of MSQPT in selectively characterizing quantum processes. Furthermore, the ability to perform full QPT using MSQPT is showcased, enabling the construction of a complete quantum process element-wise. Additionally, efficient measurement settings are introduced, where detecting only one qubit from the entire system is sufficient. It is demonstrated that the results obtained from MSQPT can be utilized to construct valid underlying quantum operations by solving constrained convex optimization problems.

The IBM quantum processor, based on superconducting technology, is freely accessible to researchers worldwide through the cloud \cite{santos-2017,harper-prl-2019,Rui2017}. The user-friendly graphical interface of the IBM computer allows researchers to easily implement 2-qubit and 3-qubit quantum operations by directly incorporating the corresponding unitary gates into the quantum circuit. In this study, the five-qubit IBM QX2 processor is utilized, initializing all qubits in the $\vert 0 \rangle$ state as the input state. This processor enables the direct implementation of fifteen single-qubit gates, twelve 2-qubit gates, and two 3-qubit gates in the quantum circuit. The final step of the circuit involves projective measurements in the Pauli $\sigma_z$ basis. By repeating the quantum circuit multiple times, Born probabilities can be computed. The important experimental parameters are given in the Table\ref{qx2} of the chapter\ref{csqpt_chap}. It is important to note that the IBM quantum circuit begins with a pure quantum state as the input state and only allows for the implementation of unitary operations. Therefore, additional ancillary qubits are required, along with the system qubits, to prepare the system in mixed states or to simulate non-unitary evolutions, which are necessary for efficiently demonstrating the MSQPT protocol.

\subsection{Initial state preparation for efficient MSQPT}\label{sub1}

From eq.\ref{e_2.11}, in order to compute $F_{mn}$ one needs to prepare quantum system in the state corresponding to $E_i$ and then pass it through quantum channel which is to be characterized and then efficiently compute expectation value of $E_k$ on output state $\Lambda(E_i)$. In order to experimentally prepare quantum system in the desired state the corresponding density operator $\rho$ should satisfy three conditions: i) $\rho = \rho^{\dagger}$, ii) $\rho \geq 0 $ and iii) \textbf{Tr}$(\rho) = 1$. However, the Pauli operators satisfy only Hermiticity condition. So we need to prepare system in the state which evolve like Pauli operators.

 The method developed in the papers\cite{ekert-prl-2002, horodecki-arxiv-2001} shows that one can construct valid density matrix from any hermitian operator $M$ as follows:
 \begin{itemize}
 \item First step is to construct the operator $M'$ s.t.
\begin{equation}
M' = M + \lambda I
\label{eq6}
\end{equation} 
  where $-\lambda$ is the smallest eigenvalue of $M$ and $I$ is the identity operator. The operator $M'$ will be positive semi definite Hermitian operator.
  \item For trace condition, divide $M'$ by its trace i.e.
\begin{equation}
\tilde{M} = \frac{M'}{\textbf{Tr}(M')}
\end{equation} 
$\tilde{M}$ will satisfy all three properties of density matrix.
 \end{itemize}
 We say that the density operator $\tilde{M}$ corresponds to a given Hermitian operator $M$ and will evolve as if the system is in the state $M$.
  
In the case of Pauli operators all $E_i$s are traceless and $\lambda$ in Eq.\ref{eq6} will be equal to 1. So the valid density operators will be of the form:
\begin{equation}
 \tilde{M}_i = \frac{E_i + I}{D} 
 \label{eq81}
\end{equation}
where $D$ is dimension of the Hilbert space. Eq.\ref{eq81} is true for all Pauli basis operators except for $E_0 = I$. Now since the matrix $\tilde{M}_i$ represents valid quantum state, one can experimentally prepare the quantum system in the state $\tilde{M}_i$ corresponding to given basis operator $E_i$ and compute expectation values $\langle E_k^i \rangle$ given in eq.\ref{e_2.11}.

 Note that for general $n$-qubit system, all density operators $\tilde{M_i}$ in Eq.\ref{eq81} represents mixed state except for $n=1$. So if the system is initialized in a pure state and only unitary evolutions are allowed, then one needs extra ancillary qubits to experimentally prepare the quantum system in the state $\tilde{M_i}$. 
\begin{figure}[!]
\centering
\includegraphics[angle=0,scale=0.85]{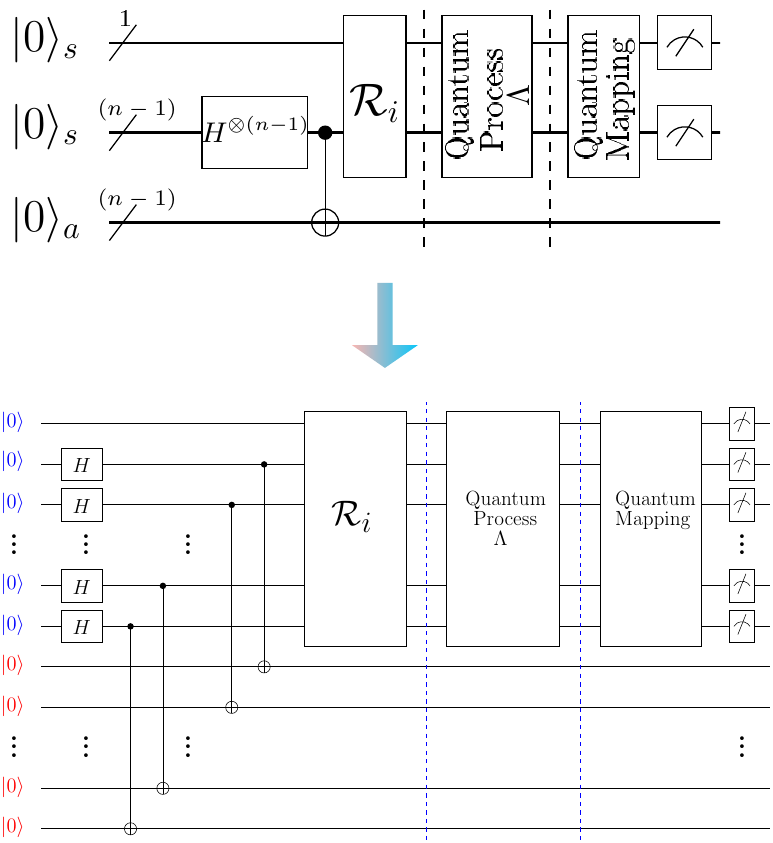} 
\caption{The generalized quantum circuit to acquire data to
perform n-qubit MSQPT is given. The symbol '/' through the
input wire represents multiqubit quantum register. 
The first quantum register contains single qubit while
second and third quantum registers comprise $n-1$ qubits.
The first and the second quantum registers collectively
represent system qubits and denoted by $\vert 0 \rangle_s$
where third quantum register represents ancilla qubits
denoted by $\vert 0 \rangle_a$. The quantum circuit below is the elaborated version where blue color denotes the system qubits and red color denotes the ancilla qubits.
}
\label{genckt}
\end{figure}
\subsubsection{Generalized algorithm for preparation of input states for n-qubit system }

It turns out that for a $n$-qubit system all non-zero eigenvalues of operator $\tilde{M_i}$ in Eq.\ref{eq81} are the same and are equal to $1/2^{n-1}$. Let $\lbrace \vert u_1^i \rangle, \vert u_2^i \rangle, \vert u_3^i \rangle,...,\vert u_{2^{n-1}}^i \rangle \rbrace $ be a complete set of normalized eigenvectors of the operator $\tilde{M_i}$ corresponding non-zero eigenvalues. The complete state of the combined system (system + ancilla) is as\cite{gaikwad-sr-2022}:
\begin{equation}\label{geneq}
\vert \Psi_i \rangle = \frac{ \vert u_1^i \rangle \vert a_1\rangle+ \vert u_2^i \rangle \vert a_2\rangle+.....+ \vert u_{2^{n-1}}^i \rangle \vert a_{2^{n-1}}\rangle}{\sqrt{2^{n-1}}}
\end{equation}
where $\vert a_i\rangle$'s are basis states (computational basis vectors) of ancillary system. Note that in general $\vert\Psi_i\rangle$ represents an entangled state. After tracing over ancillary system, the main system will be in the desired state $\tilde{M_i}$.

The procedure to find out the unitary operation $U^i$ s.t. $U^i \vert 0 \rangle ^{sys} \vert 0 \rangle ^{ancilla} = \vert \Psi_i \rangle $is given below:
\begin{enumerate}
\item[1.] Apply Hadamard gate on any ($n-1$) number of system qubits so that after application of hadmard gate there will be total $2^{n-1}$ number of states in superposition while state of the ancilla system is $\vert 0 \rangle ^{ancilla}$.
\item[2.] Utilize CNOT operations between the system qubits and ancillary qubits, with the system qubits serving as the control qubits and the ancillary qubits as the target qubits. This arrangement ensures that we achieve the transformations $\vert 0 \rangle ^{\text{ancilla}} \longrightarrow \vert a_1^i \rangle $, $\vert 0 \rangle ^{\text{ancilla}} \longrightarrow \vert a_2^i \rangle $, and so forth.
\item[3.] After the second step, map the computational basis states of the system qubits which we have got in the step 1. to eigenvectors of $\tilde{M_i}$ using unitary gate $\mathcal{R}_i$ where the columns of $\mathcal{R}_i$ are just normalized eigenvectors of $\tilde{M_i}$. Note that the column position of eigenvectors depends on which computational basis vector we want to map on to which eigenvector. This is sometimes referred to as change of basis operation. 
\item[4.] At the end of step 3, the combined system (main system + ancilla) will be in $\vert\Psi_i\rangle$ state. Repeat the procedure to prepare other states $\tilde{M_i}$.
\end{enumerate}

The generalized quantum circuit to implement MSQPT protocol is given in Fig.\ref{genckt}. 

\subsubsection*{Constructing a valid quantum process }

Note that in order to represent a valid quantum map (CPTP), the $\chi$ matrix should satisfy following conditions: i) $\chi = \chi^{\dagger}$, ii) $\chi \geq 0$ and iii) $\sum_{m,n}\chi_{mn}E_m^{\dagger}E_n = I$. Using the MSQPT method, $\chi$ matrix is Hermitian by construction but there is no guarantee that it will satisfy the last two conditions. However, one can use the constrained convex optimization (CCO) technique to obtain a valid $\chi_{\rm cco}$ matrix from $\chi_{\rm msqpt}$ as follows:
\begin{subequations}
\begin{alignat}{2}
&\!\min_{\chi_{\rm cco}}        &\qquad& \Vert \chi_{\rm msqpt}-\chi_{\rm cco}\Vert_{l_2}\label{e12}\\
&\text{subject to} &      & \chi_{\rm cco} \geq 0,\label{e12:constraint1}\\
&                  &      & \sum_{m,n}\chi_{mn}^{\rm cco}E_m^{\dagger}E_n = I.\label{e12:constraint2}
\end{alignat}
\end{subequations}
where $\chi_{\rm msqpt}$ is experimentally obtained process matrix using MSQPT protocol and $\chi_{\rm cco}$ is variable process matrix which represents a valid underlying quantum process.
 

\subsection{MSQPT of 2-qubit quantum gates on IBM QX2 processor }
\label{sub2}

In the case of trace preserving completely positive maps (CPTP) the quantum state corresponding to identity operator does not evolve, i.e. $\Lambda(I) = I$, so we will be considering only evolution of quantum states corresponding to all pauli operators $\lbrace E_i \rbrace$ except for $E_0 = I$.

In the case of 2-qubit system, one needs to prepare 15 input(mixed) states  $\tilde{M}_i$ given in the Eq.\ref{eq81} corresponding to all $E_i$. For all $\tilde{M}_i$'s, it turns out that out of 4 eigenvalues only 2 eigenvalues are non-zero ($\lambda_1 = \lambda_2 = 1/2$) and other 2 are zero ($\lambda_3 = \lambda_4 = 0$). Let say $\vert v^i_1 \rangle$ and $\vert v^i_2 \rangle$ are two normalized eigenvectors of operator $\tilde{M_i}$ corresponding to $\lambda_1$ and $\lambda_2$ respectively. So to efficiently perform MSQPT of 2-qubit system on IBM computer, one ancillary qubit is required and the input state of complete system which is to be prepared is given by:
\begin{equation}
\vert \psi_i \rangle = \frac{\vert v^i_1 \rangle \vert 0 \rangle + \vert v^i_2 \rangle \vert 1 \rangle }{\sqrt{2}}
\end{equation}

\begin{figure*}[t]
\centering
\includegraphics[angle=0,scale=0.87]{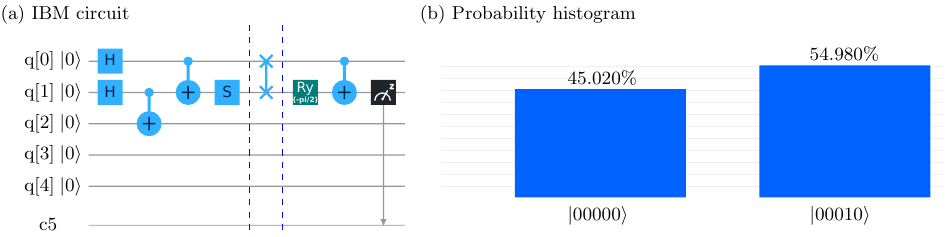} 
\caption{(a) The IBM quantum circuit to perform MSQPT of
SWAP gate is given where in the first block the 3-qubit
input state $\vert \psi_{6} \rangle $ is prepared using $U^6
= S_2. CNOT_{12} .CNOT_{23}. H_2. H_1$ and in the second
block quantum process corresponding to 2-qubit SWAP gate is
applied and in the last block quantum map $U_{13}=
CNOT_{12}.R_y(-\frac{\pi}{2})$ is applied to efficiently
compute $\textbf{Tr}(\sigma_z \bigotimes \sigma_x
\Lambda(\tilde{M_6}))$ by detecting system's second qubit in
$\sigma_z$ basis. (b) Histogram represents statistical
results after running quantum circuit given in (a) 4096
times. It gives the probability $p_0 = 0.4502$ and $p_1 =
0.5498$ of obtaining second qubit in $\vert 0 \rangle $ and
$\vert 1 \rangle $ state. } 

\label{ibm2q}
\end{figure*}

All fifteen 3-qubit pure input states $\vert \psi_i \rangle$ corresponding to $E_i$ are listed below:

\begin{align*}
       & \vert \psi_1 \rangle =[(0,1,0,1,1,0,1,0)/2]^{T},  \\&
        \vert \psi_2 \rangle=[(0,-i,0,1,-i,0,1,0)/2]^T,  \\&
         \vert \psi_3 \rangle=[(1,1,0,0,0,0,0,0)/\sqrt{2}]^T, \\&
         \vert \psi_4 \rangle=[(0,-1,1,0,0,-1,1,0)/2]^T, \\&
         \vert \psi_5 \rangle=[(1,0,0,1,0,1,1,0)/2]^T, \\&
         \vert \psi_6 \rangle=[(-i,0,0,1,0,-i,1,0)/2]^T, \\&
         \vert \psi_7 \rangle=[(0,1,-1,0,0,1,1,0)/2]^T, \\&
         \vert \psi_8 \rangle=[(0,-1,-i,0,0,-i,1,0)/2]^T, \\&
         \vert \psi_9 \rangle=[(-i,0,0,1,0,i,1,0)/2]^T, \\&
          \vert \psi_{10} \rangle=[(-1,0,0,1,0,1,1,0)/2]^T, \\&
           \vert \psi_{11} \rangle=[(0,1,1,0,0,1,1,0)/2]^T, \\&
           \vert \psi_{12} \rangle=[(1,0,0,1,0,0,0,0)/\sqrt{2}]^T, \\&
           \vert \psi_{13} \rangle=[(0,1,0,1,-1,0,1,0)/2]^T, \\&
           \vert \psi_{14} \rangle=[(0,-i,0,1,i,0,1,0)/2]^T, \\&
           \vert \psi_{15} \rangle=[(1,0,0,0,0,0,0,1)/\sqrt{2}]^T
\end{align*}
As an example, IBM quantum circuit for $SWAP$ gate MSQPT corresponding to quantum state $\vert \psi_6 \rangle$ and observable $E_{13} = \sigma_z \otimes \sigma_x$ is given in the Fig.\ref{ibm2q}, where the system qubits are denoted by $q[0]$ and $q[1]$ as first and second qubit respectively while ancilla qubit is denoted by $q[2]$. In the first block unitary operation $U^6 = S_2. CNOT_{12} .CNOT_{23}. H_2. H_1$ is applied on initial state $\vert 000 \rangle$ to prepare 3-qubit system in the pure state $\Psi_6 = \vert \psi_6 \rangle \langle \psi_6 \vert$. After that in the second block the quantum process $(\Lambda_{system} \otimes I_{ancilla})$ corresponding to 2-qubit $ SWAP$ gate is implemented on system qubits. And in the last block quantum map corresponding to unitary operation $U_{13}= CNOT_{12}.R_y(-\frac{\pi}{2})$ is used to transform system's output state to determine $ \langle \sigma_z \otimes \sigma_x \rangle $ by measuring only second qubit in the $ \sigma_{z} $ basis. In this particular example the quantity corresponding to $\textbf{Tr}(\sigma_z \otimes \sigma_x \Lambda(\tilde{M_6}))$ is experimentally computed which is equal to $\textbf{Tr}(\sigma_{2z} U_{13}(\Lambda(\tilde{M_6})){U_{13}}^{\dagger})$ and using Eq.\ref{eq81} we finally get,
{\small
\begin{align}
\textbf{Tr}(\sigma_z \otimes \sigma_x \Lambda(\sigma_x \otimes \sigma_y )) &= 4\textbf{Tr}(\sigma_z \otimes \sigma_x \Lambda(\tilde{M_6}))\\
 &= 4\textbf{Tr}(\sigma_{2z} U_{13}(\Lambda(\tilde{M_6})){U_{13}}^{\dagger})
\end{align}
}
In this way one can efficiently compute all $ \langle E_k^i \rangle$ and estimate corresponding average survival probability $F_{mn}$.

To perform efficient measurement, the list of all unitary operations $U_i$ corresponding to all quantum maps which transform output states in order to determine $\langle E_k \rangle $ just by detecting one of the system's qubit in $\sigma_z$ basis (\textit{i.e}. by measuring either $\langle \sigma_{1z} \rangle$ or $\langle \sigma_{2z} \rangle$) are given in Table.\ref{Map}.

\subsection{MSQPT of 3-qubit toffoli gate on IBM QX2 processor }
\label{sub3}

In case of 3-qubit system, 63 number input(mixed) states $\tilde{M_i}$ need to be prepared corresponding to all 3-qubit pauli operators $E_i$. It turns out that for all $\tilde{M_i}$, out of 8 eigenvalues only 4 are non zero and are equal to 1/4. Let $\vert u_1^i \rangle$, $ \vert u_2^i \rangle$, $ \vert u_3^i \rangle$ and $\vert u_4^i \rangle$ are 4 eigenvectors of $\tilde{M_i}$ corresponding to non-zero eigenvalues. In order to prepare main system in the $\tilde{M_i}$ states, one needs to prepare 5-qubit pure state in the form,
\begin{equation}
\vert \Omega_i \rangle = \frac{\vert u^i_1 \rangle \vert 00 \rangle + \vert u^i_2 \rangle \vert 01 \rangle + \vert u^i_3 \rangle \vert 10 \rangle + \vert u^i_4 \rangle \vert 11 \rangle }{2}
\label{eq14}
\end{equation}

After tracing out the last two ancillary qubits, the main system will be left in the state $\tilde{M_i}$, i.e, $\textbf{Tr}^{ancilla}(\vert \Omega_i \rangle \langle \Omega_i \vert) = \tilde{M_i}$. 
\begin{figure*}[t]
\centering
\includegraphics[angle=0,scale=0.9]{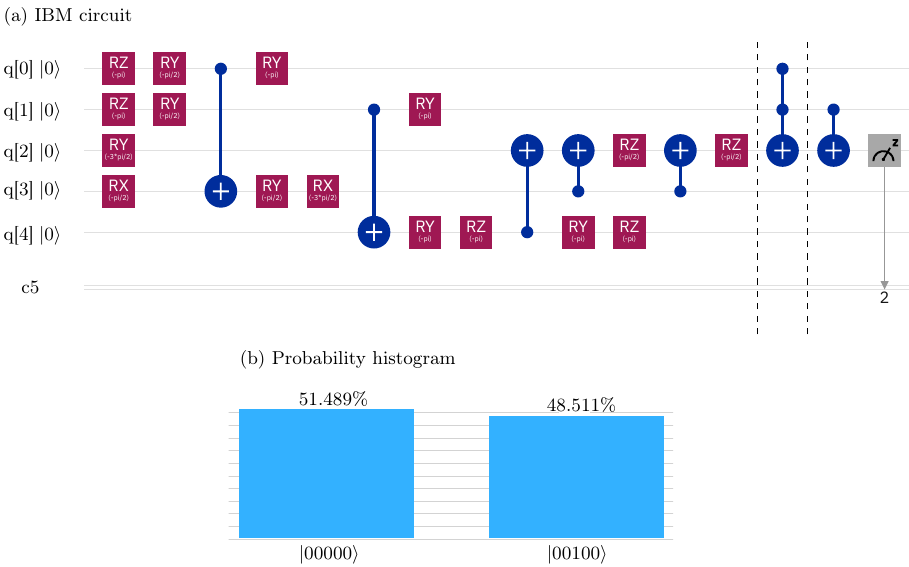} 
\caption{(a) The IBM quantum circuit to perform MSQPT of 3 qubit Toffoli gate is given where in the first block the 5-qubit input state $\vert \Omega_{50} \rangle $ is prepared and in the second block quantum process corresponding to 3-qubit  Toffoli gate is applied on the system qubits and in the last block quantum map $U_{15}= CNOT_{23}$ is applied to efficiently compute $\textbf{Tr}(I \otimes \sigma_z \otimes \sigma_y \Lambda(\tilde{M}_{50}))$ by detecting system's third qubit in $\sigma_z$ basis. (b) Histogram represents statistical results after running quantum circuit given in (a) 4096 times. It gives the probability $p_0 = 0.51489$ and $p_1 = 0.48511$ of obtaining third qubit in $\vert 0 \rangle $ and $\vert 1 \rangle $ state respectively.} 
\label{ibm4}
\end{figure*}
\subsubsection*{Preparation of 5-qubit pure input states $\vert \Omega_i \rangle $ on IBM QX2 processor }

As mentioned in the secction III, the IBM quantum processor does not allow to directly implement any arbitrary unitary quantum gate. There are very limited number of quantum gates avalaible that you can directly implement on IBM processor. In order to implement arbitrary quantum gate one has to first decompose it into set of available IBM quantum gates and then implement the appropriate sequence of set IBM quantum gates corresponding to desired quantum gate. So while preparing 5-qubit pure input state given in eq.\ref{eq14} one has to first find the correct decomposition of unitary operator $\mathcal{R}_i$ given in the Fig.\ref{genckt} in order to prepare quantum state $\vert \Omega_i \rangle$. There are several techniques available that you can use to decompose given unitary into number of CNOTs and single qubit rotation gates\cite{bergholm-pra-2005, shende-ieee-2006}. In this study, the UniversalQCompiler is used as tool to find out the optimized gate sequence of CNOT and single qubit rotation gates to prepare input state $\vert \Omega_i \rangle$ from initial state $\vert 00000 \rangle$. It is a Mathematica package developed from the techniques proposed in the papers\cite{plesch-pra-2011, colbeck-pra-2016, colbeck-arxiv-2019}.

As an example, IBM quantum circuit performing MSQPT of 3-qubit Toffoli gate is given in the Fig.\ref{ibm4} corresponding to 5-qubit pure input state $\vert \Omega_{50} \rangle$ and observable $E_{15} = I \otimes \sigma_{z} \otimes \sigma_{y} $. The system qubits are denoted by $q[0]$, $q[1]$ and $q[2]$ as first, second and third qubit while the ancilla qubits are denoted by $q[3]$ and $q[4]$. The first block in the Fig.\ref{ibm4} prepares the 5-qubit pure input state $\vert \Omega_{50} \rangle$ while the second block represents the action of 3-qubit toffoli gate on system qubits and the last block represents the action of quantum map corresponding to unitary operation $U_{15}= CNOT_{23}$ is used to transform system's output state to determine $ \langle I \otimes \sigma_{z} \otimes \sigma_{y} \rangle $ by measuring only third qubit in the $ \sigma_{z} $ basis. At the end the IBM circuit given in Fig.\ref{ibm4} computes the quantity corresponding to $\textbf{Tr}(\sigma_{3z} U_{15}(\Lambda(\tilde{M}_{50})){U_{15}}^{\dagger})$. Using eq.\ref{eq81} we finally get,
{\small
\begin{align}
\textbf{Tr}(E_{15} \Lambda(E_{50} )) &= 8 \textbf{Tr}(E_{15} \Lambda(\tilde{M}_{50}))\\
 &= 8\textbf{Tr}(\sigma_{3z} U_{15}(\Lambda(\tilde{M}_{50})){U_{15}}^{\dagger})
\end{align}
}

In the similar way, all $\textbf{Tr}(E_k \Lambda(E_i))=\langle E^i_k\rangle$ can be efficiently computed corresponding to desired average survival probability $F_{mn}$. For 3 qubit system, the list of all unitary operations $U_i$ corresponding to all quantum maps which transform output states in order to determine $\langle E_k \rangle $ just by detecting one of the system's qubit in $\sigma_z$ basis (i.e. by measuring either $\langle \sigma_{1z} \rangle$ or $\langle \sigma_{2z} \rangle$ or  $\langle \sigma_{3z} \rangle$) can be found in the papers\cite{gaikwad-pra-2018,singh-pra-2018}.

\subsection{IBM Experimental Results and analysis}\label{result}

The MSQPT protocol was implemented as described above to characterize quantum processes carried out on the IBM QX2 quantum processor for various two-qubit gates such as the Identity gate, CNOT gate, and SWAP gate, as well as the three-qubit Toffoli gate. The corresponding full $\chi$ matrices were constructed elementwise. In all cases, the experimentally constructed $\chi_{\rm msqpt}$ was used, and the CCO problem given in Eq. \ref{e12} was solved to obtain $\chi_{\rm cco}$ representing the underlying true quantum process. To ensure circuit correctness, the MSQPT protocol was also theoretically simulated on the IBM processor to obtain $\chi_{\rm sim}$. The fidelity of the experimentally implemented quantum gates was calculated using Eq. \ref{ch1_eq77}.

\subsubsection{2-qubit MSQPT results and analysis on IBM}

\begin{figure*}[t]
\centering
\includegraphics[angle=0,scale=0.9]{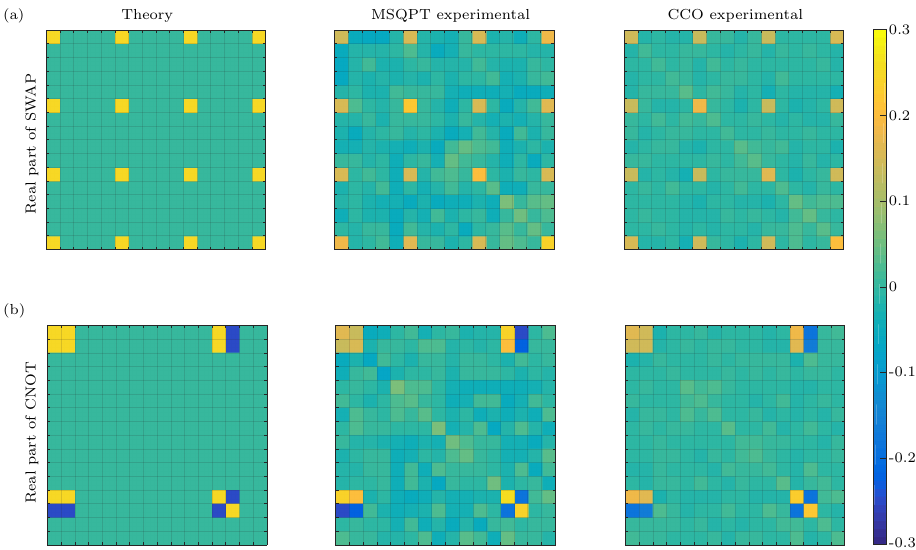} 
\caption{Matrix plots are shown corresponding to (a) real part of $\chi$ matrix
	for CNOT gate and (b) real part of $\chi$ matrix for SWAP gate. The
	first column represents theoretically constructed process matrix
	$\chi_{\rm the}$ while second and third column represent $\chi_{\rm
	msqpt}$, and $\chi_{\rm cco}$ respectively obtained by implementing
	MSQPT protocol on IBM QX2 processor.} 

	\label{plot1}
\end{figure*}

In Fig.\ref{plot1} the experimentally obtained matrix plots corresponding to 2-qubit CNOT and SWAP gates are given. The $16\times 16$ grid matrix plots given in part (a) in Fig\ref{plot1} represent the $16\times 16$ real part of $\chi$ matrix corresponding to a CNOT gate. The first yellow square (1st grid) in the matrix plot denotes the element $\chi_{11}=0.25$ of theoretically constructed process matrix $\chi_{\rm the}$ and so on. In the theoretically constructed matrix plot for the CNOT gate, out of 256 squares (formed by $16 \times 16$ grid lines) only 16 squares (10 yellow and 6 blue) has non-zero values. On the right the color bar is presented where color coding is given.  The second and third column represent matrix plots corresponding $\chi_{\rm msqpt}$, and $\chi_{\rm cco}$ respectively obtained by implementing MSQPT protocol on IBM QX2 processor. The slight deviations in color grids in experimentally obtained matrix plots with reference to theoretical matrix plot represents quantum gate error caused due to decoherence effect  and various statistical and systematic errors while preparing initial input state.  The matrix plots corresponding to imaginary part of $\chi$ matrix is not presented here as all the elements of imaginary part of $\chi$ matrix are zero and does not hold any significant information. The experimental fidelity of $\chi_{\rm msqpt}$ for CNOT gate turned out be 0.8281 while fidelity of $\chi_{\rm cco}$ turns out to be 0.9530. One can also see that the color grids in matrix plot given in the third column (CCO experimental) has smaller deviation compared to matrix plot given in the second column (MSQPT experimental). This improved fidelity shows that one can use the MSQPT data to further solve CCO problem to more accurately construct full process matrix. Note that to perform full QPT using MSQPT protocol the experimental complexity does not increase and number of experiments will be same as standard QPT. However, MSQPT allows one to gain partial knowledge about the underlying process by selectively computing desired elements of process matrix. Similar description goes for the subfigure (b) corresponding to SWAP gate. For the SWAP gate, experimental fidelity of $\chi_{\rm msqpt}$ turns out to be 0.7998 while improved fidelity of $\chi_{\rm cco}$ turns out to be 0.9299.

All three quantum gates  we obtained $\mathcal {F}(\chi_{\rm sim})\geq 0.99$ which ensures that all the quantum circuits are correct and lower values of $\mathcal {F}(\chi_{\rm msqpt})\leq 0.9$ are purely due to imperfections in initial input state preparation and imperfect quantum gates and decoherence effects. The values $\mathcal {F}(\chi_{\rm cco})\geq 0.9$ shows that one can retrieves full and true dynamics of quantum process with considerably high precision by solving optimization problem given in eq.\ref{e12} using experimentally constructed full $\chi_{\rm msqpt}$. The experimental complexity of MSQPT on IBM is the same as NMR case and is given in the Table.\ref{complex}.

\subsubsection{3-qubit MSQPT results and analysis on IBM}
\begin{figure*}
\centering
\includegraphics[angle=0,scale=0.85]{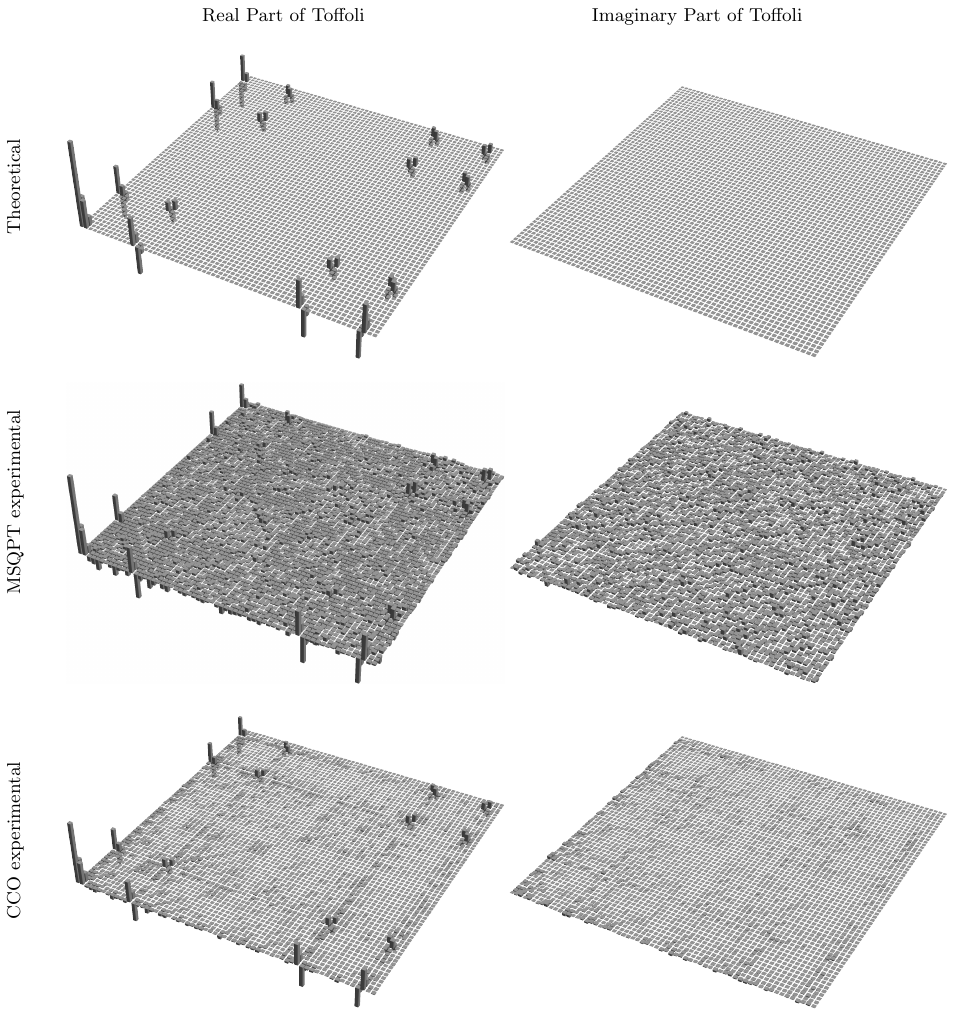} 
\caption{Tomographs are shown corresponding to 3 qubit toffoli gate. The first column represents the real part of $\chi$ matrix and second column represents the imaginary part of $\chi$ matrix corresponding to Toffoli gate. The first row denotes the theoretically constructed $\chi$ matrix while second and third row represent experimentally constructed $\chi$ matrix obtained by implementing MSQPT protocol and CCO protocol on IBM QX2 processor respectively. } 

	\label{toffoli}
\end{figure*}

In the Fig.\ref{toffoli} the experimentally obtained $ 64\times64$ dimensional $\chi$ matrix corresponding to 3-qubit Toffoli gate is represented by a $ 64\times64$ dimensional bar plot, where the first column represents the bar plot corresponding to the real part of $\chi$ matrix while the second column represents the bar plot corresponding to the imaginary part of $\chi$ matrix for Toffoli gate.   The first row denotes the theoretically constructed process matrix $\chi_{\rm the}$ while the second and third rows represent experimentally constructed process matrices $\chi_{\rm msqpt}$ and $\chi_{\rm cco}$ respectively, obtained by implementing the MSQPT protocol on the IBM QX2 processor. All the bar plots given in the Fig.\ref{toffoli} are plotted on the same scale. 
The experimental gate fidelity for case of $\chi_{\rm msqpt}$ turns out to be 0.5899 while surprisingly the improved experimental gate fidelity obtained for the case of $\chi_{\rm cco}$ turns out be 0.9457. To ensure the correctness of the circuits we have also simulated all the IBM circuits on IBM simulator. The simulation  fidelity of the Toffoli gate turns out to be 0.9804. The variation in the height of various bars in experimental tomographs and the lower fidelity for  $\chi_{\rm msqpt}$ in the experimentally constructed process matrices are due to imperfections in the initial state preparations, decoherence and various statistical and systematic errors. 
\begin{table}[h!]
\centering
\caption{\label{ibmsqpt_complexity3}
The experimental complexity and the extra ancillary qubits required for 3-qubit MSQPT is compared with linear inversion based standard QPT and standard SQPT method.}
\begin{tabular}{c c c c}
\hline \hline
&
MSQPT&
SQPT&
Standard QPT~~~\\
\hline \hline
Preparations & 63 & 288 & 63 ~~~\\
Readouts & 504 & 2016 & 3969 ~~~\\ 
No. of extra ancilla& 2 & 0 & 0 ~~~\\   
\hline
\end{tabular}
\end{table}
In the conventional QPT approach, one is required to prepare 63 linearly independent input states and conduct 63 measurements or readouts corresponding to a tomographically complete set of observables. This is equivalent to performing 63 state tomographies. The total number of readouts in this case is equal to $63 \times 63 = 3969 $. In the case of 3 qubit system the cardinality of set of quantum 2-design states is 72 (9 MUB sets each having cardinality of 8). In standard SQPT method, one needs to prepare states of the form $ (E_m \pm E_n)^\dagger \rho_j 
(E_m \pm E_n)$ and $ (E_m \pm iE_n)^\dagger \rho_j 
(E_m \pm iE_n)$ to determine real and imaginary part of $F_{mn}$ respectively requires 288 number of state preparations: 72 for $ (E_m + E_n)^\dagger \rho_j 
(E_m + E_n)$ + 72 for $ (E_m - E_n)^\dagger \rho_j 
(E_m - E_n)$ + 72 for $ (E_m + iE_n)^\dagger \rho_j 
(E_m + iE_n)$ and + 72 for $ (E_m - iE_n)^\dagger \rho_j 
(E_m - iE_n)$. Furthermore, the estimation of the overlap with the original 2-design state $\rho_j$ is necessary. To estimate a single overlap, 7 readouts are required (since the number of non-zero coefficients in the decomposition of $\rho_j$ is 7). Thus, the total number of readouts required amounts to $288 \times 7 = 2016$. However, when using the MSQPT approach, the total number of states to be prepared is 63 (corresponding to a complete set of basis operators), and the total number of required readouts is 504 (for each mutually unbiased basis set, 56 readouts are needed, resulting in a total of $9 \times 56 = 504$ readouts). The experimental complexity and the required ancilla qubits are provided in Table \ref{ibmsqpt_complexity3}.

\section{Conclusions}\label{seqpt_con}

In this study, a modified scheme based on local measurements (MSQPT), was proposed for achieving selective and efficient quantum process tomography on both NMR and IBM platforms. The scheme offers a distinct advantage by requiring significantly fewer experiments to determine the desired elements of the process matrix. Successful experimental implementation of the scheme was demonstrated for various cases, including 'no operation', a controlled-NOT gate, and a controlled-Hadamard gate on two NMR qubits. Additionally, MSQPT was implemented using ancilla qubits on the IBM processor for SWAP, CNOT gates, and the 3-qubit Toffoli gate. This method combines selectivity and efficiency, making it particularly valuable for scenarios where complete experimental characterization is not necessary. The results presented in this chapter have been 
published in
\href{https://journals.aps.org/pra/abstract/10.1103/PhysRevA.97.022311}{\rm
Phys. Rev. A \textbf{97}, 022311 (2018)} and \href{https://www.nature.com/articles/s41598-022-07721-3}{\rm
Sci. Rep \textbf{12}, 1-11 (2022)}.

\chapter{Direct tomography of quantum states and processes using weak measurements on NMR}

\section{Introduction}
\label{weakqpt_sec1}

The task of selectively estimating elements of the process matrix using quantum 2-design states was addressed in the previous chapter. Its effectiveness was demonstrated using both NMR and the IBM quantum processor. However, a large number of experiments are still required by the MSQPT protocol, and the ancilla system grows exponentially with the system size. In this chapter, the task of selectively estimating density and process matrix elements is revisited, which are sometimes referred to as direct quantum state tomography (DQST) and direct quantum process tomography (DQPT) respectively. The weak measurement method has been employed, and an even more efficient scheme than MSQPT for performing DQST and DQPT has been proposed.

Over the past few decades, the concept of weak measurement and weak value has garnered significant attention from both a fundamental and an applied perspective in the field of quantum theory \cite{yokota-njp-2009, palacios-np-2010, avella-prl-2016, lundeen-prl-2009}. The weak value of a given observable obtained through weak measurement is generally a complex number \cite{lev-prl-1990, lev-prl-1988}, and it allows us to sequentially measure incompatible observables to extract useful information from the quantum system without collapsing it, unlike projective measurement where the system collapses into one of the eigenstates, resulting in maximum disturbance to the state \cite{lundeen-prl-2016, lim-prl-2014}. This unique characteristic of weak measurement provides an elegant way to address important and fundamental issues in the foundations of quantum theory, such as the reality of the wave function \cite{lundeen-nat-2011, bhati-pla-2022}, the observation of a quantum Cheshire Cat in a matter-wave interferometer experiment \cite{yuji-natcom-2014}, observing single photon trajectories in a two-slit interferometer \cite{sacha-science-2011}, the three-box paradox \cite{lundeen-pla-2004}, Leggett-Garg inequality \cite{andrew-prl-2008, groen-prl-2013}, and more. Furthermore, weak measurements are actively utilized in the field of quantum information processing, covering a wide range of applications. Some of these applications include quantum state and process tomography \cite{kim-natcom-2018, lundeen-prl-2012, wu-sr-2013, bolduc-natcom-2016}, quantum state protection from decoherence \cite{kim-np-2012, xiao-epjd-2013, wang-pra-2014, wang-pra-2014_2}, quantum state manipulation \cite{blok-np-2014}, performing minimum disturbance measurements \cite{baek-pra-2008}, precision measurements and quantum metrology \cite{zhang-prl-2015}, sequential measurement of two non-commuting observables \cite{tosi-prl-2016, avella-pra-2017}, and tracking the precession of single nuclear spins by weak measurements \cite{cujia-nature-2019}, among others.

This study primarily focuses on the problem of direct quantum state and process tomography using the weak measurement (WM) technique. The task of directly estimating quantum states and processes has been extensively studied in the past, and various techniques have been described. For the task of DQST, one of the earliest protocols based on a controlled-SWAP quantum network was reported in \cite{ekert-prl-2022}. This protocol allows for the determination of selective elements of the density matrix using linear and nonlinear functionals of a quantum state. Additionally, the paper \cite{salvail-natpho-2013} presents an experiment demonstrating the direct measurement of polarization states of light. In a recent paper \cite{feng-pra-2021}, a new method based on phase-shifting techniques for DQST is proposed and experimentally demonstrated on a photonic chip \cite{li-pra-2022}. Similarly, for the task of direct quantum process tomography (DQPT), selective and efficient protocols based on quantum 2-design states have been reported in \cite{pears-pra-2021, perito-pra-2018} and successfully demonstrated on various physical platforms \cite{paz-prl-2011, gaikwad-pra-2018, gaikwad-sr-2022}. However, most of the proposed direct tomography methods are still not experimentally efficient in terms of complexity and resource requirements.

On the other hand, WM-based tomography techniques have shown superiority and efficiency in terms of directness and ease of implementation compared to existing direct tomography protocols, as they do not require full reconstruction. However, most WM-based tomography protocols have been implemented using optics, involving projective measurements for post-selection, and are yet to be demonstrated in other physical setups involving ensemble quantum systems such as NMR. Efforts have been made in this direction, and the first successful experimental demonstration of circuit-based WM with post-selection on an NMR ensemble quantum information processor has been reported in \cite{lu-njp-2014}. However, the proposed technique is experimentally expensive in terms of resources, as it requires two additional qubits to compute the weak value of a single qubit observable and involves the implementation of a multi-qubit controlled phase gate, which is challenging in practice.
\begin{figure*}  
\centering
{\includegraphics[scale=0.75]{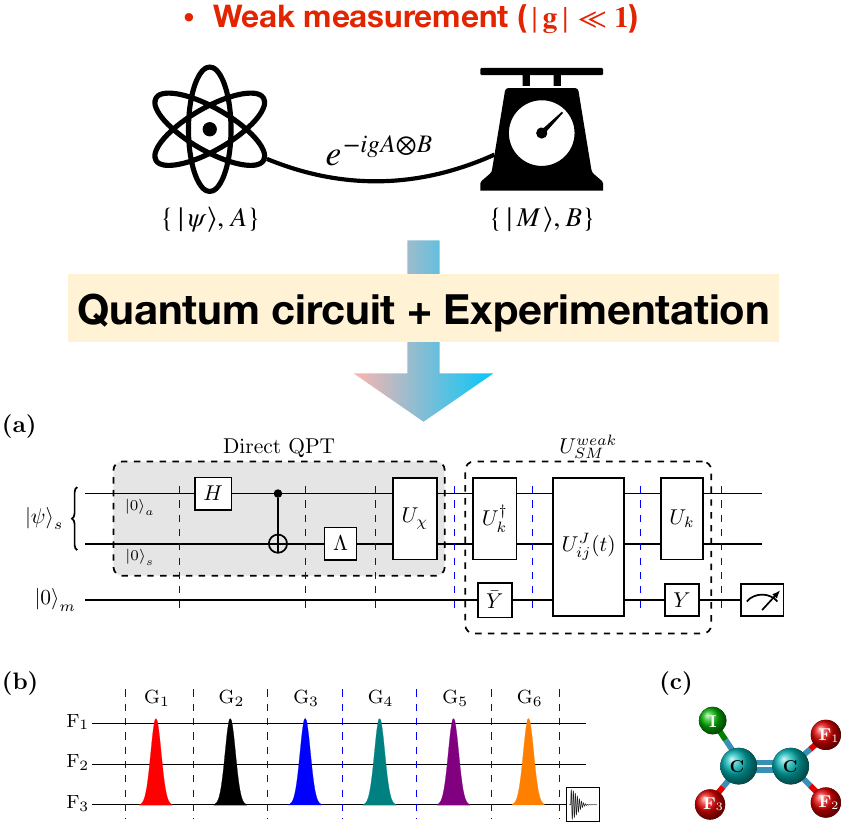}}
\caption{Schematic of WM based DQST and DQPT. (a) General quantum circuit for DQST of initial unknown state $\vert \psi \rangle_s$ and DQPT of quantum channel $\Lambda$ using WM method. (b) GRAPE optimised pulses to implement blocks of quantum circuit. (c) 3-qubit quantum system. }
\label{weakqpt_ckt}
\end{figure*}

In this study, an experimentally efficient scheme for performing DQST and DQPT is proposed using WM of Pauli spin operators on an NMR ensemble quantum information processor\cite{gaikwad-epjd-2023}. The scheme enables the computation of weak values of any observable and is designed in such a way that it eliminates the need for an ancillary qubit system and reduces complexity compared to recently proposed WM-based direct QST and QPT methods. The scheme possesses the following advantages over existing methods: i) Sequential weak measurements are not required, ii) The implementation of complex quantum gates like three-qubit multi-control phase gates is not involved, and iii) Projective measurements are not necessary. Furthermore, the proposed method is experimentally feasible, requiring only a single experiment to determine multiple selective elements of the density (process) matrix. The scheme is general and can be applied to any circuit-based implementation. To demonstrate its effectiveness, the scheme is experimentally implemented to characterize several two-qubit quantum states and single-qubit quantum processes with high fidelity. Additionally, experimental results are utilized as input for a convex optimization algorithm \cite{gaikwad-qip-2021} to reconstruct valid underlying states and processes. A comparison between the results and theory, as well as numerical simulations in terms of state and process fidelity, is performed to assess the efficacy of the method.

\section{General scheme for direct QST and QPT via weak measurement}\label{weakqpt_sec2}

Consider the initial preparation of the system and the measuring device in states $\vert \psi \rangle$ (pre-selection) and $\vert M \rangle$ respectively. To perform a weak measurement of the observable ${A}$, the joint state $\vert \psi, M \rangle$ evolves under the action of the evolution operator ${U}_{SM} = e^{-ig {A}\otimes {B}}$, where $g$ represents the coupling strength ($\vert g \vert \ll 1$) between the system and the measuring device. The operator ${B}$ associated with the measuring device is chosen such that $\langle M \vert {B} \vert M \rangle =0$. In the limit of weak measurement ($\vert g \vert \ll 1$), the evolution operator can be approximated to first order in $g$ as ${U}^{weak}_{SM} = ({I} -ig {A}\otimes {B})$. The joint state (system+measuring device) evolves according to this operator, and subsequently, a projective measurement is performed on the system using the projector $\vert \phi \rangle \langle \phi \vert$ (post-selection). This leads to the following expression:

\begin{eqnarray}
\vert \psi, M \rangle   & = &  e^{-ig {A}\otimes {B}}\vert \psi, M \rangle \nonumber  \\
 & \approx &  (I -ig {A}\otimes {B} ) \vert \psi, M \rangle \nonumber \\
 & = &  \vert \psi, M \rangle - ig {A} \vert \psi \rangle  \otimes {B} \vert M \rangle \label{weakqpt_eq1} \\
 & = &  \langle \phi \vert \psi \rangle \vert \phi, M \rangle - ig \langle \phi \vert {A} \vert \psi \rangle \vert \phi \rangle \otimes {B} \vert M \rangle \nonumber
\end{eqnarray}
At the end, up to some normalization constant the final state $\vert M_f \rangle$ of measuring device is given as,
\begin{equation}
\vert M_f \rangle   = \langle \phi \vert \psi \rangle \lbrace {1} - ig \langle {A} \rangle_w^{\phi} {B} \rbrace \vert M \rangle
\end{equation}
where 
\begin{equation}
\langle {A} \rangle_w^{\phi}  = \frac{\langle \phi \vert {A} \vert \psi \rangle}{\langle \phi \vert \psi \rangle}
\label{weakqpt_eq3}
\end{equation}  
is called the weak value of observable ${A}$ corresponding to post selected state $\phi$. Since the weak value $\langle {A} \rangle_w^{\phi}$ is registered on the final state $\vert M_f \rangle$ of the measuring device, it can be estimated by measuring expectation values of certain observables.

If we multiply Eq.\ref{weakqpt_eq3} by post-selection probability $\Pi_{\psi}^{\phi} = \vert \langle \phi \vert \psi \rangle \vert^2$ we get,
\begin{equation}
\langle {A} \rangle_w^{\phi} \Pi_{\psi}^{\phi}  = \frac{\langle \phi \vert {A} \vert \psi \rangle}{\langle \phi \vert \psi \rangle} \langle \phi \vert \psi \rangle \langle \psi \vert \phi \rangle = \langle \phi \vert {A} \vert {\rho} \vert \phi \rangle
\label{weakqpt_eq4}
\end{equation}
where ${\rho} = \vert \psi \rangle \langle \psi \vert$ is the density matrix representing the initial state of the system. If we choose the observable ${A}$ and post selected state $ \vert \phi \rangle $ such that, $\vert \phi \rangle = \vert n \rangle $ and ${A}^\dagger \vert n \rangle = \vert m \rangle $ where $\vert n \rangle$ and $\vert m \rangle$ are some basis ket vectors then Eq.\ref{weakqpt_eq4} becomes,
\begin{equation}
\langle {A} \rangle_w^{ \phi  } \Pi_{\psi}^{\phi} = \langle m \vert {\rho} \vert  n \rangle
\label{weakqpt_eq5}
\end{equation}
which is nothing but density matrix element $\rho_{mn}$ in chosen basis vectors. And estimating desired element $\rho_{mn}$ is referred as direct quantum state tomography.    

Eq.\ref{weakqpt_eq5} also can be extended to perform DQPT of unknown quantum channel. Generally quantum processes are either represented via corresponding i) $\chi$ matrix (also be referred as process matrix) by means of Kraus operator decomposition\cite{kraus-book-83} or ii) Choi-Jamiolkowski state using channel-state duality theorem\cite{jiang-pra-2013}. In the case of $N$-qubit system, the $\chi$ matrix and the Choi-Jamiolkowski state corresponding to quantum channel $\Lambda$ are given in Eq.\ref{weakqpt_eq6} and \ref{weakqpt_eq7} respectively as follows,
\begin{eqnarray}
\Lambda({\rho}_{in}) &=& \sum_{i=0}^{4^N-1} {K}_i {\rho}_{in} K_i^{\dagger} = \sum_{m,n=0}^{4^N-1} \chi_{mn} {E}_m \hat{\rho}_{in} {E}_n^{\dagger} \label{weakqpt_eq6} \\ 
\left|\Phi_{\Lambda}\right\rangle &=& (I \otimes \Lambda )|\Phi\rangle=\frac{1}{2^{N / 2}} \sum_{m=0}^{2^{N}-1}| m \rangle \otimes \Lambda| m \rangle  \label{weakqpt_eq7}
\end{eqnarray}
where $\chi_{mn}$ in Eq.\ref{weakqpt_eq6} are elements of $\chi$ matrix and $\vert \Phi_{\Lambda} \rangle$ in Eq.\ref{weakqpt_eq7} is Choi-Jamiolkowski state.  $\lbrace K_i \rbrace$'s and the $\lbrace {E}_i \rbrace$'s in Eq.\ref{weakqpt_eq6} are Kraus operators and fixed basis operators respectively while quantum state $\vert \Phi \rangle$ given in Eq.\ref{weakqpt_eq7} is pure maximally entangled state of $2N$ qubits given as, $|\Phi\rangle=2^{-N / 2} \sum_{m=0}^{2^{N}-1}|m\rangle|m\rangle$. The density matrix ${\rho}_{\Lambda} = \vert \Phi_{\Lambda} \rangle \langle \Phi_{\Lambda} \vert $ corresponding to Choi-Jamiolkowski state can be mapped to $\chi$ matrix using appropriate unitary transformation $U_{\chi}$ as, $\chi = U_{\chi} \rho_{\Lambda} U^{\dagger}_{\chi}$. Note that the unitary transformation matrix $ U_{\chi}$ only depends on the fixed set of basis operators $\lbrace E_i\rbrace$ given in Eq.{\ref{weakqpt_eq6}} and does not depend on quantum channel to be tomographed. To perform DQPT of given quantum channel $\Lambda$ in terms of $\chi$ matrix, one needs to apply unitary transformation $ U_{\chi}$ on $\vert \Phi_{\Lambda} \rangle$ and then follow the direct QST protocol and estimate desired element $\chi_{mn}$ using Eq.\ref{weakqpt_eq5} as,
\begin{equation}
\langle {A} \rangle_w^{\phi} \Pi_{\psi}^{\phi} = \langle m \vert U_{\chi} \rho_{\Lambda} U^{\dagger}_{\chi} \vert n \rangle = \langle m \vert \chi \vert n \rangle = \chi_{mn}
\label{weakqpt_eq8}
\end{equation}  
Note that performing direct QPT of $N$ qubit quantum channel is equivalent to performing direct QST of $2N$ qubit quantum state. 

Let's consider the operators $O^{\phi}_x = \vert \phi \rangle \langle \phi \vert \otimes \sigma_x$ and $O^{\phi}y = \vert \phi \rangle \langle \phi \vert \otimes \sigma_y$, where $\vert \phi \rangle$ represents the state for post-selection, and $\sigma_{x(y)}$ refers to the Pauli spin operators for a single qubit. Upon performing some calculations, the expectation values of the $O^{\phi}_x$ and $O^{\phi}_y$ operators can be computed in the weakly evolved joint state, as provided in Eq.\ref{weakqpt_eq1}. This results in the following expressions:
\begin{eqnarray}
\langle O^{\phi}_x \rangle & = & ig \Big[ \langle \psi \vert A^{\dagger} \vert \phi \rangle \langle \phi \vert \psi \rangle - \langle \phi \vert A \vert \psi \rangle \langle \psi \vert \phi \rangle \Big] \label{weakqpt_eq9}\\
\langle O^{\phi}_y \rangle & = & -g \Big[ \langle \psi \vert A^{\dagger} \vert \phi \rangle \langle \phi \vert \psi \rangle + \langle \phi \vert A \vert \psi \rangle \langle \psi \vert \phi \rangle \Big] \label{weakqpt_eq10}
\end{eqnarray}

If Eq. \ref{weakqpt_eq9} is multiplied by $i$ and then subtracted from Eq. \ref{weakqpt_eq10}, the result is obtained as follows:
\begin{equation}
\frac{\langle O^{\phi}_y \rangle - i \langle O^{\phi}_x \rangle }{-2g} = \langle \phi \vert A \vert \psi \rangle \langle \psi \vert \phi \rangle = \langle A \rangle_w^{\phi} \Pi_{\psi}^{\phi} \label{weakqpt_eq11}
\end{equation}

So using Eq.\ref{weakqpt_eq5} and Eq.\ref{weakqpt_eq11} the following equation can be obtained,
\begin{equation}
\frac{\langle O^{\phi}_y \rangle - i \langle O^{\phi}_x \rangle }{-2g} = \rho_{mn} = \langle m \vert \rho \vert n \rangle \label{weakqpt_eq12}
\end{equation}
Similarly using Eq.\ref{weakqpt_eq8} and Eq.\ref{weakqpt_eq11} one can have,
 \begin{equation}
\frac{\langle O^{\phi}_y \rangle - i \langle O^{\phi}_x \rangle }{-2g} =  \chi_{mn} = \langle m \vert \chi \vert n \rangle  \label{weakqpt_eq13}
\end{equation}

So using Eq.\ref{weakqpt_eq12} and \ref{weakqpt_eq13} one can perform direct QST and QPT by measuring $\langle O^{\phi}_x \rangle $ and  $ \langle O^{\phi}_y \rangle $ for appropriate choice of $A$ and $\vert \phi \rangle$. In the following section the detailed procedure and experimental demonstration are given for both direct QST and direct QPT tasks. 

\section{Efficient experimental implementation of weak measuremnt scheme on NMR}\label{weakqpt_sec3}

\begin{figure*}
\centering
{\includegraphics[scale=0.83]{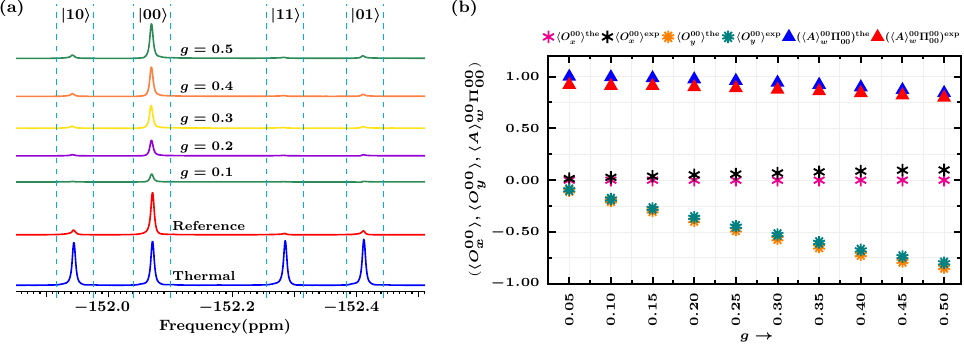}}
\caption{In the part (a), all the NMR spectra are obtained by acquiring third nuclear spin $\rm F_3$ corresponding to meter qubit. The thermal spectra is depicted in blue color while reference spectra is depicted in red color. The first five spectrums (from top) are obtained by implementing quantum circuit given in Fig.\ref{weakqpt_ckt} for different values of $g$ corresponding to initially prepared input state $\vert 00 \rangle_s \vert 0 \rangle_m$ and weak interaction $U^{\rm weak}_{\rm SM}$ corresponding to Pauli operator $\sigma_{1z}$ followed by $90^{\circ}$ phase shift on $\rm F_3$. In part the (b), the theoretically and experimentally obtained quantities $\langle O^{00}_x \rangle, \langle O^{00}_y \rangle$ and $\langle A \rangle^{00}_w \Pi^{00}_{00}$ are compared for different values of $g$. }
\label{fig1_zi}
\end{figure*}
\subsection{Weak measurement of Pauli spin operators on NMR} \label{sec3a}

For experimental implementation three $\rm ^{19}F$ nuclei in the molecule trifluoroiodoethylene, dissolved in acetone-D6 are used as the three-qubit system. 
The first, second, and third qubits are denoted as $\rm F_1$, $\rm F_2$, and $\rm F_3$ nuclear spins, respectively. Among these, $\rm F_1$ and $\rm F_2$ spins serve as system qubits, while $\rm F_3$ acts as the meter qubit.
 The NMR Hamiltonian for given three spin-1/2 nuclei system in the rotating frame is given as,
\begin{equation}
\mathcal{H}=-\sum_{i=1}^{3} \nu_{i} I_{i z}+\sum_{i, j=1, i>j}^{3} J_{i j} I_{i z} I_{j z}
\end{equation}
where, $\nu_i$ is the chemical shift of the $i$th nucleus, $J_{ij}$ is the scaler coupling strength between $i$th and $j$th nuclei and $I_{iz}$ is the $z$-component of spin angular momentum of the $i$th nucleus. Various experimental parameters like chemical shifts, $J$ couplings and relaxation rates characterising given system can be found in the paper \cite{gulati-arxiv-2022}.

The initial state of meter qubit is set to be $\vert M \rangle = \vert 0 \rangle_m $ and observable is set to be $B = \sigma_x$. In this case, the weak interaction evolution operator ${U}^{weak}_{SM}$ will be of the following form,
\begin{equation}
{U}^{weak}_{SM} = {I} -ig {P_k}\otimes {\sigma_x} \label{weakqpt_eq15}
\end{equation}
where operator $I$ is $8 \times 8$ dimensional identity matrix and $P_k = \lbrace I, \sigma_x, \sigma_y, \sigma_z \rbrace^{\otimes 2} $ is two-qubit Pauli spin operator. Note that the operator ${U}^{weak}_{SM}$ given in Eq.\ref{weakqpt_eq15} can be decompose as follows:
\begin{eqnarray}
\footnotesize	
{I} -ig {P_k}\otimes {\sigma_x} &=& {I} -ig (U_{k}\sigma_{iz} U_{k}^{\dagger}) \otimes (R_y(\frac{\pi}{2}) \sigma_z R_y^{\dagger}(\frac{\pi}{2})) \nonumber \\
 &=& \mathcal{U}_k ({I} -ig {\sigma_{iz}}\otimes {\sigma_z}) \mathcal{U}_k^{\dagger} \label{weakqpt_eq16}          
\end{eqnarray}
where $\sigma_{iz}$ is either $\sigma_{1z} = \sigma_z \otimes I$ or $\sigma_{2z} = I \otimes\sigma_z $ and $\mathcal{U}_k = U_k \otimes R_y(\frac{\pi}{2}) $ where $U_k$ is two-qubit unitary operator acting on system qubits and is constructed such that $P_k = U_{k}\sigma_{iz} U_{k}^{\dagger} $. To further simplify Eq.\ref{weakqpt_eq16}, consider the $J$-evolution operator $U_{ij}^{J}(t)$ between $i$th and $j$th qubit,

\begin{equation}
U_{i j}^{J}(t)=e^{-i 2 \pi J_{ij} I_{i z} I_{j z} t }  \label{weakqpt_eq17}
\end{equation}
If the evolution time $t$ is sufficiently small such that, $g = \frac{\pi J_{ij} t}{2} \ll 1$ then Eq.\ref{weakqpt_eq17} can be approximated as,
\begin{equation}
U_{i j}^{J}(t) \approx I - i g \sigma_{iz} \otimes \sigma_{jz}  \label{weakqpt_eq18}
\end{equation}
Using Eq.\ref{weakqpt_eq16} and \ref{weakqpt_eq18} we will have,
\begin{equation}
{U}^{weak}_{SM}  \approx  \mathcal{U}_{k} U_{ij}^J(t) \mathcal{U}_k^{\dagger} \label{weakqpt_eq19}
\end{equation}
where $t = \frac{2g}{\pi J_{ij}}$, $i = 1,2$ and $j=3$.

So the WM of given Pauli operator $P_k$ can be performed by applying sequence of unitary operations given in Eq.\ref{weakqpt_eq19} on initial joint state of three-qubit system followed by the measurement of $O^{\phi}_x$ and $O^{\phi}_y$. The list of all $\mathcal{U}$s corresponding to all $P_k$s is given in the Table.\ref{weakqpt_table1}.

\begin{table}[h!]
\centering
\caption{ The list of all unitary operators $\mathcal{U}_k$ and $U_{i j}^{J}(t)$ to implement weak interaction operation ${U}^{weak}_{SM}$ corresponding to all 2-qubit Pauli operators $P_k$. }
\setlength{\tabcolsep}{15pt} 
\renewcommand{\arraystretch}{1.3}
\footnotesize{
\begin{tabular}{c c c}
\hline \hline $P_k$ &
~~~$\mathcal{U}_k$~~~& ~~~$U_{i j}^{J}(t)$~~~\\
\hline \hline $ \sigma_{2x} $ & $Y_2$ & $U_{23}^{J}(t)$   \\ 
$\sigma_{2y} $ & $\bar{X}_2 $ & $U_{23}^{J}(t)$  \\ 
$\sigma_{2z} $ & $I$ & $U_{23}^{J}(t)$  \\ 
$\sigma_{1x}$ & $Y_1 $ & $U_{13}^{J}(t)$  \\ 
$\sigma_{1x}\sigma_{2x} $ & $\bar{Z}_1 Y_2 U_{12}^{J}(\frac{1}{2J_{12}}) Y_1 $ & $U_{13}^{J}(t)$  \\ 
$\sigma_{1x}\sigma_{2y}$ & $\bar{Z}_1 \bar{X}_2 U_{12}^{J}(\frac{1}{2J_{12}}) Y_1 $ & $U_{13}^{J}(t)$  \\ 
$\sigma_{1x}\sigma_{2z}$ & $\bar{Z}_1 U_{12}^{J}(\frac{1}{2J_{12}}) Y_1 $ & $U_{13}^{J}(t)$ \\ 
$\sigma_{1y}$ & $\bar{X}_1 $ & $U_{13}^{J}(t)$  \\
$\sigma_{1y}\sigma_{2x}$ & $ Y_2 U_{12}^{J}(\frac{1}{2J_{12}}) Y_1 $ & $U_{13}^{J}(t)$  \\
$\sigma_{1y}\sigma_{2y}$ & $ \bar{X}_2 U_{12}^{J}(\frac{1}{2J_{12}}) Y_1 $ & $U_{13}^{J}(t)$   \\
$\sigma_{1y}\sigma_{2z}$ & $ U_{12}^{J}(\frac{1}{2J_{12}}) Y_1 $ & $U_{13}^{J}(t)$  \\
$\sigma_{1z}$ & $I$ & $U_{13}^{J}(t)$  \\
$\sigma_{1z}\sigma_{2x}$ & $X_1 Y_2 U_{12}^{J}(\frac{1}{2J_{12}}) Y_1 $ & $U_{13}^{J}(t)$  \\
$\sigma_{1z}\sigma_{2y}$ & $X_1 \bar{X}_2 U_{12}^{J}(\frac{1}{2J_{12}}) Y_1$ & $U_{13}^{J}(t)$ \\
$\sigma_{1z}\sigma_{2z}$ & $X_1 U_{12}^{J}(\frac{1}{2J_{12}}) Y_1 $ & $U_{13}^{J}(t)$ \\
\hline \end{tabular} }
\label{weakqpt_table1}
\end{table}
\subsubsection{Measurement of $O_x^{\phi}$ and $O_y^{\phi}$ on NMR}

To simplify the analysis, let us consider the post-selected state $\vert \phi \rangle$ to be one of the computational basis vectors: $\lbrace \vert 00\rangle, \vert 01 \rangle, \vert 10 \rangle, \vert 11 \rangle \rbrace$. These basis vectors are essential for performing QST or QPT, as indicated in Eq.\ref{weakqpt_rho}. In this scenario, it is found that the observables $O_{x(y)}^{\phi}$ can be conveniently measured by capturing the NMR signal from the third nuclear spin, denoted as $\rm F_3$ (meter qubit). The NMR signal obtained from the $\rm F_3$ nucleus exhibits four distinct peaks (shown as thermal spectra in blue color in Fig.\ref{fig1_zi}(a)). These peaks correspond to four transitions and are associated with specific elements of the density matrix, referred to as readout elements. Specifically, these readout elements are denoted as $\rho_{56}$, $\rho_{12}$, $\rho_{78}$, and $\rho_{34}$. In the NMR spectra depicted in Fig.\ref{fig1_zi}(a), the first peak from the left (identified using dashed lines) corresponds to the post-selected state $\vert \phi \rangle = \vert 10 \rangle$, while the second, third, and fourth peaks correspond to $\vert 00 \rangle$, $\vert 11 \rangle$, and $\vert 01 \rangle$, respectively. Furthermore, the order of the peaks (from left to right) corresponds to the readout elements $\rho_{56}$, $\rho_{12}$, $\rho_{78}$, and $\rho_{34}$. In the provided spectra, the absorption mode ($x$-magnetization) is directly proportional to the real part of the corresponding readout element, whereas the dispersion mode ($y$-magnetization) is proportional to the imaginary part of the readout element. Upon conducting calculations, the following expressions can be obtained.:
\begin{equation}
 \langle O_{x}^{\phi} \rangle \propto {\rm Re}(\rho_{ij}) \quad {\rm and} \quad \langle O_{y}^{\phi} \rangle \propto {\rm Im}(\rho_{ij})
\end{equation}
where $\rho_{ij}$ is readout element of 3-qubit density matrix on which observables $O_{x(y)}^{\phi}$ are being measured. The complete list of quantities $\langle O_{x(y)}^{\phi} \rangle$ with corresponding energy transitions and readout elements are listed in the Table.\ref{weakqpt_table2}.
\begin{table}[h!]
\centering
\caption{ The list of $\langle O_{x(y)}^{\phi} \rangle$ corresponding to energy transitions and readout elements for post selected state $\vert \phi \rangle$ being one of the computation basis vector.}
\setlength{\tabcolsep}{4pt} 
\footnotesize{
\begin{tabular}{c c c c}
\hline \hline 
$\langle O_{x(y)}^{\phi} \rangle$ &~~~Transitions~~~&~~~elements~~~&~~~Peak (from left)~~~\\
\hline \hline 
$\langle O_{x(y)}^{00} \rangle$ & $\vert 000 \rangle \leftrightarrow \vert 001 \rangle$ & $\rho_{12}$ & second   \\ 
$\langle O_{x(y)}^{01} \rangle$ & $\vert 010 \rangle \leftrightarrow \vert 011 \rangle$ & $\rho_{34}$ & fourth  \\ 
$\langle O_{x(y)}^{10} \rangle$ & $\vert 100 \rangle \leftrightarrow \vert 101 \rangle$ & $\rho_{56}$ & first   \\ 
$\langle O_{x(y)}^{11} \rangle$ & $\vert 110 \rangle \leftrightarrow \vert 111 \rangle$ & $\rho_{78}$ &  third  \\ 

\hline \end{tabular} }
\label{weakqpt_table2}
\end{table}
It should be noted that when dealing with an arbitrary post-selected state $\vert \phi \rangle$, it becomes necessary to decompose the observables $O_{x(y)}^{\phi}$ into operators based on the Pauli basis, given by $O_{x(y)}^{\phi} = \sum_i a_i^{x(y)}P_i$. Subsequently, one needs to measure $\langle P_i \rangle$ for those basis operators $P_i$ that have non-zero coefficients $a_i^{x(y)}$. By doing so, it becomes possible to compute the expectation value $\langle O_{x(y)}^{\phi} \rangle$ for the given post-selected state $\vert \phi \rangle$. The efficient method for measuring the expectation value of any Pauli observable is detailed in the paper by Singh et al. \cite{singh-pra-2018}.

\begin{figure}
\centering
{\includegraphics[scale=1.1]{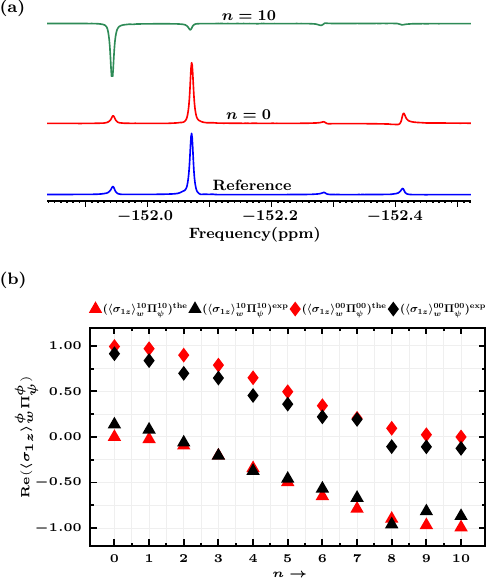}}
\caption{ NMR spectra shown in part (a) are obtained by implementing quantum circuit given in Fig.\ref{weakqpt_ckt}, for initial state of the form $\vert \psi \rangle_s = Cos(\frac{n \pi}{20}) \vert 00 \rangle + Sin(\frac{n \pi}{20}) \vert 10 \rangle $ for fixed value of $g = 0.1$ and observable $\sigma_{1z}$. The spectrum green and red correspond to $n=10$ and $n=0$ respectively. In part (b), experimental quantity $\langle \sigma_{1z} \rangle_w^{\phi} \Pi_{\psi}^{\phi}$ is compared with theoretical value as function of initial state $\vert \psi \rangle_s = Cos(\frac{n \pi}{20}) \vert 00 \rangle + Sin(\frac{n \pi}{20}) \vert 10 \rangle$.}
\label{fig2_zi}
\end{figure}

\subsection{Experimentally demonstrating WM of $\sigma_{1z}$ }

To demonstrate the methodology, an experimental calculation of several pertinent quantities (detailed in Section \ref{weakqpt_sec2}) was carried out for the 2-qubit Pauli operator $\sigma_{z}\otimes I_{2 \times 2} = \sigma_{1z}$. The WM of $\sigma_{1z}$ allows to measure all the diagonal elements of the density matrix in a single experiment. Fig.\ref{fig1_zi} showcases the implementation of proposed scheme on NMR and the measurement of $\langle O_{x(y)}^{\phi} \rangle$ and $A_w^{\phi} \Pi_{\psi}^{\phi}$ quantities, with varying strengths of the weak interaction parameter $g$. For this specific case, following setting are chosen: $A = P_k = \sigma_{1z}$, an initial state of $\vert \psi \rangle = \vert 00 \rangle$, and $\vert \phi \rangle$ as the computational basis vectors. All the NMR spectra depicted in Fig.\ref{fig1_zi}(a) pertain to the $\rm F_3$ nucleus. The bottom spectrum in blue is referred as thermal spectrum obtained by applying a readout pulse on the thermal state, followed by detection on the third spin, $\rm F_3$. The peaks in this spectrum yield valuable information. Referring to Table \ref{weakqpt_table2}, the first peak (from the left) in the thermal spectrum corresponds to $\langle O_{x(y)}^{10} \rangle$, while the second, third, and fourth peaks correspond to $\langle O_{x(y)}^{00} \rangle$, $\langle O_{x(y)}^{11} \rangle$, and $\langle O_{x(y)}^{01} \rangle$, respectively.

The red-colored reference spectrum, depicted as the second spectrum from the bottom, is obtained by applying a readout pulse to an experimentally prepared pseudo-pure state (PPS) using the spatial averaging technique, followed by detection on $\rm F_3$. The reference peak is assigned a value of 1. In reference to this baseline spectrum, the measurement of $\langle O_{x}^{\phi} \rangle$ can be directly obtained by calculating the spectral intensity through integration of the corresponding peak area. Conversely, the measurement of $\langle O_{y}^{\phi} \rangle$ involves applying a $90^{\circ}$ phase shift first, followed by intensity computation. It is important to note that the quantity $\langle O_{y}^{\phi} \rangle$ is equal to the negative value of the spectral intensity. For instance, the green-colored third spectrum (from the bottom) in Fig.\ref{fig1_zi}(a), corresponding to $g = 0.1$, is obtained by implementing the quantum circuit illustrated in Fig.\ref{weakqpt_ckt} and subsequently applying a $90^{\circ}$ phase shift. The peak intensity relative to the reference spectrum is determined to be $0.1821 \pm 0.003$, resulting in $\langle O_{y}^{00} \rangle = -0.1821 \pm 0.003$, while the experimental value of $\langle O_{x}^{00} \rangle$ (intensity prior to the $90^{\circ}$ phase shift) is measured as $0.0272 \pm 0.0066$. Similarly, the other four spectra (from the top) are obtained and correspond to various values of $g$, as shown in the figure. The error bars on the various quantities are calculated by repeating the experiments multiple times.
From Fig.\ref{fig1_zi}(a), it is evident that for all values of $g$, the spectral intensity of the first, third, and fourth peaks, which correspond to the post-selected states $\vert 10 \rangle$, $\vert 11 \rangle$, and $\vert 01 \rangle$, respectively, is negligible compared to the reference peak. This suggests that the quantities $\langle O_{x(y)}^{10} \rangle$, $\langle O_{x(y)}^{11} \rangle$, and $\langle O_{x(y)}^{01} \rangle$ are nearly zero, as expected since theoretically we have $\langle \phi \vert \sigma_{1z} \vert \rho \vert \phi \rangle = 0$, except when $\vert \phi \rangle = \vert 00 \rangle$. Conversely, the spectral intensity of the second peak, corresponding to $\langle O_{y}^{00} \rangle$, is non-zero and increases with $g$.

In Fig.\ref{fig1_zi}(b), the experimental values of $\langle O_{x(y)}^{00} \rangle$ and ${\langle \sigma_{1z} \rangle}_w^{00} \Pi_{00}^{00}$ are compared with their corresponding theoretical values for different values of $g$. Only the real part of ${\langle \sigma_{1z} \rangle}_w^{00} \Pi_{00}^{00}$, denoted as ${\rm Re}({\langle \sigma_{1z} \rangle}_w^{00} \Pi_{00}^{00}) = \frac{\langle O_{y}^{00} \rangle}{-2g}$, is plotted, as the imaginary part is found to be negligible, as observed from the $\langle O_{x}^{00} \rangle$ values. The theoretical and experimental values of various quantities in Fig.\ref{fig1_zi}(b) are computed by simulating the quantum circuit presented in Fig.\ref{weakqpt_ckt} both theoretically and experimentally. The experimental quantity ${\langle \sigma_{1z} \rangle}_w^{00} \Pi_{00}^{00}$ is evaluated using Eq. \ref{weakqpt_eq11}, although it can also be computed by rescaling the spectrum by a factor of $\left\lvert \frac{1}{2g} \right\rvert$ with respect to the reference spectrum. It is important to note that the expected value of ${\langle \sigma_{1z} \rangle}_w^{00} \Pi_{00}^{00}$ is equal to 1, which corresponds to the density matrix element $\rho_{11}$ of the initial state $\vert \psi \rangle = \vert 00 \rangle$. Moreover, as the value of $g$ increases, both the experimental and theoretical values of $\rho_{11}$ deviate more and more from 1. This deviation occurs because the weak interaction approximation no longer holds for relatively large values of $g$. At $g = 0.05$, the experimental value of $\rho_{11}^{\rm exp}$ is obtained as $(0.9198 \pm 0.0057) + i (0.0525 \pm 0.0399)$, while at $g = 0.5$, it is $(0.7971 \pm 0.0021) + i (0.0989 \pm 0.0439)$. Furthermore, it should be noted that in practical experiments, very small values of $g$ may not be effective, as the signal strength after weak interaction could be too weak to detect, leading to significant errors in the respective quantity.

Moreover, the scheme was employed for different initial states. The experimental results, depicted in Fig.\ref{fig2_zi}, are obtained by experimentally implementing the WM scheme with a fixed interaction strength of $g = 0.1$ for various initial states of the form $\vert \psi \rangle = \cos(\frac{n \pi}{20})\vert 00 \rangle + \sin(\frac{n \pi}{20}) \vert 10 \rangle$. In Fig. \ref{fig2_zi}(a), the first NMR spectrum (from the bottom) shown in blue represents the reference spectrum, while the other two spectra in red and green correspond to the states $n = 0$ and $n = 10$, respectively. These spectra are obtained by implementing the quantum circuit shown in Fig.\ref{weakqpt_ckt}, followed by a $90^{\circ}$ phase shift. It is worth noting that since $g = 0.1$ is fixed, the spectra corresponding to all values of $n$ are rescaled by a factor of $\frac{1}{2(0.1)} = 5$ with respect to the reference spectrum. This rescaling directly provides the real and imaginary parts of the corresponding density matrix elements, ${\rm Re}( {\langle \sigma_{1z} \rangle}_w^{\phi} \Pi_{\psi}^{\phi} )$ and ${\rm Im}( {\langle \sigma_{1z} \rangle}_w^{\phi} \Pi_{\psi}^{\phi} )$, respectively. For the case of $n = 0$ (red spectrum), the quantities $\langle O_{y}^{10} \rangle$ (1st peak), $\langle O_{y}^{00} \rangle$ (2nd peak), $\langle O_{y}^{11} \rangle$ (3rd peak), and $\langle O_{y}^{01} \rangle$ (4th peak) are found to be $-0.1370 \pm 0.0008$, $-0.9137 \pm 0.0038$, $-0.0267 \pm 0.0010$, and $-0.1243 \pm 0.0085$, respectively. On the other hand, for $n = 10$ (green spectrum), the corresponding quantities are $0.8687 \pm 0.0054$, $0.1255 \pm 0.0033$, $0.0294 \pm 0.0003$, and $0.0237 \pm 0.0030$, respectively. In Fig.\ref{fig2_zi}(b), the experimentally obtained ${\rm Re}(\langle {\sigma_{1z}}\rangle_w^{\phi} \Pi_{\psi}^{\phi} )$ is compared with the theoretical value for $\vert \phi \rangle = \vert 00 \rangle$ and $\vert \phi \rangle = \vert 10 \rangle$ for various initial states. It is evident that the experimental values align closely with the theoretical values in both Fig.\ref{fig1_zi} and \ref{fig2_zi}, demonstrating the successful implementation of the WM of the $\sigma_{1z}$ operator. In the subsequent subsection, the element-wise complete reconstruction of density and process matrices for various states and quantum gates is demonstrated using the proposed scheme of WM-based DQST and DQPT.

\begin{figure}[t]
\centering
{\includegraphics[scale=1]{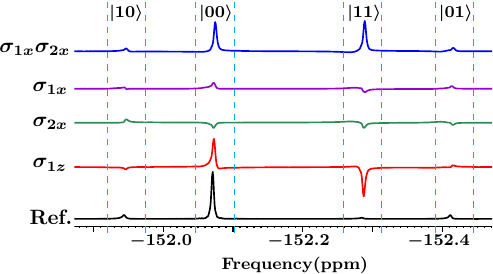}}
\caption{Experimental readouts demonstrating DQST of Bell state $\vert \psi \rangle_s = \frac{1}{\sqrt 2} (\vert 00 \rangle + \vert 11 \rangle) $ via WM technique on NMR. NMR spectra are obtained by implementing quantum circuit given in Fig.\ref{weakqpt_ckt} followed by $90^{\circ}$ phase shift for initial state $\vert \psi \rangle_s = \frac{1}{\sqrt 2} (\vert 00 \rangle + \vert 11 \rangle) $, value of $g = 0.2$ and observables $\sigma_{1z}$ (red), $\sigma_{1x}$(purple), $\sigma_{2x}$(green), and $\sigma_{1x}\sigma_{2x}$ (blue).}
\label{bell_spectra}
\end{figure}
\subsection{Experimental DQST and DQPT using WM scheme}

The proposed WM scheme is experimentally demonstrated for the performance of DQST and DQPT on two-qubit quantum states and single-qubit quantum processes, respectively. To estimate desired element $\rho_{mn}$ of density matrix or $\chi_{mn}$ of process matrix, one of the possible choices of the post selected state $\vert \phi \rangle$ together with the Pauli operator $P_k$ is depicted as $(\vert \phi \rangle, P_k)$ in the corresponding matrix position as given below:

\begin{equation}  \label{weakqpt_rho}
\footnotesize	
\begin{pmatrix}
(\vert 00 \rangle, \sigma_{1z}) & (\vert 01 \rangle, \sigma_{2x}) & (\vert 10 \rangle, \sigma_{1x}) & (\vert 11 \rangle, \sigma_{1x}\sigma_{2x}) \\
\rho_{12}^* & (\vert 01 \rangle, \sigma_{1z}) & (\vert 10 \rangle, \sigma_{1x}\sigma_{2x}) & (\vert 11 \rangle, \sigma_{1x}) \\
\rho_{13}^*  & \rho_{23}^* & -(\vert 10 \rangle, \sigma_{1z}) & (\vert 11 \rangle, \sigma_{2x}) \\
 \rho_{14}^* & \rho_{24}^* & \rho_{34}^* & -(\vert 11 \rangle, \sigma_{1z}) \\
\end{pmatrix}
\end{equation}

In this scenario, conducting a complete QST of a two-qubit quantum state only requires Weak Measurement (WM) of four Pauli operators: $\lbrace \sigma_{1z}, \sigma_{1x}, \sigma_{2x}, \sigma_{1x} \sigma_{2x} \rbrace$. The WM of $\sigma_{1z}$ allows us to directly estimate all the diagonal elements, which represent the population of energy eigenstates and can also be interpreted as the probability of finding the particle in the respective eigenstate. On the other hand, WM of $\sigma_{1x}$, $\sigma_{2x}$, and $\sigma_{1x} \sigma_{2x}$ provides two off-diagonal elements each, representing single and multiple quantum coherences in a selective manner. To demonstrate this, the experimental DQST is performed on maximally entangled Bell states: $\vert \psi_1 \rangle = (\vert 00 \rangle + \vert 11 \rangle)/\sqrt{2} $ and $\vert \psi_2 \rangle = (\vert 01 \rangle + \vert 10 \rangle)/\sqrt{2} $. Additionally, the DQPT is conducted on two quantum gates: $H$ (Hadamard gate) and $R_x(\frac{\pi}{2})$ (rotation gate around the x-axis). In both cases of DQST and DQPT, the value of $g$ has been set to $0.2$

\begin{table}[h!]
\centering
\caption{ Experimental state and process fidelities obtained using WM based DQST and DQPT method respectively.}
\setlength{\tabcolsep}{15pt} 
\renewcommand{\arraystretch}{1.3} 
\footnotesize{
\begin{tabular}{c c c}
\hline \hline 
state($\vert \psi \rangle$)/process($\Lambda$) &~~~$\mathcal{F}(\rho^{\rm dqst}_{\rm weak})$~~~&~~~$\mathcal{F}(\rho^{\rm true}_{\rm weak})$~~~\\
\hline \hline 
$\vert \psi_1 \rangle =(\vert 00 \rangle + \vert 11 \rangle)/\sqrt{2} $ & $0.9511 \pm 0.0065$ &  $0.9791$   \\ 
$\vert \psi_2 \rangle =(\vert 01 \rangle + \vert 10 \rangle)/\sqrt{2} $ & $0.9266 \pm 0.0075$ &  $0.9739$  \\ 
$\Lambda_1 = H $ & $0.9447 \pm 0.0060$ &  $0.9703$   \\ 
$\Lambda_2 = R_x(\frac{\pi}{2}) $ & $0.9476 \pm 0.0029$ & $0.9729$   \\ 

\hline \end{tabular} }
\label{weakqpt_table3}
\end{table}

In Fig.\ref{bell_spectra}, the NMR readouts demonstrating the DQST of Bell state $\vert \psi_1 \rangle = (\vert 00 \rangle + \vert 11 \rangle)/\sqrt{2}$ are shown where WM of four Pauli operators are carried out. The NMR readouts corresponding to WM of $\sigma_{1z}$, $\sigma_{2x}$, $\sigma_{1x}$ and $\sigma_{1x}\sigma_{2x}$ are depicted in color red, green, purple and blue respectively, while bottom most spectrum given in black color represents the reference spectrum. As mentioned earlier the 1st, 2nd, 3rd and 4th peaks (regions are specified using dashed lines) correspond to the post-selected state $\vert 10 \rangle$, $\vert 00 \rangle$, $\vert 11 \rangle$ and $\vert 01 \rangle$ respectively. One can clearly see that, the non-zero spectral intensity of NMR peaks corresponding to $(\vert 00 \rangle, \sigma_{1z})$, $(\vert 11 \rangle, \sigma_{1z})$, $(\vert 11 \rangle, \sigma_{1x}\sigma_{2x})$ yield three density matrix elements $\rho_{11}$, $\rho_{44}$ and $\rho_{14}$ respectively whereas other peak intensities corresponding to respective elements as shown in Eq.\ref{weakqpt_rho} tend to zero as compared to reference spectra. The experimentally obtained real and imaginary parts of the density matrix corresponding to the Bell state $\vert \psi_1 \rangle $ is given in Eq.\ref{bell_re} and \ref{bell_im} respectively. All the elements are measured with considerably high accuracy and precision. Note that the experimental density matrix is Hermitian by construction (the imaginary part of all diagonal elements can be ignored) but may not satisfy positivity and trace conditions as all the independent elements $\lbrace \rho_{ij}, i \leq j \rbrace$ are computed individually and independently. In the case of DQST of Bell state $\vert \psi_1 \rangle$, the trace turns out to be $1.3435$ and eigenvalues are $1.2836$, $0.2825$, $-0.1553$ and $-0.0673$. However, the underlying true quantum state satisfying all the three properties of density matrix can be recovered from experimental density matrix by solving the constrained convex optimization problem as follows,
\begin{subequations}
\begin{alignat}{2}
&\!\min_{\overrightarrow{\rho}^{\rm true}_{\rm weak}}        &\qquad& \Vert \overrightarrow{\rho}^{\rm true}_{\rm weak}-\overrightarrow{\rho}^{\rm dqst}_{\rm weak}\Vert_{l_2}\label{weakqpt_eq122}\\
&\text{subject to} &      & \rho^{\rm true}_{\rm weak} \geq 0,\label{weakqpt_eq122:constraint1}\\
&                  &      & { Tr}(\rho^{\rm true}_{\rm weak}) =1.\label{weakqpt_eq122:constraint2} 
\end{alignat}
\end{subequations}
where ${\rho}^{\rm true}_{\rm weak}$ is the variable density matrix corresponding to the true quantum state to be constructed while ${\rho}^{\rm dqst}_{\rm weak}$ is experimentally obtained density matrix using weak measurement DQST scheme. The overrightarrow denotes the vectorized form of the corresponding matrix and $\Vert . \Vert_{l_2}$ represents $l_2$ norm, also referred as Euclidean norm of vector.  The valid density matrix ${\rho}^{\rm true}_{\rm weak}$ representing underlying the true quantum state is recovered from ${\rho}^{\rm dqst}_{\rm weak}$  and is given in Eq.\ref{bell_true}.  Also note that the experimentally obtained density matrices ${\rho}^{\rm dqst}_{\rm weak}$ (or ${\rho}^{\rm true}_{\rm weak}$) corresponding to states $\vert \psi_1 \rangle$ and $\vert \psi_2 \rangle$ can be interpreted as Choi-6Jamiolkowski state corresponding to identity gate ($\Lambda = I$) and bit flip gate ($\Lambda = \sigma_x$) respectively.
{\footnotesize
\begin{equation} \label{bell_re}
\rm{Re}(\rho^{\rm dqst}_{\rm weak})=\begin{pmatrix}
 0.6089\pm 0.0015 & -0.0104\pm 0.0007 & 0.0774\pm 0.0047 & 0.6314\pm 0.0074 \\
 -0.0104\pm 0.0007 & 0.0155\pm 0.0059 & 0.0633\pm 0.0073 & -0.0534\pm 0.0183 \\
 0.0774\pm 0.0047 & 0.0633\pm 0.0073 & 0.0764\pm 0.0019 & -0.0956\pm 0.0292 \\
 0.6314\pm 0.0074 & -0.0534\pm 0.0183 & -0.0956\pm 0.0292 & 0.6425\pm 0.0011 \\
\end{pmatrix}
\end{equation}
\begin{equation} \label{bell_im}
\rm{Im}(\rho^{\rm dqst}_{\rm weak})=\begin{pmatrix}
 0 & -0.0566\pm 0.0085 & 0.0352\pm 0.0123 & 0.1126\pm 0.0374 \\
 0.0566\pm 0.0085 & 0 & 0.0860\pm 0.0217 & -0.1384\pm 0.0036 \\
 -0.0352\pm 0.0123 & -0.0860\pm 0.0217 & 0 & 0.1367\pm 0.0139 \\
- 0.1126\pm 0.0374 & 0.1384\pm 0.0036 & 0.1367\pm 0.0139 & 0 \\
\end{pmatrix}
\end{equation}
\begin{equation} \label{bell_true}
\rho^{\rm true}_{\rm weak}=\begin{pmatrix}
 0.4667 & -0.0300 - 0.0333i & -0.0217 - 0.0618i  & 0.4858 - 0.0811i \\
  -0.0300 + 0.0333i &  0.0043  &  0.0058 + 0.0024i & -0.0255 + 0.0399i \\
  -0.0217 + 0.0618i  &  0.0058 - 0.0024i &  0.0092  &  -0.0118 + 0.0681i \\
   0.4858 + 0.0811i &  -0.0255 - 0.0399i &  -0.0118 - 0.0681i  &  0.5198  \\
\end{pmatrix}
\end{equation}
}
\begin{figure}[t]
\centering
{\includegraphics[scale=1.1]{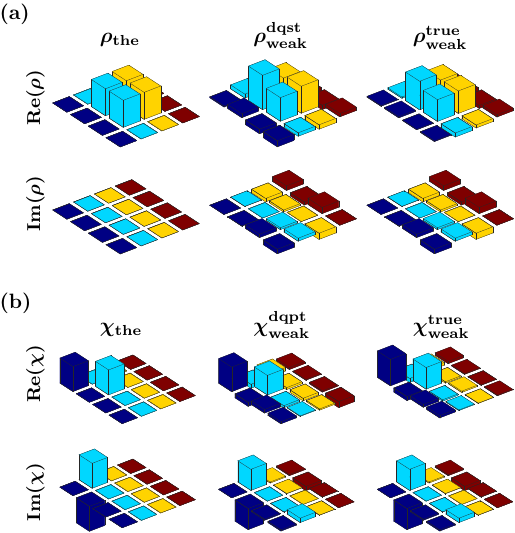}}
\caption{In part (a) theoretical and experimental density matrices corresponding to quantum state $\vert \psi_2 \rangle = (\vert 01 \rangle + \vert 10 \rangle)/\sqrt{2} $  are given whereas in part (b) theoretical and experimental process matrix corresponding to rotation operation $R_{x}(\frac{\pi}{2})$ is given.}
\label{tomo}
\end{figure}
For the DQPT case, the unitary operator $U_{\chi}$ serves as a change of basis operation that transforms the Choi-Jamiolkowski state to the process matrix $\chi$ in the selected basis operators. In this context, the desired unitary operator $U_{\chi}$ is defined as follows:
\begin{equation} \label{uchi}	
U_{\chi}=\frac{1}{\sqrt{2}}
\begin{pmatrix}
1 & 0 & 0 & 1 \\
0 & 1 & 1 & 0 \\
0 & -i & i & 0 \\
1 & 0 & 0 & -1
\end{pmatrix}
\end{equation}
The unitary operator $U_{\chi}$, as defined in Eq.\ref{uchi}, enables to estimate the process matrix $\chi$ in the Pauli basis.
For the Hadamard gate $H$, the real and imaginary parts of the process matrix $\chi^{\text{dqst}}_{\text{weak}}$ in the Pauli basis, obtained through the WM-based DQPT protocol, are provided in Eq.\ref{had_re} and \ref{had_im}, respectively. The trace of the process matrix is found to be 0.9589, and the eigenvalues are -0.1646, 0.0795, 0.1427, and 0.9014. However, the true quantum process $\chi^{\text{true}}_{\text{weak}}$ can be recovered from $\chi^{\text{dqst}}_{\text{weak}}$ by solving a similar convex optimization problem outlined in Eq.\ref{weakqpt_eq122}, with an additional constraint: $\sum_{m, n} \chi_{m n} E_n^{\dagger} E_m = I$. The recovered true quantum process is given in Eq.\ref{had_true}. The theoretical process matrix for the Hadamard gate contains only four non-zero elements, specified as $\lbrace \rho_{ij} = 0.5 \vert i,j=2,4 \rbrace$. From Eq.\ref{had_re} and \ref{had_im}, it is evident that the WM scheme accurately determines all these elements. The graphical representation of the theoretical and experimental density and process matrices corresponding to the quantum state $\vert \psi_2 \rangle$ and the gate $R_x(\frac{\pi}{2})$ is shown in Fig.\ref{tomo}.
In all cases, the experimental state (process) fidelity $\mathcal{F}$ is computed by normalizing the trace distance between the experimental and theoretical density (process) matrices. The experimental fidelity for various quantum states and processes obtained via the WM scheme is presented in Table \ref{weakqpt_table3}.
{\footnotesize
\begin{equation} \label{had_re}
\rm{Re}(\chi^{\rm dqst}_{\rm weak})=\begin{pmatrix}
 -0.0010 \pm 0.0005 &0.0422\pm 0.0041&0.0635\pm 0.0012&-0.0854\pm 0.0004\\
0.0422\pm 0.0041&0.3964 \pm 0.0099&-0.0827\pm 0.0004 &0.4406 \pm 0.0097\\
0.0635\pm 0.0012&-0.08269\pm 0.0004&0.0789 \pm 0.0036&0.0429\pm 0.0019\\
-0.0854\pm 0.0004&0.4406\pm 0.0097 &0.0429\pm 0.0019&0.4846\pm 0.0035\\
\end{pmatrix}
\end{equation}
  
\begin{equation} \label{had_im}
\rm{Im}(\chi^{\rm dqst}_{\rm weak})=\begin{pmatrix}
 0&0.0243\pm 0.0033&0.0554\pm 0.0155 &-0.0195\pm 0.0012\\
-0.0243\pm 0.0033&0&-0.0664\pm 0.0081&0.0754\pm 0.0445\\
-0.0554\pm 0.0155&0.0664\pm 0.0081&0&0.0633\pm 0.0445\\
0.0195\pm 0.0012&-0.0754\pm 0.0045&-0.0633\pm 0.0004&0\\
\end{pmatrix}
\end{equation}

 \begin{equation} \label{had_true}
\chi^{\rm true}_{\rm weak}=\begin{pmatrix}
 0.0319  &  -0.0145 + 0.0072i  & 0.0144 + 0.0246i  & -0.0389 + 0.0077i \\
  -0.0145 - 0.0072i  &  0.4021  &  -0.0380 - 0.0542i   & 0.3992 + 0.0653i \\
   0.0144 - 0.0246i  & -0.0380 + 0.0542i  &  0.0831  &   0.0008 + 0.0646i \\
  -0.0389 - 0.0077i  &  0.3992 - 0.0653i  &  0.0008 - 0.0646i  &  0.4829  \\
\end{pmatrix}
\end{equation}
}
\subsubsection{Extension to $n$-qubit system}

In the case of $n$-qubit density (or process) matrix, all the independent elements $\lbrace \rho_{ij}, i \leq j \rbrace$ can be obtained in a similar manner as described in Eq.\ref{weakqpt_rho}. The $2^n$ diagonal elements can be determined using the WM of the $\sigma_{1z} = \sigma_z \otimes I^{\otimes n-1}$ operator. On the other hand, the $2^{n-1}(2^n-1)$ off-diagonal elements ($\lbrace \rho_{ij}, i < j \rbrace$) can be obtained through the WM of $n$-qubit Pauli operators, specifically those of the form $\lbrace I, \sigma_x \rbrace^{\otimes n}$ (excluding $I^{\otimes n}$), with each operator yielding $2^{n-1}$ elements. For example, the operator $\sigma_x^{\otimes n}$ will provide the off-diagonal elements $\lbrace \rho_{ij}, 1 \leq i \leq 2^{n-1}, j = 2^n+1-i \rbrace$. The complete density matrix necessitates the WM of $2^n$ Pauli operators, in contrast to the standard protocol, which requires measurements of $4^n-1$ operators. Thus, even in the case of full reconstruction, the WM-based tomography protocol proves to be significantly more efficient than both the standard protocol and other selective tomography protocols. In terms of experimental implementation, the quantum circuit depicted in Fig.\ref{weakqpt_ckt} can be extended to an $n$-qubit system. For the DQST case, an additional qubit will be required as a meter qubit, while for the DQPT case, $n$ extra ancillary qubits, along with the meter qubit, will be necessary.

\section{Conclusions}\label{weakqpt_sec4}

An efficient scheme was introduced, and a generalized quantum circuit was constructed for the execution of DQST and DQPT using the WM technique. This scheme was successfully demonstrated on an NMR quantum information processor. Through the utilization of $J$-coupling, control over the interaction strength between spin qubits was attained, facilitating the simulation of the WM process on the NMR platform. This simulation allowed for the computation of pertinent quantities with good accuracy and precision. The proposed protocol facilitates the direct determination of multiple selective elements within the density and process matrices of an unknown quantum state/process, in a single experiment. This efficiency enhancement places it ahead of other direct tomography methods. To showcase its efficacy, DQST and DQPT were conducted on various quantum states and processes, resulting in high fidelities.

Furthermore, a convex optimization method was employed to recover the underlying true quantum states and processes using the data obtained through the WM scheme. In all instances, a substantial enhancement in experimental fidelities was achieved. To the best of our knowledge, this is the first experimental demonstration of WM-based DQST and DQPT on an ensemble quantum computer like NMR, without necessitating any projective measurements. It should be emphasized that, unlike other WM-based DQST (or DQPT) methods that demand projective measurements on system qubits, potentially leading to significant disruption of the state of the system, the proposed protocol avoids any measurements on system qubits, thereby minimizing disturbance to the state of the system. The experimental demonstration opens up new avenues for exploring diverse and intriguing WM experiments that were previously constrained by the limitations of projective measurements on ensemble quantum systems.

\chapter{Experimental simulation of non-unitary quantum processes on NMR}

\section{Introduction}
\label{snd_sec1}

In 1982, Richard Feynman introduced the concept of utilizing a universal quantum computer to simulate quantum systems\cite{feynman-ijtp-1982}, which garnered significant attention within the scientific community\cite{lloyd-science-1996,kassal-arpc-2011,franco-rmp-2014}. Subsequently, extensive efforts were made over the ensuing decades to construct quantum computers capable of exponentially outperforming classical counterparts in solving computational problems\cite{nielsen-book-10}. The fundamental component of a quantum computer is the underlying physical system and its temporal evolution under a specific Hamiltonian\cite{divi-fdp-2000}. However, the primary challenge in constructing such a quantum computer lies in mitigating its inevitable interaction with the surrounding environment, commonly referred to as decoherence. As a result, investigations into open quantum dynamics were pursued, encompassing various methodologies for studying the time evolution of quantum systems\cite{Breuer2007,Rotter-rpp-2015}.

In practical scenarios, the physical system being considered continuously interacts with its environment, resulting in a non-unitary time evolution. This interaction introduces noise from open dynamics, which can significantly contribute to errors in the computational output, leading to a decrease in experimental fidelity and a decline in the quality of the quantum device\cite{zuniga-pra-2012}. To address these challenges, several approaches have been proposed. One such approach is a duality quantum algorithm that simulates the Hamiltonian evolution of an open quantum system using Kraus operators\cite{Wei-sr-2016,Zheng-sr-2021}. This algorithm allows for the realization of time evolution while considering the non-unitary nature of the system. Another quantum algorithm was developed to simulate general finite-dimensional Lindblad master equations without the need for specific system-environment engineering\cite{Candia-sr-2015}. This provides a versatile method for simulating a wide range of quantum systems. Additionally, a recent proposal introduced a method for efficiently simulating open quantum dynamics for various Hamiltonians and spectral densities\cite{Zhang-fp-2021}. This approach aims to improve the computational efficiency and accuracy of simulations in the presence of open system dynamics.

Numerous techniques have been suggested for simulating specific types of quantum channels, and these techniques have been successfully demonstrated in experiments utilizing various physical platforms. For instance, an optical setup was utilized to demonstrate a control technique that transitions an open quantum system from the Markovian to the non-Markovian regime\cite{Liu-np-2011}. In another study, a model was developed that enables precise control over non-Markovian effects by adjusting the degree of correlation and the interaction time between qubits and their environment in an NMR system\cite{Bernardes-sr-2016}. The decoherence dynamics of a qubit were experimentally demonstrated using photons through the implementation of non-positive dynamical maps\cite{Liu-natcom-2018}. Additionally, a technique for simulating both Markovian and non-Markovian dynamics was proposed in the context of a cavity-QED setup\cite{patsch-prr-2020}. Multiqubit open dynamics were simulated on an IBM quantum processor for various quantum processes, including unital and non-unital dynamics, as well as Markovian and non-Markovian evolution\cite{garcia-npj-2020}. Furthermore, a dilation procedure employing ancilla qubits was employed to simulate non-Hermitian Hamiltonian dynamics\cite{dogra-commphy-2021}. These studies highlight the versatility of different experimental setups and methods in simulating and understanding quantum phenomena.

Recently, promising quantum algorithms have emerged for simulating arbitrary non-unitary evolutions on quantum devices. These algorithms are primarily based on the dilation technique, specifically the Stinespring dilation algorithm\cite{shirokov-jmp-2020} and Sz.-Nagy's dilation algorithm\cite{prineha-prr-2021}. The fundamental principle underlying these algorithms is the construction of a unitary operation in a higher-dimensional Hilbert space, which effectively simulates the desired non-unitary evolution in a lower-dimensional Hilbert space. It should be noted that the Stinespring dilation algorithm requires a larger Hilbert space dimension, resulting in higher computational and experimental costs compared to the Sz.-Nagy algorithm. Nevertheless, the Sz.-Nagy algorithm has been successfully utilized to experimentally simulate the single-qubit amplitude damping channel on the IBM quantum processor\cite{Hu-sr-2020}. These advancements highlight the potential of these algorithms in facilitating the simulation of non-unitary quantum processes on practical quantum computing platforms.

This chapter demonstrates the experimental implementation of the Sz.-Nagy quantum algorithm on an ensemble NMR quantum information processor to simulate open quantum dynamics\cite{gaikwad-pra-2022}. The Sz.-Nagy algorithm requires knowledge of the complete set of corresponding Kraus operators to simulate the given quantum dynamics of an open quantum system. However, in realistic scenarios, these Kraus operators may not be readily available. In such cases, it is necessary to first compute the complete set of Kraus operators before proceeding with the implementation of the Sz.-Nagy algorithm. To address this, QPT is utilized to compute the process matrix, which characterizes the specific quantum process under investigation\cite{gaikwad-pra-2018}. By employing unitary diagonalization and Lindblad generators, the complete set of Kraus operators corresponding to a general quantum channel are computed. To showcase the effectiveness of the Sz.-Nagy algorithm, the experimental simulations of three non-unitary quantum processes acting on a two-qubit system have been performed. These processes included a phase damping channel acting independently on the two qubits, a correlated amplitude damping channel with known or theoretically obtainable Kraus operators, and a magnetic field gradient pulse (MFGP) with unavailable Kraus operators that needed to be computed. Furthermore, to validate the quality of the experimentally simulated quantum channel, the convex optimization-based full quantum process tomography\cite{gaikwad-ijqi-2020} is conducted.


\begin{figure}[h]
\centering
\includegraphics[angle=0,scale=1]{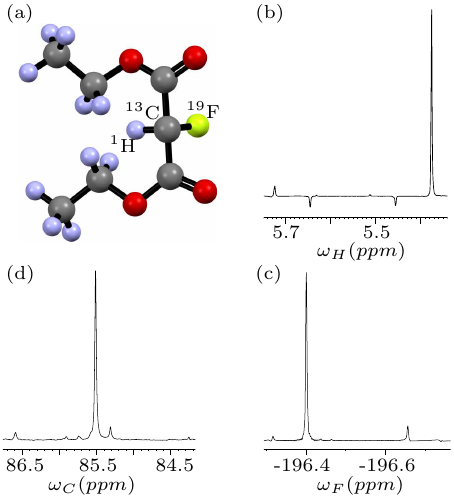} 
\caption{
(a) Molecular structure of diethyl fluoromalonate labeled with ${}^{13}$C, used as the three-qubit quantum system.
The spectra presented in (b), (c), and (d) correspond to the $^1$H, $^{19}$F, and $^{13}$C spins, respectively, obtained after applying a $90^{\circ}$ readout pulse on the initial state $\vert 000 \rangle$. The $J$-couplings between different nuclei are as follows: $J_{HF} = 47.5$ Hz, $J_{HC} = 161.5$ Hz, and $J_{FC} = -191.7$ Hz. The spin-lattice relaxation times measured for the different nuclei are $T^{H}_1 = 3.0 \pm 0.34$ s, $T^{F}_1 = 3.3 \pm 0.15$ s, and $T^{C}_1 = 3.2 \pm 0.38$ s, while the spin-spin relaxation times are $T^{H}_2 = 1.3 \pm 0.24$ s, $T^{F}_2 = 1.4 \pm 0.22$ s, and $T^{C}_2 = 1.2 \pm 0.18$ s.
} 
\label{fig1}
\end{figure}

\begin{figure*}[t] 
\centering
\includegraphics[angle=0,scale=0.83]{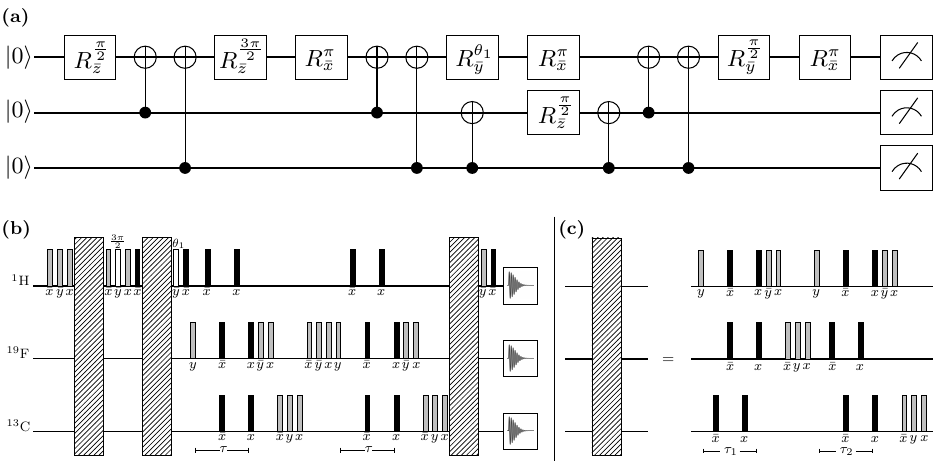} 
\caption{
(a) Quantum circuit diagram illustrating the simulation of the Kraus operator $A_1$ of the phase damping channel on the initial state $\vert 00 \rangle$ using the Sz.-Nagy algorithm. The input state $\vert 00 \rangle$ is encoded as $\vert 000 \rangle_{\rm HFC} = \vert 0 \rangle_{\rm H} \otimes \vert 00 \rangle_{\rm FC}$ to comply with the requirements of the algorithm.
(b) NMR-based implementation of the quantum circuit depicted in (a).
(c) The corresponding pulse phases are indicated below. The intervals of free evolution time are specified as $\tau = 0.0078$ s, $\tau_1 = 0.0105$ s, and $\tau_2 = 0.0031$ s, respectively.}  
\label{snd_fig2}
\end{figure*}

\section{Time evolution of open quantum systems and the Sz.-Nagy algorithm}
\label{snd_sec2} 
The Sz.-Nagy algorithm enables the evolution of the system's density matrix, denoted as $\rho$, to a density matrix $\rho(t)$ at time $t$ according to a specified evolution model. To mathematically describe the framework of the Sz.-Nagy algorithm, consider the Kraus operator-sum representation for the time evolution of the density matrix:

\begin{equation}
\rho(t) = \sum_i A_i \rho A_i^{\dagger} \label{eq1}
\end{equation}

Here, the $A_i$ operators are Kraus operators that satisfy the condition $\sum_i A_i^{\dagger}A_i=I$, where $I$ represents the identity operator. To effectively implement the Sz.-Nagy algorithm and simulate any desired open quantum dynamics, it is essential to possess the complete set of Kraus operators described in Eq.\ref{eq1}.  

The Sz.-Nagy algorithm proposes that for any contraction operator $W$ acting on a vector $v$ in a Hilbert space ${\cal H}_1$, it is possible to construct a corresponding unitary dilation operator $U_w$ in a larger Hilbert space ${\cal H}_2$. This construction is such that the action of $W^m$ can be simulated by the unitary dilation operator as follows:

\begin{equation}
W^m = P_{{\cal H}_1} U_w^m P_{{\cal H}_1}, \quad m \leq N \label{eq4}
\end{equation}

Here, $P_{{\cal H}_1}$ is the projection operator that maps the output vector back into the space ${\cal H}_1$. It should be noted that the dimension of ${\cal H}_2$ is greater than the dimension of ${\cal H}_1$ (dim($ {\cal H}_2$) > dim($ {\cal H}_1 $)). The parameters $m$ and $N$ are integers. It is important to consider that a contraction operator $W$ is one that either preserves or reduces the norm of any vector, denoted as $\Vert W \Vert = \text{sup} \frac{\Vert Wv \Vert }{\Vert v \Vert} \leq 1$. Eq.\ref{eq4} suggests that the repeated action of the contraction operator $W$ in ${\cal H}_1$, up to $N$ times, can be simulated by the corresponding unitary dilation operator $U_w$ acting up to $N$ times in the larger space ${\cal H}_2$, provided that the input vector lies in ${\cal H}_1$ and the output vector is projected back into ${\cal H}_1$.

Consider the set of Kraus operators $\lbrace A_i \rbrace$ (Eq.~\ref{eq1}) that correspond to a given quantum process, which evolves the initial density matrix to $\rho(t)$. It is necessary for the Kraus operator $A_i$ to be a 'contraction operator' in order to determine its corresponding unitary dilation operator. The a comprehensive proof that any Kraus operator satisfies all the properties of being a 'contraction operator,' is given in the Reference\cite{Hu-sr-2020}. For simplicity, consider an $n$-qubit system with a corresponding Hilbert space $\mathcal{H}$ of dimension $2^n$, where the initial density matrix $\rho$ is in a pure state, i.e., $\rho = \vert \phi \rangle \langle \phi \vert$. In this case, the steps to implement the Sz.-Nagy algorithm for simulating Eq.~\ref{eq1} are as follows: 
 
\begin{enumerate}
\item Prepare the pure input state $ \vert \Phi \rangle  =
\vert 0 \rangle \otimes \vert \phi \rangle$ in a larger
Hilbert space of dimension $2^{n+1}$.
\item Apply the unitary operation $U_{A_i}$ on the input state $\vert \Phi \rangle$, where
$U_{A_i}$ is the minimal unitary dilation of $A_i$ (with
$N=1$) given by:
\begin{equation}
{U}_{{A_i}}=\left(\begin{array}{cc} {A}_i &
{D}_{A_i^{\dagger}} \\ {D}_{{A_i}} & -A_i^{\dagger}
\end{array}\right) \label{snd_eq5} \end{equation} where
${D}_{{A_i}} = \sqrt{I-A_i^{\dagger}A_i}$.  
\item 
Project the output vector
$U_{A_i}\vert \Phi \rangle$ into a smaller Hilbert space
$\mathcal{H}$, using the appropriate projection operator
$P_{\mathcal{H}}$, the dimension of $\mathcal{H}$ being $2^n$.  
\item Repeat the above steps for the remaining
Kraus operators, and 
sum over all output density matrices
obtained after
Step~3, in order  
to compute the effect of the given quantum process on the input state
$\rho$.  
\end{enumerate}
 From Eq.\ref{snd_eq5}, one can see that for $n$-qubit system, the Kraus operator will be $2^n \times 2^n$ dimensional matrix and the corresponding unitary dilation operator  given in Eq.\ref{snd_eq5} will be $2^{n+1} \times 2^{n+1}$ dimensional operator which need to be implemented on $n+1$ qubit system to simulate the action of given $n$-qubit kraus operator. However, as the system size increases, the number of Kraus operators also grows exponentially. For an $n$-qubit system, the maximum number of Kraus operators increases as $4^n$. In such scenarios, the Sz.-Nagy algorithm remains efficient in terms of experimental resources when selectively simulating the action of a desired Kraus operator, as it requires only one extra qubit. In comparison to existing duality simulation algorithms presented in \cite{Zheng-sr-2021} and \cite{xin-pra-2017}, where an $n$-qubit system with $4^n$ Kraus operators necessitates an ancilla system of dimension $4^n$ (i.e., $2^n$ ancilla qubits) to simulate $n$-qubit open quantum dynamics, the Sz.-Nagy algorithm requires just one additional qubit. Consequently, Sz.-Nagy's algorithm remains highly efficient for larger systems in terms of experimental resources, especially when the number of experiments to be performed equals the total number of Kraus operators.

Please note that if the initial density matrix is in a mixed state, represented as $\rho = \sum_j p_j \vert \phi_j \rangle \langle \phi_j \vert$, it is necessary to repeat the Sz.-Nagy algorithm for each individual state $\vert \phi_j \rangle$ in order to determine the impact of the given quantum process on the initial density matrix that is in a mixed state.

\section{Experimentally simulating two-qubit 
non-unitary quantum processes}
\label{snd_sec3}
The focus now shifts to the experimental implementation of the Sz.-Nagy algorithm. The objective is to simulate a two-qubit pure phase damping channel, a correlated amplitude damping channel, and an MFGP process on an NMR quantum information processor. To achieve this goal, a three-qubit system will be employed. In the case of the pure phase damping channel and MFGP, the experiments were carried out utilizing a three-qubit system comprised of ${}^{13}$C-labeled diethyl fluoromalonate dissolved in acetone-D6. The assignment was made such that the $^1$H, $^{19}$F, and $^{13}$C spins represented the first, second, and third qubits, respectively. The experimental parameters are depicted in Fig.\ref{fig1} whereas for the correlated amplitude damping channel, a three-qubit system was established using three $^{19}$F nuclei within the trifluoroiodoethylene molecule, dissolved in acetone-D6.

\subsection{Simulating a two-qubit phase damping channel}
\label{pd}
The phase damping is a well-known phenomenon that plays a crucial role in solution NMR, where it contributes to the relaxation of the spin ensemble. In certain real-life scenarios, the limited experimental fidelity of quantum gates with long implementation times can be attributed to the detrimental effects of the phase damping channel. Numerous studies have focused on preserving delicate quantum coherences in the presence of phase damping \cite{singh-epl-2017, singh-pra-2017, singh-pra-2018, singh-pramana-2020}. In the paper \cite{souza-qic-2010}, the authors demonstrate circuit-based quantum simulations of single-qubit phase damping, generalized amplitude damping channels, and relaxation processes in spin 3/2 systems. However, the method used to construct the circuits is not general and does not enable the simulation of arbitrary open quantum dynamics in spin-3/2 systems. Specifically, the simulation of spin-3/2 quadrupole relaxation requires a total of a 7-qubit system, with the first two qubits representing the system qubits (spin-3/2 system) and the last five qubits serving as ancilla qubits (acting as the environment). This clearly illustrates the costly nature of experimental implementation. In contrast, Sz.-Nagy's algorithm only requires one additional qubit to simulate open dynamics of arbitrary dimensions.

In this case, the superoperator representation is utilized to depict the open quantum dynamics of a system that underging a phase damping channel, with the known generator of the phase damping process. The generators associated with the phase damping channel acting independently on qubit 1 and qubit 2 are denoted as $\mathcal{Z}_1$ and $\mathcal{Z}_2$ respectively. The matrix forms of $\mathcal{Z}_1$ and $\mathcal{Z}_2$ are given by \cite{childs-pra-2001,singh-pramana-2020}:

\begin{equation} \begin{aligned} \mathcal{Z}_1=&
\operatorname{diag}\left[0,0,-\gamma_{1},-\gamma_{1},0,
0,-\gamma_{1},-\gamma_{1},-\gamma_{1},-\gamma_{1},0,0,\right.\\
&\left.-\gamma_{1},-\gamma_{1},0, 0\right] \nonumber \\
\mathcal{Z}_2=&
\operatorname{diag}\left[0,-\gamma_{2},0,-\gamma_{2},-\gamma_{2},0
-\gamma_{2},0,0,-\gamma_{2},0-\gamma_{2},\right.\\
&\left.-\gamma_{2},0,-\gamma_{2},0\right] \\ \end{aligned}
\end{equation} 

Here, $\gamma_1$ and $\gamma_2$ represent the phase damping rates for qubit 1 and qubit 2 respectively. The resulting process is denoted by the superoperator $\Xi$, which consists of the simultaneous independent action of the phase damping channel on qubit 1 and qubit 2, with a generator given by $\mathcal{Z} = \mathcal{Z}_1+\mathcal{Z}_2$. The time evolution of the initial density matrix $\rho$ for the two-qubit system can be expressed as \cite{childs-pra-2001}:

\begin{equation}
\rho(t)=\Xi(\rho)=e^{\mathcal{Z}t}(\vec{\rho}) \label{snd_eq6}
\end{equation}

To simulate Eq.~\ref{snd_eq6} using the Sz.-Nagy algorithm, it is necessary to have the complete set of Kraus operators associated with the phase damping process. The standard technique of QPT is employed to calculate the Kraus operators in the following manner:
\begin{enumerate} 
\item Construct
the complete set of linearly independent initial input
density matrices.  
\item Estimate output density
matrices by evolving each input density matrix using
Eq.~\ref{snd_eq6}.  
\item From knowledge of the input and
output density matrices, compute the process matrix $\chi$ using
the standard QPT protocol.  
\item Using unitary diagonalization of
$\chi$ matrix as: $\chi = VDV^{\dagger}$, compute the complete
set of Kraus operators as: 
\begin{equation}
A_{i}=\sqrt{d_{i}} \sum_{j} V_{j i} E_{j} \label{snd_eq7}
\end{equation} where $A_i$s are the Kraus operators, $d_i$s
are diagonal elements of the matrix $D$, $V_{ji}$s are elements
of the matrix $V$ and the $E_j$s form a fixed operator basis. The
diagonal elements of matrix $D$ are eigenvalues of the $\chi$
matrix and the columns of matrix $V$ are the corresponding
normalized eigenvectors of the $\chi$ matrix.  
\end{enumerate}

The unitary dilation operators $\lbrace
U_{A_i} \rbrace$ corresponding to each Kraus operator
$\lbrace A_i \rbrace $ were computed using Eq.~\ref{snd_eq5}. 
The values of the phase damping rates are set to be
$\gamma_1=1.4$ and $\gamma_2=1.5$, and evolved the initial
density matrix for a time $t=2$ s using Eq.~\ref{snd_eq6}. 
It should be noted that the time required to implement a unitary dilation operator on our two NMR qubits depends significantly on the execution times of the CNOT gates, which range from 3 to 11 milliseconds in our system. Consequently, the total time needed to implement all four unitary dilation operators required for simulating the phase damping channel is approximately 80 milliseconds.  The spin-spin relaxation times ($T_2$) of the three NMR qubits, which characterize the natural phase damping channel in the NMR system, are as follows: $T_{2}^{H}=1.3$ s, $T_{2}^{F}=1.4$ s, and $T_{2}^{C}=1.2$ s. Given that the time required to implement the unitary dilation operators is much shorter than the natural phase damping rates of the system, the experimental implementation of the simulated phase damping channel is largely unaffected by the inherent noise in the NMR setup. The subsection~\ref{pd_kraus} provides the comprehensive collection of Kraus operators that govern the evolution of the initial density matrix when subjected to individual phase damping channels on each qubit, with specific values of $\gamma_1$, $\gamma_2$, and $t$. Notably, for the given parameter values, there are four distinct Kraus operators that accurately describe the characteristics of the phase damping channel.

Fig.~\ref{snd_fig2} demonstrates the implementation of the Sz.-Nagy algorithm to
simulate the action of Kraus operator $A_1$ (see subsection~\ref{pd_kraus}) on
the two-qubit initial input state $\vert \phi \rangle \langle \phi \vert = \vert
00 \rangle \langle 00 \vert$.  The initial two-qubit state $\vert 00 \rangle $
is encoded in a three-qubit input state as $\vert 000 \rangle = \vert 0
\rangle_{\rm H} \otimes \vert 00 \rangle_{\rm FC} $.  The quantum circuit given in
Fig.\ref{snd_fig2} (a) represents the action of the unitary dilation operator
$U_{A_1}$ on the input state $\rho_{000}$, followed by measurement. The
column-by-column (CBC) decomposition method has been used to decompose three-qubit unitary dilation operators $\lbrace
U_{A_i} \rbrace $\cite{iten-pra-2016,
iten-arxiv-2021}. Using the CBC  method, $U_{A_1}$ is realized using eight CNOT
gates and eight single-qubit rotation gates $R_{\phi}^{\theta}$ (where $\phi$
denotes the axis of rotation and $\theta$ denotes the angle of rotation).  The
CBC decompositions of the other unitary dilation operators are given in
subsection~\ref{pd_kraus}. 

It is worth mentioning that the same quantum circuit can also be employed to simulate the effect of $A_1$ on arbitrary initial two-qubit states $\rho = \vert \phi \rangle \langle \phi \vert$. In this case, one simply needs to prepare the three-qubit system in the state $\vert 0 \rangle_{\rm H} \otimes \vert \phi \rangle_{\rm FC}$. Furthermore, the arbitrary initial input state $\vert \phi \rangle$ of the two-qubit system exists within a smaller Hilbert space spanned by the vectors: $\vert 000 \rangle $, $\vert 001 \rangle $, $\vert 010 \rangle $, and $\vert 011 \rangle $. The process of projecting the higher-dimensional output state onto this smaller Hilbert space is equivalent to estimating a $4 \times 4$ dimensional partial density matrix, which corresponds to the first four rows and columns of the higher-dimensional output density matrix.

 The NMR pulse sequence required to implement the quantum circuit is illustrated in Fig.\ref{snd_fig2}(b). To achieve single-qubit rotation gates, spin-selective high-power rf pulses were employed. Filled gray and black rectangles i Fig.~\ref{snd_fig2}(b) represent $\pi/2$ and $\pi$ pulses respectively, while unfilled rectangles represent pulses with their corresponding flip angles given above each pulse; the value of $\theta_1$ was set to $0.3737*\frac{\pi}{2}$. The three dashed boxes consist of a set of pulses which have been expanded and depicted in Fig.~\ref{snd_fig2}(c). The phase of each pulse is indicated below its corresponding rectangle. The various time periods for free evolution were set as follows: $\tau = 0.0078$ s, $\tau_{1}=0.0105$ s, and $\tau_2=0.0031$ s.  The measurement box in the diagram represents the decaying time domain NMR signal (FID), which is Fourier transformed to obtain the NMR spectrum. Finally, tomographic measurements were conducted on all three nuclei to calculate the density matrix elements $\lbrace \rho_{ij}$, $0 \leq i,j \leq 4 \rbrace$. The normalized trace distance between the experimentally obtained output Hermitian matrix $(A_1 \rho_{00} A_1^{\dagger}){{\rm exp}}$ and the theoretically expected matrix $(A_1 \rho_{00} A_1^{\dagger})_{{\rm the}}$ was found to be $0.9885$.  A similar quantum circuit and NMR pulse sequence are utilized to simulate the MFGP. In this case, the action of the Kraus operator $A_1$ (subsection \ref{mfgp_kraus}) on the state $\vert 00 \rangle$ can be simulated using the unitary dilation operator $U$ (Eq.\ref{c1} subsection \ref{mfgp_kraus}) through the CBC method, employing 9 CNOT gates and 18 local rotations. However, in the case of a correlated amplitude damping channel, the implementation of unitary dilation operators involves NMR shaped pulses that are optimized using the GRAPE algorithm.
\begin{figure}[t]
\centering
\includegraphics[angle=0,scale=1.2]{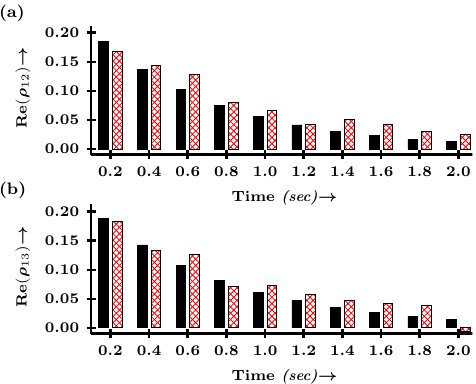} 
\caption{The impact of an independent phase damping channel on the state $\vert ++ \rangle$ is illustrated through an experimental simulation using the Sz-Nagy algorithm. Figure (a) demonstrates the damping effect on the real part of the off-diagonal element $\rho_{12}$, while Figure (b) depicts the damping of the real part of $\rho_{13}$. The $y$-axis represents the values of the real part of $\rho_{12(13)}$, and the $x$-axis represents the time intervals in seconds. The black bars correspond to theoretical values, while the red bars correspond to experimental values.} 
\label{0++}
\end{figure}
\begin{figure*}[t]
\centering
\includegraphics[angle=0,scale=0.81]{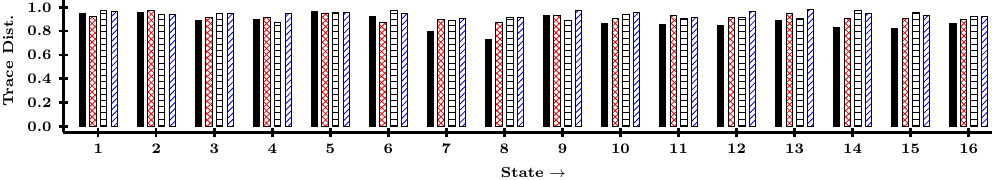} 
\caption{The bar plot illustrates the average normalized trace distance between the experimentally obtained output Hermitian matrix $(A_i \rho_j A_i^{\dagger}){{\rm exp}}$ and the theoretically expected matrix $(A_i \rho_j A_i^{\dagger}){{\rm the}}$ for the phase damping channel. The x-axis represents the numbered two-qubit states $\rho_j$. The bars are color-coded, with black, red, gray, and blue bars corresponding to Kraus operators $A_1$, $A_2$, $A_3$, and $A_4$, respectively.} 
\label{pdbar}
\end{figure*}
\begin{figure}[t]
\centering
\includegraphics[angle=0,scale=1.2]{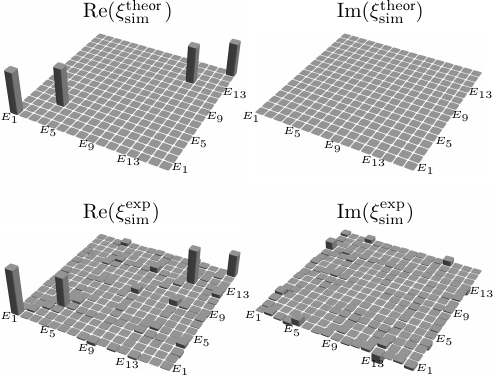} 
\caption{The process matrices obtained from theoretical simulations $(\xi_{\rm sim}^{\rm the})$ and experimental simulations $(\xi_{\rm sim}^{\rm exp})$ of phase damping channels acting independently on each qubit in a two-qubit NMR system. The first column of bar plots displays the real part of the process matrices for the theoretically simulated channel ($\rm{Re}(\xi_{\rm sim}^{\rm the})$) and the experimentally obtained phase damping channel computed using convex optimization-based QPT ($\rm{Re}(\xi_{\rm sim}^{\rm exp})$, respectively). The second column of bar plots represents the imaginary part of the respective process matrices. The average process fidelity of the experimentally simulated channel is found to be $0.8883 \pm 0.0375$.
   } 
\label{fig3} 
\end{figure}

The bar plot shown in Fig.\ref{pdbar} illustrates the normalized trace distance between the experimentally obtained output Hermitian matrix ($A_i \rho_j A_i^{\dagger}$) and the theoretically expected matrix for the phase damping channel. High values of the normalized trace distance indicate successful experimental simulation of the action of a given Kraus operator on a particular initial state. However, when observing the x-axis of Fig.\ref{pdbar} that represents the numbered initial quantum states $\rho_j$, it is observed that the black bar corresponding to $A_1$ has relatively smaller values compared to the red, gray, and blue bars corresponding to Kraus operators $A_2$, $A_3$, and $A_4$ respectively. This discrepancy arises because the experimental implementation of $U_{A_1}$ is more complex than that of $U_{A_2}$, $U_{A_3}$, and $U_{A_4}$ in terms of the number of CNOT gates used, resulting in more experimental errors during the implementation of $U_{A_1}$.
Additionally, it is noteworthy that for certain quantum states $\rho_{7}$, $\rho_{8}$, $\rho_{14}$, $\rho_{15}$, and $\rho_{16}$ indicated on the x-axis of Fig.\ref{pdbar}, which correspond to $\vert 1+ \rangle$, $\vert 1- \rangle$, $\vert -1 \rangle$, $\vert -+ \rangle$, and $\vert -- \rangle$ respectively, the trace distance values are relatively small compared to other states. This behaviour can be attributed to errors in the preparation of the initial states.

It is important to note that in order to obtain the final density matrix $\rho(t)$ evolved under a desired quantum channel, it is necessary to combine the results from each Kraus operator. Specifically, for the phase damping channel, four quantum circuits corresponding to each $U_{A_i}$ must be implemented to obtain the final $\rho(t)$. In Fig.\ref{fig3}, the process matrices representing the experimentally and theoretically simulated 2-qubit independent phase damping channel are compared. The first column displays the real part of the process matrices, while the second column displays the imaginary part. The process fidelity of the experimentally simulated phase damping channel is found to be $0.8883 \pm 0.0375$. To provide a comprehensive analysis, the action of all Kraus operators corresponding to the phase damping channel  were experimentally simulated for 16 linearly independent two-qubit density matrices. The fidelities between the experimentally simulated states using the Sz.-Nagy algorithm and the theoretically simulated states for the phase damping channel are listed in Table~\ref{table3}.
\begin{table}[h] 
\centering
\caption{\label{table3} Fidelity between the experimentally
and theoretically simulated two-qubit states evolving under
independent phase damping channels.}
\renewcommand{\arraystretch}{1.3}
\begin{tabular}{c| c| c| c}
\hline \hline ~~State~~ &
~~~~~~~Fidelity~~~~~~~& ~~~State~~~& ~~~~~~~Fidelity~~~~~~~\\
\hline \hline $\vert 00 \rangle $ & $ 0.9734 \pm 0.0286 $ & $\vert +0 \rangle $ & $0.9377 \pm 0.0270 $  \\ 
$\vert 01 \rangle $ & $ 0.9719 \pm 0.0327 $ & $\vert +1 \rangle $ & $ 0.9031  \pm 0.0754 $  \\ 
$\vert 0+ \rangle $ & $ 0.9506 \pm 0.0322 $ & $\vert ++ \rangle $ & $0.9385 \pm 0.0407 $  \\ 
$\vert 0- \rangle $ & $ 0.9478 \pm 0.0308 $ & $\vert +- \rangle $ & $0.9194 \pm 0.0584 $  \\ 
$\vert 10 \rangle $ & $ 0.9786 \pm 0.0141 $ & $\vert -0 \rangle $ & $0.9274 \pm  0.0240 $ \\ 
$\vert 11 \rangle $ & $ 0.9638 \pm 0.0258 $ & $\vert -1 \rangle $ & $0.9159 \pm 0.0261 $ \\ 
$\vert 1+ \rangle $ & $ 0.9444 \pm 0.0109 $ & $\vert -+ \rangle $ & $0.9078 \pm 0.0473 $ \\ 
$\vert 1- \rangle $ & $ 0.9546 \pm 0.0193 $ & $\vert -- \rangle $ & $ 0.9204 \pm 0.0239 $ \\ 
\hline \end{tabular}
\end{table}
The significant fidelities presented in Table~\ref{table3} serve as evidence for the successful experimental simulation of the phase damping channel on various initial quantum states. As the provided set of 16 states constitutes a complete basis set, it is possible to simulate the effect of the phase damping channel on any arbitrary quantum state with fidelities ranging from $0.9031$ to $0.9786$.

Moreover, an experimental simulation of the dynamics of the $\vert ++ \rangle$ state was conducted under the phase damping channel for various time intervals, allowing us to observe the progression of the decoherence process. Initially, the system is prepared in the state $\vert 0 \rangle \otimes \vert ++ \rangle$ with an experimental state fidelity of $0.9800 \pm 0.0004$. Subsequently, four required unitary dilation operators are applied according to the algorithm and constructed the final output density matrix for each time interval. The decay of the off-diagonal elements $\rho_{12}$ (part (a)) and $\rho_{13}$ (part (b)) is clearly evident in Fig.\ref{0++}, with the theoretical values represented by black bars and the experimental values by red bars. The $y$-axis represents the real part of $\rho_{12(13)}$, while the $x$-axis corresponds to the time interval in seconds. Notably, the imaginary part of $\rho_{12(13)}$ is not displayed, as the theoretical values are consistently zero for all time intervals, providing no additional insight into the phase damping process. 
\subsection{Kraus operators \& unitary dilation operators for
phase damping channel}\label{pd_kraus}

The complete set of Kraus operators corresponding to
an independent phase damping channel, acting on the two-qubit system
with parameter values $\gamma_1 = 1.4$, $\gamma_2 = 1.5$ and
$t=2$ sec, is given below:

{\footnotesize
\[  
 A_1= \left(
\begin{array}{cccc}
 -0.4723+0. i & 0.\, +0. i & 0.\, +0. i & 0.\, +0. i \\
 0.\, +0. i & 0.4723\, +0. i & 0.\, +0. i & 0.\, +0. i \\
 0.\, +0. i & 0.\, +0. i & 0.4723\, +0. i & 0.\, +0. i \\
 0.\, +0. i & 0.\, +0. i & 0.\, +0. i & -0.4723+0. i \\
\end{array}
\right)
\]

\[  
 A_2= \left(
\begin{array}{cccc}
 0.0181\, -0.4961 i & 0.\, +0. i & 0.\, +0. i & 0.\, +0. i \\
 0.\, +0. i & 0.0181\, -0.4961 i & 0.\, +0. i & 0.\, +0. i \\
 0.\, +0. i & 0.\, +0. i & -0.0181+0.4961 i & 0.\, +0. i \\
 0.\, +0. i & 0.\, +0. i & 0.\, +0. i & -0.0181+0.4961 i \\
\end{array}
\right)
\]

\[  
 A_3= \left(
\begin{array}{cccc}
 -0.0085-0.5019 i & 0.\, +0. i & 0.\, +0. i & 0.\, +0. i \\
 0.\, +0. i & 0.0085\, +0.5019 i & 0.\, +0. i & 0.\, +0. i \\
 0.\, +0. i & 0.\, +0. i & -0.0085-0.5019 i & 0.\, +0. i \\
 0.\, +0. i & 0.\, +0. i & 0.\, +0. i & 0.0085\, +0.5019 i \\
\end{array}
\right)
\]

\[  
 A_4= \left(
\begin{array}{cccc}
 -0.5276-0.007 i & 0.\, +0. i & 0.\, +0. i & 0.\, +0. i \\
 0.\, +0. i & -0.5276-0.007 i & 0.\, +0. i & 0.\, +0. i \\
 0.\, +0. i & 0.\, +0. i & -0.5276-0.007 i & 0.\, +0. i \\
 0.\, +0. i & 0.\, +0. i & 0.\, +0. i & -0.5276-0.007 i \\
\end{array}
\right)
\]
}

The decomposition of the unitary dilation operators $U_{A_i}$
corresponding to the Kraus operators  for the phase damping channel
are given below. The column-by-column decomposition method is used to decompose a given
unitary into single-qubit rotation gates and two-qubit CNOT gates.

\begin{enumerate}
\item  {\footnotesize $U_{A_1}$: 
$^1R_{\bar{x}}^{\pi}$.$^1R_{\bar{y}}^{\frac{\pi}{2}}$.$U_{\rm{CNN}}$.${\rm
CNOT}_{32}$.$^2R_{\bar{z}}^{\frac{\pi}{2}}$.${\rm
CNOT}_{32}$.$^1R_{\bar{x}}^{\pi}$.$^1R_{\bar{y}}^{\theta_1}$.$U_{\rm{CNN}}$.$^1R_{\bar{x}}^{\pi}$.$^1R_{\bar{z}}^{\frac{3\pi}{2}}$.$U_{\rm{CNN}}$.$^1R_{\bar{z}}^{\frac{\pi}{2}}$ }
\\ where $U_{\rm{CNN}} = {\rm CNOT}_{31}.{\rm CNOT}_{21}$ 
and $\theta_1 = 0.5870$

\item  $U_{A_2}$ =
$^1R_{\bar{x}}^{\pi}$.$^1R_{\bar{y}}^{\frac{
\pi}{2}}$.${\rm CNOT}_{21}$.$^1R_{\bar{x}}^{\theta_3}
$.$^1R_{\bar{y}}^{\theta_2}$.
${\rm CNOT}_{21}$.$^1R_{\bar{x}}^{\theta_1}$.$^1R_{\bar{z}}^{\frac{
\pi}{2}}$.${\rm CNOT}_{21}$.$^2R_{\bar{z}}^{\frac{3 \pi}{2}}$
\\ where $\theta_1 = 3.0803 $, $\theta_2 = 0.5329 $, and
$\theta_3 = 1.6059 $

\item  $U_{A_3}$ =
$^1R_{\bar{x}}^{\pi}$.$^1R_{\bar{y}}^{\frac{
\pi}{2}}$.${\rm CNOT}_{31}$.$^1R_{\bar{x}}^{\theta_3}$.
$^1R_{\bar{y}}^{\theta_2}$.
${\rm CNOT}_{31}$.$^1R_{\bar{x}}^{\theta_1}$.$^1R_{\bar{z}}^{\frac{
\pi}{2}}$.${\rm CNOT}_{31}$.$^3R_{\bar{z}}^{\frac{3 \pi}{2}}$
\\ where $\theta_1 = 3.1711$, $\theta_2 = 0.5193 $, and
$\theta_3 = 1.5536 $

\item $U_{A_4}$ =
$^1R_{\bar{z}}^{\theta_3}$.$^1R_{\bar{y}}^{\theta_2}$.$^1R_{\bar{z}}^{\theta_1}$
where $\theta_1 = 3.1549$, $\theta_2 = 2.0299 $, and
$\theta_3 = 0.0133 $

\end{enumerate}
where $^iR_{\phi}^{\theta}$ represents a single-qubit rotation
gate acting on the $i$th qubit with the rotation angle $\theta$ and
the rotation axis is denoted by $\phi$ and ${\rm CNOT}_{ij}$
represents a two-qubit CNOT gate with the 
$i$th qubit being the control and the $j$th qubit 
being the target qubit.
\subsection{Simulation of 2-qubit correlated amplitude damping channel}\label{cadsec}
The amplitude damping (AD) channel is a widespread phenomenon that naturally occurs in many physical systems. It plays a crucial role in processes such as spin-lattice relaxation in NMR\cite{childs-pra-2001} and spontaneous emission in optical systems\cite{srikant-pra-2008}, where energy exchange between the system and its environment takes place. Specifically, it leads to population decay from the excited state to the ground state of the system. In the case of a single qubit system, the amplitude damping channel can be described by two Kraus operators, namely $A_1$ and $A_2$, as given in previous work\cite{Hu-sr-2020}: 
\begin{equation}
A_1 = \left(\begin{array}{cc}
1 & 0 \\
0 & \sqrt{1-p}
\end{array}\right) \quad \text{and} \quad A_2 = \left(\begin{array}{cc}
0 & \sqrt{p} \\
0 & 0
\end{array}\right)
\label{eq9}
\end{equation}

\begin{figure}[t]
\centering
\includegraphics[angle=0,scale=1.2]{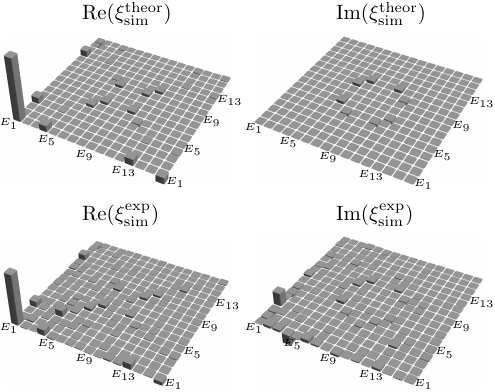} 
\caption{The process matrices obtained from theoretical simulations $(\xi_{\rm sim}^{\rm the})$ and experimental simulations $(\xi_{\rm sim}^{\rm exp})$ of a 2-qubit fully correlated amplitude damping channel on a two-qubit NMR system.
The first column of bar plots depicts the real part of the process matrices for the theoretically simulated channel ($\rm{Re}(\xi_{\rm sim}^{\rm the})$) and the experimentally obtained fully correlated amplitude damping channel computed using convex optimization-based QPT ($\rm{Re}(\xi_{\rm sim}^{\rm exp})$, respectively). The second column of bar plots represents the imaginary part of the respective process matrices. The average process fidelity of the experimentally simulated channel is determined to be $0.9216 \pm 0.0017$.'   } 
\label{cad} 
\end{figure}
here, the parameter $p$ represents the amplitude damping parameter. In this study, the Sz.-Nagy algorithm was employed to simulate a 2-qubit correlated amplitude damping channel. In a fully correlated amplitude damping channel, the amplitude damping operation affects all the qubits simultaneously. The two Kraus operators, denoted as $A_1^{CAD}$ and $A_2^{CAD}$, which characterize the 2-qubit fully correlated amplitude damping channel, can be constructed as follows:,
\begin{equation}
A_1^{\small CAD} = \sqrt{p} (\sigma_{+} \otimes \sigma_{+})
\end{equation}
\begin{equation}
A_2^{\small CAD} = \sqrt{I - {A_1^{CAD}}^{\dagger} A_1^{CAD} } 
\end{equation}
where $\sigma_+ = \left(\begin{array}{cc}
0 & 1 \\
0 & 0
\end{array}\right)$ and the damping parameter $p$ can be expressed as $p = 1-e^{- \gamma t}$ where $\gamma$ is amplitude damping rate and $t$ is time such that as $t \rightarrow 0 \Rightarrow p \rightarrow 0$ and as $t \rightarrow \infty \Rightarrow p \rightarrow 1$ for some given finite value of $\gamma$.

For the purpose of experimental demonstration, the value damping parameter is set to be $p = 0.6$. The corresponding Kraus operators can be found in subsection~\ref{cad_kraus}. The unitary dilation operators, denoted as $\lbrace U_{A_i^{\small CAD}} \rbrace$ for each Kraus operator $\lbrace A_i^{\small CAD} \rbrace$, were computed using Eq.~\ref{snd_eq5} and implemented on NMR using a pulse sequence optimized with the GRAPE algorithm. The implementation time for individual $U_{A_1^{\small CAD}}$ and $U_{A_2^{\small CAD}}$ was approximately 15 ms [5000 (number of time intervals) $\times$ 3 $\mu$s (length of each time interval)], which is much shorter than the system's natural relaxation times $T_1$ and $T_2$. The simulation of the correlated amplitude damping channel is performed using a similar quantum circuit as shown in Fig.\ref{snd_fig2}, where the desired initial state is prepared and the unitary dilation operators are implemented using shaped NMR pulses optimized through the GRAPE algorithm. This is in contrast to the phase damping and MFGP cases where the unitary dilation operators are implemented using spin-selective hard pulses. 

\begin{figure}[t]
\centering
\includegraphics[angle=0,scale=1.1]{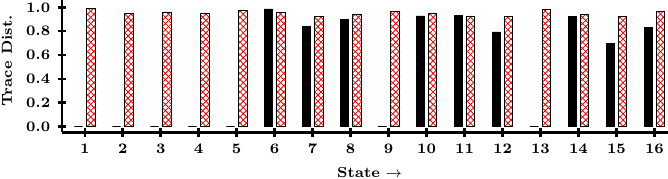} 
\caption{Bar plot showing the average normalized trace distance between the experimentally obtained output Hermitian matrix $(A_i^{CAD} \rho_j {A_i^{CAD}}^{\dagger})_{{\rm exp}}$ and the theoretically expected matrix $(A_i^{CAD} \rho_j {A_i^{CAD}}^{\dagger})_{{\rm the}}$ for a correlated amplitude damping channel. The x-axis represents the numbered two-qubit states $\rho_j$. The black and red bars represent the corresponding Kraus operators $A_1^{CAD}$ and $A_2^{CAD}$. } 
\label{cadbar}
\end{figure}

In Fig.~\ref{cad}, the process matrices of the 2-qubit fully correlated amplitude damping channel obtained from experimental and theoretical simulations are compared. The first column displays the real part of the process matrices, while the second column displays the imaginary part. The process fidelity of the experimentally simulated correlated amplitude damping channel is found to be $0.9216 \pm 0.0017$. Table~\ref{cadtable} provides the fidelities between the experimentally simulated states using the Sz.-Nagy algorithm and the theoretically simulated states for the correlated amplitude damping channel, covering the complete set of linearly independent 16 states. The higher state fidelities indicate the successful simulation of the correlated amplitude damping channel. Additionally, the normalized trace distance between the experimentally obtained output Hermitian matrix $(A_i^{CAD} \rho_j {A_i^{CAD}}^{\dagger})_{\rm exp}$ and the theoretically expected matrix $(A_i^{CAD} \rho_j {A_i^{CAD}}^{\dagger})_{\rm the}$ is depicted using a bar plot in Fig.\ref{cadbar}. The black and red bars correspond to the Kraus operators $A_2^{CAD}$ and $A_3^{CAD}$, respectively. The absence of black bars for the states $\rho_{1}-\rho_5$, $\rho_9$, and $\rho_{13}$, which correspond to the states $\vert 00 \rangle$, $\vert 01 \rangle$, $\vert 0+ \rangle$, $\vert 0- \rangle$, $\vert 10 \rangle$, $\vert +0 \rangle$, and $\vert -0 \rangle$ in Fig.~\ref{cadbar}, signifies that $(A_1^{CAD} \rho_j {A_1^{CAD}}^{\dagger})_{\rm the} = 0$, indicating an output matrix with all entries equal to zero.

\begin{table}[h] 
\centering
\caption{\label{cadtable} Fidelity between the experimentally
and theoretically simulated two-qubit states evolving under
correlated amplitude damping channels.}
\renewcommand{\arraystretch}{1.3}
\begin{tabular}{c| c| c| c}
\hline \hline ~~State~~ &
~~~~~~~Fidelity~~~~~~~& ~~~State~~~& ~~~~~~~Fidelity~~~~~~~\\
\hline \hline $\vert 00 \rangle $ & $0.9927  \pm  0.0001 $ & $\vert +0 \rangle $ & $0.9747   \pm 0.0011$   \\ 
$\vert 01 \rangle $ & $0.9527  \pm  0.0023$ & $\vert +1 \rangle $ & $0.9515 \pm  0.0013$ \\ 
$\vert 0+ \rangle $ & $0.9486  \pm  0.0008$ & $\vert ++ \rangle $ & $0.9284  \pm  0.0035$ \\ 
$\vert 0- \rangle $ & $0.9386  \pm  0.0012$ & $\vert +- \rangle $ & $0.9292  \pm  0.0003 $ \\ 
$\vert 10 \rangle $ & $0.9781  \pm  0.0036$ & $\vert -0 \rangle $ & $ 0.9859 \pm   0.0030$ \\ 
$\vert 11 \rangle $ & $0.9702  \pm  0.0033$ & $\vert -1 \rangle $ & $0.9397  \pm  0.0056 $ \\ 
$\vert 1+ \rangle $ & $0.9179  \pm  0.0052$ & $\vert -+ \rangle $ & $0.9286  \pm  0.0088 $\\ 
$\vert 1- \rangle $ & $0.9347  \pm  0.0026$ & $\vert -- \rangle $ & $0.9476  \pm  0.0046 $\\ 
\hline \end{tabular}
\end{table} 

It is important to note that the actual experimental quality of GRAPE-optimized unitary dilation operators depends on the chosen set of GRAPE parameter values during the optimization process. The lower fidelity observed can be attributed to various factors related to the specific parameter values used, such as rf-distribution, number of time-steps, length of each time step, soft pulse buffering delay, and others. Therefore, even if the GRAPE algorithm theoretically converges to the desired targeted gate fidelity of 0.999, the actual experimental quality of the optimized pulse sequence may turn out to be lower. To ensure accurate results, caution must be exercised when finding the optimized pulse sequence of a given unitary operator using the GRAPE algorithm. Improving the experimental quality of unitary dilation operators can be achieved by selecting GRAPE parameter values that closely resemble realistic values for the specific physical system under consideration.

\begin{figure}[t]
\centering
\includegraphics[angle=0,scale=1.1]{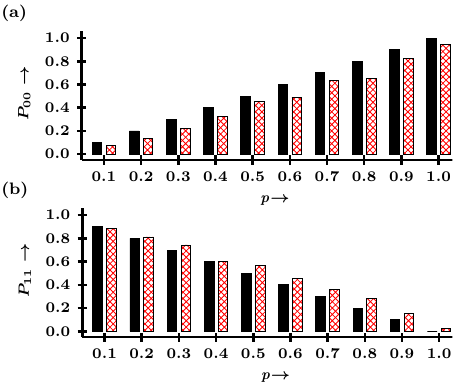} 
\caption{Experimental simulation of the correlated amplitude damping channel's effect on the $\vert 11 \rangle$ state using the Sz-Nagy algorithm. Part (a) represents the ground state population ($P_{00}$), and part (b) represents the excited state population ($P_{11}$). The y-axis represents population, and the x-axis represents the damping parameter ($p$). The black bars indicate theoretical values, while the red bars indicate experimental values.} 
\label{011}
\end{figure}

In this study, a 3-spin half homo-nuclear system consisting of three $^{19}$F nuclei (3-qubits) in the molecule trifluoroiodoethylene, dissolved in d6-acetone is utilized for the correlated amplitude damping case.
Furthermore, an experimental simulation of the dynamics of the $\vert 11 \rangle$ state under a correlated amplitude damping channel was conducted for various time intervals. Following the algorithm's requirements, the initial state $\vert 0 \rangle \otimes \vert 11 \rangle$ was prepared with a fidelity of $0.9806 \pm 0.0025$. The state was then subjected to two unitary dilation operators, followed by measurement and the construction of the final output density matrix for each time interval. In Fig.\ref{011}, the results show that as the damping parameter $p$ (or time $t$) increases, the population of the excited state $\vert 11 \rangle$, denoted as $P_{11}$ in part (b), decays, while the population of the ground state $\vert 00 \rangle$, denoted as $P_{00}$, increases. As $p$ approaches 1, $P_{11}$ approaches 0, and $P_{00}$ approaches 1. The black bars represent the theoretical values, while the red bars indicate the experimental values. From Fig.\ref{011}, it is evident that the experimental values align well with the theoretical values, demonstrating the successful simulation of the $\vert 11 \rangle$ state under a correlated amplitude damping channel using the Sz.-Nagy algorithm.

\begin{figure*}[t]
\centering
\includegraphics[angle=0,scale=0.85]{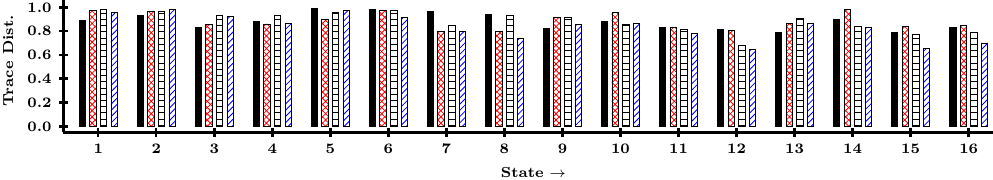} 
\caption{The bar plot illustrates the average normalized trace distance between the experimentally simulated output Hermitian matrix $(A_i \rho_j A_i^{\dagger})^{\rm{exp}}_{\rm{sim}}$, obtained using the SND algorithm, and the experimentally obtained matrix $(A_i \rho_j A_i^{\dagger})^{\rm{exp}}_{\rm{qpt}}$ through quantum process tomography. The simulation involves the implementation of MFGP on two qubits. The x-axis represents the numbered two-qubit states $\rho_j$, while the black, red, gray, and blue bars correspond to the Kraus operators $A_1$, $A_2$, $A_3$, and $A_4$, respectively.  } 
\label{gradbar}
\end{figure*}


\subsection{Kraus operators \& unitary dilation operators for
fully correlated amplitude damping channel}\label{cad_kraus}
\label{cad_append}
The complete set of Kraus operators corresponding to
a fully correlated amplitude damping channel, acting on the two-qubit system
with damping parameter values $p = 0.6$ is given below:
{\footnotesize
\[
A_1^{CAD}=\left(
\begin{array}{cccc}
 0    &     0    &     0  &  0.7746 \\
         0   &      0      &   0     &    0\\
         0    &     0     &    0    &     0\\
         0    &     0     &    0    &     0\\
\end{array}
\right)
\quad {\text {\rm and} } \quad
A_2^{CAD}=\left(
\begin{array}{cccc}
 1.0000   &      0     &    0    &     0 \\
         0  &  1.0000    &     0    &     0\\
         0   &      0  &  1.0000     &    0\\
         0    &     0     &    0  &  0.6325\\
\end{array}
\right)
\] }
The unitary dilation operators $\lbrace
U_{A_i^{\small CAD}} \rbrace$ are implemented using GRAPE algorithm. The time required to implement individual $
U_{A_1^{\small CAD}} $ and $
U_{A_2^{\small CAD}} $ turns out to be $\approx$ 15 ms [5000 (no. of time intervals) $\times$ 3 $\mu s$ (length of each time interval)].
\subsection{Simulating a magnetic field gradient pulse}
\label{mfgp}
Magnetic Field Gradient Pulses (MFGPs) find extensive use in NMR and magnetic resonance imaging experiments, encompassing various applications such as molecular diffusion studies and spatial encoding for imaging\cite{denis-neuro-2012,paga-analyst-2017,han-chemmat-2021}. The behavior of an MFGP is analogous to that of a phase damping channel, as it selectively suppresses the off-diagonal elements (coherences) of the density matrix in a controlled manner. Recently, a method based on time and space discretization was proposed to efficiently simulate shaped gradient pulses\cite{laflamme-arxiv-2020}. The authors present guidelines for selecting suitable discretization values to achieve precise and rapid simulation of gradient pulses, which can further be utilized for implementing specific non-unitary channels. However, this approach has limitations and may not be capable of simulating arbitrary open quantum dynamics using alternative physical setups. In contrast, the Sz.-Nagy algorithm, employed in this study, does not require any specific hardware requirements for simulating non-unitary operations. It can be implemented on any physical system solely using unitary gates. Here, the Sz.-Nagy algorithm is used to simulate the dynamics of a two-qubit system subjected to a shaped MFGP applied for a given duration.

The MFGP process
is typically implemented using gradient coils in NMR hardware where the magnetic
field gradient is along the $z$-axis.  The MFGP process to be tomographed is a
sine-shaped pulse of a duration of 1000$\mu$s, the number of time intervals equal to 100, and an
applied gradient strength of 15\%.  For the intrinsic non-unitary quantum
processes D1 D2, and the MFGP, the FFNN is able to predict the corresponding
process matrix with average fidelities of $\mathcal{\bar{F}} = 0.8373 \pm
0.0381, 0.7607 \pm 0.0690$ and $0.7858 \pm 0.0703$ respectively, using a reduced
data set of size 160. It is evident from the computed fidelity values that the
FFNN performs better if the process matrix is sparse

To simulate the desired MFGP using the Sz.-Nagy algorithm, several steps were followed. Firstly, the MFGP needed to be characterized and its corresponding Kraus operators computed. This was achieved through convex optimization-based QPT, which experimentally characterized the MFGP. By preparing a complete set of linearly independent initial two-qubit quantum states, including $\vert 0\rangle$, $\vert 1\rangle$, $\vert +\rangle$, and $\vert -\rangle$, the MFGP was applied to these states using gradient coils. Full QST was then performed on all output states to compute the process matrix $\chi$, which characterizes the MFGP. Using Eq.~(\ref{snd_eq7}), the complete set of Kraus operators was calculated. The Sz.-Nagy algorithm was subsequently employed to simulate the MFGP, utilizing only unitary operations. Finally, the process fidelity was computed by comparing the experimental process matrix characterizing the MFGP with the experimental process matrix of the simulated MFGP. The complete set of Kraus operators and their corresponding unitary dilation operators for the shaped MFGP can be found in Appendix\ref{mfgp_kraus}.

 It is worth noting that Sz.-Nagy's algorithm could potentially be used to simulate the MFGPs employed in medical imaging and diffusion experiments. For instance, in the paper \cite{mansfield-bjr-1977}, a linear MFGP with strengths ranging from 0.30 G cm$^{-1}$ to 0.90 G cm$^{-1}$ is utilized for medical imaging purposes. Additionally, diffusion experiments discussed in \cite{tanner-jcp-1965} employ time-dependent MFGPs with amplitudes as high as 100 G cm$^{-1}$ to calculate the diffusion coefficient of dry glycerol. To simulate such MFGPs, the procedure mentioned earlier for computing Kraus operators needs to be followed, followed by implementing Sz.-Nagy's algorithm to simulate the diffusion and imaging experiments.

\begin{table}[h] 
\centering
\caption{\label{table4} Fidelity between experimentally 
simulated and experimentally implemented
two-qubit state under 
the action of a magnetic field gradient pulse.}
\renewcommand{\arraystretch}{1.3}
\begin{tabular}{c| c| c| c}
\hline \hline ~~State~~ &
~~~~~~~Fidelity~~~~~~~& ~~~State~~~& ~~~~~~~Fidelity~~~~~~~\\
\hline \hline $\vert 00 \rangle $ & $0.9904 \pm 0.0121 $ & $\vert +0 \rangle $ &  $0.9104 \pm 0.0050$  \\ 
$\vert 01 \rangle $ & $0.9866 \pm 0.0025$ & $\vert +1 \rangle $ & $0.9446 \pm 0.0008$ \\ 
$\vert 0+ \rangle $ & $0.9525 \pm 0.0521$ & $\vert ++ \rangle $ & $0.8960 \pm 0.0193$ \\ 
$\vert 0- \rangle $ & $0.9208 \pm 0.0581$ & $\vert +- \rangle $ & $0.8118 \pm 0.1088$ \\ 
$\vert 10 \rangle $ & $0.9794 \pm 0.0106$ & $\vert -0 \rangle $ & $0.8924 \pm 0.0030$ \\ 
$\vert 11 \rangle $ & $0.9870 \pm 0.0164$ & $\vert -1 \rangle $ & $0.9368 \pm 0.0206$ \\ 
$\vert 1+ \rangle $ & $ 0.9506 \pm 0.0100$ & $\vert -+ \rangle $ & $0.8533 \pm 0.1101$ \\ 
$\vert 1- \rangle $ & $0.9201 \pm 0.0319$ & $\vert -- \rangle $ & $0.8530 \pm 0.0801$\\ 
\hline \end{tabular}
\end{table}

In the case of the MFGP operation, the normalized trace distance between the output Hermitian matrix $(A_i \rho_j A_i^{\dagger})^{\text{exp}}_{\text{sim}}$ obtained through the Sz.-Nagy algorithm and the experimentally derived matrix $(A_i \rho_j A_i^{\dagger})^{\text{exp}}_{\text{qpt}}$ using quantum process tomography is illustrated in Fig~\ref{gradbar}. The bars in black, red, gray, and blue correspond to the Kraus operators $A_1$, $A_2$, $A_3$, and $A_4$, respectively. Regarding the MFGP process, the Kraus operators that require experimental simulation are themselves computed from the experimentally constructed process matrix. Additionally, implementing the unitary dilation operators for the MFGP process involves a relatively high experimental complexity, specifically the number of CNOT gates required. These factors could potentially account for the smaller trace distance values shown in the bar plot presented in Fig~\ref{gradbar} for the MFGP process, compared to the independent phase damping and correlated amplitude damping channels.

To simulate the MFGP operation, it is necessary to implement four quantum circuits corresponding to each $U_{A_i}$ in order to obtain the final $\rho(t)$. In order to provide a comprehensive analysis, the action of all Kraus operators were experimentally simulated on 16 linearly independent two-qubit density matrices for the MFGP process. The fidelities between the experimentally simulated state using the Sz.-Nagy algorithm and the experimentally tomographed state corresponding to the MFGP operation are presented in Table~\ref{table4} for all 16 states. Since the set of 16 states in Table~\ref{table4} forms a complete basis set, the MFGP can be simulated on any arbitrary quantum state with fidelities ranging from 0.8118 to 0.9904.

\begin{figure}[h] 
\centering
\includegraphics[angle=0,scale=1.2]{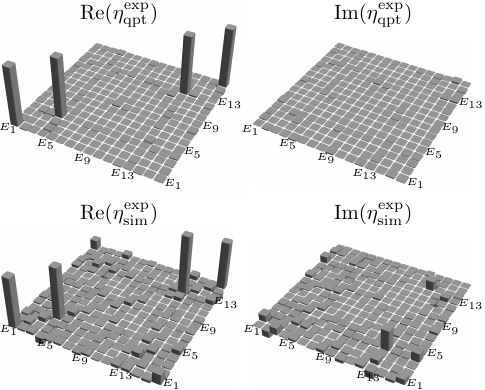} 
\caption{In the top panel, $\rm{Re}(\eta^{\rm exp}_{\rm qpt})$ and $\rm{Im}(\eta^{\rm exp}_{\rm qpt})$ represent the real and imaginary parts of the experimental process matrix obtained from QPT of the shaped MFGP applied on two qubits. In the bottom panel, $\rm{Re}(\eta^{\rm exp}_{\rm sim})$ and $\rm{Im}(\eta^{\rm exp}_{\rm sim})$ depict the real and imaginary parts of the process matrix for the same MFGP, experimentally simulated using the Sz.-Nagy algorithm. The process fidelity between $\eta^{\rm exp}_{\rm qpt}$ and $\eta^{\rm exp}_{\rm sim}$ is found to be $0.8687 \pm 0.0192$.} 
\label{fig4} 
\end{figure}

The process fidelity is computed for the independent phase damping and fully correlated amplitude damping channels by comparing the experimentally simulated channel $\xi^{\rm exp}_{\rm sim}$ with the theoretically simulated channel $\xi^{\rm the}_{\rm sim}$. For the shaped MFGP process, the process fidelity is computed by comparing the experimentally simulated shaped MFGP $\eta^{\rm exp}_{\rm sim}$ with the result of quantum process tomography $\eta^{\rm exp}_{\rm qpt}$ performed on the experimentally implemented shaped MFGP. The computed process fidelity for the phase damping channel is $0.8883 \pm 0.0375$, while for the fully correlated amplitude damping channel it is $0.9216 \pm 0.0017$. The corresponding tomographs can be compared in Fig.\ref{fig3} and Fig.\ref{cad}, respectively. In the case of the shaped MFGP process, the computed process fidelity is $0.8687 \pm 0.0192$, and the respective tomographs can be compared in Fig.~\ref{fig4}. Error bars in the process and state fidelities represent the standard deviation obtained from repeating the experiments and computing the fidelities in each case. It is important to note that all the tomographs are plotted on the same scale.

For both the phase damping channel and the MFGP process, it can be observed that the real part of the process matrix only has four non-zero elements, corresponding to the set of Kraus operators: $E_1 = I \otimes I$, $E_4 = I \otimes \sigma_z$, $E_{13} = \sigma_z \otimes I$, and $E_{16} = \sigma_z \otimes \sigma_z$. The imaginary part of the process matrix is nearly zero for both processes. Fig.\ref{fig3} and \ref{fig4} demonstrate that the action of the shaped MFGP and the phase damping channel is similar in terms of effectively eliminating the off-diagonal elements of the density matrix. The deviations observed in the simulated process matrix from the desired process matrix are attributed to experimental errors in state preparation, implementation of unitary dilation operators, and inevitable systematic errors. These errors can be reduced through the use of appropriate optimization protocols. Specifically for the MFGP process, the experimental implementation of all four unitary dilation operators requires 9 CNOT gates (i.e., 9 CNOT gates $\times$ 4 Kraus operators = 36 CNOT gates in total for simulating the MFGP process). On the other hand, for the phase damping channel, the experimental implementation of the unitary dilation operators $U_{A_1}$, $U_{A_2}$, $U_{A_3}$, and $U_{A_4}$ requires 8, 3, 3, and 0 CNOT gates, respectively (i.e., 14 CNOT gates in total for simulating the phase damping channel). Consequently, the experimental errors are higher when simulating the MFGP process compared to the phase damping channel, resulting in lower values of the process fidelities as indicated in Tables~\ref{table3} and \ref{table4}. In the case of the fully correlated amplitude damping channel, rather than decomposing the unitary dilation operators into CNOT and single-qubit gates, GRAPE-optimized pulses were utilized, which require less implementation time. This approach yields higher process and state fidelities compared to the phase damping and MFGP processes.

\subsection{Kraus operators and unitary dilation operators for shaped MFGP}\label{mfgp_kraus}
The complete set of Kraus operators corresponding to 
the desired shaped gradient
pulse with parameter values given in Sec.\ref{mfgp} applied on 
a two-qubit 
system were experimentally computed via 
the convex optimization based QPT method. The Kraus operators are 
given by:

{\footnotesize
\[
A_1=\left(
\begin{array}{cccc}
 0.1231\, -0.0877 i & -0.0038+0.0026 i & -0.0077+0.0085 i & 0.0023\, +0.0004 i \\
 0.0122\, -0.0279 i & -0.1899-0.1181 i & 0.0101\, +0.0085 i & 0.0097\, +0.006 i \\
 -0.0174+0.0165 i & -0.0073+0.0042 i & -0.3573+0.4876 i & 0.0167\, -0.0073 i \\
 -0.0036-0.0034 i & -0.0056+0.0133 i & -0.0009+0.0275 i & 0.5454\, +0.4572 i \\
\end{array}
\right)
\]

\[
A_2=\left(
\begin{array}{cccc}
 -0.0434-0.4568 i & 0.0061\, +0.0085 i & 0.0095\, +0.0121 i & -0.0055-0.0064 i \\
 0.0329\, +0.0096 i & 0.181\, -0.4594 i & -0.0029+0.0105 i & -0.0003+0.0002 i \\
 0.0017\, -0.0235 i & 0.0036\, -0.003 i & -0.35-0.3762 i & 0.0141\, -0.0184 i \\
 -0.0055-0.0042 i & 0.0124\, -0.007 i & 0.012\, +0.0275 i & 0.3231\, -0.3787 i \\
\end{array}
\right)
\]

\[
A_3=\left(
\begin{array}{cccc}
 -0.4842-0.5645 i & 0.0305\, +0.0057 i & 0.027\, -0.0027 i & -0.0011+0.0033 i \\
 -0.0206+0.0166 i & -0.327+0.0929 i & 0.0007\, -0.0019 i & 0.0034\, -0.0026 i \\
 0.0102\, +0.0216 i & -0.0024+0.0064 i & 0.3035\, -0.2407 i & 0.0096\, +0.0199 i \\
 -0.0005-0.0058 i & 0.0024\, +0.0041 i & 0.015\, +0.006 i & -0.0094+0.4166 i \\
\end{array}
\right)
\]

\[
A_4=\left(
\begin{array}{cccc}
 0.4475\, +0.0416 i & -0.0139+0.0256 i & -0.0099+0.0021 i & 0.0055\, +0.0044 i \\
 -0.0239-0.0035 i & -0.7081+0.2924 i & -0.0143-0.0018 i & 0.0063\, -0.0201 i \\
 0.0027\, -0.0084 i & 0.0055\, +0.0079 i & -0.1662-0.4034 i & 0.0107\, -0.0154 i \\
 0.0045\, -0.0062 i & 0.0167\, -0.0093 i & -0.0253+0.0106 i & 0.1022\, -0.1527 i \\
\end{array}
\right)
\]
}
The decomposition of unitary dilation operators $U_{A_i}$ corresponding to
respective Kraus operators are given below for a shaped gradient pulse.  The column-by-column decomposition method is used to decompose a given unitary into
single-qubit rotations and CNOT gates. It turns out that in the case of a shaped
gradient pulse, the form of decomposition of unitary dilation operators
corresponding to all Kraus operators is the same. The general form of the
decomposition of unitary dilations is denoted by $U$ and given below.
{\footnotesize
\begin{equation}
\label{c1}
\begin{split}
U  =
{}^1R_{\bar{x}}^{\theta_{17}}.^1R_{\bar{y}}^{\theta_{16}}.\rm{CNOT}_{31}.^1R_{\bar{x}}^{\theta_{15}}.^1R_{\bar{y}}^{\theta_{14}}.\rm{CNOT}_{21}.^1R_{\bar{x}}^{\theta_{13}}.^1R_{\bar{y}}^{\theta_{12}}.\rm{CNOT}_{31}.^1R_{\bar{x}}^{\theta_{11}}.^1R_{\bar{y}}^{\theta_{10}}.\rm{CNOT}_{21}.^1R_{\bar{x}}^{\theta_9}.\\
^1R_{\bar{y}}^{\theta_8}.
\rm{CNOT}_{31}.^1R_{\bar{x}}^{\theta_7}.^1R_{\bar{y}}^{\theta_6}.\rm{CNOT}_{31}.^1R_{\bar{x}}^{\theta_5}.^1R_{\bar{z}}^{\theta_4}.\rm{CNOT}_{21}.^1R_{\bar{z}}^{\theta_3}.\rm{CNOT}_{31}.^1R_{\bar{z}}^{\theta_2}.\rm{CNOT}_{21}.^1R_{\bar{z}}^{\theta_1}.^3R_{\bar{z}}^{\theta_0}
\end{split}
\end{equation}
}

\begin{table}[h!] 
\centering
\caption{\label{table6} The values of $\theta_i$s
(Eq.\ref{c1}) required 
to the implement unitary dilation operators $U_{A_j}$.}
\begin{tabular}{c | c | c | c | c}
\hline \hline ~~~~~&~~~~~ $U_{A_1}$ ~~~~~& ~~~~~~$U_{A_2}$~~~~~~&~~~~~ $U_{A_3}$~~~~~~&~~~~~~$U_{A_4}$~~~~~\\
\hline \hline
$\theta_{0} $ &  1.5708   &  4.7124   &  4.7124  & 1.5708  \\
$\theta_{1} $ &  6.2759   &  0.0486   &  6.1354  & 0.1079  \\
$\theta_{2} $ &  5.7332   &  0.0306   &  0.1041  & 0.8518  \\
$\theta_{3} $ &  0.5359   &  0.1169   &  5.7425  & 5.6472  \\
$\theta_{4} $ &  4.2067   &  1.3599   &  2.4207  & 4.9160  \\
$\theta_{5} $ &  2.8192   &  3.0589   &  3.4918  & 2.6544  \\
$\theta_{6} $ &  1.8641   &  1.5181   &  1.2934  & 1.2327  \\
$\theta_{7} $ &  2.2842   &  1.0045   &  5.3556  & 1.0113  \\
$\theta_{8} $ &  0.4323   &  0.0979   &  0.4432  & 0.5851  \\
$\theta_{9} $ &  3.1416   &  3.1416   &  3.1416  & 3.1416  \\
$\theta_{10} $ &  0.4323   &  0.0979   &  0.4432  & 0.5851 \\
$\theta_{11} $ &  2.2856   &   0.6158  &  5.0076  & 3.7509  \\
$\theta_{12} $ &  1.0560   &  0.1859   &  0.6384  & 1.1481  \\
$\theta_{13} $ &  2.3100   &  2.9610   &  2.8007  & 5.0366  \\
$\theta_{14} $ &  0.6701   &  0.4389   &  0.9546  & 1.3207  \\
$\theta_{15} $ &  1.1972   &  0.1460   &  3.6109  & 4.3664  \\
$\theta_{16} $ &  1.6675   &  1.4041   &  2.5217  & 1.8259  \\
$\theta_{17} $ &  2.9373   &  2.1115   &  2.4411  & 3.8623  \\

\hline \end{tabular}
\end{table}
where $^iR_{\phi}^{\theta}$ represents a single-qubit rotation gate acting on
the $i$th qubit with rotation angle $\theta$ and axis of rotation 
$\phi$; $\rm{CNOT}_{ij}$ represents a standard two-qubit CNOT gate with 
$i$ being the control qubit 
and $j$ being the target qubit.
\pagebreak
\section{Conclusions}
\label{sec5}
The experimental implementations of the Sz.-Nagy algorithm was conducted on an NMR quantum information processor to simulate an independent phase damping channel, a fully correlated amplitude damping channel, and a shaped MFGP acting on two qubits using one ancilla qubit. The approach involved designing a protocol to compute the complete set of Kraus operators using quantum process tomography and the unitary diagonalization technique. To assess the quality of the experimentally simulated quantum process, quantum process tomography based on the constrained convex optimization technique was employed. The findings demonstrate the feasibility of implementing the Sz.-Nagy algorithm experimentally, as it only necessitates one ancilla qubit to simulate open quantum dynamics of arbitrary dimensions. The protocol is general and applicable to any quantum process, and it can be adapted for other physical platforms to simulate complex quantum processes using the Sz.-Nagy algorithm. Additionally, GRAPE optimization algorithm\cite{khaneja-jmr-2005} and genetic programming\cite{Devra-qip-2018}, along with other techniques\cite{iten-pra-2016, iten-arxiv-2021}, have been well developed. These methods aid in implementing high-dimensional unitary dilation operators, which is the main challenge in simulating the action of a given Kraus operator in the Sz.-Nagy algorithm. However, the current need is to develop computationally efficient algorithms for decomposing a given unitary dilation operator into a universal set of quantum gates. The results presented in this chapter have been published
in \href{https://journals.aps.org/pra/abstract/10.1103/PhysRevA.106.022424}{Phys. Rev. A 106, 022424 (2022)}.

\pagebreak

\chapter{Summary and Future Outlook }

\section{Summary}

In this thesis, the primary focus revolves around the design and practical implementation of diverse quantum tomography protocols. These protocols play a vital role in efficiently characterizing and reconstructing unknown quantum states and processes. The study leverages both nuclear magnetic resonance (NMR) quantum processors based on spin ensembles as well as superconducting technology-based IBM quantum processors. The core objective encompasses two distinct but interconnected tasks: the reconstruction of quantum states using quantum state tomography (QST) protocols and the characterization of quantum processes through quantum process tomography (QPT) protocols. These QST and QPT techniques are pivotal, serving to assess the reliability and performance of quantum processors. However, both methods grapple with a significant challenge- computational complexity that escalates exponentially with the system's size- rendering their experimental implementation infeasible, even for relatively smaller systems. Additionally, the finite size of ensembles and unavoidable systematic errors introduce issues leading to unphysical density and process matrices.

To combat these challenges, various QST and QPT protocols have been proposed, yet many remain untested in practical applications. The central aim of the study carried out in this thesis is to devise experimental strategies that enable the efficient implementation of tomography protocols on both NMR and IBM quantum processors. In this pursuit, generalized quantum circuits are introduced as a means to efficiently gather experimental data for QST and QPT. These circuits have been shown to be effective in both two- and three-qubit quantum states and processes.

Addressing the challenge of unphysical reconstructions of experimentally determined quantum states and processes through conventional tomography techniques, the study introduces an innovative approach. QST and QPT tasks are reformulated as constrained convex optimization (CCO) problems, with solutions that yield valid quantum states and processes. Importantly, in the case of QPT, this approach facilitates the computation of a complete set of Kraus operators associated with a given quantum process. Additionally, the study employs compressed sensing (CS) and artificial neural network (ANN) techniques for efficient tomography, achieving accurate results with significantly reduced data compared to traditional methods. The implementation of CS and ANN-based tomography holds promise for managing complexity challenges in characterizing higher-dimensional quantum gates.

Moreover, the study delves into the task of selectively and directly estimating specific elements of the process matrix that characterizes quantum processes. This exploration results in the development of a novel protocol termed selective and efficient quantum process tomography (SEQPT). The study presents a generalized quantum algorithm and corresponding circuit for SEQPT, successfully demonstrating its functionality on both NMR and IBM quantum processors. Additionally, the thesis introduces an efficient scheme for direct QST and QPT rooted in the weak measurement approach, illustrated through experimentation on a three-qubit NMR system. Further, the thesis investigates the practical simulation of open quantum system dynamics using dilation techniques.

To validate the effectiveness of the proposed quantum tomography and simulation protocols, the study compares experimental outcomes with theoretically predicted results for several instances of two- and three-qubit quantum systems. This comparison underscores the practical viability and utility of the strategies developed in this thesis.

\section{Future Scope}

The research conducted in this thesis not only contributes to the current understanding of QST and QPT but also lays the foundation for several promising avenues of future exploration within these fields.

Firstly, an area ripe for further investigation involves the refinement and optimization of the CCO method. By extending its application beyond the NMRQC platform, researchers can strive to elevate the accuracy and fidelity of density and process matrix characterization across diverse experimental contexts. This expansion promises to unlock novel insights and advancements in quantum state and process tomography.

Secondly, the realm of scalability presents an exciting challenge to address. Future studies can delve into alternative CS algorithms and techniques, all while considering varying operator bases and measurement strategies. The goal is to establish methods that can efficiently tackle QST and QPT with larger and more complex quantum systems, thereby enhancing the applicability and efficiency of these essential techniques.

Moreover, the continued development of ANN techniques holds significant potential for overcoming scalability hurdles in QST and QPT. Researchers can explore diverse architectures and training methods to ensure the effectiveness of these approaches. Additionally, selective and efficient tomography schemes like SEQPT and WM-based direct QST/QPT could benefit from further refinement and validation across different quantum systems and processors.

Furthermore, embracing new experimental platforms and techniques remains an avenue brimming with possibilities. By considering various physical realizations of quantum processors, researchers can explore innovative ways to implement quantum state and process tomography, potentially unlocking new insights and expanding the horizon of quantum technologies. Lastly, the exploration of novel algorithms and approaches for simulating and characterizing the dynamics of open quantum systems, particularly through experimental dilation techniques, offers the potential to unravel previously uncharted territories of quantum behaviour.

Overall, the thesis provides a solid foundation for future research in experimental QST and QPT, offering opportunities for improving fidelity, scalability, and efficiency in characterizing quantum states and processes.



\renewcommand{\bibname}{References} 
\bibliographystyle{phreport}
\bibliography{master}

\end{document}